\documentclass[letterpaper,twocolumn,10pt]{article}
\usepackage{usenix}
\usepackage{tikz}
\newcommand*\circled[1]{\tikz[baseline=(char.base)]{
            \node[shape=circle,draw,inner sep=0.2pt] (char) {#1};}}

\usepackage{tabularx}
\usepackage{array}
\usepackage{booktabs}
\usepackage{epsfig,endnotes}
\usepackage[ruled, lined, linesnumbered,longend]{algorithm2e}
\usepackage{multirow}
\usepackage{url}
\newcommand{\Orca}{{\scshape Orca}\xspace}
\usepackage{subfig}
\usepackage{wrapfig}
\newcommand{\DELF}[1]{\iffalse #1 \fi}
\newcommand{\IdeatoRead}[1]{\iffalse #1 \fi}

\newcommand{\DELMayNeed}[1]{\iffalse #1 \fi}

\newcommand\bl[1]{{\color{blue}#1}}

\usepackage{amsthm}
\theoremstyle{definition}

\usepackage{color}
\usepackage{amsfonts}
\newtheorem{thm}{\textbf{Observation}}
\usepackage[most]{tcolorbox}
\definecolor{light-gray}{gray}{0.95}
\tcolorboxenvironment{thm}{
   center,
   boxsep=0.5pt,
   top=1.0pt,
   bottom=0.0pt,
width=\linewidth,
    colframe=light-gray,
    colback=light-gray
}

\usepackage{mathtools}      
\usepackage{cleveref}
\crefname{section}{§}{§§}
\Crefname{section}{§}{§§}
\crefformat{section}{§#2#1#3}

\newcommand{\sh}[1]{\textcolor{blue}{[??: #1]}}

\newcommand{\DEL}[1]{\iffalse #1 \fi}

\usepackage[normalem]{ulem}
\usepackage{mathtools}      

\newcommand{\squishlist}{
\begin{list}{$\bullet$}
  { \setlength{\itemsep}{0pt}
     \setlength{\parsep}{0pt}
     \setlength{\topsep}{0pt}
     \setlength{\partopsep}{0pt}
     \setlength{\leftmargin}{0em}
     \setlength{\labelwidth}{0em}
     \setlength{\labelsep}{0.2em} } }
 
\newcommand{\squishlisttwo}{
\begin{list}{$\bullet$}
  { \setlength{\itemsep}{0pt}
     \setlength{\parsep}{0pt}
    \setlength{\topsep}{0pt}
    \setlength{\partopsep}{0pt}
    \setlength{\leftmargin}{2em}
    \setlength{\labelwidth}{1.5em}
    \setlength{\labelsep}{0.5em} } }
 
\newcommand{\squishend}{
  \end{list}  }

\newcommand{\sys}{\textsc{EconoServe}\xspace}
	
\begin{document}
\DEL{Schedule 
Both analysis and experiment

1. Figures for 13B for the alpaca and ShareGPT dataset. 04/21/2024 - 3pm (Analysis included, exp- collected, will include 04/29 5:30pm)

2. Figures for 13B for the book corpus dataset. 04/22/2024 - noon  ---> 04/23/2024 --- 11:00AM ---> 5:30PM> 11:30PM (analysis included, exp-collected, will include 04/29 5:30 pm)

3. Figures like vllm paper for 13B for the alpaca and ShareGPT, bookcorpus dataset. 04/23/2024 - 8:30 AM (data collected, will include 04/29/2024 5:30pm)
\\

4. Figures for 175B for alpaca (done) and ShareGPT dataset  04/24/2024 - noon ---- midnight------ 04/26/2024 --- 10:00 PM (analysis data collected: will be included 04/29: 10:00 pm, exp data collected, will be included 04/30, 8:00 am)

5. Figures for 175B for the book corpus dataset. 04/24/2024 - midnight ------ 04/25/2024 - 10:00pm -----4/26/2024 - midnight -> 04/27/2024 - morning -- 3:00pm -- midnight (analysis data collected: will be included 04/29: 10:00 pm, exp data collected, will be included 04/30, 8:00 am)

6. Figures for the Llama 13B (alpaca done) for shareGPT 04/27/2024 - morning (need to receive update from zafar) (done) ---- bookcorpus  ---> 04/30--> noon (experiment only)

7. Figures for the Oracle for all the trace data. 04/27/2024 - 5 pm  ---- 04/28/2024 9pm (got the data) ----> 04/30, 8:00 am

8. Figures for the pipelining method. 04/27/2024 --- midnight  ---> 04/29/2024 -- 10pm ----> 04/30, 3:00pm

9. Varying the parameters ----> 
-------
\\
In the subsequent exp., record and give me these values too
•	the number of preemptions in pipelining, the number of its KVC blocks
•	the number of preemptions without swamping KV values due to underprediction, the number of its KVC blocks
•	the number of the times to use of the reserved KVC and the number of the used blocks due to underprediction, 
•	the number of switching due to urgency of waiting GTs, the number of the times to use of the reserved KVC and the number of the used blocks due to urgency of waiting GTs.
\\

To Do:
1. How FastGen handle preemption? 
--- free KV cache and do the

2. Seconds/ token in the analysis 

3. Check MultiRes with vLLM, it should be higher}

\title{EconoServe: Maximizing Multi-Resource Utilization with SLO Guarantees in LLM Serving}
\author{Haiying Shen and Tanmoy Sen\\ University of Virginia}
\maketitle

\vspace{-0.0in}

\IdeatoRead{(??in your analsysis and exp., you need to vary the response length}

\begin{abstract}
\DEL{Large Language Model (LLM) inference faces an inherent problem of GPU memory bottleneck, leading to GPU under-utilization, significantly hindering throughput. 
To enhance the performance of LLM inference, researchers have proposed iteration-level scheduling. However, it contributes to high scheduling time and GPU idle time (11\% in our measurement) for advanced schedulers that enhance throughput and/or reduce latency. Addressing the concurrent challenges of minimizing scheduling time and resolving the inherent problem is essential. In this paper, 
we conduct comprehensive experimental analysis, and propose a system that Decouples Prompt and Generation task (GT) processing (\sys). \sys keeps a prompt waiting queue and a GT waiting queue, and conducts iteration-level scheduling for prompt tasks to fully utilize GPU, and 
batches GTs with similar predicted RLs to avoid iteration-level GT scheduling. To enhance Key-Value (KV) cache utilization, \sys has a novel KV cache (KVC) pipelining method that allows sharing allocated but unused KVC space. Experimental results demonstrate that \sys reduces inference time by up to ??\% and increases throughput by up to ??\% compared to the state-of-the-art scheduler.} 

As Large Language Models (LLMs) continue to grow, reducing costs and alleviating GPU demands has become increasingly critical. However, existing schedulers primarily target either GPU compute or Key-Value Cache (KVC) utilization, failing to fully optimize both GPU compute and KVC usage during each iteration or guarantee timely KVC allocations when needed. To address these challenges, we conducted a trace-based experimental analysis and made insightful observations, leading to the design of a system called \sys. \sys maximizes multi-resource utilization while ensuring service-level objective (SLO) guarantees in LLM serving. To enable adding prompts to a batch to maximize GPU utilization in each iteration, \sys maintains separate waiting queues for prompt processing tasks (PTs) and generation tasks (GTs). It batches GTs with the same predicted response lengths (RL) to save scheduling time and allocates KVC space for the predicted RL to avoid KVC allocation failures. It further has a novel KVC pipelining method, allowing sharing allocated but unused KVC space to enhance KVC utilization. In addition, it prioritizes queued requests that occupy more KVC to release KVC earlier and satisfy request service-level-objective (SLO). Experimental results demonstrate that \sys increases throughput by up to 4$\times$ with the same level of latency, generates up to 91\% lower job completion time and up to 91\% higher SLO satisfaction ratio compared to vLLM. It also reduces the number of GPUs used in DistServe by up to 78\% while maintaining the same level of goodput. 


\end{abstract}



\DEL{{\sh{check if can reduce 1200 reserved tokens considering GT group will finish and reserved tokens can be from there}}

{\sh{offload-free preemption: testing
with and without borrow reserved tokens}}-- add the figure to~\ref{fig:memory-percentage}1) request preemption only 3) Preemption using reserved KVC only 

{\sh{I guess for SYND, 16-token block may work better. you use this setting too and choose the better one, and show 1-2 figs for this block size variance.
Final exp. Results need to include KVC allocation failures
}}}



\vspace{-0.0in}
\section{Introduction}
\label{sec:intro}
Large Language Models (LLMs) have emerged as transformative tools widely used in various applications such as natural language understanding, machine translation, text summarization, question answering, and sentiment analysis. In LLM inference, a user's request undergoes two primary stages: computation-intensive prompt processing task (PT) and memory-intensive token generation task (GT). 
The first token is generated by the PT in the first iteration, and then GT utilizes the newly generated token as input to produce the next token in the response sequence in each iteration. During these iterations, each token's key-value (KV) pairs are stored in the KV cache (KVC) in GPU memory and are accessed during the calculation of subsequent tokens. This process continues iteratively until the request is completed, and then its occupied KVC is released. 







To meet the capacity requirements of different applications, LLMs have experienced significant growth in size. Early models like BERT~\cite{devlin2018bert} had around 340 million parameters, while GPT-3 and GPT-4~\cite{mann2020language,bubeck2023sparks} reached 175 billion. GPT-4o~\cite{islam2024gpt} exceeds 200 billion parameters and LLama 3.1~\cite{llama,touvron2023llama} reaches 405 billion parameters. The increasing size and widespread use of machine learning (ML) inference applications have led to the reliance on expensive, power-hungry GPUs, making model inference a significant operational cost for production clusters~\cite{hongzhang0025SHEPHERDServingDNNs2023}. For instance, Facebook processes over 2 quadrillion inference requests daily~\cite{hongzhang0025SHEPHERDServingDNNs2023}, inference accounts for more than 90\% of ML production costs on AWS\cite{AmazonCost}, and operating ChatGPT incurs approximately \$700,000 per day in compute hardware costs for 28,936 GPUs~\cite{ChatGPTCost}. Small companies such as AnyScale and Crayon also provide LLM inference services. Projections suggest that LLM-related costs for data centers could exceed \$76 billion by 2028, driven by the rapid expansion of GPU deployments~\cite{ProjectCost}. 
Consequently, maximizing resource utilization in LLM serving systems is crucial for reducing costs and alleviating GPU demands, particularly as LLM deployment scales.



For this purpose, in this paper, our goal is to design a scheduler that \emph{fully utilizes both GPU compute and memory resources (dual-resources) in \underline{each iteration} while meeting service-level objectives (SLOs), minimizing scheduling time (i.e., batch formation time) and preventing KVC overflow.} 
To the best of our knowledge, no existing schedulers have achieved this goal (see Table~\ref{tab:compareMethods}). 


\begin{table}[t]
\centering
\caption{Comparison of \sys and current methods.}
\label{tab:compareMethods}\vspace{-0.1in}
\resizebox{\columnwidth}{!}{%
\begin{tabular}{|l|c|c|c|c|c|}
\hline
Method & Avoid KVC & Increase GPU uti. & High GPU uti. & High KVC uti. & Low schedu-\\ 
 & allo. failures & when KCV allows& in each iteration& in each iteration & ling time \\ \hline
\Orca~\cite{280922} & $\checkmark$ & $\times$ & $\times$ & $\times$ & $\checkmark$ \\ \hline
FastServe~\cite{Wu2023FastDI}& $\checkmark$ & $\times$ & $\times$ & $\times$ & $\times$ \\ \hline
vLLM~\cite{vllm} & $\times$ & $\times$ & $\times$ & $\checkmark$ & $\checkmark$ \\ \hline
FastGen~\cite{deepspeed-fastgen}  & $\times$ & $\checkmark$ & $\times$ & $\times$ & $\checkmark$ \\ \hline
Sarathi-Serve~\cite{Agrawal2023SARATHIEL} & $\times$ & $\checkmark$ & $\times$ & $\times$ & $\checkmark$ \\ \hline
\sys & $\checkmark$ & $\checkmark$ &$\checkmark$ & $\checkmark$ & $\checkmark$ \\ \hline
\end{tabular}%
}\vspace{-0.1in}
\end{table}

Different schedulers have been proposed and many schedulers use first-come-first-serve (FCFS). To improve request-level scheduling~\cite{fang2021deployment,46801}, \DEL{Initially, LLM inference employs request-level scheduling, in which all requests within a batch are processed collectively until they are completed~\cite{fang2021deployment,46801}. However, it introduces delays for requests that complete significantly earlier than others within a batch and prolongs the waiting time for queued requests. To address this problem, }\Orca~\cite{280922} employs iteration-level scheduling that returns completed requests after each iteration and selects waiting requests to form a new batch. It allocates a request with KVC space for the maximum total sequence length, i.e., the sum of the prompt length and response length (RL), to prevent cache overflow~\cite{fastertransformer}. 
However, this \emph{max-allocation} approach leads to low KVC utilization due to unutilized allocated KVC and limits batch size, resulting in low GPU utilization (potentially as low as 0.4\% \cite{jin2023s}) and throughput. FastServe~\cite{Wu2023FastDI} employs the traditional Multi-Level Feedback Queue (MLFQ)~\cite{skq} to address issues of head-of-line blocking and prolonged JCT, but also uses max-allocation.  Subsequent work has focused on improving either KVC utilization or GPU utilization, as outlined below.

\noindent\textbf{Improve KVC utilization.} To mitigate this KVC bottleneck, vLLM~\cite{vllm} allocates fixed-size KVC blocks (e.g., 32 tokens) to a request that starts execution or uses up its previously allocated block (we call it \emph{block-allocation}), and preempts requests upon KVC allocation failures in execution by swapping their KV values to CPU memory or conducting recomputation. However, it cannot guarantee KVC allocations upon request 
in execution, generating preemptions and delay. In addition, it may not maximize GPU compute utilization. \looseness=-1

\noindent\textbf{Improve GPU utilization.} FastGen~\cite{deepspeed-fastgen} and Sarathi-Serve~\cite{Agrawal2023SARATHIEL} chunk long prompts and batch PTs with GTs 
to reach the target forward size (TFS) that maximizes GPU compute utilization. 
Forward size is the number of tokens in a batch~\cite{megatron,megaequ}. 
However, they may not fully utilize KVC and inherit block-allocation's shortcomings mentioned above.




\DEL{To enhance throughput, user requests are typically processed in batches with a predetermined batch size.} 





\DEL{In \Orca, the scheduling delay typically amounts to ?? seconds, which corresponds to ??\% of one iteration time, making it an acceptable level of delay. However, more intricate schedulers can introduce significantly higher delays and inefficient use of GPU resources.}



\DEL{Theoretically, the GPU utilization depends on the forward size (i.e., the number of tokens in a batch)~\cite{megatron,megaequ}, and the KVC utilization depends on the total sequence lengths of active requests. We refer to the forward size that fully utilize GPU 
as \emph{Target Forward Size (TFS)}. 
}


\looseness=-1 


\DEL{To avoid KVC allocation failures and fully utilize dual-resources concurrently, 
we could adopt the approach in~\ref{}\sh{add the reference}that allocate the KVC equal to the estimated RL\sh{check if it is sequence length or outputlength}, referred to as \emph{exact-allocation}. Exact-allocation enables a scheduler to estimate each waiting prompt's memory-demand, and hence form a batch to fully allocate dual-resources. However, fully allocated KVC cannot be fully utilized due to reserved memory waste~\cite{vllm} and also prevent adding more requests to a batch to fully utilize the GPU compute resource until a request in the batch completes. }

To prevent KVC allocation failures and fully utilize dual-resources concurrently, we could adopt the approach in~\cite{Zheng2023ResponseLP} that allocates KVC based on the estimated sequence length, referred to as \emph{exact-allocation}. This method allows the scheduler to estimate each request's memory demand for its sequence length and form a batch that fully allocates dual-resources, a strategy we refer to as \emph{MultiRes}. However, the fully allocated KVC cannot be fully utilized, and it also prevents additional requests from being added to the batch to fully utilize GPU compute resource until a request in the batch completes. Specifically, PTs
will become GTs in the next iteration and continue occupying the KVC with the sequence length, so KVC won’t be available for adding more requests to the batch until a request completes. 
We refer to this problem as the \emph{GT domination issue}. In this paper, we conducted trace-based experiment analysis, and made observations:\looseness=-1


\DEL{With exact-allocation, each waiting prompt's memory-demand can be estimated, and we can choose prompts to fully allocate the dual-resources. Note that the allocated compute resource will be fully utilized while the allocated KVC won't be fully utilized. A batch is formed by selecting requests until the KVC is fully allocated. After each iteration, the allocated KVC for a prompt with length $L_p$ is retained until completion, while the compute resources for $(L_p-1)$ tokens are released. However, no new requests can be added to the batch to utilize the released compute resources, leading to low GPU utilization. We call such an issue \emph{GT domination issue}. We verified this issue through trace-based experimental analysis, which also revealed that:}
\squishlist
\item [1)] Existing methods fail to fully utilize dual-resources concurrently. While \emph{MultiRes} enhances utilization, further improvements are needed for the reasons discussed above and its coupled scheduling mechanism, which focuses on fully
utilizing both resources per request. Additionally, it incurs high scheduling time to identify the requests.
\item [2)] To address the KVC under-provisioning using LLM-based RL prediction~\cite{Zheng2023ResponseLP}, there exists a sweetspot padding ratio to the predicted value that balances execution time and waiting time. 
In addition, upon a KVC allocation failure, using the reserved KVC and preemption without KVC offloading to CPU memory (offload-free preemption) may be more efficient than vLLM's offload-based preemption.
\squishend

\DEL{\begin{figure}[h]\vspace{-0.0in}
\centering
    \subfloat[W/o decoupling: cannot add PTs until a GT completes.\vspace{-0.0in}\label{fig:DecoupleDemo11}]{{\includegraphics[width=0.45\linewidth,height=0.08\textheight]{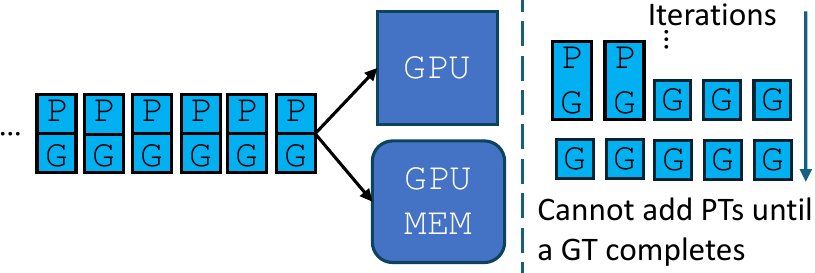}  }}
    \hfill
    \subfloat[W/ decoupling: can add PTs in each iteration. \vspace{-0.0in}\label{fig:DecoupleDemo12}]{{\includegraphics[width=0.45\linewidth,height=0.08\textheight]{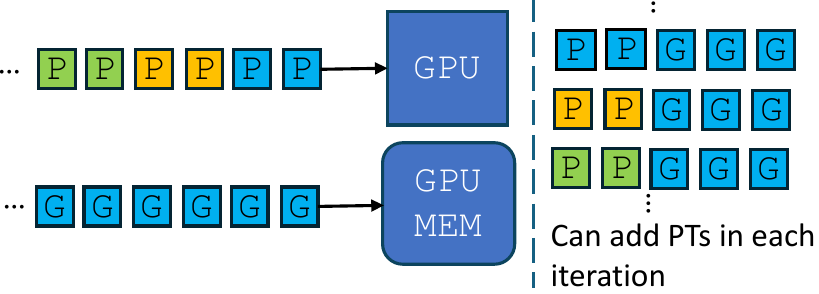} }}
    \hfill
    \vspace{-0.0in}
   \caption{{Decoupling enables adding PTs to the batch in each iteration and avoiding iteration-level scheduling.
\vspace{-0.0in}}}
    \label{fig:DecoupleDemo}
\end{figure}
}

\DEL{In this paper, we aim to \emph{achieve the aforementioned goal with low scheduling time}.
Our trace-based experimental analysis indicates that same-RL requests widely exist. Thus, to avoid heavyweight iteration-level scheduling, we can use \emph{time-sync batching (SyncCoupled)} that batches the requests with the same predicted RL to fully utilize KVC. 
}

\DEL{due to the earlier completion of some requests in the batch. However, \emph{SyncCoupled} cannot maximize GPU utilization in each iteration. In addition to the aforementioned reason, it is also because the added prompts may not complete at the same time as the current requests in the batch, contradicting the time-sync principle of \emph{SyncCoupled}.} 

\DEL{To fully utilize GPU in each iteration, prompts need to be added to the batch after each iteration, but it is almost infeasible as explained in the above. 
In addition, the added prompts may not complete at the same time as the current requests in the batch, contradicting the time-sync principle of \emph{SyncCoupled}. 
}

\DEL{To maximize GPU utilization, we augment \emph{SyncCoupled} with a Decoupled prompt and generation approach (\emph{SyncDecoupled}). \emph{SyncDecoupled} innovatively maintains separate waiting queues for PTs and GTs. The allocated KVC size of a prompt equals to its prompt length and that of a GT equals to its predicted RL. It reserves a small space in KVC for PTs. 
As shown in Figure~\ref{fig:DecoupleDemo12}, PTs will become GTs in the next iteration, and then enter the GT queue. Thus, queuing PTs can always be added to the batch in each iteration. 
}




\DEL{Basically, \sys maintains two waiting queues: one for prompts and one for TG tasks. After a prompt is processed, its TG tasks are entered to the TG waiting queue. The prompts are responsible for fully utilizing the GPU resource and the TG tasks are responsible for fully utilize the KVC resource. To reduce the scheduling times during a request's processing, \sys predicts the response lengths and batches the requests with the same response lengths so that no scheduling is needed until all requests complete processing. However, since a batch consists of all GTs, each task only contains one input token and the GPU workload linearly depends on the number of input tokens, this approach exacerbates the GPU under-utilization problem that inherently exist in LLM inference. To address this problem, since prompt processing tasks are computation heavy while token GTs are memory heavy, \sys novelly decouples the prompt processing and the generation processing of requests, and then finely batches prompt processing tasks and generation processing tasks in order not to overload or underload both the GPU and memory resources, improving the response latency and throughput performance. Further, though a request is allocated with a large KVC, it gradually fills up the KVC space. Not using assigned KVC wastes the resource especially considering it is the bottleneck. Our experiments show that ?? and ?? cache space is not in use at a time on average in the maximum space allocation method and the block-based cache allocation method, respectively. To address this problem, based on the maximum sequence length allocation method, \sys incorporates a memory pipelining method to share the allocated but unused memory among multiple requests, ensuring that a request can always continuously use its KVC without any interruption.}



To leverage \emph{MultiRes} to achieve the goal while addressing its issues, 
we propose the \sys system. \sys consists of the following components: 
\squishlist

\vspace{0.05in}\item  
\textbf{KVC Pipelining.} Each GT lends its unused allocated KVC to another GT, carefully selecting a recipient to ensure that by the time the original GT requires the lent KVC, the recipient has completed and released it. The recipient GT follows the same process and so on. This process continues iteratively until the original GT’s allocated KVC is fully utilized.

\IdeatoRead{If a request has not been completed by the predicted time, additional KVC space is allocated from a reserved KVC pool designated for such requests.? }


\vspace{0.05in}\item \textbf{Resource Responsibility Decoupling with Time-Synced Batching.} Instead of letting each request to contribute to the utilizations of the dual-resources, it assigns PTs and GTs the responsibility of fully utilizing the GPU and KVC, respectively, in each iteration, maintaining separate queues for both. This allows PTs and GTs to be fetched directly from their individual queues to be added to the batch, eliminating the need to identify specific requests for dual-resource utilization. To further reduce scheduling time, it synchronizes request completion times within a batch by grouping GTs with similar predicted RL, thereby avoiding iteration-level scheduling. To handle RL under-prediction, it employs padding on predicted values, reserves KVC, and uses offload-free preemption based on our observation.\looseness=-1

\vspace{0.05in} \item 
\textbf{Prompt and Generation Task Ordering.} 
It orders the PTs and GTs in their queues for the goal. It first tries to ensure compliance with JCT service-level-objective (SLO) requirements, then prioritizes the tasks that occupy a larger KVC space so they can release KVC earlier, and finally prioritizes the tasks with longer prompt lengths or predicted RLs to quickly find tasks to fully utilize the resources.
\squishend

\DEL{
\squishlist

\item Since some queued requests have similar output lengths, grouping same-RL requests is viable in batching (O\ref{thm-decoupling}). 

\item\emph{SyncDecoupled} holds potential to achieve the aforementioned goal, but its exact-allocation results in a large portion of allocated KV going unused (e.g., 37\%) (O\ref{thm2}).


\item To address the KVC under-provisioning using LLM-based RL prediction~\cite{Zheng2023ResponseLP}, there exists a sweetspot padding ratio to the predicted value that balances execution time and waiting time. 
In addition, upon a KVC allocation failure, using the reserved KVC and preemption without KVC offloading to CPU memory (offload-free preemption) may be more efficient than vLLM's offload-based preemption (O\ref{RLprediction}).

\item Queued requests exhibit varying occupied KVC sizes (O\ref{thm2-sort}).



\squishend
}

\DEL{\begin{figure}[h]
\centering
{{\includegraphics[width=0.8\linewidth,height=0.13\textheight]{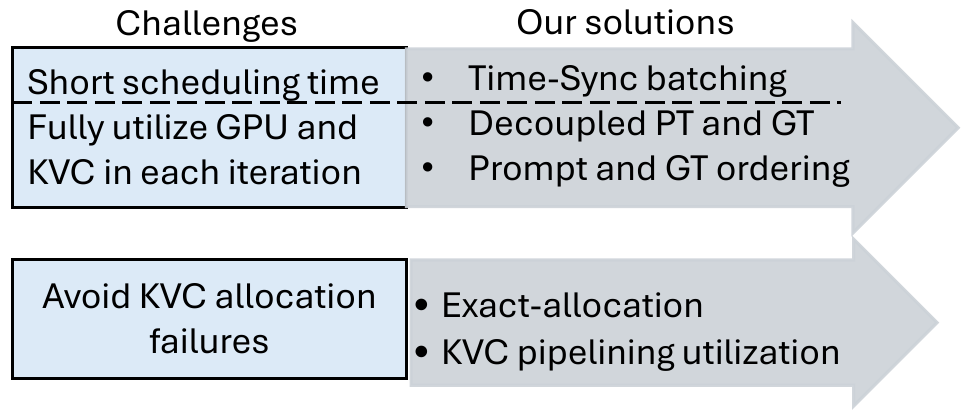}  }}
  \vspace{-0.0in} \caption{\sys's solutions to the challenges.\vspace{-0.0in} 
\vspace{-0.0in}}
    \label{fig:SolutionDemo}
\end{figure}}


Unlike methods that disaggregate GTs and PTs of a request across different GPUs~\cite{zhong2024distserve,patel2024splitwise,liu2023deja}, which are suitable for high node-affinity clusters with Infiniband (for KV value transfer) and abundant GPU resources (for hosting multiple model copies), \sys operates without such constraints, recognizing that resource efficiency is crucial and some clusters lack Infiniband, such as those found in smaller organizations and companies. In contrast to these methods, which may lead to low resource utilization, \sys focuses on fully utilizing GPU dual-resources to achieve resource-efficiency, essential in today's LLM serving systems where GPUs are both scarce and costly.



The contribution of this work includes:
\squishlist 
\vspace{0.02in}\item An in-depth trace-based experimental analysis that lays the foundation of the system design;

\vspace{0.02in} \item The EcoServe system that achieves the goal outlined, which previous methods, as shown in Table~\ref{tab:compareMethods}, failed to accomplish.


\vspace{0.02in} \item Real implementation of \sys and a comprehensive trace-driven performance evaluation. 
\squishend

Our experiments show that \sys increases throughput by up to 4$\times$ with the same level of latency, generates up to 91\% lower job completion time (JCT) and up to 91\% higher SLO satisfaction ratio compared to the vLLM state-of-the-art scheduler. Compared with DistServe~\cite{zhong2024distserve}, 
it increases GPU utilization by 16\%, reduces the number of used GPUs by 78\% for the same level of goodput 
for a cluster with Ethernet connections.\looseness=-1 
\begin{figure*}[t]
\centering
  \DEL{  \subfloat[GPU utilization. 
    \vspace{-0.0in}\label{fig:gpu-schedulers}]{{\includegraphics[width=0.32\linewidth,height=0.112\textheight]{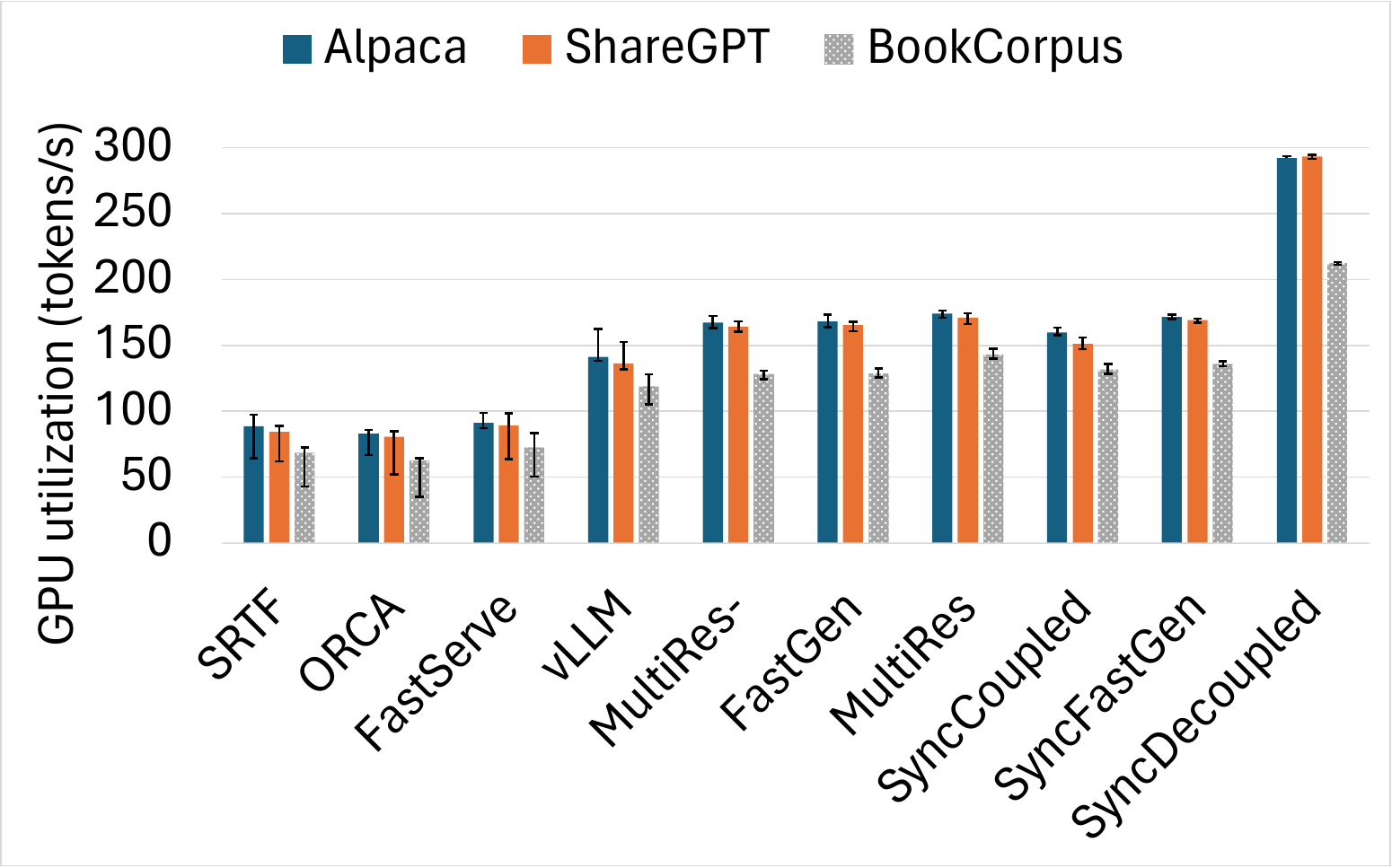} }}
    \hfill}
    \subfloat[Throughput.\vspace{-0.0in}\label{fig:throughput-schedulers}]{{\includegraphics[width=0.32\linewidth,height=0.112\textheight]{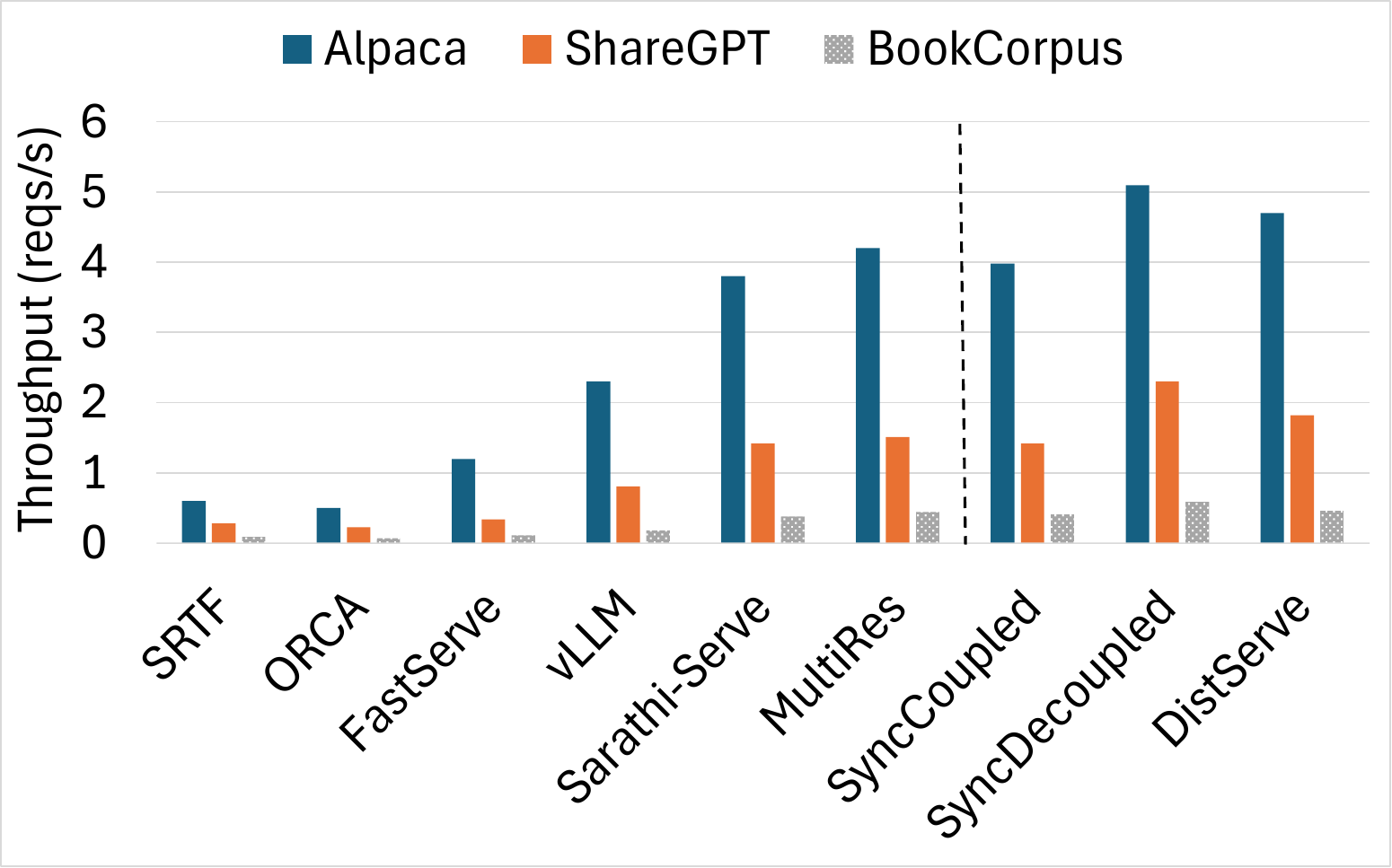} }}
    \hfill
    \DEL{\subfloat[TG latency.\vspace{-0.0in}\label{fig:tgl-schedulers}]{{\includegraphics[width=0.32\linewidth,height=0.112\textheight]{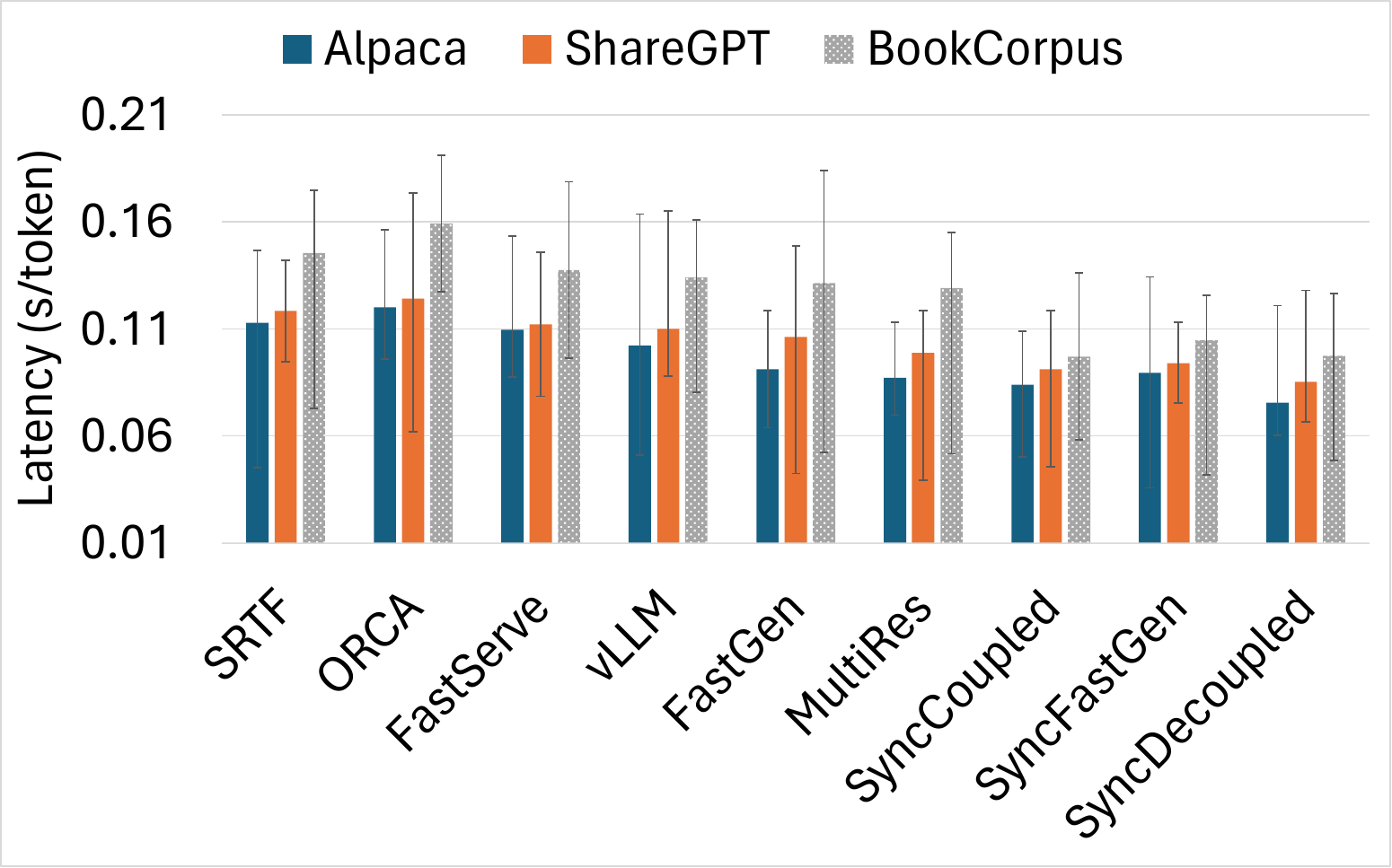} }}
    \hfill}
    \subfloat[KVC utilization.\vspace{-0.0in}\label{fig:KVC-schdulers}]{{\includegraphics[width=0.32\linewidth,height=0.112\textheight]{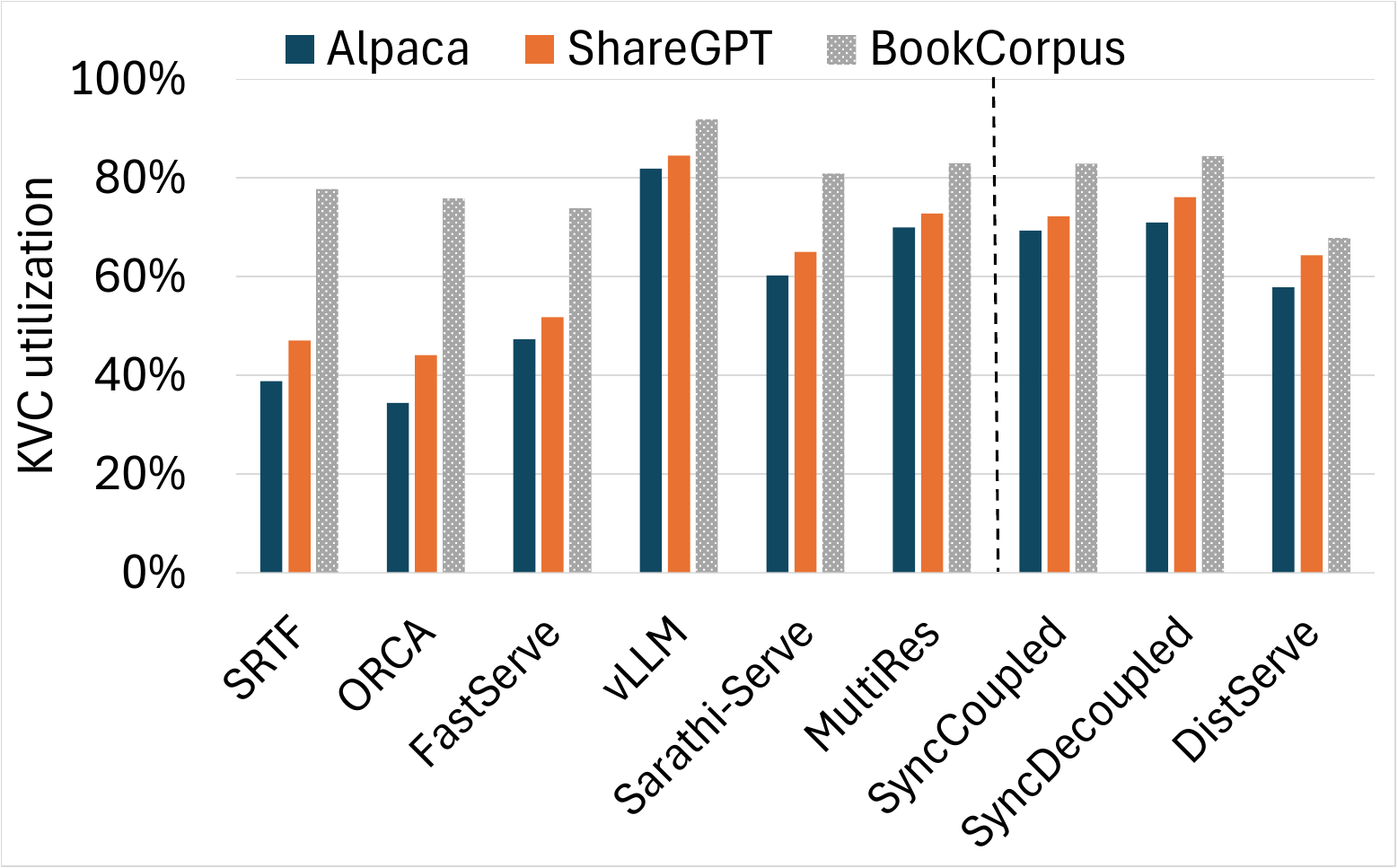} }}
    \hfill
    \subfloat[Forward size.\vspace{-0.0in}\label{fig:forwardsize-schedulers}]{{\includegraphics[width=0.32\linewidth,height=0.112\textheight]{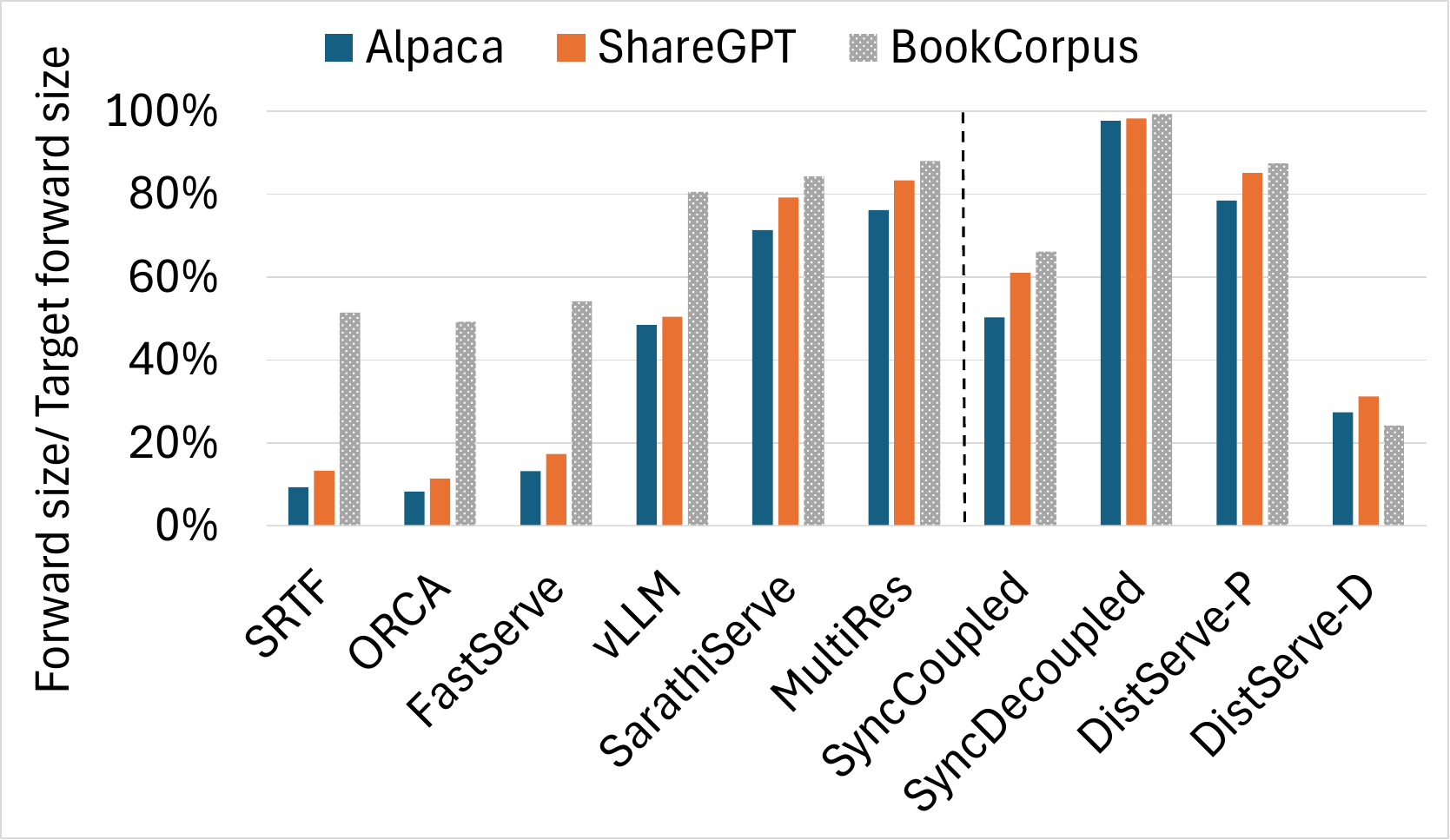} }}
    \hfill
    \subfloat[KVC allocation failure percentage.\vspace{-0.0in}\label{fig:alloc-schedulers}]{{\includegraphics[width=0.32\linewidth,height=0.112\textheight]{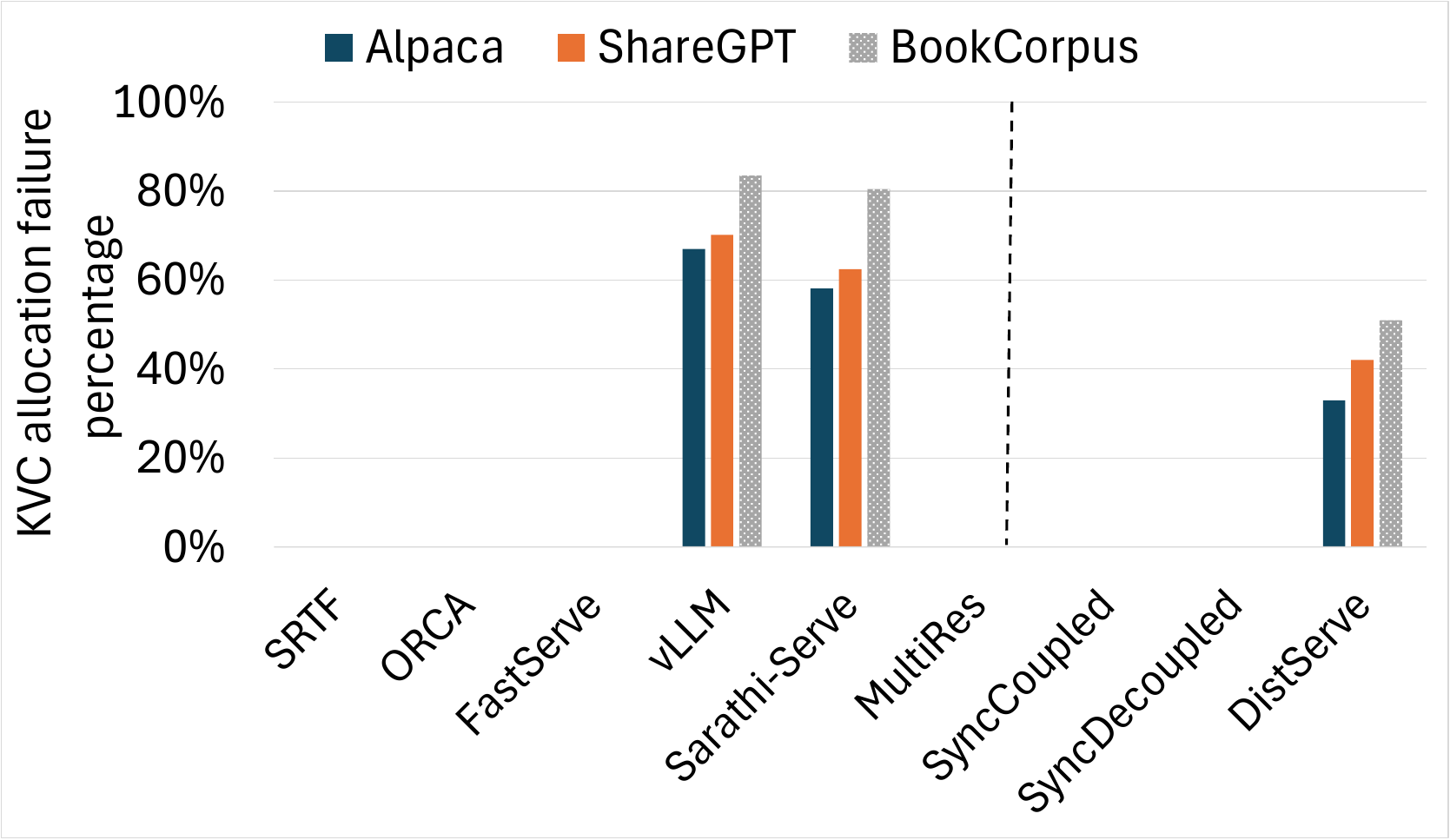}  }}
    \hfill
    \subfloat[JCT for Alpaca.\vspace{-0.0in}\label{fig:overhead-schedulers}]{{\includegraphics[width=0.32\linewidth,height=0.112\textheight]{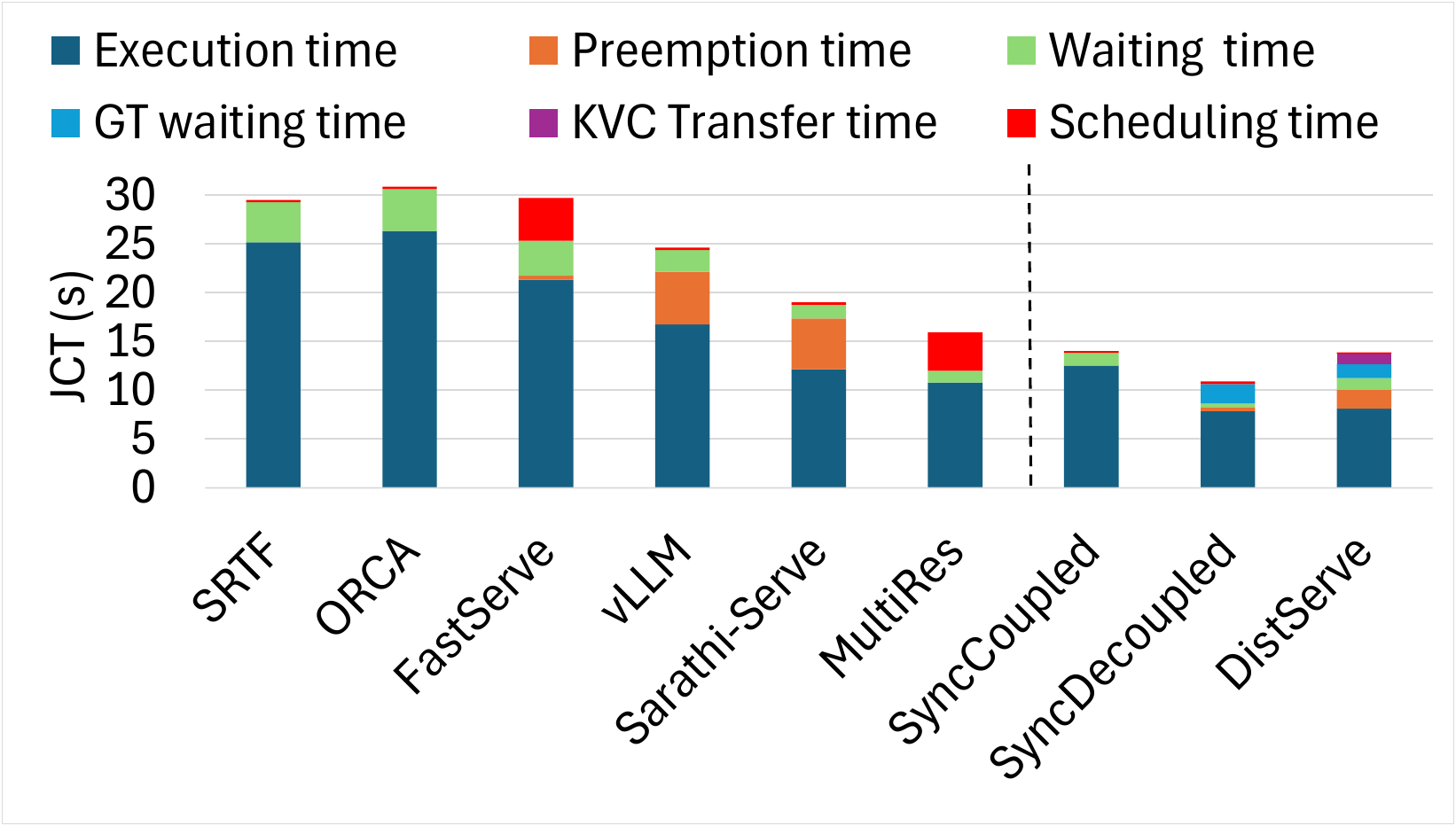}  }}
    \hfill
   \DEL{ \subfloat[JCT for ShareGPT.\vspace{-0.0in}\label{fig:overhead-schedulers}]{{\includegraphics[width=0.32\linewidth,height=0.112\textheight]{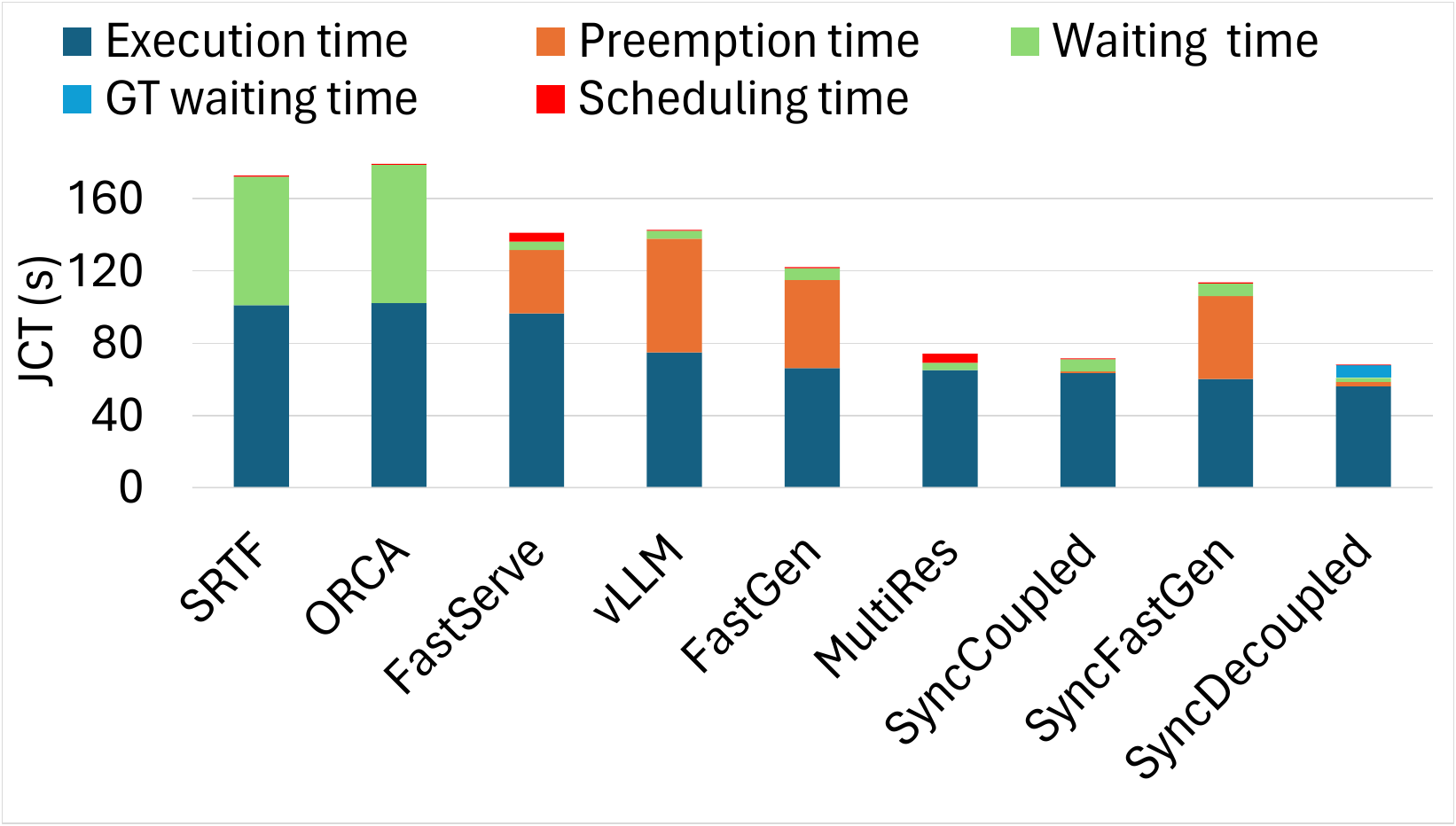}  }}
    \hfill
    \subfloat[JCT for BookCorpus.\vspace{-0.0in}\label{fig:overhead-schedulers-b}]{{\includegraphics[width=0.32\linewidth,height=0.112\textheight]{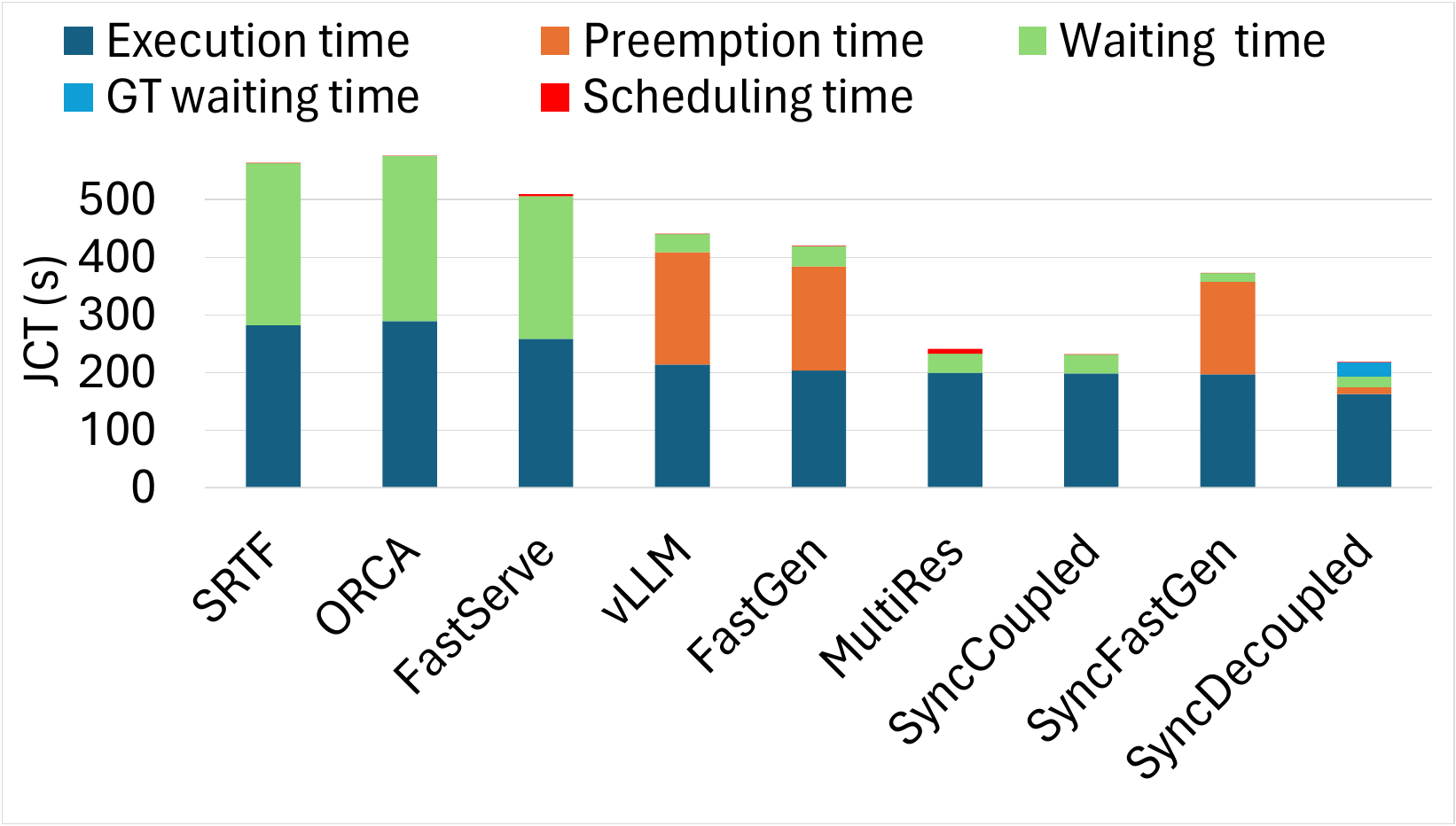}}}
    \hfill}
    \DEL{\subfloat[Scheduling time.\vspace{-0.0in}\label{fig:overhead-schedulers}]{{\includegraphics[width=0.32\linewidth,height=0.112\textheight]{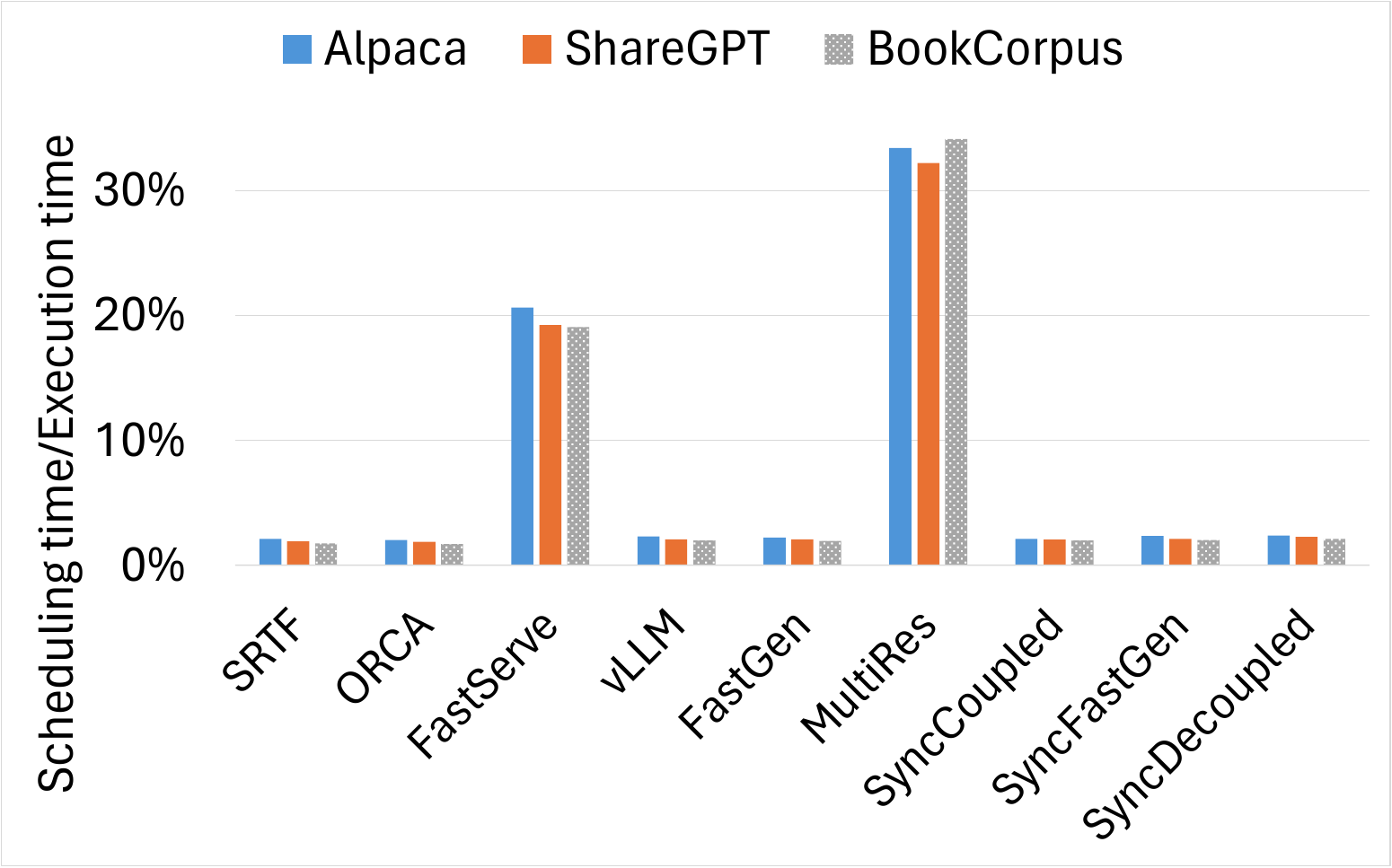}}}
    \hfill}
    \subfloat[Distribution of iterations with a certain num. of requests completed after an iteration for Alpaca.\vspace{-0.0in}\label{fig:req-schedulers}]{{\includegraphics[width=0.32\linewidth,height=0.112\textheight]{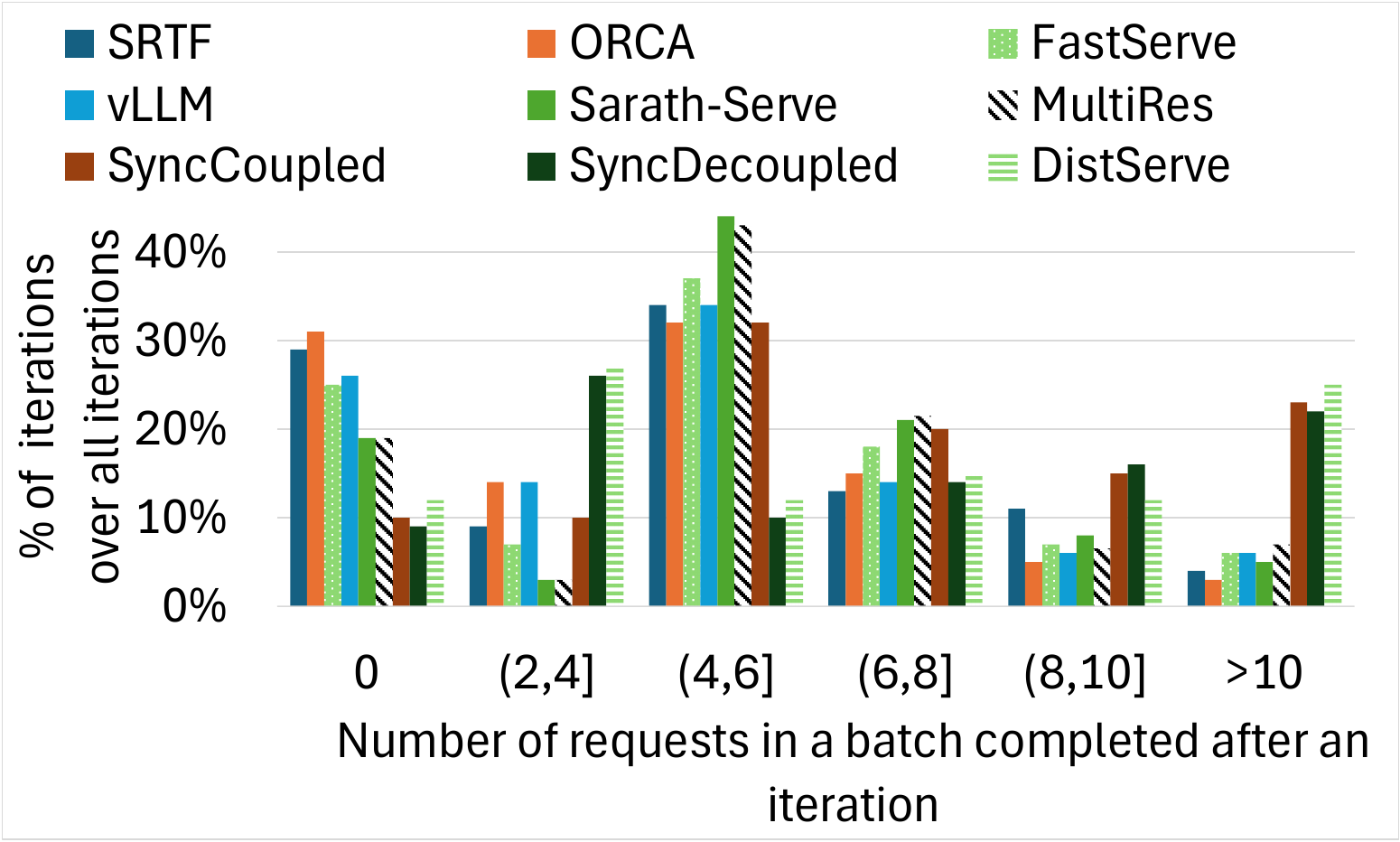}}}
    \hfill
    \DEL{\subfloat[Distribution of iterations with a certain num. of requests completed after an iteration for ShareGPT.\vspace{-0.0in}\label{fig:req-schedulers-sharegpt}]{{\includegraphics[width=0.32\linewidth,height=0.112\textheight]{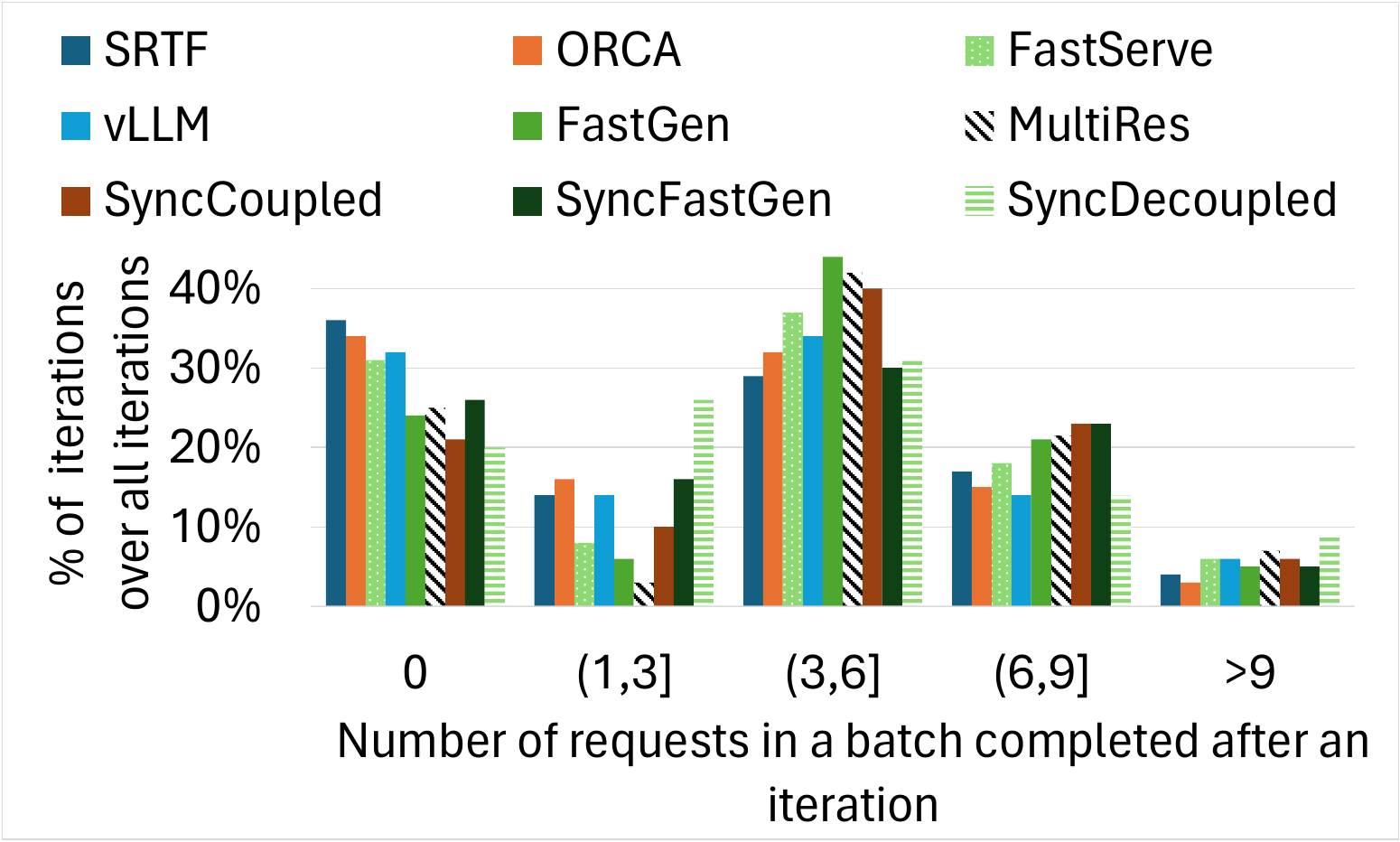}}}
    \hfill
     \subfloat[Distribution of iterations with a certain num. of requests completed after an iteration for BookCorpus.\vspace{-0.0in}\label{fig:req-schedulers-bookcprpus}]{{\includegraphics[width=0.32\linewidth,height=0.112\textheight]{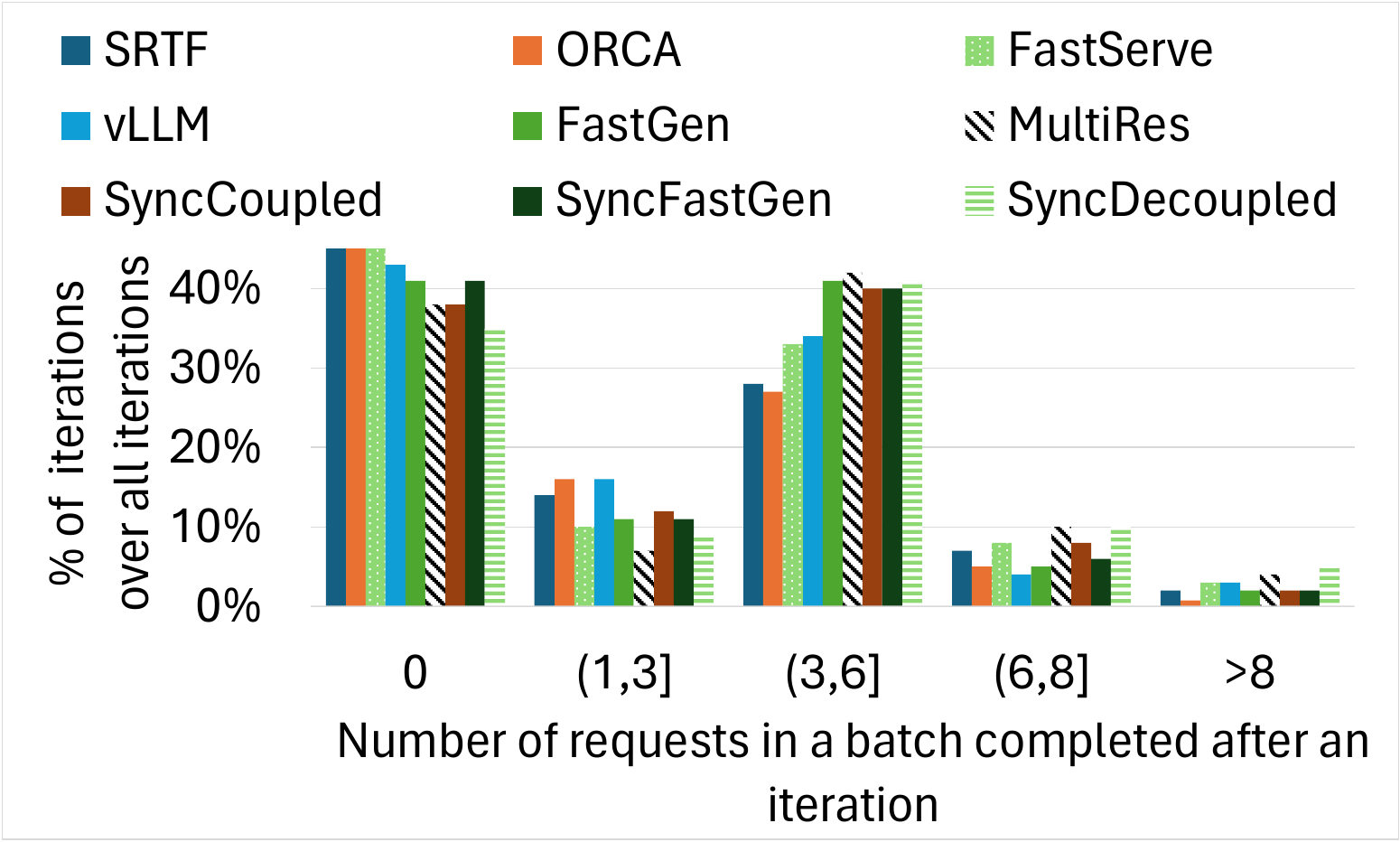}}}
    \hfill}
    \vspace{-0.15in}
   \caption{{Comparison of different schedulers.
\vspace{-0.1in}}}%
    \label{fig:schedulers-measurement}
\end{figure*}
\DEL{\vspace{-0.0in}
\section{Preliminaries}\vspace{-0.0in}}

\DEL{\begin{figure}
    \centering
\includegraphics[width=1\columnwidth,height=0.145\textheight]{fig/transformer-updated.png}
    \caption{Transformer layer of GPT.}
    \label{fig:transformer}\vspace{-0.0in}
\end{figure}
}

\DEL{As a representative of the transformer model, we describe the GPT structure. 
As shown in Figure~\ref{fig:transformer}, before the attention layer, GPT transformer consists of the layer normalization operation (LayerNorm) and the QKV Linear (linear and split operations to get the query, key, and value). Operations performed after Attention are, in order, a linear operation (Attn Out Linear), an add operation for residual connection (Add), layer normalization operation (LayerNorm), the fully connected layers (FC) operations, and the other residual
connection operation (Add). These layers can be divided into two categories: 1) forward computation layers (FCL), and 2) attention layer (marked within a dashed box) in Figure~\ref{fig:transformer}. Rest of the layers marked within the solid box are part of the FCL.

The three input components to the attention layer, query, key and value, are the projections of the embeddings using three learned metrics $\mathbf{W_q},\mathbf{W_k}$ and $\mathbf{W_v}$ of the transformer. The embedded matrix for the whole token sequence input is $\mathbf{E}$, where $e_i$ is the embedding of the each token in the sequence. 
Then, the projection of  is created by
\begin{equation*}
    \mathbf{Q} = \mathbf{W_q} \mathbf{E} \text{, }
    \mathbf{K} = \mathbf{W_K} \mathbf{E} \text{, }
    \mathbf{V} = \mathbf{W_v} \mathbf{E}
\end{equation*}
The projection calculation involves matrix multiplications between the learned matrices and the embedded matrix. At the layer level, each matrix multiplication is a fully connected layer with input and output neurons. Then for each matrix multiplication the number of operations is $2 \times \# tokens \times in\_neurons \times out\_neurons$. 

For each token in the input, the Attention operation 
takes three inputs represented as  Query ($\mathbf{Q}$), Key ($\mathbf{K}$), and Value ($\mathbf{V}$). The attention module includes the dot products of the query tokens in $\mathbf{Q}$  with all the keys of previous tokens in $\mathbf{K}$ to measure the similarity of previous tokens from the new tokenâ€šÃ„Ã´s perspective:
\begin{equation}
    \label{eq:3}
    Similarity(\mathbf{Q,K}) = \mathbf{QK}^T
\end{equation}
It then applies the Softmax to the dot products to get weights and produces the output ($Self-attention(\mathbf{Q,K,V}$)) as a weighted sum of the values. Here, the weights are called attention score of the query tokens with respect to the previous tokens. 

\DEL{\begin{equation}
\label{eq:4}
    \mathbf{S} = softmax(\mathbf{QK}^T)
\end{equation}
\begin{equation}
\label{eq:5}
Self-attention(\mathbf{Q,K,V}) = [WeightedSum(\mathbf{V_1,S}),\cdots, WeightedSum(\mathbf{V_n,S})]
\end{equation}
All these mathematical operations can be summarized as follows:}

\begin{equation}
    \label{eq:6}
    Self-attention(\mathbf{Q,K,V}) = softmax(\frac{\mathbf{QK}^T}{\sqrt{d_k}}) \mathbf{V},
\end{equation} where 
$d_k$ is the dimension of the $\mathbf{K}$ or $\mathbf{Q}$.


The computational complexity at the layer level is $4 \times (\# \text{ of tokens})^2 \times model\_dim$.


The output of the transformer layer is the probability of each word to be present in the output. 
The term ``logitsâ€šÃ„Ã¹ refers to the raw outputs of a model before they are transformed into probabilities. Specifically, logits are the output of the FC2 layer, which is passed through the Add to generate the probabilities and then the new token.
}

\DEL{\tannsdi{Let us mathematically explain the attention score calculation process in detail. Given a sequence of input tokens $\mathbf{x_1,} \cdots \mathbf{x_n}$ where any $x_i \in \mathbb{R}^d$, 
its self-attention outputs a sequence of the same length $y_1,\cdots,y_n$, where function mapping to learn is as follows.
\begin{equation}
\label{eq:1}
 y_i = f(x_i,(x_1,x_1),\cdots, (x_n,x_n)\in \mathbb{R}^d
\end{equation}
Now, as mentioned above, the attention contains three major components, which are Query ($\mathbf{Q}$), Key($\mathbf{K}$), and Value($\mathbf{V}$) matrices, which are derived from three learned matrices $\mathbf{W_Q},\mathbf{W_K},$ and $\mathbf{W_V}$. A projection of the mentioned $\mathbf{Q}$, $\mathbf{K}$ and $\mathbf{V}$ can be created as follows from the embedding of the embedding of each input token $\mathbf{e_i}$.
\begin{equation}
\label{eq:2}
    \mathbf{Q_i}=\mathbf{W_Q}\mathbf{e_i},
\mathbf{K_i}=\mathbf{W_K}\mathbf{e_i},
\mathbf{V_i}=\mathbf{W_V}\mathbf{e_i}
\end{equation}
Then the dot-product similarity between the Query ($\mathbf{Q}$) and Key ($\mathbf{K}$) is being calculated.
\begin{equation}
    \label{eq:3}
    Similarity(\mathbf{Q,K}) = \mathbf{QK}^T
\end{equation}
After the computation the similarity between the $\mathbf{Q}$ and $\mathbf{K}$, the softmax function is applied to obtain the weights. The softmax function ensures that the weights for each Value are positive and sum to 1.
\begin{equation}
\label{eq:4}
    \mathbf{S} = softmax(\mathbf{QK}^T)
\end{equation}
Then, the final output matrix (\mathbf{y}) is obtained by concatenating the Weighted Sum matrices computed for each embedding in the input sequence:
\begin{equation}
\label{eq:5}
\mathbf{y} = [WeightedSum(\mathbf{V_1,S}),\cdots, WeightedSum(\mathbf{V_n,S})]
\end{equation}
Finally, the  Self-attention score calculation mechanism can be summarized with the following formula:
\begin{equation}
    \label{eq:6}
    Self-attention(\mathbf{Q,K,V}) = softmax(\frac{\mathbf{QK}^T}{\sqrt{d_k}}) \mathbf{V}
\end{equation}, where $\mathbf{Q}$, $\mathbf{K}$, and $\mathbf{V}$ are the Query weighted matrix, Key weighted matrix and Value weighted matrix, respectively. $d_k$ is the dimension of the $\mathbf{K}$ or $\mathbf{Q}$.}}


\DEL{In the Transformer model, the above operations are executed by each attention head in parallel. Each attention head uses its own set of Query, Key, and Value parameters. This allows the model to capture various aspects of the context and focus on different relationships among words simultaneously~\cite{Vaswani2017AttentionIA}.
Finally, the context-aware representations (results) from all attention heads are concatenated. 
This new matrix is later used to produce the final attention score.

(?could delete)The GPT model receives a sequence of input tokens and then completes the sequence by generating all subsequent output tokens. 
The input tokens are processed sequentially by the layers of the GPT model. {The whole process can be divided into prompt processing and token generation. In the prompt processing during the first iteration, the input tokens from the given prompt are processed, generating one output token. This is the step, where all the attention scores of the tokens from the step is calculated and stored in the KVC.}
While in the next iterations in the token generation phase, only the key-value tensors of the newly generated token require computation, and others are loaded from the KV-cache. One run of all layers is an iteration or a forward pass of the model.
--- content from connor's email, will need to update figure accordingly---

In terms of the KVC size~\cite{KVC}, the amount of bytes stored by each token is: $2*2* \# \text{ of transformer layers} \times hidden\_size$. The first '2' is to account for the two vectors,
$\mathbf{K}$ and  $\mathbf{V}$.  Then multiply by 2 again for the number of bytes as we consider 16-bit precision formats. $\# of transformer layers$ are involved as the subsequent tokens are passed through transformer layers until the generation is complete. 

Now, for a single forward pass of the model, weâ€šÃ„Ã´ll pass through all the blocks shown in the Figure~\ref{fig:transformer}. From a throughput perspective, the transformer layers will typically dominate the end-to-end (E2E) runtime and can be used as a FLOPs proxy for the whole GPT model.
}

\DEL{As a representative of the transformer model, we describe the GPT structure. As shown in Figure~\ref{fig:transformer}, before the attention layer, GPT transformer consists of the layer normalization operation (LayerNorm) and the QKV Linear (linear and split operations to get the query, key, and value). Operations performed after Attention are, in order, a linear operation (Attn Out Linear), an add operation for residual connection (Add), layer normalization operation (LayerNorm), the fully connect layers (FC) operations, and the other residual
connection operation (Add). These layers can be divided into two categories: 1) forward computation layers (FCL), and 2) attention layer (marked within a dashed box) in Figure~\ref{fig:transformer}. Rest of the layers marked within the solid box are part of the FCL.}


\DEL{\noindent\textbf{Throughput.} For a transformer layer, there are two primary throughput contributors, the forward computation layers (FCL) layers and attention. \DEL{The other layers Add and LayerTransform have negligible operations with respect to these so that we can ignore their count of operations.}
The total operations for a single transformer laye (assuming only one sequence) is:\vspace{-0.0in} 
\begin{equation}
\label{eq:totop}
24 \times tokens \times model\_dim^2 \times (1 + \dfrac{tokens} {6 \times model\_dim}).\vspace{-0.0in}
\end{equation}
We derived this equation from the equation in~\cite{megaequ}.
For short sequence lengths ($tokens$$<<$$6 \times model\_dim$), it is easy to proxy this by $2 \times tokens \times model\_dim^2$.
\DEL{The first addend comes from the FCL operations, and the second number is derived from the attention operations.} Therefore, the GPU utilization depends on the forward size. \DEL{We refer to the forward size that fully utilize GPU but does not greatly compromise JCT
as \emph{Target Forward Size (TFS)}.} 
}






%


\DEL{\noindent{\textbf{Processing steps of a Transformer Model.}}
The GPT model receives a sequence of input tokens and then completes the sequence by generating all subsequent output tokens. 
The input tokens are processed sequentially by the layers of the GPT model. \tannsdi{The whole process can be divided into prompt processing and token generation. In the prompt processing during the first iteration, the input tokens from the given prompt are processed, generating one output token. This is the step, where all the attention scores of the tokens from the step is calculated and stored in the KVC.}}


\DEL{Then, in the token generation part, the generated output token is fed back into the model to generate the next output token. This autoregressive procedure of generating a single token is done by running all the layers of the GPT model with the input, which is either a sequence of input
tokens from the client or a previously generated output token. One run of all layers is an iteration of the model. }

\DEL{\noindent\textbf{KVC Allocation.} In an iteration, the KV values of all the tokens from a step are calculated and stored in the KVC. \DEL{In the subsequent iteration, only the KV tensors of the newly generated token need computation, while others are loaded from the KVC.} 
The byte storage per token in KVC is equal to~\cite{kv-cache}: \vspace{-0.0in}
\begin{equation}\label{equ:cache} 2\times 2\times \# \text{ of transformer layers} \times model\_dim. \vspace{-0.0in}\end{equation}
\DEL{In addition to the max-allocation method used in \Orca, there are two other KVC allocation approaches. 
The Transformers library by Huggingface \cite{Face3} allocates new memory at each token generation but it causes delays associated with memory allocation. The vLLM approach~\cite{vllm} uses a block-based approach and one block hosts a certain number of tokens, say 128 tokens. 
When a token is generated, if this token has not been allocated with KVC space, a block is created for this token and the next 127 tokens.\looseness=-1}
}

\DEL{{\color{red}\noindent\textbf{KV-cache occupancy calculation.} We derive the following equation to calculate the KV-cache utilization. Let us assume, 
$L$ : average number of tokens stored in GPU memory per second for all $n$ prompts,
$l$ The number of the final output tokens (including the prompt and the generated tokens),
$p$: The number of tokens in the prompt,
$t$: The average time required to generate one token,
$T$: sampling time window.

If there are multiple prompts at the $T$ time-window, then following the arithmetic sequence algorithm. 
\begin{multiline}
 p + (p+1) + (p+2) + ... + (l-1) + l = [(p + l) * (l - p + 1)] / 2 = (l^2 - p^2 + p + l) / 2
\end{multiline}
This leads to the following equation:
\begin{equation}
\footnotesize
    L = \frac{(l_1^2+l_2^2+\cdots l_n^2)-(p_1^2+p_2^2+\cdot p_n^2)+(p_1+p_2+\cdots p_n)+ (l_1+l_+\cdots l_n)}{2\times T}\times t
    \label{eq:1}
\end{equation}
Finally, since the memory size of a token for a specific model is:
\begin{equation*}
\footnotesize
m = 2 \times 2 \times n_{layer} \times d_{model}    
\end{equation*}

where $n_{layer}$ is the number of layers, $d_{model}$ is the dimension of the model.
Thus the average memory usage of those L tokens is
$L \times m$.

}}

\DEL{This eliminates the need to determine the output sequence length but comes at the expense of prolonged inference latency. On the other hand, the FasterTransformer library by NVIDIA \cite{fastertransformer} pre-assigns memory for the maximum sequence length, leading to excessive memory allocation, which limits batch size 
and throughput. The maximum sequence lengths can be from 8K to 32K~\cite{OpenAI5}.}

\DEL{For an input sequence, the VLLM approach allocate a certain number of blocks. For example, 4 blocks are allocated to a 512-token input sequence. Then, when a token is generated, a block is created for this token and the next 127 tokens. Next, when the $129^{th}$ token is generated, another 128-token block is created.
This is an improvement over both the above solutions. First, since we pre-allocate, we do not have to incur per-token allocation costs (which also has high occupancy costs (??Connor, what is occupancy cost?)). Second, the maximum over-use of KVC in our system (??Connor, what is this system, who are the users? which are the input sequences?) is not that of the maximum sequence length, but that of a single KV-block, which is like to be 64-128 tokens. So there is much less wasted memory.
}


\DEL{To support the growing KVC size,
Huggingfaceâ€šÃ„Ã´s Transformers library \cite{Face3} constantly allocates new memory at each token generation and incurs latency overhead associated with memory allocation. This improves usability since users
do not have to know the output sequence length but suffers long inference latency. Alternatively,
NVIDIAâ€šÃ„Ã´s FasterTransformer library \cite{fastertransformer} pre-allocates memory for the maximum sequence length,
which ends up wasting memory by allocating more than is necessary. This approach limits batch
size and prioritizes latency over throughput. The trend towards long maximum sequence lengths
in recent models (e.g., 8K or 32K) \cite{OpenAI5} amplifies the KVC overhead, demanding more efficient
memory utilization in serving Transformer-based text generation models."

vLLM approach~\cite{vllm}, in which we allocate a large number of blocks initially and then map to them to sequences as is necessary. This is an improvement over both the solutions. First, since we pre-allocate, we do not have to incur per-token allocation costs (which also has high occupancy costs). Second, the maximum over-use of KVC in our system is not that of the maximum sequence length, but that of a single KV-block, which is like to be 64-128 tokens. So there is much less wasted memory.}

\vspace{-0.1in}
\section{Experiment Analysis}\vspace{-0.1in}
\label{sec:analysis}
\subsection{Experiment Settings}\vspace{-0.05in}
\noindent{\textbf{Machine settings.}} We ran our experiments on an AWS p4d.24xlarge instance, equipped with 8 NVIDIA A100
GPUs and 1152GB of total CPU and GPU memory. Each GPU is equipped with 80GB of memory. The GPUs are connected with a 600 GB/s NVSwitch. We ran the OPT model with 13B parameters~\cite{Zhang2022OPTOP} on one GPU. The memory for KVC is 12GB.

\noindent{\textbf{Request settings.}} 
We used
the Alpaca~\cite{alpaca}, ShareGPT~\cite{sharegpt} and BookCorpus~\cite{soskkobayashi2018bookcorpus} traces. Table~\ref{tab:trace-table} shows the properties of the traces and our settings. \DEL{Alpaca and ShareGPT have 52K and 90K requests\sh{add these value to the table, and call it "Reqs \#"}. 
BookCorpus
has 11K unpublished books, with longer prompt lengths up
to 461K.}For each trace, we used 10K requests for fine-tuning the RL prediction model, and the rest of the requests for experiments. 
The batch size of \Orca and FastServe was set to 8~\cite{280922}. 
We set TFS 
by finding the forward size that saturates GPU utilization as in~\cite{deepspeed-fastgen}. As in~\cite{vllm}, the block size was set to 32 and the request arrival rate followed a Poisson distribution as indicated in Table~\ref{tab:trace-table}. We set the arrival rates to create our scenario that some requests are queued when a batch is processing. 
We divided the prompts in BookCorpus~\cite{soskkobayashi2018bookcorpus} into 2048-token chunks to meet the requirement of our used LLM model. 
The KVC utilization (measured using the gpustat library~\cite{gpustat}) was collected at a 1s time interval. 



\DEL{\begin{table}[]
\centering
\caption{Trace Statistics??check the values and update the statistics}
\label{tab:trace-table}
\resizebox{\columnwidth}{!}{%
\begin{tabular}{l|lll|l|l|l|l|l|}
\hline
\multicolumn{1}{|l|}{Trace} & \multicolumn{3}{l|}{Input length} & Output length &  &  & TFS & Arrival rate (reqs/s) \\ \hline
                                 & \multicolumn{1}{l|}{avg}    & \multicolumn{1}{l|}{min} & max & avg    & min & max &  &    \\ \hline
\multicolumn{1}{|l|}{Alpaca}     & \multicolumn{1}{l|}{191.31} & \multicolumn{1}{l|}{}    &     & 58.41  &     &     &  & 36 \\ \hline
\multicolumn{1}{|l|}{ShareGPT}   & \multicolumn{1}{l|}{161.31} & \multicolumn{1}{l|}{}    &     & 337.99 &     &     &  & 4  \\ \hline
\multicolumn{1}{|l|}{BookCorpus} & \multicolumn{1}{l|}{}       & \multicolumn{1}{l|}{}    &     &        &     &     &  &    \\ \hline
\end{tabular}%
}
\end{table}}
\DEL{\begin{table}[!htbp]
\centering
\caption{Trace properties and experiment settings.\sh{fix the right bar}}\vspace{-0.0in}
\label{tab:trace-table}
\resizebox{\columnwidth}{!}{%
\begin{tabular}{|c|ccc|ccc|c|c|}
\hline
\multirow{2}{*}{Trace} &
  \multicolumn{3}{c|}{Input length} &
  \multicolumn{3}{c|}{Output length} &
   \multirow{2}{*}{Req \#} &
 \multirow{2}{*}{Arrival rate} \\ \cline{2-7}
 &
  \multicolumn{1}{c|}{avg} &
  \multicolumn{1}{c|}{min} &
  max &
  \multicolumn{1}{c|}{avg} &
  \multicolumn{1}{c|}{min} &
  max &
   \\ \hline
Alpaca &
  \multicolumn{1}{c|}{19.31} &
  \multicolumn{1}{c|}{9} &
  2.47K &
  \multicolumn{1}{c|}{58.41} &
  \multicolumn{1}{c|}{13} &
  292 & 52k &
  36 reqs/s \\ \hline
ShareGPT   & \multicolumn{1}{c|}{161.31}  & \multicolumn{1}{c|}{16} & 3.2K & \multicolumn{1}{c|}{337.99} & \multicolumn{1}{c|}{19} & 991 & 90k & 28 reqs/s \\ \hline
BookCorpus & \multicolumn{1}{c|}{1952.11} & \multicolumn{1}{c|}{18} & 461K & \multicolumn{1}{c|}{681.2}  & \multicolumn{1}{c|}{32} & 1041 & 11k &1.2 reqs/s\\ \hline
\end{tabular}%
}
\end{table}}


\begin{table}[h]\vspace{-0.1in}
\centering
\caption{Trace properties and experiment settings.}\vspace{-0.15in}
\label{tab:trace-table}
\resizebox{\columnwidth}{!}{%
\begin{tabular}{|c|ccc|ccc|c|c|}
\hline
\multirow{2}{*}{Trace} &
  \multicolumn{3}{c|}{Input length} &
  \multicolumn{3}{c|}{Output length} &
  \multirow{2}{*}{Req\#} &
  \multirow{2}{*}{Arrival rate} \\ \cline{2-7}
 &
  \multicolumn{1}{c|}{avg} &
  \multicolumn{1}{c|}{min} &
  max &
  \multicolumn{1}{c|}{avg} &
  \multicolumn{1}{c|}{min} &
  max &
   &
   \\ \hline
Alpaca &
  \multicolumn{1}{c|}{19.31} &
  \multicolumn{1}{c|}{9} &
  2.47K &
  \multicolumn{1}{c|}{58.41} &
  \multicolumn{1}{c|}{13} &
  292 &
  52K &
  36 reqs/s \\ \hline
ShareGPT &
  \multicolumn{1}{c|}{161.31} &
  \multicolumn{1}{c|}{16} &
  3.2K &
  \multicolumn{1}{c|}{337.99} &
  \multicolumn{1}{c|}{19} &
  991 &
  90K &
  28 reqs/s \\ \hline
BookCorpus & \multicolumn{1}{c|}{1952.11} & \multicolumn{1}{c|}{18} & 461K & \multicolumn{1}{c|}{681.2} & \multicolumn{1}{c|}{32} & 1041 & 11K & 1.2 reqs/s \\ \hline
\end{tabular}%
}\vspace{-0.1in}
\end{table}

\noindent{\textbf{Schedulers.}} We assumed that the RL of each request is pre-known in our first measurement. We evaluated the following schedulers: 1) \Orca; 2) The shortest-remaining-time-first in iteration-level scheduling using max-allocation (SRTF); 3) FastServe~\cite{Wu2023FastDI} using 5-level queues; 4) vLLM~\cite{vllm} with the KVC swapping strategy; and 5) Sarathi-Serve. 
We also adopt the scheduler for the cloud, \emph{MultiRes}~\cite{227623}. 
After each iteration, when there are available GPU and KVC resources, \emph{MultiRes} calculates the Euclidean distance between each request's demands on GPU and KVC and the available resources, and chooses the request with the minimum distance, and repeats this process after each request selection until no requests can be added to the batch. Its time complexity for scheduling is $O(n^2)$, where $n$ is the number of requests in the queue. 

\DEL{Since previous work focuses either on improving KVC utilization~\cite{vllm} or GPU utilization~\cite{deepspeed-fastgen}, to improve both, we naturally adopt \emph{MultiRes}~\cite{227623} as our iteration-level and exact-allocation scheduler. 
After each iteration, when there are available GPU and KVC resources, \emph{MultiRes} calculates the Euclidean distance between each request's demands on GPU and KVC and the available resources and chooses the request with the minimum Euclidean distance, and repeats this process after each request selection until no requests can be added to the batch. Its time complexity for scheduling is $O(n^2)$, where $n$ is the number of requests in the queue.} 

We implemented
\Orca and FastServe ourselves since their source codes are not publicly available. We used the source code of Sarathi-Serve~\cite{sarathi-code} and DistServe~\cite{distserve-code}, 
and implemented the other schedulers based on vLLM's source code~\cite{vllm}. The error bars in our reported results are the 5th percentile and the 95th percentile. For figures with three datasets, unless specified otherwise, we discuss the average value of the three datasets.

\DEL{\begin{table}[h]
\centering
\caption{Exploring an advanced scheduler.}
\label{tab:offline-time}\vspace{-0.0in}
\resizebox{\columnwidth}{!}{%
\begin{tabular}{|l|l|l|l|l|}
\hline
Method & Scheduling time & GPU and Mem utilization  \\ \hline
Un\emph{SyncCoupled} & High & Relatively high  \\ \hline
\emph{SyncCoupled}  & Low & Low GPU, hard to increase    \\ \hline
SyncDecoupled  & Low\  & High\    \\ \hline
\end{tabular}%
}\vspace{-0.0in}
\end{table}
}


\DEL{we schedule the requests based on the distance between the demanded and available resources of a request~\cite{227623}\sh{which KV cache allocation method is used}.  In this scheduling, we consider a request to have a priority based on the distance based on the available resource and demanded resource. The \emph{MultiRes-} only considers the total available GPU and total available memory at the beginning for selecting the requests while calculating the distance. The \emph{MultiRes} considers reduces the available GPU and memory by the demanded memory after every request is selected and allocated the resource.}

\subsection{Motivation and Exploration for a New Scheduler}

\DEL{??Analysis section: need to show the advantage of decoupling and the pre-sorting instead of selecting prompts one by one to reach TFS. For example, we need to fill up 1000 prompt tokens, and checking requests in the queue sequentially takes time. if you group them based on prompt length, then you can quickly find the group with 1000 tokens. the same for the GT group below--need analysis figs to show this)}





\DEL{\begin{figure}[t]
    \centering
\begin{minipage}[t]{0.48\linewidth}
\includegraphics[width=\linewidth,height=0.112\textheight]{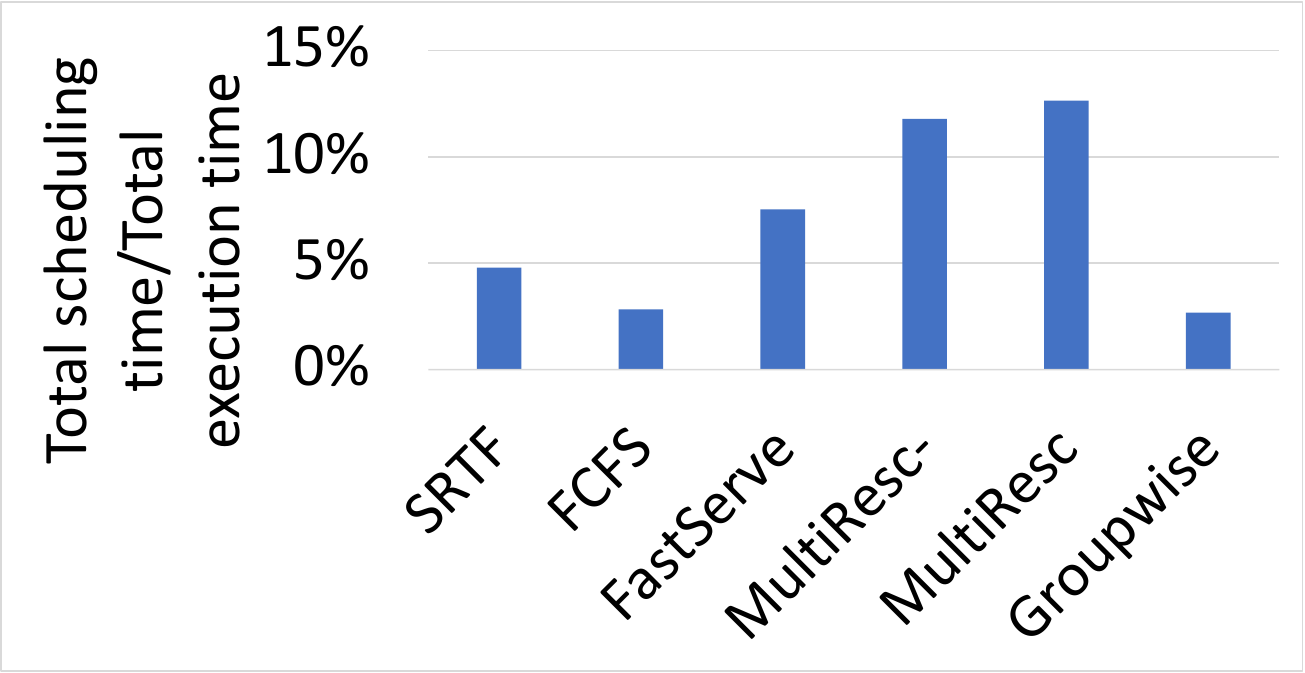} 
\vspace{-0.0in}    \caption{Scheduling time.\vspace{-0.0in} }
    \label{fig:overhead-schedulers}
\end{minipage} 
\begin{minipage}[t]{0.48\linewidth}
\includegraphics[width=\linewidth,height=0.112\textheight]{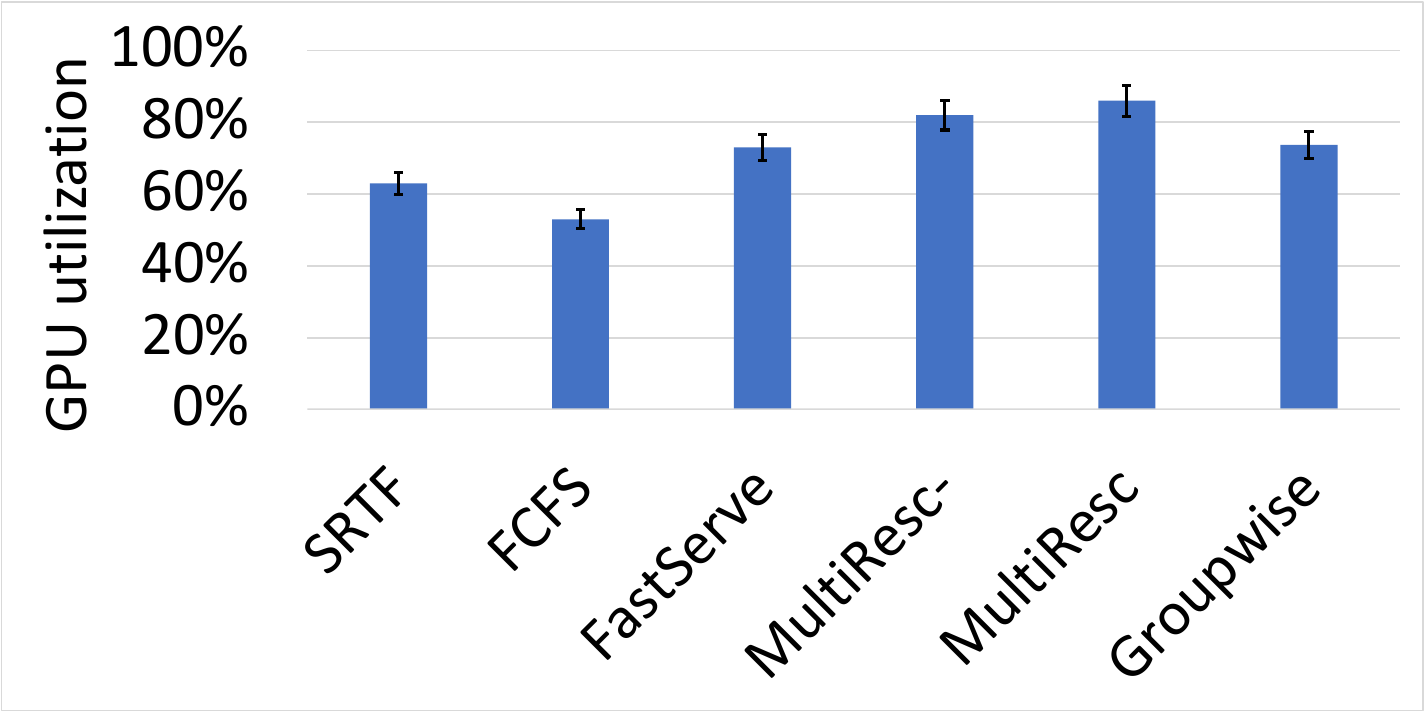} 
\vspace{-0.0in}    \caption{GPU utilization.\vspace{-0.0in}}
    \label{fig:gpu-schedulers}
\end{minipage} 
\DEL{\begin{minipage}[t]{0.48\linewidth}
\includegraphics[width=\linewidth,height=0.112\textheight]{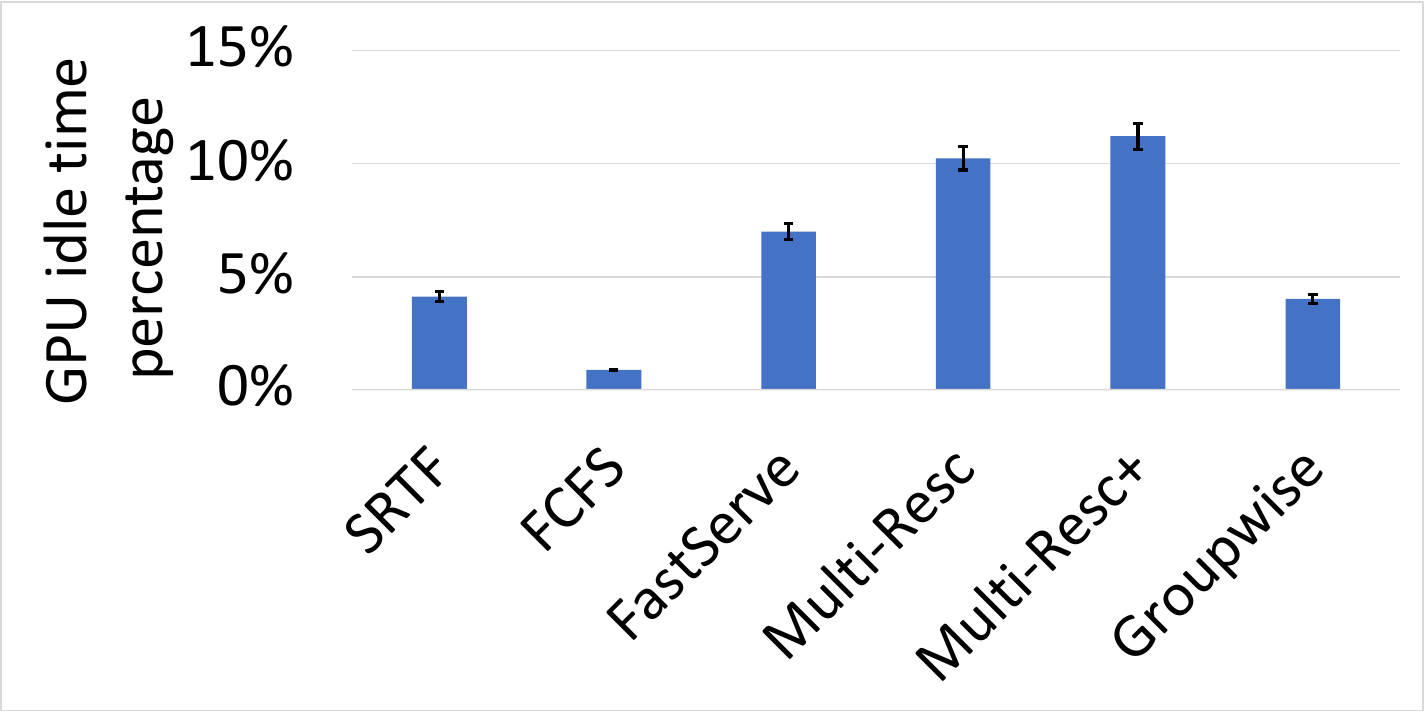}
 \vspace{-0.0in}   \caption{GPU idle time. \vspace{-0.0in}}
    \label{fig:idle-time-schdulers}
\end{minipage} }
\end{figure}

\begin{figure}[t]
\begin{minipage}[t]{0.48\linewidth}
\includegraphics[width=\linewidth,height=0.112\textheight]{Fig/unused-KVC-schedulers.pdf}%
 \vspace{-0.0in}   \caption{KVC wastage. \vspace{-0.0in}}
    \label{fig:KVC-schdulers}
\end{minipage} 
\DEL{\begin{minipage}[t]{0.48\linewidth}
\includegraphics[width=\linewidth,height=0.112\textheight]{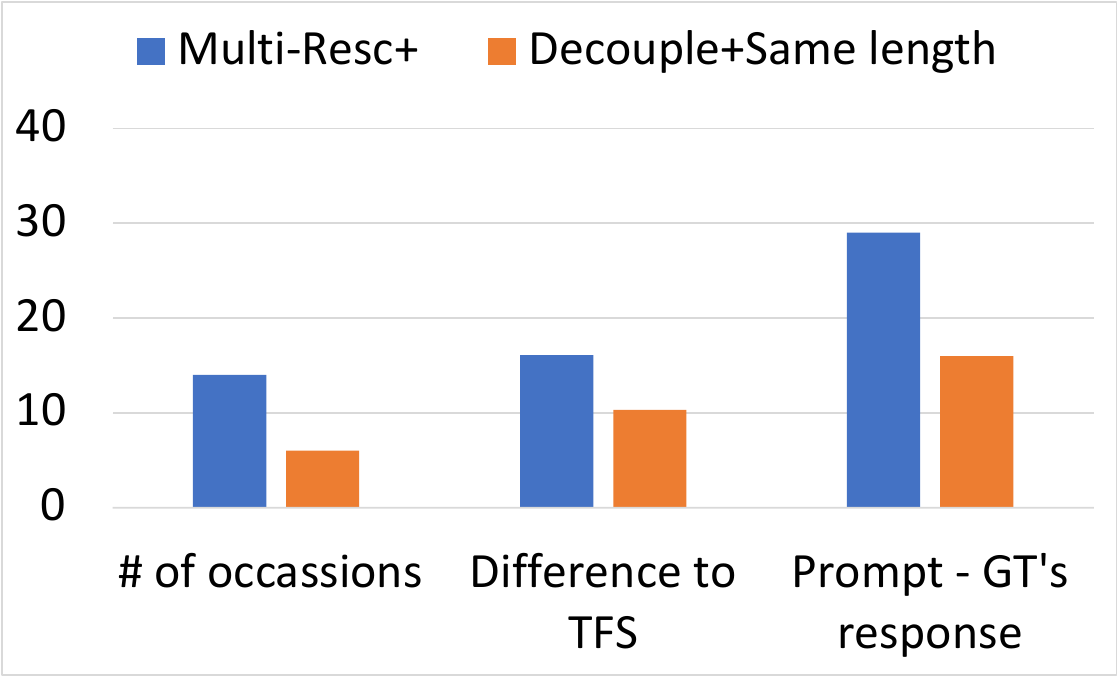}
    \caption{Diffrence of MultiResource and decoupling??Figure 6: Y to 30 , Same length - GT-30 \sh{in the fig, the name should be "Decouple+\emph{SyncCoupled}". Coupled should be \emph{SyncCoupled}+prompt--it is the method after Observation 2. Add "GPU utilization" and "Unused KVC percentage". 3. Change Y in previous  fig from ""Wasted KVC wasted/Allocated KVC" "to "Unused KVC percentage". 4. Add scheduling time. }
}
    \label{fig:tfs}
\end{minipage} }
\end{figure}}


A request's \emph{waiting time} is the period its prompt waits in the queue before it starts to execute, its \emph{preemption time} is the period it pauses running, and its \emph{execution time} is the period from when it is dispatched to the execution engine to when it is completed, excluding the scheduling time and preemption time, and GT queuing time in \sys. Figure~\ref{fig:schedulers-measurement} shows the 
performance in different metrics of different schedulers (we will introduce the right three schedulers later). The JCT is decomposed into different stages. 
Due to space limits, we only present the last two figures for Alpaca. Results for the other two datasets show similar patterns and are presented in the supplemental material.

\DEL{{MultiRes} and {MultiRes-} is {\color{red}162.8 and 153.8 tokens/s {\sh{MultiRes cannot reach TFS every iteration, only when requests complete, it can add requests, this value seems high}}, respectively, and that of  \Orca, SRTF, and FastServe are only 75.3, 80.6, and 84.3 tokens/s, respectively. } \sh{shouldn't FastGen generate higher CPU utilizaiton than vLLM, can you check FastGan blog and see if the results are the same?}
}

SRTF, \Orca, and FastServe underperform vLLM in terms of throughput, KVC utilization, forward size, and JCT.
Their worse performance stems from the max-allocation, which results in KVC bottleneck, and limits the batch size and GPU utilization. 
However, as shown in Figure~\ref{fig:alloc-schedulers}, using block-allocation, vLLM and Sarathi-Serve generate 74\% and 67\% KVC allocation failure percentages. The resultant preemptions or recomputations cause additional delays, constituting 20\% and 16.5\% of the JCT in vLLM and Sarathi-Serve, respectively, as shown in Figure~\ref{fig:overhead-schedulers}.

In batching, vLLM fully allocates KVC while Sarathi-Serve aims to fully utilize GPU by reaching TFS, so vLLM generates 20\% higher KVC utilization, 41\% lower throughput, 27\% lower forward size, and finally 29\% higher JCT than Sarathi-Serve. 
We see that even though Sarathi-Serve aims to reach TFS, it cannot achieve it in each iteration due to the limit of KVC.

\DEL{Sarathi-Serve batches prompts and prompt chunks to reach the TFS, but their required KVC may be much less than the available KVC. In addition, freeing KVC when KVC is full reduces KVC utilization. }

Instead of focusing on a single resource, \emph{MultiRes} aims to fully utilize the dual-resources and it uses the exact-allocation to eliminate the KVC allocation failures as shown in Figure~\ref{fig:alloc-schedulers}. Consequently, it
improves the throughput of Sarathi-Serve by 10\% and its forward size by 5\%, and its JCT is 42.63\% and 35\% lower than vLLM and Sarathi-Serve, respectively. 
However, since the allocated KVC is not always used, it generates similar KVC utilization as Sarathi-Serve.
Both \emph{MultiRes} and FastServe generate high scheduling time occupying 34\% and 17\% of their JCT, respectively, comparing to the negligible scheduling time of the FCFS method used in other schedulers. 
Comparing the results of the three datasets, we found BookCorpus generates lower throughput, higher KVC utilization, more KVC allocation failures due to its longer prompts.\looseness=-1 

\DEL{SyncCoupled $\approx$ MultiRes, both fully utilize KVC

SyncSarathi-Serve $<$ SyncCoupled, Sarathi-Serve only use 1 block, SyncCoupled use exact-allocation.

SyncSarathi-Serve $<$ SyncDecoupled $<$ SyncCoupled, SyncDecoupled can more fully utilize KVC with decouping but because of exact-allocation, it has higher unused KVC than block-allocation.?
}

\DEL{Similarly, Figure~\ref{fig:KVC-schdulers} shows the unused KVC percentage (including the allocated but unused KVC) per request. The percentage of vLLM, Sarathi-Serve, MultiRes, \Orca, SRTF, and FastServe is {\color{red} 31.2\%, 49.3\%, 40.7\%, 60.5\%, 57.3\% and 56.3\%, respectively on average for three traces. MultiRes improves the GPU utilization of \Orca and FastServe by 20\% and 16\%\sh{fig shows they are similar} \sh{add comparison with vLLM}, respectively, and improves their unused KVC by 41.1\% and 33.61\%, respectively.} {\color{red}Figure~\ref{fig:alloc-schedulers} shows the percentage of KV cache allocation failures in execution. The schedulers using the max-allocation method or exact-allocation method do not need KV cache allocation in execution. vLLM and Sarathi-Serve, which use block-based allocation methods, encounter 66\% and 72\% failures\sh{these failures seem have no influence on JCT?}.} \sh{JCT and component times need discussion} Figure~\ref{fig:overhead-schedulers} shows the scheduling time percentage in execution time per request. This percentage also is the GPU time that is not processing the requests (i.e., GPU idle time) {\sh{Masahiro, can we say this?}}. The result follows: {\color{red} \Orca (1.8\%)$<$SRTF (2.9\%)$<$FastServe (19.6\%)$<$MultiRes- (21.6\%)$<$MultiRes (34.2\%)}{\sh{how many queries in the queue? }}. }


\DEL{Surprisingly, \emph{MultiRes} does not improve the throughput and JCT greatly and fails to maximize GPU utilization in each iteration. The reasons are explained in \cref{sec:intro}. First, as in Sarathi-Serve, PTs only run one iteration but other PTs cannot be added to the batch until there is available both GPU and KVC resources (i.e., the inherent GT domination issue shown in Figure~\ref{fig:DecoupleDemo11}). Since PTs only run one iteration, GPU is not maximally utilized. Second, since \emph{MultiRes} uses each request to help fully utilize both GPU and KVC resources, it is unlikely to find prompts that can exactly fully utilize GPU and KVC, it leads to 76\%, 83\%, and 88\% 
TFS and 59\%, 63\%, and  75\% 
KVC utilization for Alpaca, ShareGPT, and BookCorpus datasets, respectively.}

To validate the GT domination issue explained in \cref{sec:intro}, we measured the distribution of iterations with varying numbers of completed requests, as shown in Figure~\ref{fig:req-schedulers}. We observe that not every iteration completes requests, and the number of completed requests in an iteration is limited. \DEL{SRTF, \Orca, FastServe, vLLM, Sarathi-Serve and \emph{MultiRes} have 27\%-35\% iterations with 0 completed requests for the three datasets, and have 4\%-14\% iterations with (2,4] completed requests. }As a result, zero or few PTs may be added to the batch, leading to underutilization of GPU compute resources.


\DEL{Second, it is difficult to find requests that can exactly fully utilize the dual-resources.

In addition, there is room to further improve the effectiveness of MultiRes. First, like Sarathi-Serve, MultiRescan add PTs only when some requests complete and release their KVC space. Since PTs only run one iteration, GPU is not maximally utilized. Second, since {MultiRes} cannot find prompts that can exactly fully utilize GPU and KVC, it leads to  14\% GPU unused resource and 6\% KVC unused resource on average.}


\DEL{\begin{thm}\label{release}
The percent of iterations that can add more requests is low, so there are few chances to add prompts to increase GPU utilization in previous methods. 
\end{thm}}



\begin{thm}\label{thm-decoupleMotivation}
\DEL{Advanced task schedulers generate higher time overhead and GPU idle time. For a schedule to fully utilize GPU and memory resources, the scheduling time and GPU idle time can constitute 11\% of an iteration time.}
\vspace{-0.05in}

\DEL{Though \emph{MultiRes} uses exact-allocation and attempts to fully utilize the dual-resources in each iteration,  it still cannot fully utilize GPU due to the inherent GT domination issue, 
and difficulty in finding prompts that can fully utilize the dual-resources simultaneously. In addition, it incurs high scheduling time, which can constitute a maximum of 34.2\%\sh{?? {\color{red}(15\% on average for three datasets)}} of JCT. 
}

Although \emph{MultiRes} enhances the utilizations of dual resources, there is critical need for further improvement due to the following challenges: 1) the GT domination issue, 2) the coupled scheduling mechanism, which focuses on fully utilizing both resources per request, 3) KVC underutilization, and 4) high scheduling time overhead. 


\end{thm}



\DEL{{\sh{ Pipelining's improvement should be much higher. It should improve ~50
vLLM's poor performance is due to cache allocation failures, this metric needs to be measured. }}
}


 \DEL{Therefore, the search for a scheduler that is both time-efficient and highly effective is imperative.However, we encounter challenges in this pursuit, which we elaborate on in the following sections.  }

Note that while the simple FCFS method does not lead to long scheduling times, \emph{MultiRes} designed to \begin{wrapfigure}{c}{4.5cm}\vspace{-0.15in}
\centering
{{\includegraphics[width=1\linewidth,height=0.112\textheight]{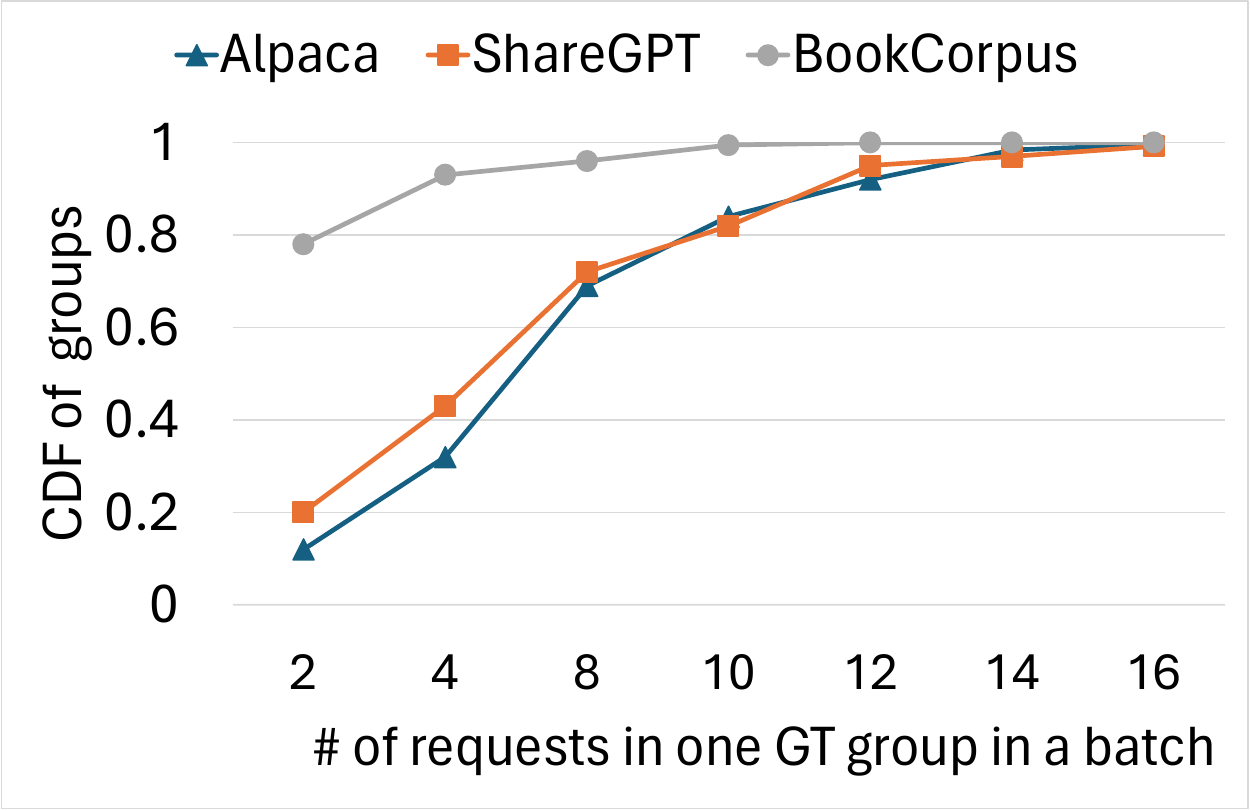} }}%
    \hfill \vspace{-0.25in}
   \caption{{Num. of requests in a same-RL group in experiment.}\vspace{-0.0in}}
\vspace{-0.2in}%
    \label{fig:cdf-gt-group}
\end{wrapfigure} fully utilize dual resources incurs significant scheduling delays. To reduce scheduling time, we aim to avoid heavyweight  iteration-level scheduling by grouping same-RL requests and allocating groups
to the batch until the
KVC is fully utilized. Since their completion times are synchronized, we refer to this method as \emph{SyncCoupled}, whereas \emph{MultiRes} can be viewed as \emph{UnsyncCoupled}.  
\DEL{\emph{SyncCoupled} avoids the need to add more requests due to the completion of some requests in the batch.} A question here is whether there are some same-RL queued requests that can be grouped. Figure~\ref{fig:cdf-gt-group} shows the  CDF of groups versus the number of requests in one GT group in a batch for all three traces in \emph{SyncCoupled}. 
For Alpaca and ShareGPT, 20\% and 23\% of the groups have at least 12 requests and 61\% and 59\% of the requests have at least 4 requests. For BookCorpus, 22\% of the groups have at least 2 requests in a batch.\looseness=-1 

\begin{figure}[t]\vspace{-0.0in}
    \centering
\includegraphics[width=0.65\columnwidth,height=0.05\textheight]{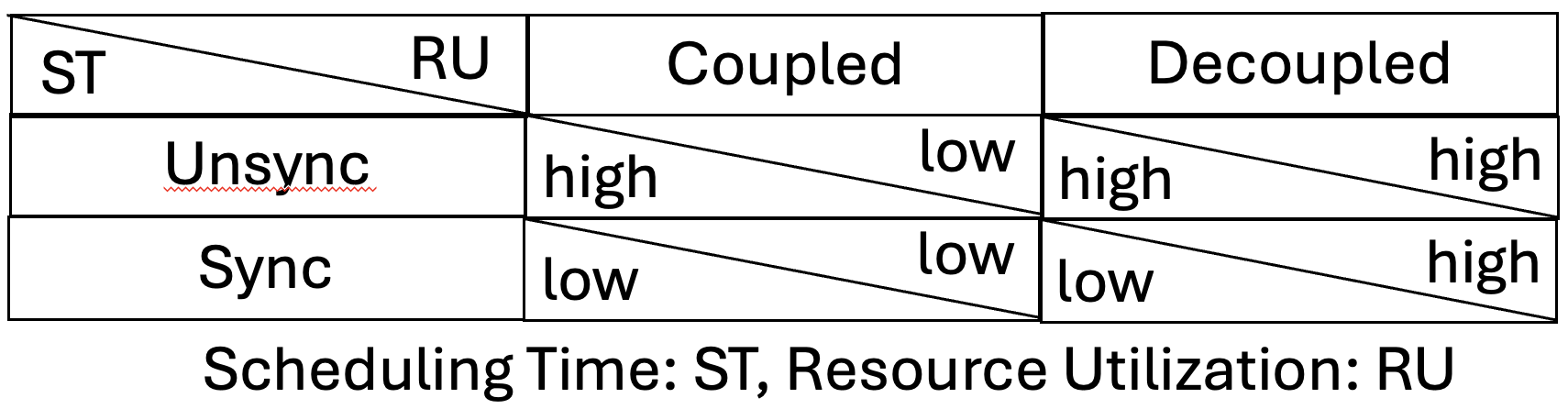}\vspace{-0.1in}
    \caption{Different combinations.}\vspace{-0.3in}
    \label{fig:MethodSummary}
\end{figure}

\vspace{-0.0in}
\begin{thm}\label{thm-decoupling}
Same-RL requests exist in the waiting queue that can be grouped into a batch to reduce the need for iteration-level scheduling.

\end{thm}
\vspace{-0.05in}




\DEL{Figure~\ref{fig:cdf-gt-group-q} shows the CDF of groups versus the number of requests in a batch on queue for all three traces. For the BookCorpus dataset, 12\% of the groups have at least 4 requests in a batch while in the queue. For the Alpaca and ShareGPT datasets, 45\% and 55\% of the batches have at least 4 requests in the queue, respectively. From these two figures, we observe that there are always requests with similar RLs in the batch while running and in the queue. The number of requests with the same length in the queue is smaller because they are already scheduled\sh{not clear}.
}

\DEL{\begin{figure}[t]
\centering
   \DEL{ \subfloat[Unused KVC. \sh{change "Prompt" to "New GTs". this fig is wrong. Y is "\% of ...", denominator is always the KVC full size}\vspace{-0.0in}\label{fig:KVC}]{{\includegraphics[width=0.48\linewidth,height=0.112\textheight]{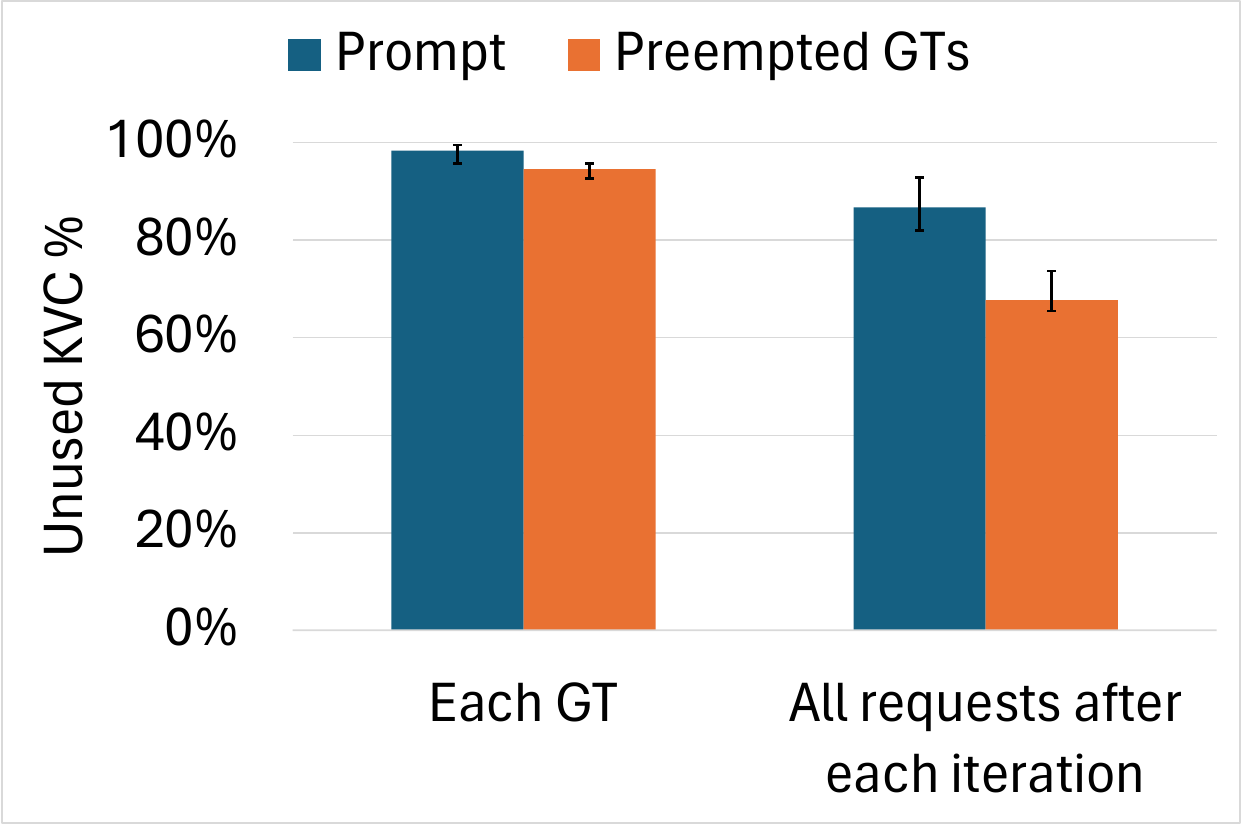}  }}
    \hfill
    \subfloat[CDF of GT. \sh{remove -, change T to t.-done what is the purpose of this fig?}\vspace{-0.0in}\label{fig:cdf-gt}]{{\includegraphics[width=0.48\linewidth,height=0.112\textheight]{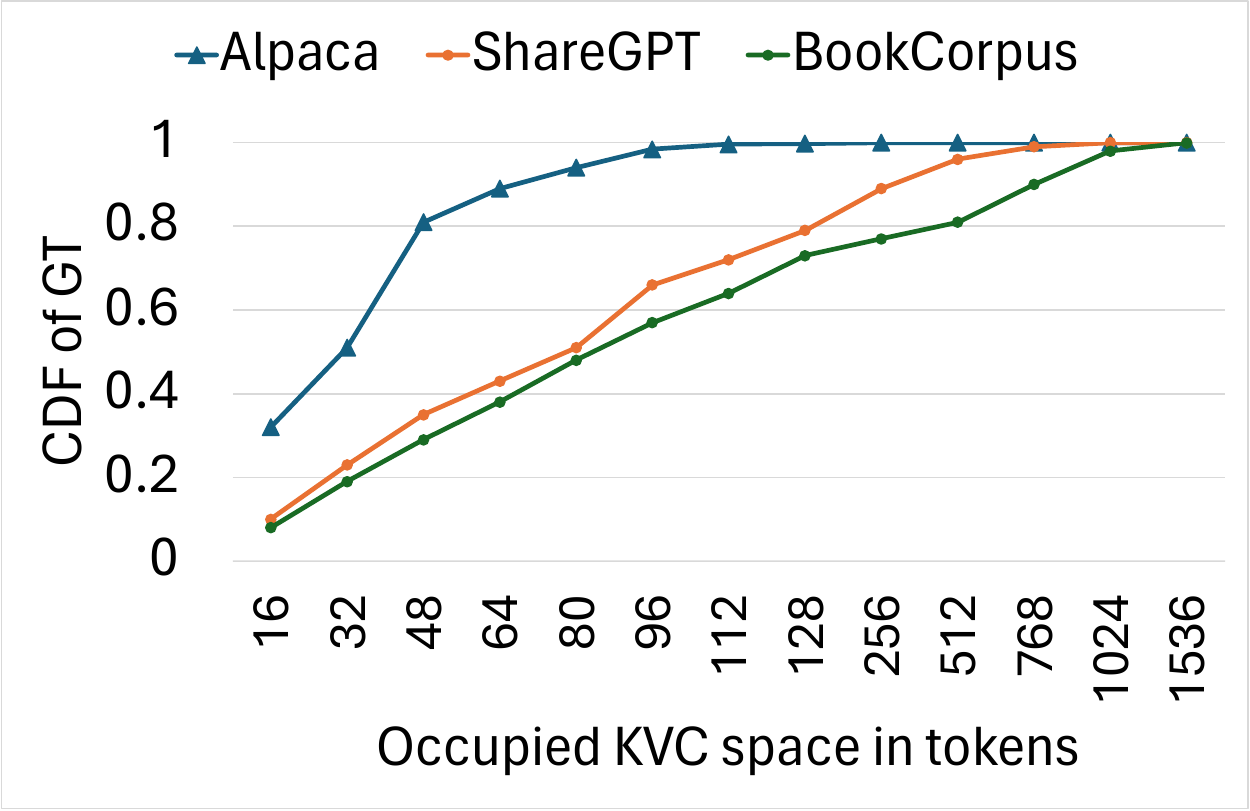} }}%
    \hfill
    \hfill}
    \subfloat[Requests in a batch during whole experiment. \sh{Y values are in \%, Y name is "CDF of groups", same for (b)}. \vspace{-0.0in}\label{fig:cdf-gt-group}]{{\includegraphics[width=0.48\linewidth,height=0.112\textheight]{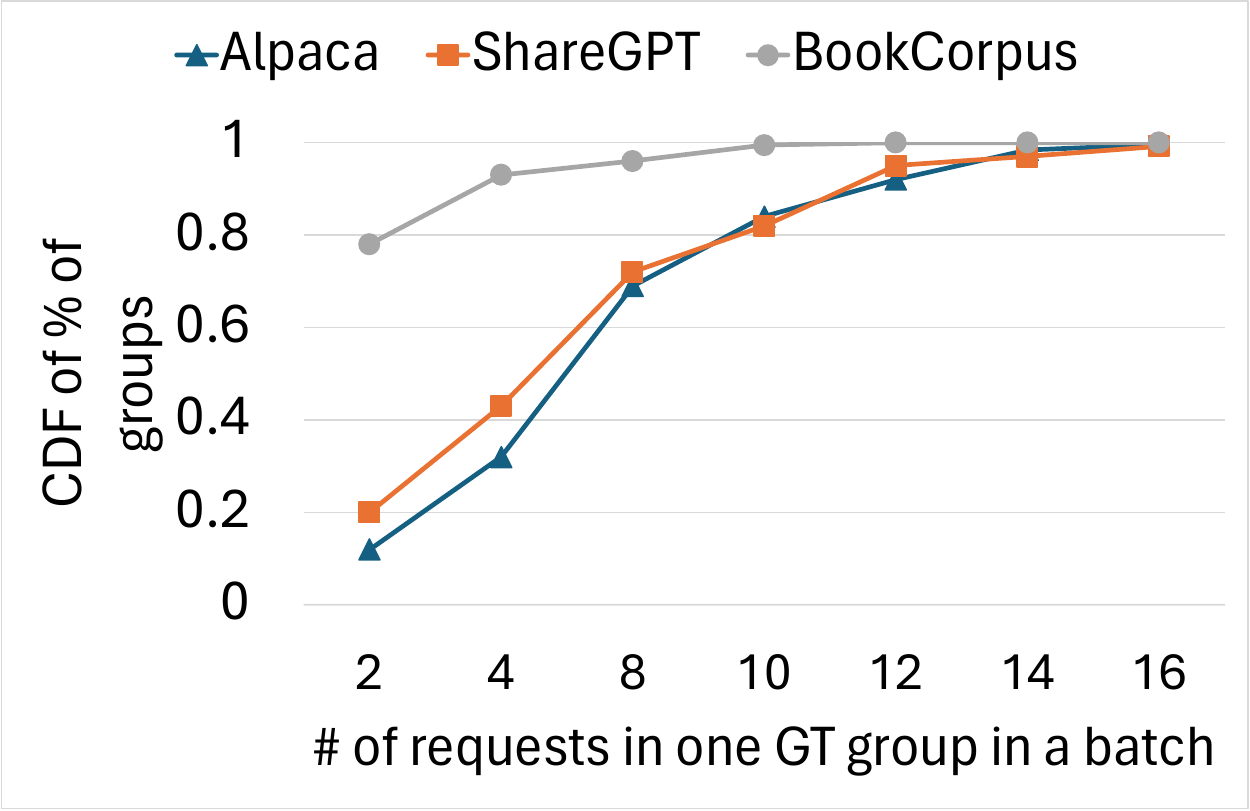} }}%
    \hfill
    \subfloat[Requests in a batch before each iteration. \vspace{-0.0in}\label{fig:cdf-gt-group-q}]{{\includegraphics[width=0.48\linewidth,height=0.112\textheight]{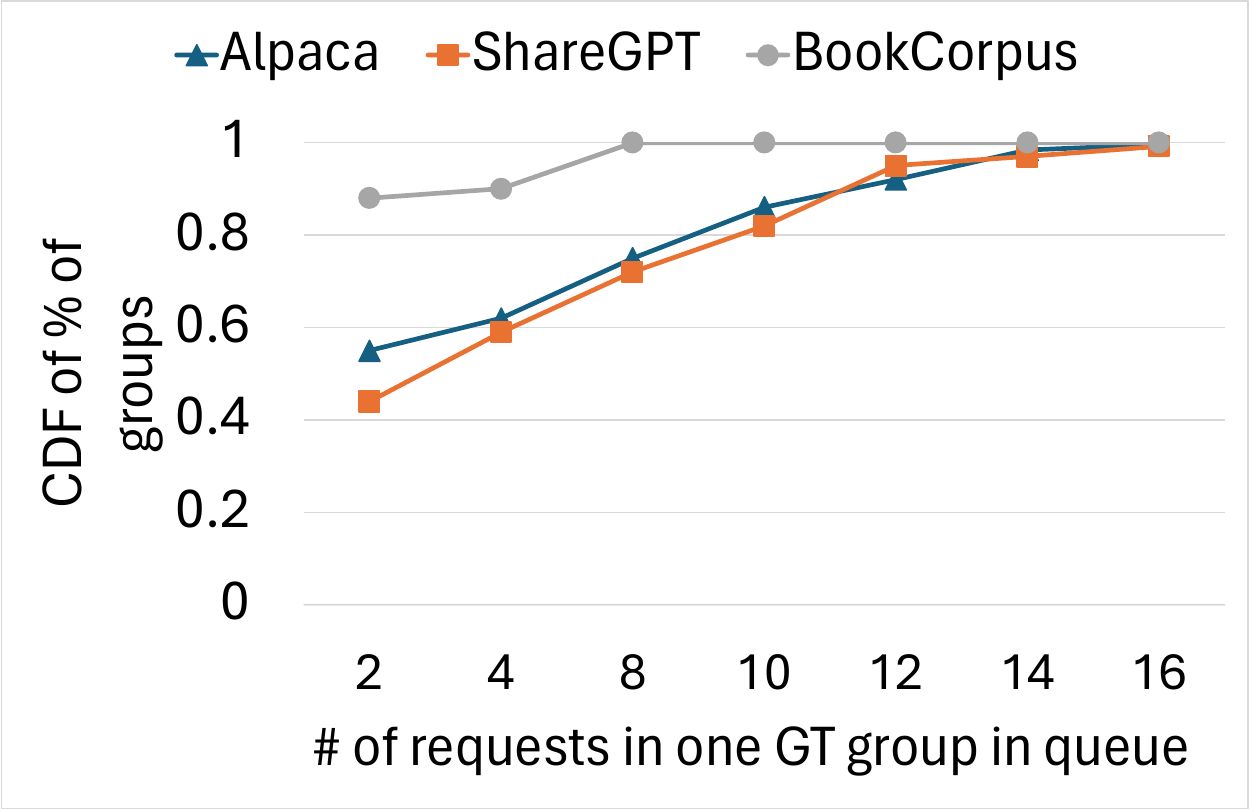} }}%
    \hfill
\vspace{-0.0in}
   \caption{\small{Requests in a group.}\vspace{-0.0in}}
\vspace{-0.0in}%
    \label{fig:request-measurement-2}
\end{figure}
}


\DEL{\vspace{-0.0in}
\begin{thm}\label{thm-groupGT}
If we group same-RL prompts to execute, it can save 78\% scheduling time per request compared to MultResc (from 10.65\% to 2.68\%) on average, but achieves 17\% lower GPU utilization {\sh{why?}}.
\end{thm}
\vspace{-0.0in}
}



The performance of \emph{SyncCoupled} is also included in Figure~\ref{fig:schedulers-measurement}. 
Compared to \emph{MultiRes}, 
\emph{SyncCoupled} generates scheduling time constituting only 2.14\% of JCT, 
due to fewer opportunities to include computation-intensive prompts in the batch. 

\DEL{However, due to the inherent GT domination issue, KVC is unlikely to have the space to enable adding prompts to increase GPU utilization. Even if we can add prompts, they are unlikely to complete at the same time as the current requests in the batch, which conflicts with the time-sync principle of \emph{SyncCoupled}. 
To mitigate the KVC limitation, we could use the block-allocation as in Sarathi-Serve~\cite{deepspeed-Sarathi-Serve}  (we call this method \emph{SyncSarathi-Serve}). The results of \emph{SyncSarathi-Serve} are included in Figure~\ref{fig:schedulers-measurement}. However, \emph{SyncSarathi-Serve} inherits the shortcoming of high KVC allocation failures from the block-allocation. 
It even increases the KVC allocation failures of vLLM and Sarathi-Serve by 19\% and 26\%, respectively. This is because prompts in a group use up their assigned block and request a new block at the same time, exacerbating the KVC bottleneck. 
Finally, \emph{SyncSarathi-Serve} 
has 24\% higher JCT and 28\% lower throughput than the \emph{SyncCoupled} due to preemptions.  }

\DEL{we see that for \emph{SyncSarathi-Serve}, the percentage of allocation failures is 92.7\%, the GPU utilization is 159 {?why so high?}, the unused KVC is 7.7\%, and the scheduling time is 2.15\%. \emph{SyncSarathi-Serve} chooses the request in FCFS manner and tries to schedule at each iteration.}



\DEL{\begin{figure}[t]\vspace{-0in}
\includegraphics[width=0.7\columnwidth,height=0.09\textheight]{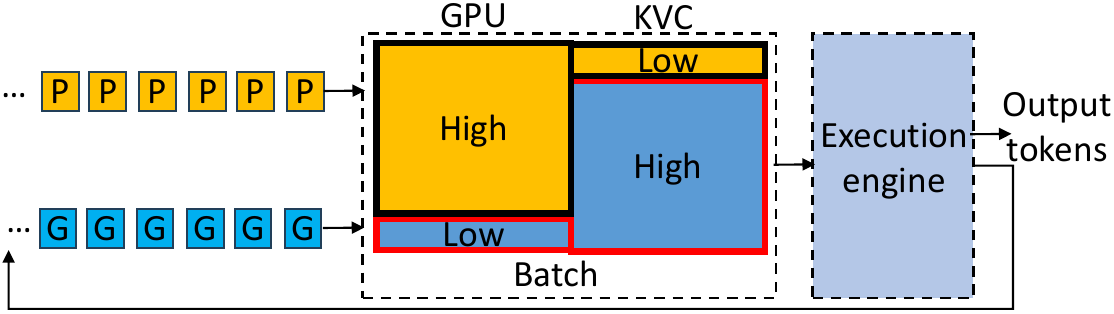}\vspace{-0.0in}
\caption{Decoupling PTs and GTs to fully utilize resources.}
\label{fig:combine}\vspace{-0.0in}
\end{figure}
}



As a result, to handle the problems in \emph{SyncCoupled} and \emph{MultiRes} (or \emph{UnsyncCoupled}) (O\ref{thm-decoupleMotivation}), we propose \emph{SyncDecoupled}, which augments \emph{SyncCoupled} with decoupling PT and GT processing. 
\emph{SyncDecoupled} maintains two separate waiting queues for PTs and GTs. The KVC allocated to a PT equals to its prompt length and that for a GT equals to its predicted RL. PTs are responsible for fully utilizing the GPU while GTs are  responsible for fully utilizing the KVC. After a prompt is processed, its GT is entered into the GT waiting queue, and the GTs form time-synced groups to be scheduled. A small amount of KVC space is reserved for adding PTs at each iteration. The reserved space for Alpaca, ShareGPT and BookCorpus was set to 1.2\%, 3\%, 5\% of the total size. 
\DEL{Consequently, \emph{SyncDecoupled} can address the aforementioned challenges. First, the added PTs require much less KVC space than their sequence lengths 
so PTs can always be added in each iteration as there are always available GPU and KVC resources. Second, computation-intensive PTs and memory-intensive GTs are responsible for fully utilizing each resource, respectively, thus avoiding the problem due to coupled scheduling. Third, GTs are scheduled only after a group completes so heavyweight iteration-level scheduling is avoided. 
Note that though PTs must be added in each iteration to reach TFS, the scheduler simply fetches the PTs in sequence from the queue, generating low scheduling time. 
}



\begin{figure*}[t]
    \centering
  \subfloat[ Response latency.\vspace{-0.0in}\label{fig:response-latency-buffer}]{{\includegraphics[width=0.325\linewidth,height=0.112\textheight]{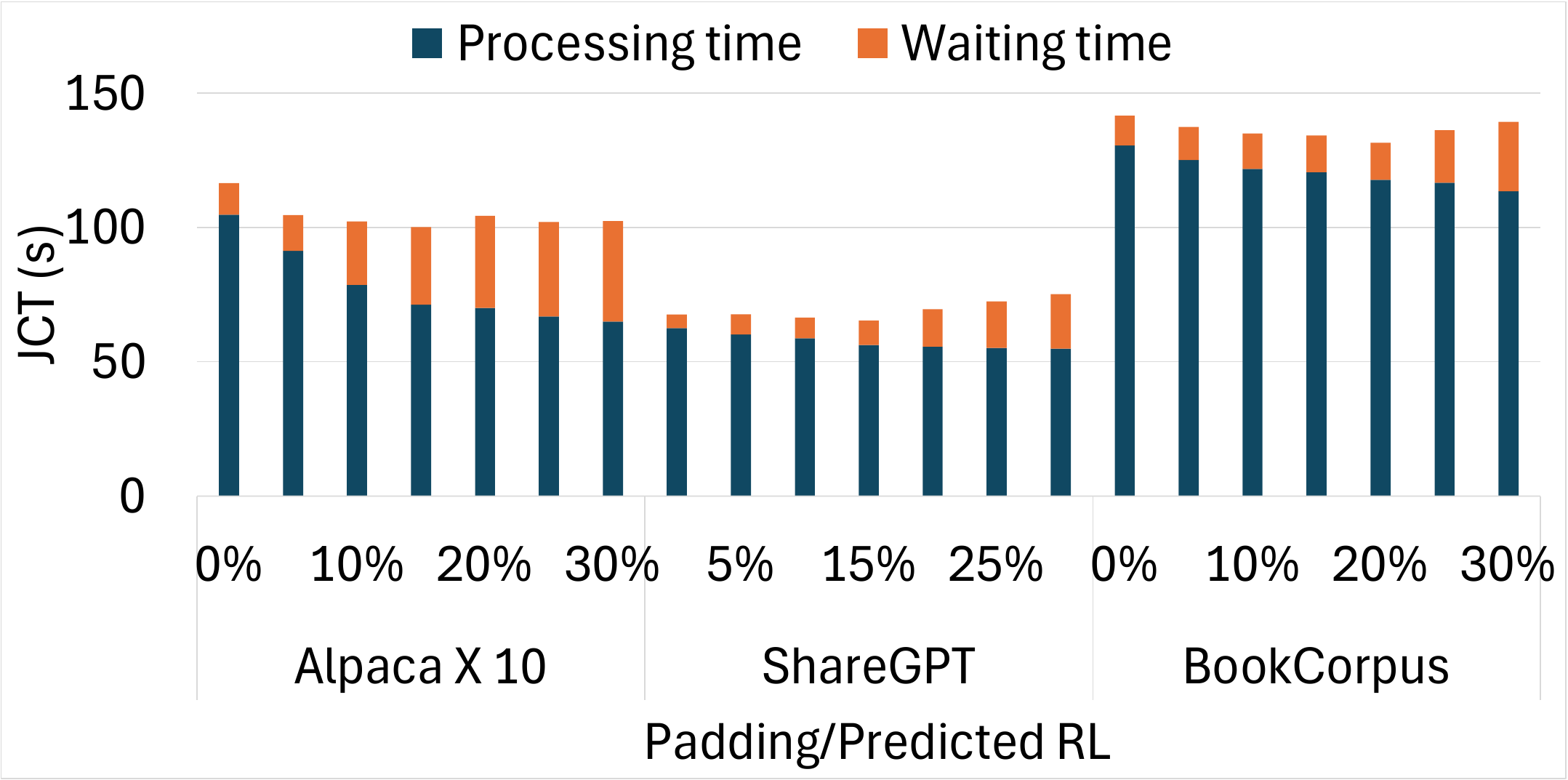}  }}
    \hfill
    \subfloat[KVC utilization.\vspace{-0.0in}\label{fig:KVC-wsate-buffer}]{{\includegraphics[width=0.325\linewidth,height=0.112\textheight]{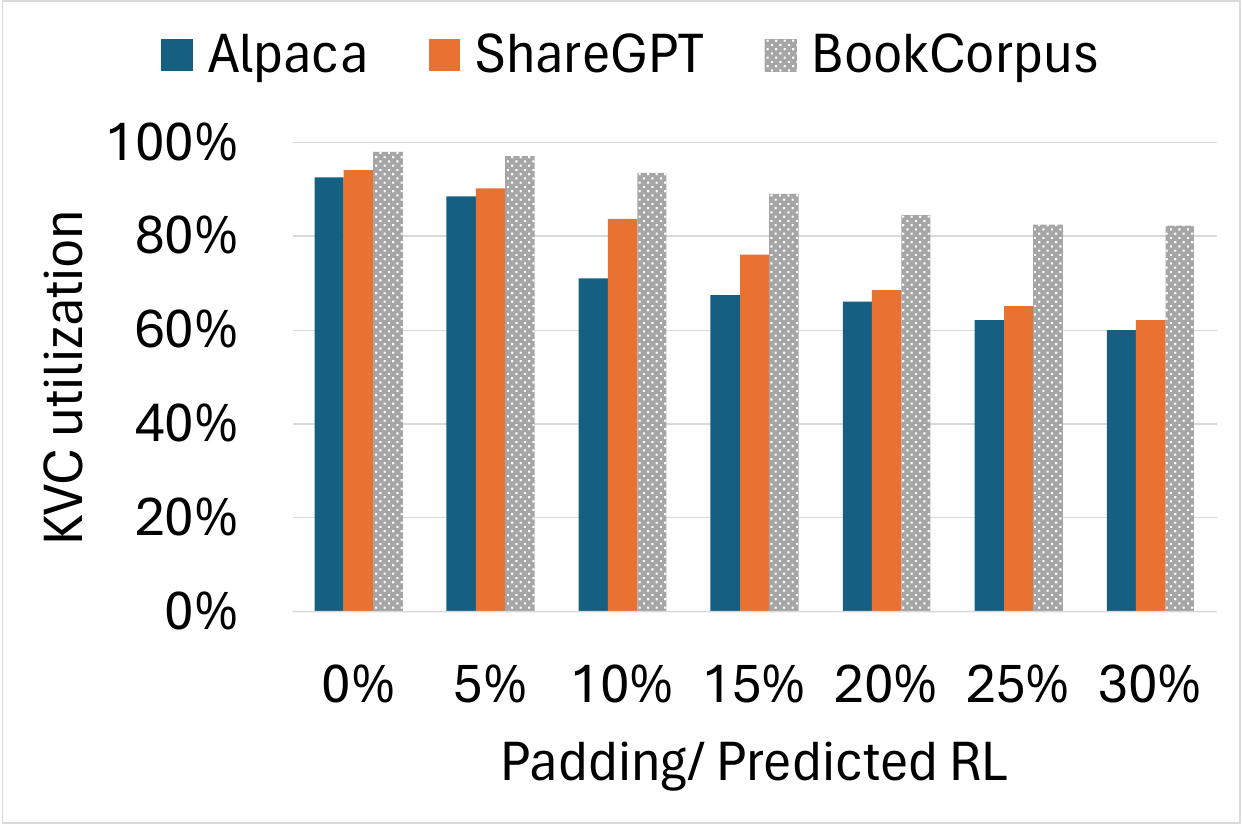} }}
    \hfill
    \subfloat[Under-provisioned requests.\vspace{-0.0in}\label{fig:under-provision-buffer}]{{\includegraphics[width=0.325\linewidth,height=0.112\textheight]{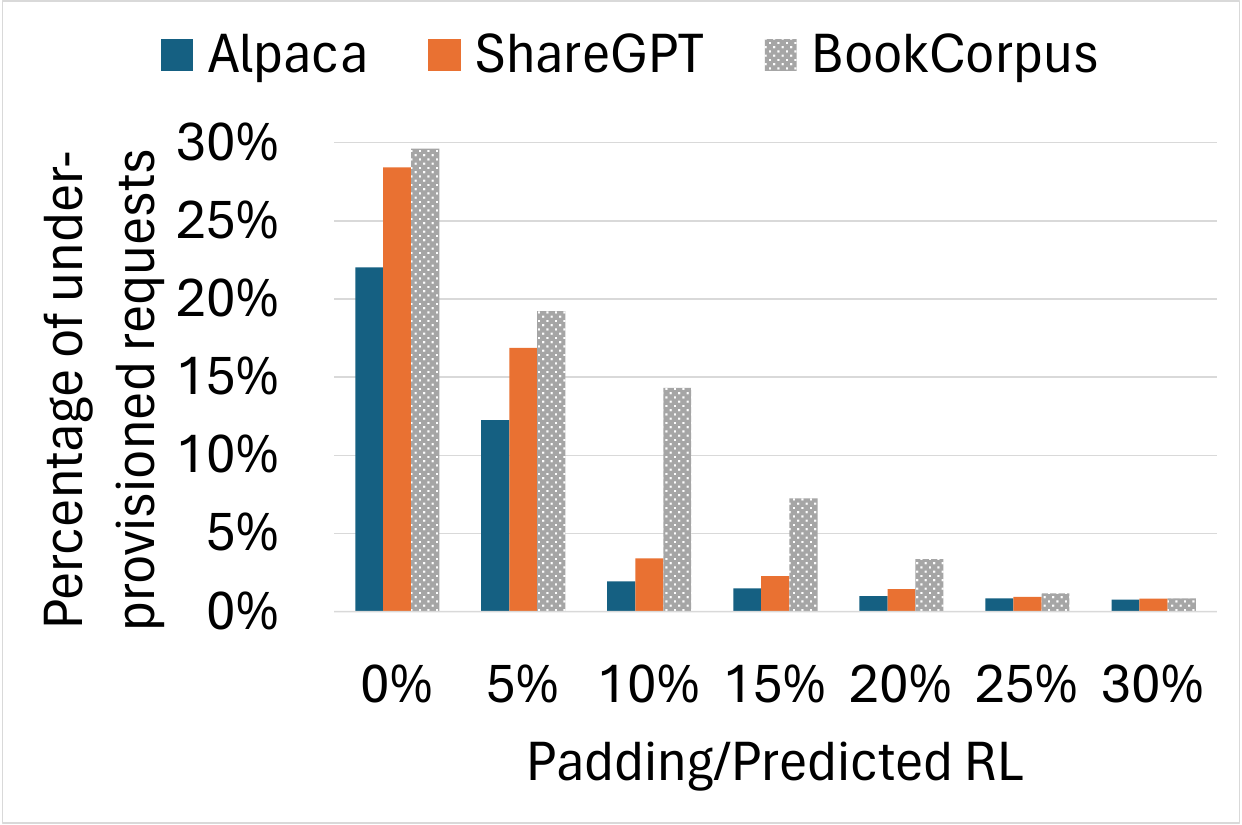} }}
    \hfill
\vspace{-0.1in}\caption{Impact of adding padding to the predicted response length (RL).}\vspace{-0.2in}
\label{fig:padding-assignment}
\end{figure*}

\DEL{\begin{figure}
    \centering
    \subfloat[KVC utilization.\vspace{-0.0in}\label{fig:KVC-wsate-buffer}]{{\includegraphics[width=0.48\linewidth,height=0.112\textheight]{NewFigs/kvc-utilization-padding.pdf} }}
    \hfill
    \subfloat[Under-provisioned requests.\vspace{-0.0in}\label{fig:under-provision-buffer}]{{\includegraphics[width=0.48\linewidth,height=0.112\textheight]{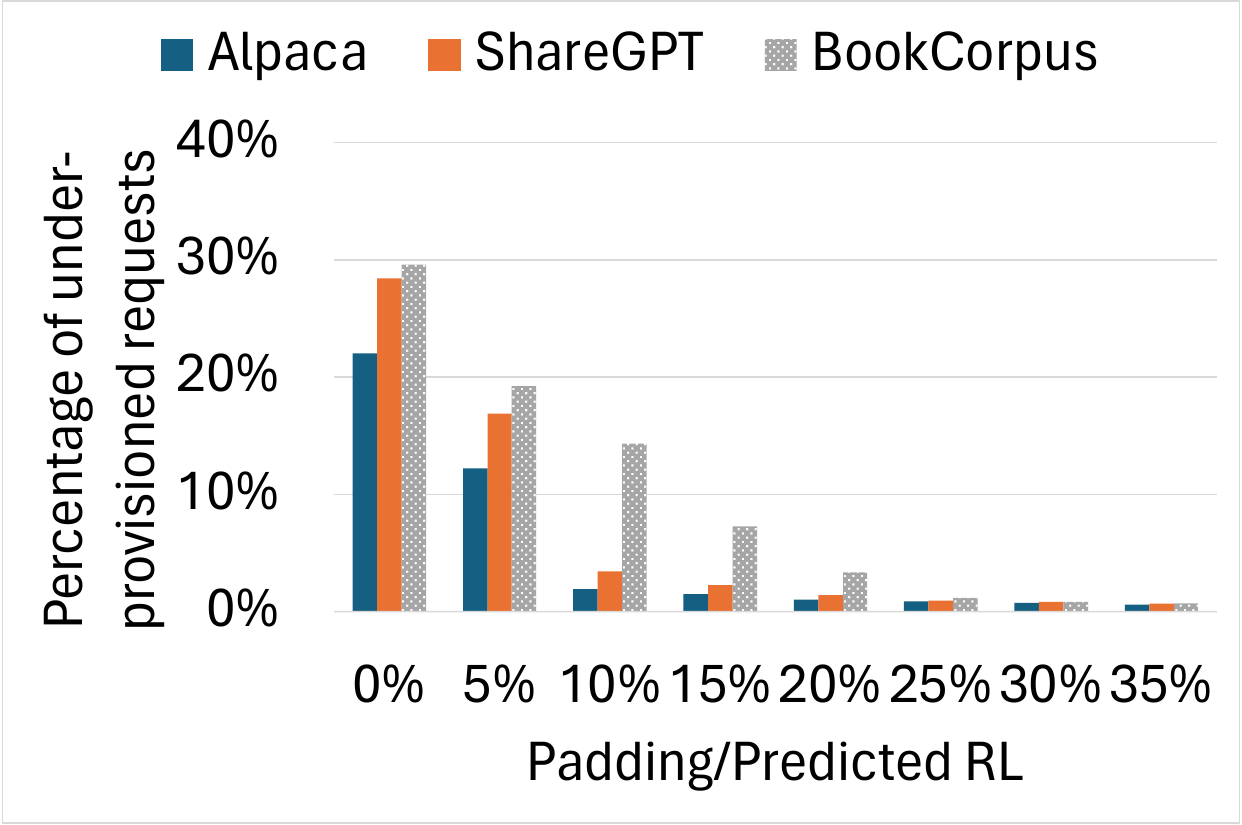} }}
    \hfill
  \vspace{-0.0in}\caption{\small{Impact of adding padding to the predicted response length.}}\vspace{-0.0in}
\label{fig:layer-assignment}
\end{figure}}

\begin{figure}[t]
\centering
    \DEL{\subfloat[{Under- and over-prediction.}\vspace{-0.0in}\label{fig:requests-pie}]{{\includegraphics[width=0.48\linewidth,height=0.112\textheight]{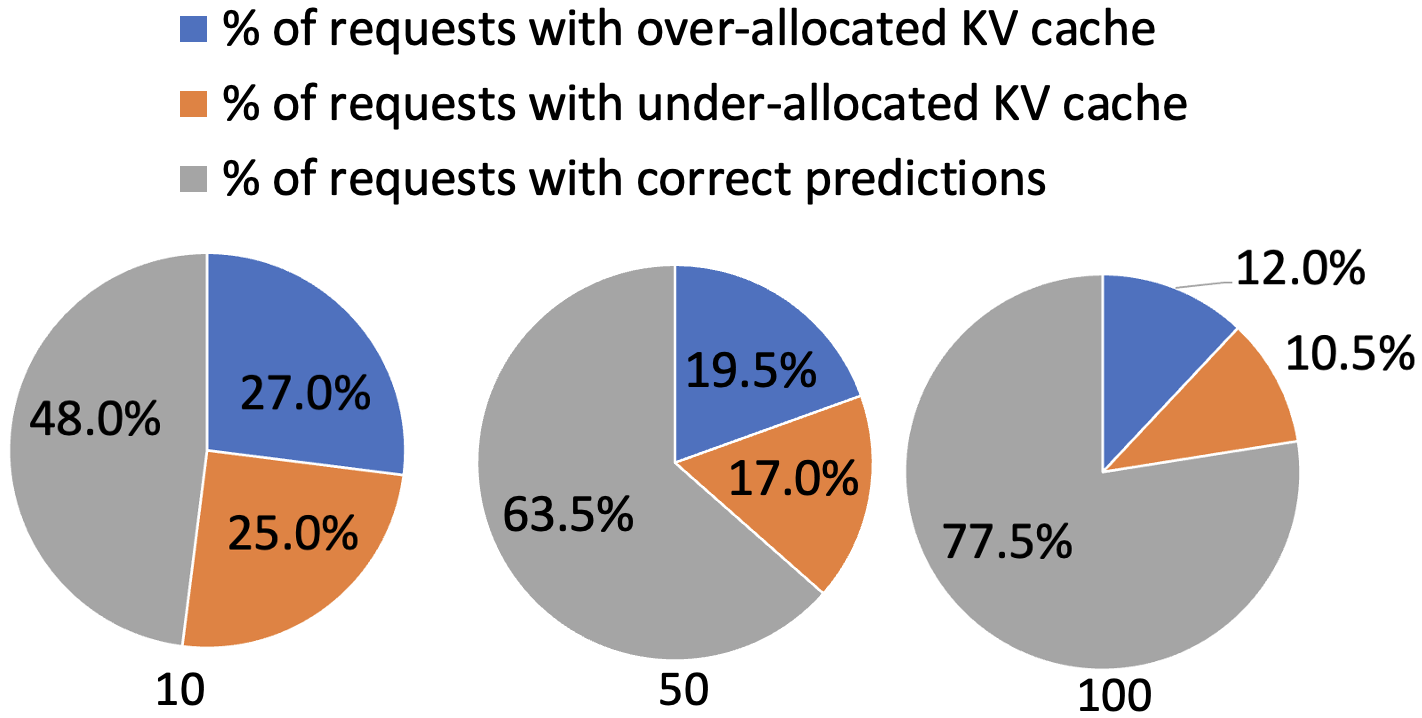} }}
    \hfill
    }
    \subfloat[Over/under KVC allocation.\vspace{-0.0in}\label{fig:KVC-wasted}]{{\includegraphics[width=0.48\linewidth,height=0.112\textheight]{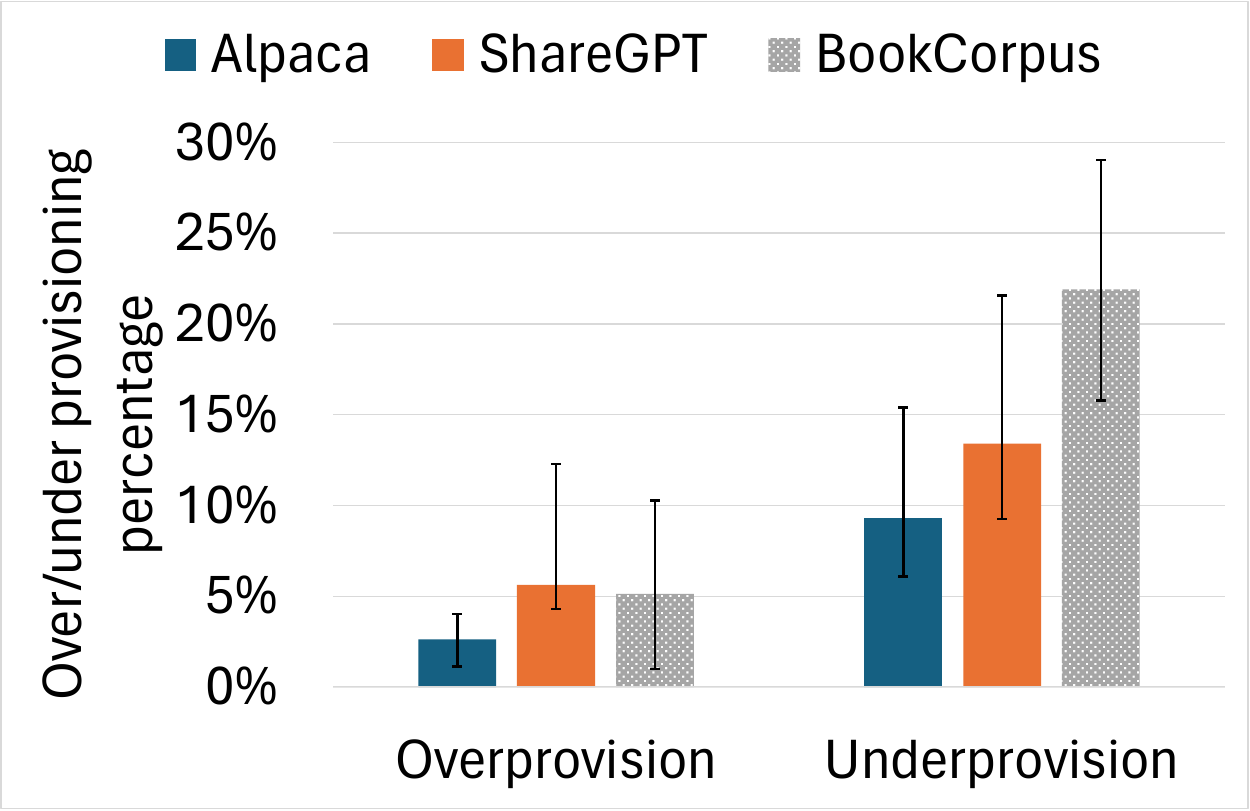} }}
    \hfill
   \DEL{\subfloat[Iterations for under-prediction \sh{add x=1, 2, 3}.\vspace{-0.0in}\label{fig:under-prediction-iter}]{{\includegraphics[width=0.48\linewidth,height=0.112\textheight]{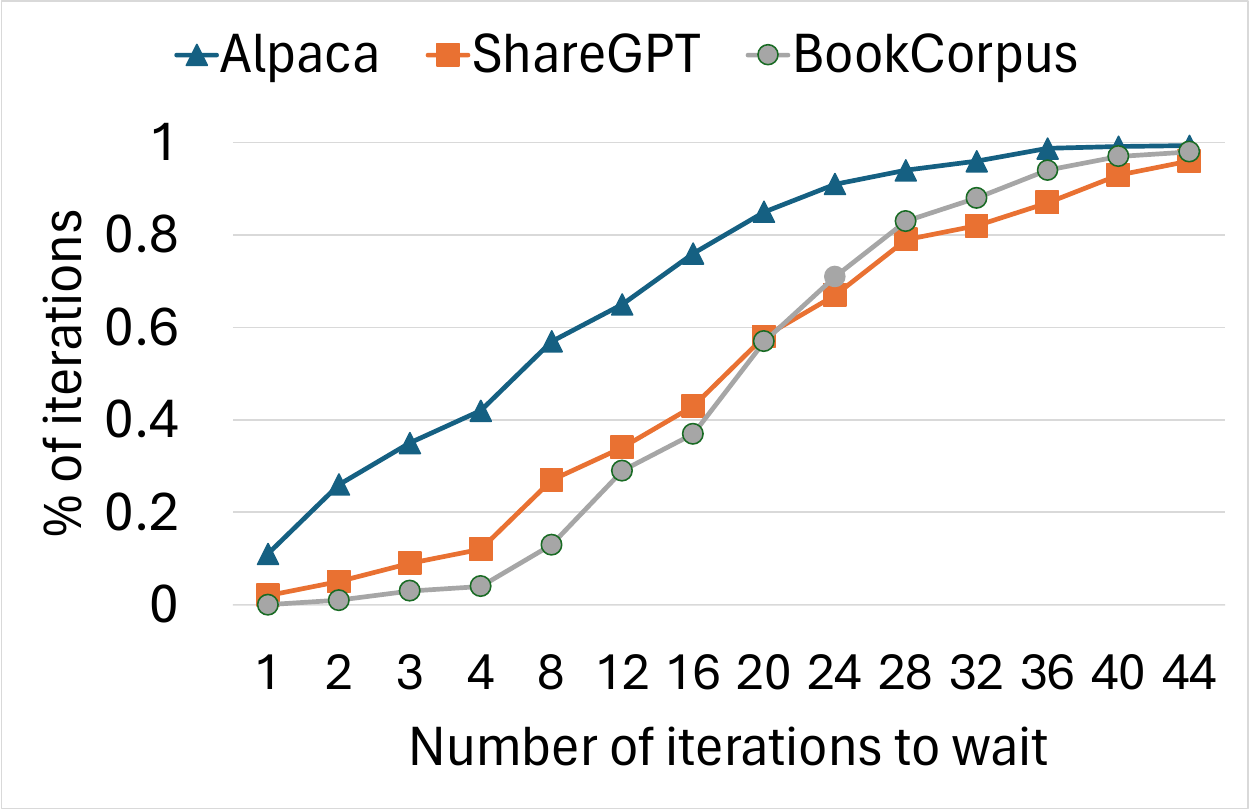} }}
    \hfill
    }
\subfloat[Preemption time. \vspace{-0.0in}\label{fig:memory-percentage}]{{\includegraphics[width=0.48\linewidth,height=0.112\textheight]{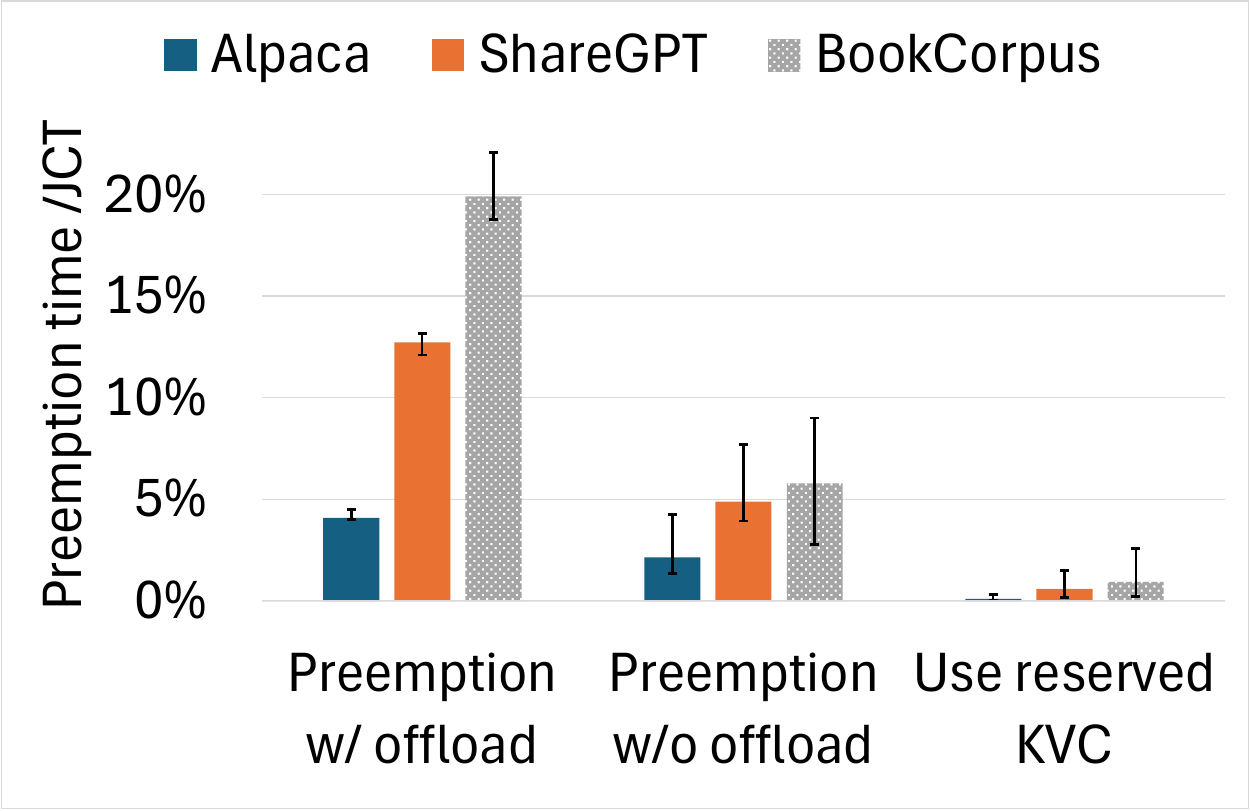} }}
    \DEL{\subfloat[Time for reserved KVC. \vspace{-0.0in}\label{fig:preemption-reserved}]{{\includegraphics[width=0.48\linewidth,height=0.112\textheight]{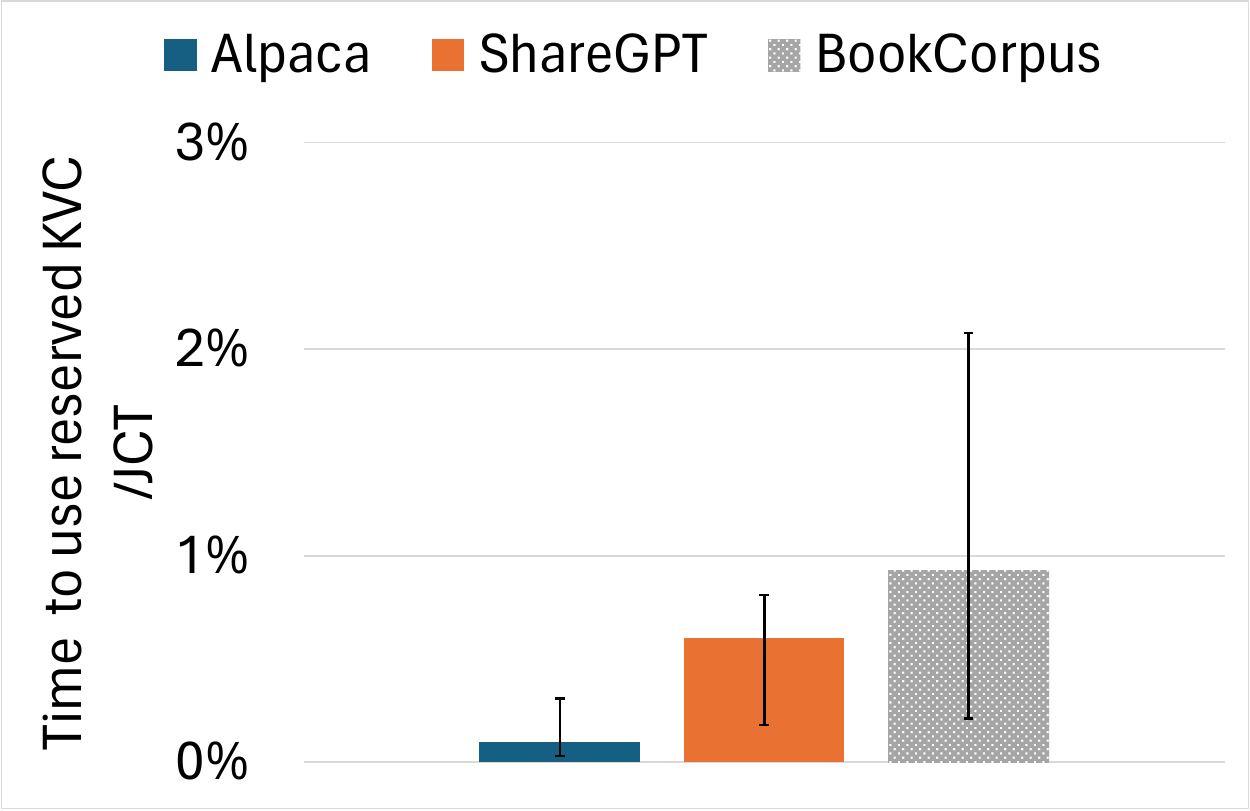} }}
    \hfill}    
    \vspace{-0.15in}
   \caption{{Impact of RL misprediction.\vspace{-0.25in}}}%
    \label{fig:mispredict}
\end{figure}

The performance of \emph{SyncDecoupled} is included in Figure~\ref{fig:schedulers-measurement}. Compared to \emph{SyncCoupled}, it increases the throughput by 1.31$\times$, and reduces the JCT by 35\%, respectively.\DEL{achieves 1.31$\times$ higher throughput. 
Compared to \emph{SyncSarathi-Serve}, it reduces the number of allocation failures by 91\%, 
increase the throughput by 2$\times$.} It is intriguing to see that it reaches TFS since PTs can be added to the batch in each iteration, 
and reduces 
JCT by 35.3\% compared to \emph{SyncCoupled}. Finally, we see that it 
only achieves 63\%-79\%  KVC utilization due to its employment of exact-allocation.\looseness=-1  

\vspace{-0.0in}
\begin{thm}\label{thm2}
\DEL{Compared to \emph{MultiRes}, time-synced batching (\emph{SyncCoupled}) greatly saves scheduling time but reduces GPU utilization. 
Augmenting \emph{SyncCoupled} with decoupling prompt and GT processing (\emph{SyncDecoupled}) maximizes GPU utilization in each iteration. However, the exact allocation they employ generates reserved waste, which can amount to as much as 37\% of KVC.}

Combining time-synced batching and decoupling request processing (\emph{SyncDecoupled}) addresses issues 1), 2), and 4) outlined in O\ref{thm-decoupleMotivation}, but it still fails to resolve issue 3) due to the use of exact-allocation.

\end{thm}
\vspace{-0.0in}

\DEL{\begin{figure}[t]
\begin{minipage}[t]{0.48\linewidth}
\includegraphics[width=\linewidth,height=0.112\textheight]{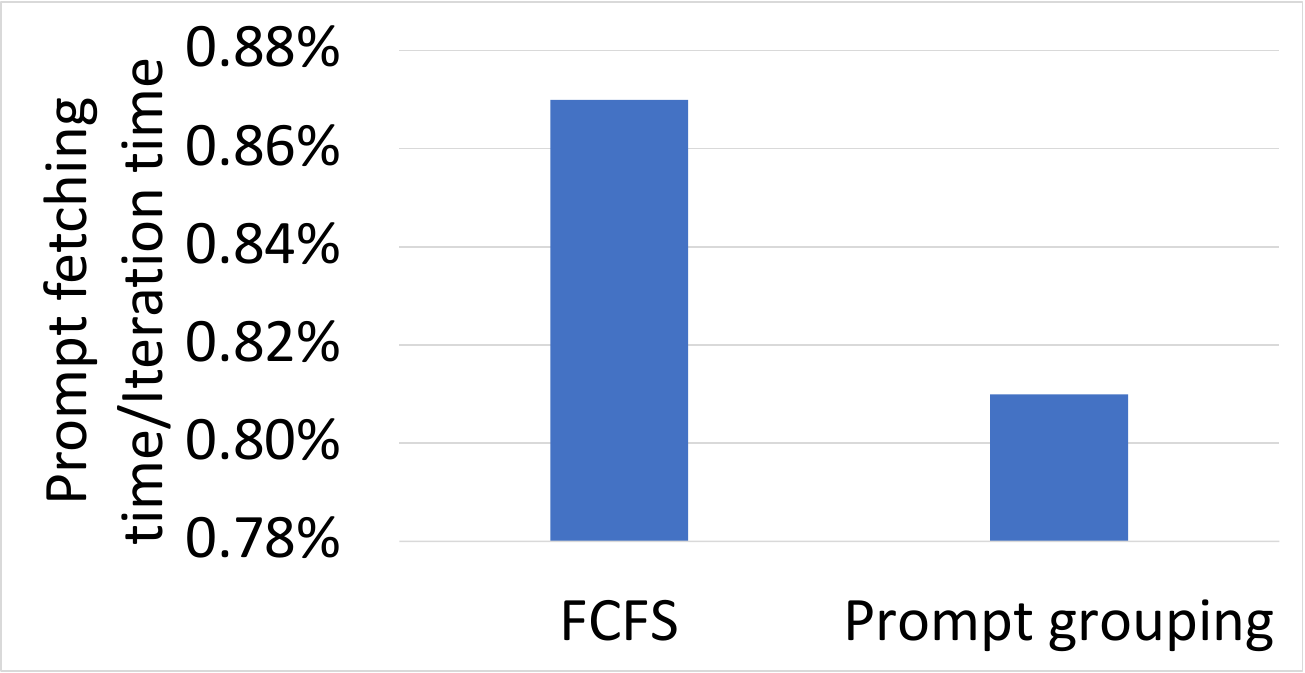} 
\vspace{-0.0in}    \caption{Prompt fetching.\vspace{-0.0in}}
    \label{fig:target-prompt}
\end{minipage} 
\begin{minipage}[t]{0.48\linewidth}
\includegraphics[width=\linewidth,height=0.112\textheight]{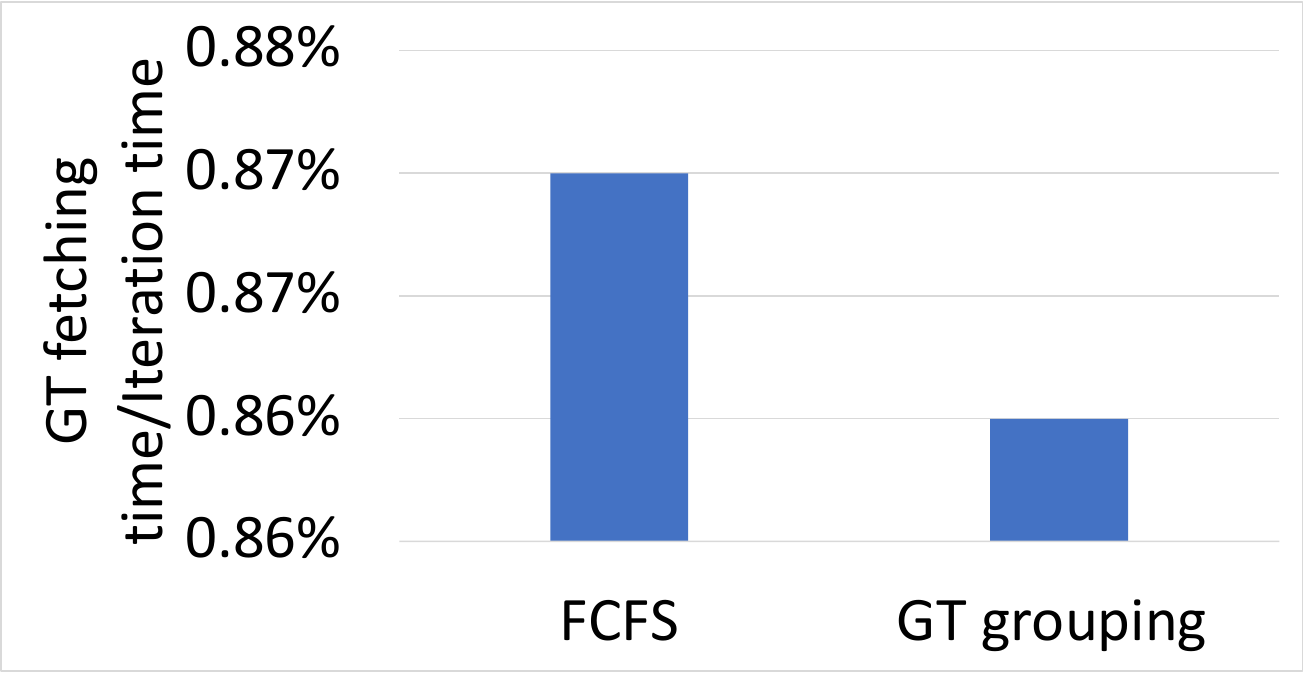}
 \vspace{-0.0in}   \caption{GT fetching. \vspace{-0.0in}}
    \label{fig:target-GT}
\end{minipage} 
\end{figure}
}

Figure~\ref{fig:MethodSummary} summarizes the relative performance on the scheduling time and both resource utilizations of different approach combinations. Time-synced batching helps reduce scheduling time and the decoupled method facilitates more fully utilizing the dual-resources simultaneously (indicated in red words). \emph{SyncDecoupled} achieves the best performance. The results of \emph{UnsyncDecoupled} will be presented in~\cref{sec:evaluation}.




\DEL{\begin{figure}[htb]
    \centering
    \includegraphics[width=0.65\columnwidth,height=0.112\textheight]{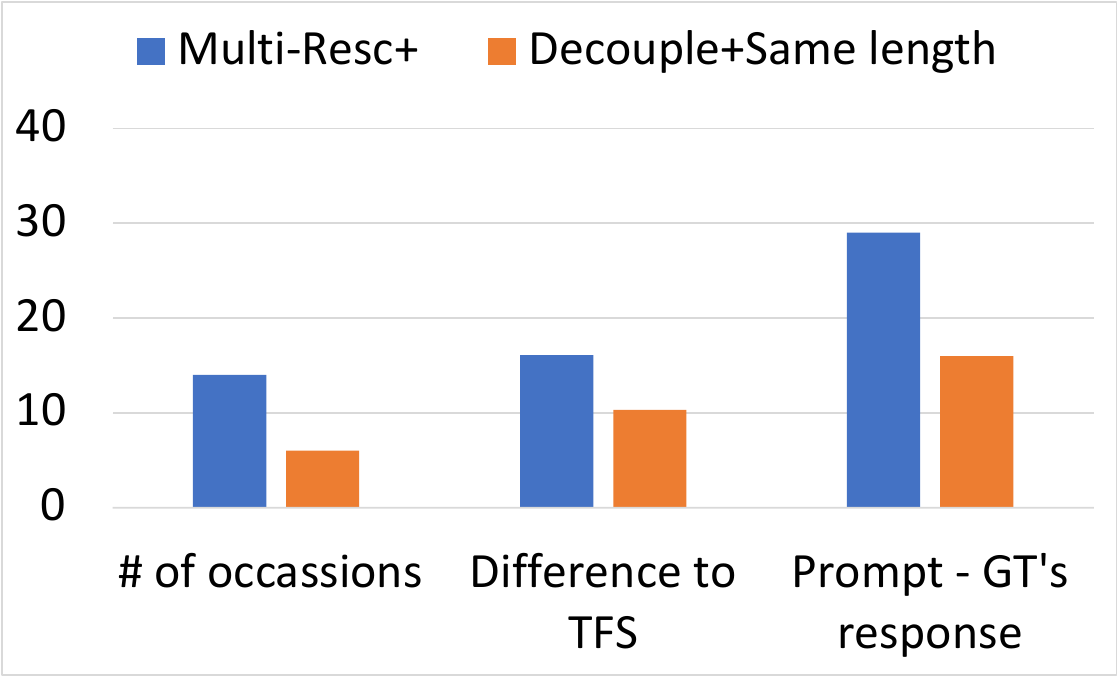}
    \caption{Diffrence of MultiResource and decoupling??Figure 6: Y to 30 , Same length - GT-30 \sh{in the fig, the name should be "Decouple+\emph{SyncCoupled}". Coupled should be \emph{SyncCoupled}+prompt--it is the method after Observation 2. Add "GPU utilization" and "Unused KVC percentage". 3. Change Y in previous  fig from ""Wasted KVC wasted/Allocated KVC" "to "Unused KVC percentage". 4. Add scheduling time. }
}
    \label{fig:tfs}
\end{figure}
}

\DEL{\vspace{-0.0in}
\begin{thm}\label{thm-decoupling2}
If we group same-RL prompts to execute (\emph{SyncCoupled}), it can save 78\% scheduling time per request compared to \emph{MultiRes} (from 10.65\% to 2.68\%) on average, but reduces  GPU utilization by 17\%. Though adding long-prompts to the batch can increase the GPU utilization, it contradicts the endtime-synchronization principle of \emph{SyncCoupled}. Decoupling prompt and GT processing (SyncDecoupled) can facilitate increasing GPU (28\%) and KVC utilization (32\%), as well as reducing scheduling time (79\%). 
\end{thm}
\vspace{-0.0in}
}


\DEL{\begin{figure}[t]
\centering
    \subfloat[Absolute value metrics.\vspace{-0.0in}\label{fig:tfs}]{{\includegraphics[width=0.48\linewidth,height=0.112\textheight]{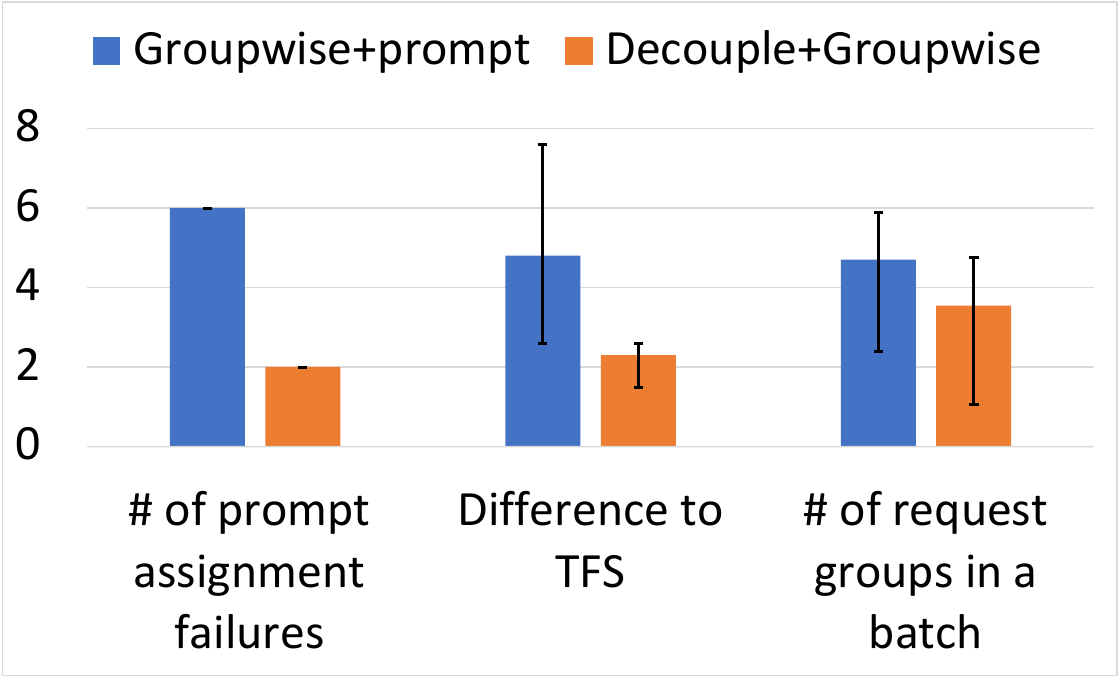} }}
    \hfill
    \subfloat[Percentage value metrics.\vspace{-0.0in}\label{fig:tfs-2}]{{\includegraphics[width=0.48\linewidth,height=0.112\textheight]{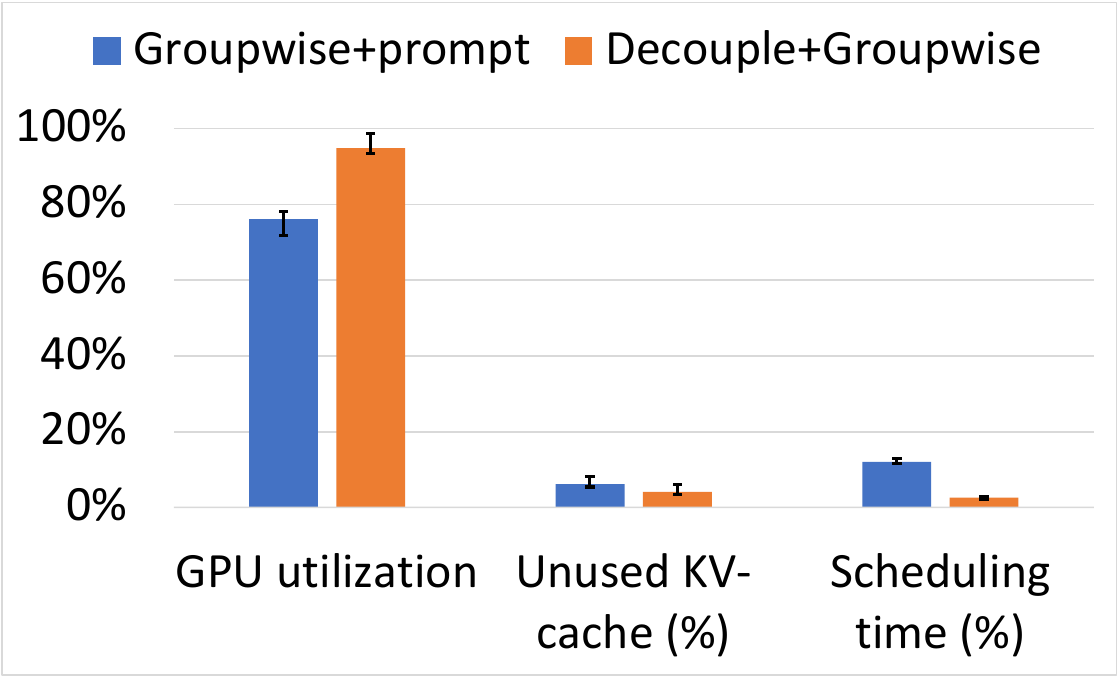} }}
    \hfill
    \vspace{-0.0in}
   \caption{\small{Advantage of decoupling. 
\vspace{-0.0in}}}%
    \label{fig:decoupling}
\end{figure}
}

\DEL{\begin{thm}\label{3factorsSortingPrompt}
When we select prompts to be added to the batch, we need to consider SLO, prompt length, and their occupied KVC space. The priority of the ?? factor should be the highest, followed by ?? factor, and then ?? factor.
\end{thm}

\begin{thm}\label{3factorsSortingGT}
When we select GTs to be added to the batch, we need to consider SLO, prompt length, and their occupied KVC space. The priority of the ?? factor should be the highest, followed by ?? factor, and then ?? factor.
\end{thm}
}

\vspace{-0.1in}
\subsection{Response Length Prediction}
\DEL{Figure~\ref{fig:requests-pie} depicts the percentages of over-predicted requests and under-predicted requests when considering predictions within a 10, 50, and 100 token difference as correct predictions.} 



We fine-turned the OPT-13B model~\cite{Zhang2022OPTOP} to predict the RL~\cite{Zheng2023ResponseLP} based on the prompt. 
We used 10K requests from each trace for fine-tuning; 70\% of the data for training and 30\% for testing. 
All other remaining requests from each trace were used in our experiments. To optimize the training process 
resources, we employed the efficient training method LoRA~\cite{Hu2021LoRALA} and trained the model for three epochs with a learning rate of $0.01$. 
We used a separate LLM running on another server (with four A100 GPUs) for RL prediction to avoid the interference on the request processing. For more details of the RL predictor, please refer to~\cite{Zheng2023ResponseLP}.\looseness=-1





To avoid under-prediction, we could add a certain ratio of the predicted value as padding. 
Figures~\ref{fig:response-latency-buffer}-\ref{fig:under-provision-buffer}
show the JCT decomposed to waiting time and processing time, the KVC utilization, and the percentage of under-provisioned requests for each padding ratio in \emph{SyncDecoupled}. 
The figure scales Alpaca's results by a factor of 10 (marked by ``$\times 10$'') to make them visible. 
As the padding ratio keeps increasing, the processing time decreases but the waiting time increases, leading to decrease and then increase in JCT. This is because adding padding also decreases the KVC utilization and the percentage of under-provisioned requests, as shown in Figures~\ref{fig:KVC-wsate-buffer} and \ref{fig:under-provision-buffer}. 
The sweetspot padding ratios for Alpaca, ShareGPT and BookCorpus are 10\%, 15\% and 20\%, which lead to 77.5\%, 73.2\%, and 69.8\% prediction accuracy, respectively, in our experiments.\looseness=-1



\DEL{As long as the deviation of the predicted RL is within the mentioned percentage of the ground-truth RL for the specific datasets, 
we consider the prediction as accurate. This approach leads to 77.5\%, 73.2\%, and 69.8\% accuracy for the Alpaca, ShareGPT, and BookCorpus datasets for the padding of 10\%, 15\%, and 20\%, respectively. 
In the following, we use these padding values.}






\DEL{\begin{figure}[t]
\centering
    \subfloat[Iterations for under-prediction .\vspace{-0.0in}\label{fig:under-prediction-iter}]{{\includegraphics[width=0.48\linewidth,height=0.112\textheight]{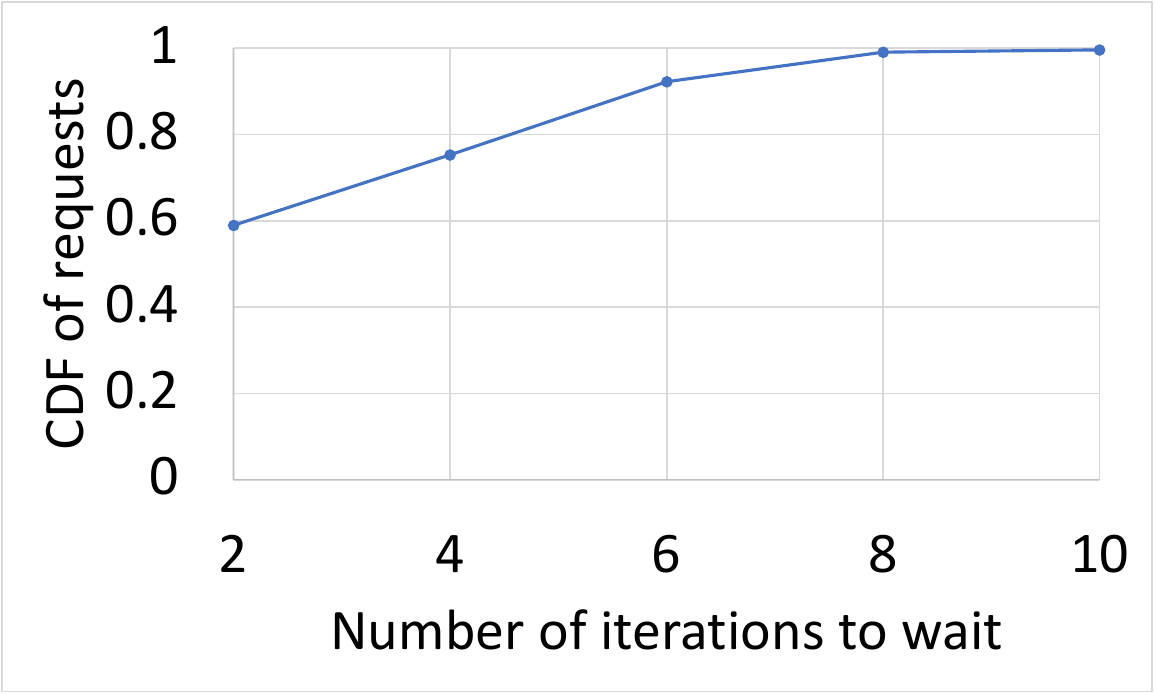} }}
    \hfill
    \subfloat[CDF of overpredicted requests.\vspace{-0.0in}\label{fig:cdf-over-requests}]{{\includegraphics[width=0.48\linewidth,height=0.112\textheight]{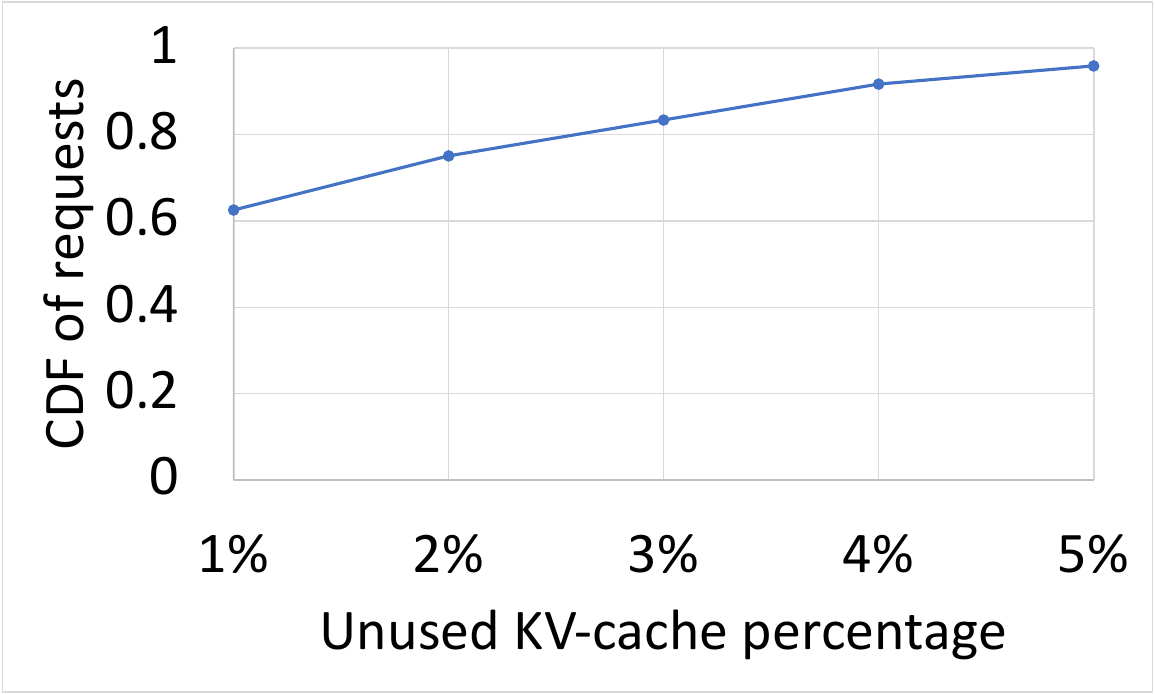} }}
    \hfill
    \vspace{-0.0in}
   \caption{{Measurement of mispredicted requests.
\vspace{-0.0in}}}%
    \label{fig:mispredict}
\end{figure}
}

\DEL{--our purpose is to show no need to wait for many iterations, but this fig is opposite. 
Fig3f shows almost each iteration has completed requests. Do you see the conflict?
}

\DEL{Figure~\ref{fig:under-prediction-iter} shows the Cumulative Distribution Function (CDF) of under-predicted requests versus the number of iterations a request needs to wait in the batch before receiving the required KVC.  When we do not move the KVC data to CPU, we observe that 39\% of the under-predicted requests that need 4 iterations to receive the required KVC space for the Alpaca dataset. For the ShareGPT and the BookCorpus dataset, 15\% of the requests need to wait for 8 iterations and 14\% of the requests need to wait for 24 iterations, respectively. {\sh{these values are too high. vLLM won't pick up new request from the queue until all preempted requests are done, did you do it? if 24 iterations, how long will be delayed? is it reasonable?}} 
\DEL{Figure~\ref{fig:cdf-over-requests} shows the CDF of the over-predicted requests versus the unused KVC percentage in KVC. We observe that 99\% and 59\% of the over-predicted requests produce within 5\% and 1\% of the unused KVC percentage, respectively.}
\sh{need to double check the results and add discussion}
}

\DEL{\vspace{-0.0in}
\begin{thm}\label{misprediction}
Using LLM for response length prediction can lead to 77.5\% accuracy. Because of the response length misprediction, 2.35\% KVC can be wasted per request and 10.5\% of requests in the workload encounter KVC {\sh{encounter KVC allocation failure?}}.
Therefore, it is important to handle misprediction.
\end{thm}
\vspace{-0.0in}
}

\DEL{\begin{figure}[t]
\begin{minipage}[t]{0.48\linewidth}
\includegraphics[width=\linewidth,height=0.112\textheight]{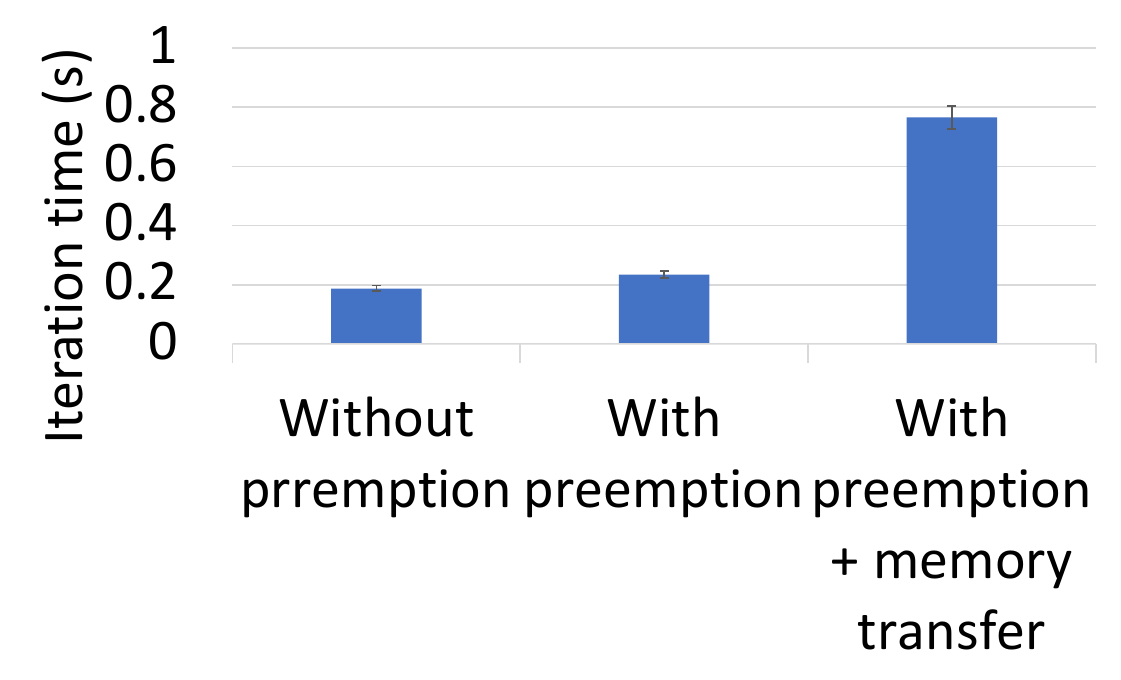} 
\vspace{-0.0in}    \caption{Memory transfer time{?remove the 3rd bar}.\vspace{-0.0in}}
    \label{fig:memory-time}
\end{minipage} 
\begin{minipage}[t]{0.48\linewidth}
\includegraphics[width=\linewidth,height=0.112\textheight]{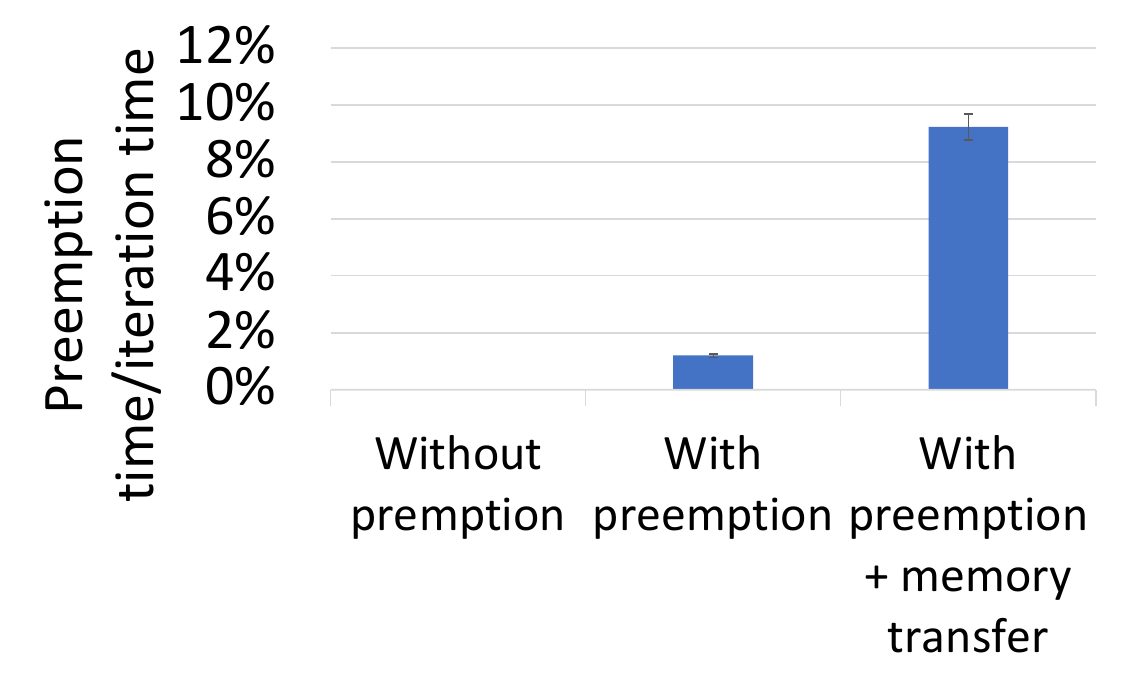}
 \vspace{-0.0in}   \caption{Preemption time. \sh{combine to 1 fig using 2 subfigs}\vspace{-0.0in}}
    \label{fig:memory-percentage}
\end{minipage} 
\end{figure}}

\DEL{\begin{figure}[t]
\centering
\DEL{   \subfloat[Iteration time (??remove this fig).\vspace{-0.0in}\label{fig:memory-time}]{{\includegraphics[width=0.48\linewidth,height=0.112\textheight]{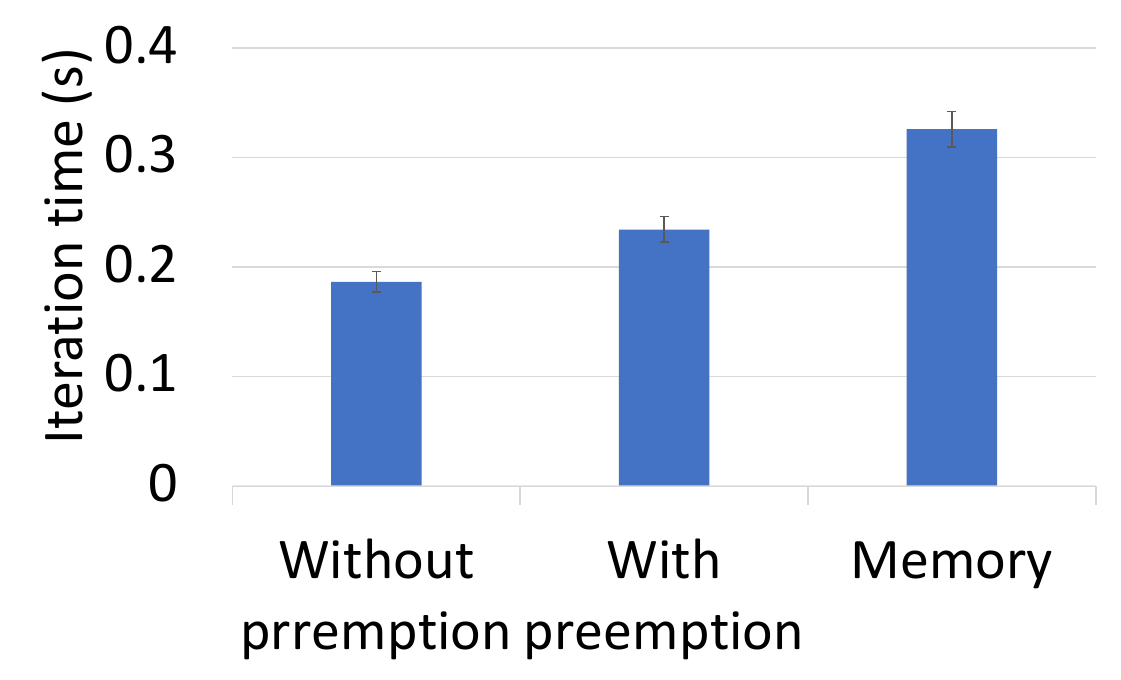}  }}
    \hfill}
    \subfloat[Preemption time. \vspace{-0.0in}\label{fig:memory-percentage}]{{\includegraphics[width=0.48\linewidth,height=0.112\textheight]{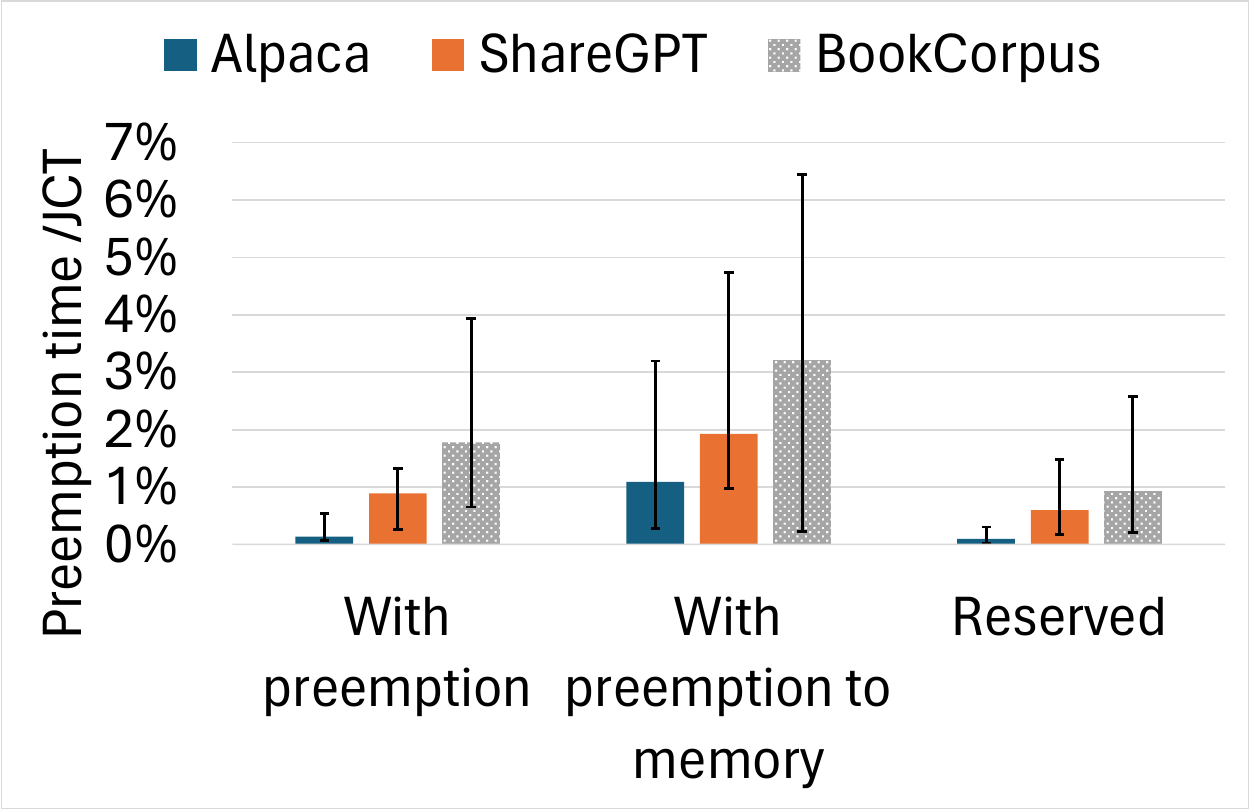} }}
    \DEL{\subfloat[Time for reserved KVC. \vspace{-0.0in}\label{fig:preemption-reserved}]{{\includegraphics[width=0.48\linewidth,height=0.112\textheight]{NewFigs/preemption-time-reserved.pdf} }}
    \hfill}
    \vspace{-0.0in}
   \caption{\small{Preemption and memory transfer time.\sh{add another data result for using reserved KVC for preempted requests}
\vspace{-0.0in}}}%
    \label{fig:memory-measurement}
\end{figure}
}



 \DEL{\begin{figure}
  \centering
\includegraphics[width=0.85\columnwidth,height=0.112\textheight]{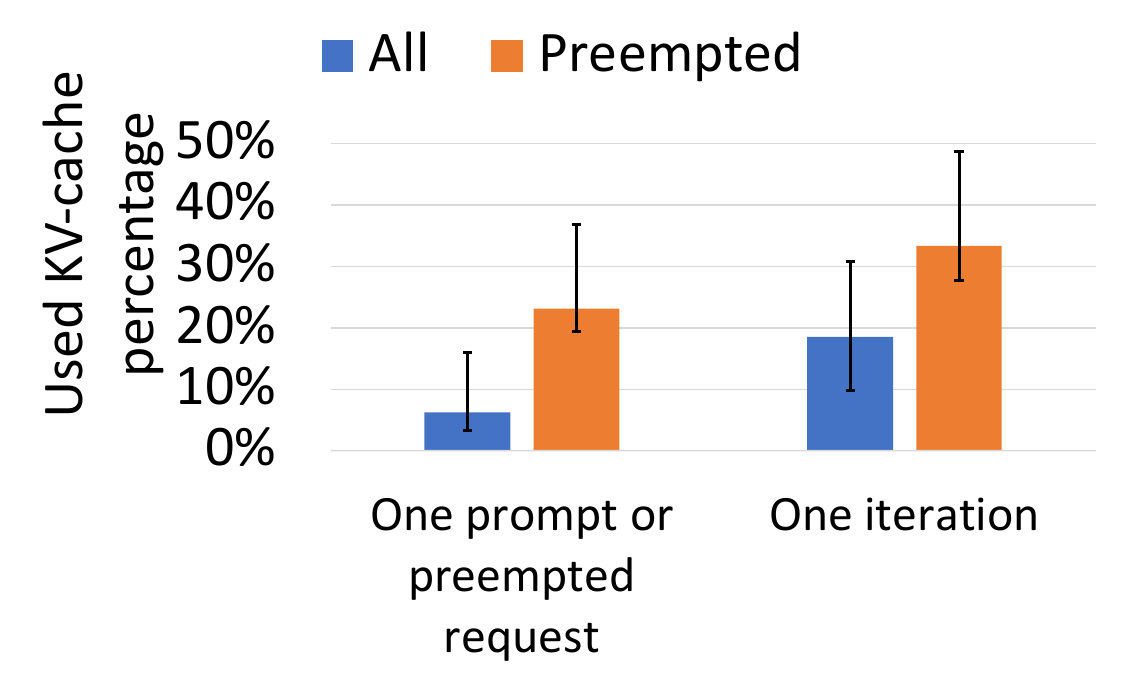}
    \caption{Used KVC.{\sh{Change names: Prompts (blue) and preempted GTs (orange). The left two bars:  each request, right two bars: all requests after each iteration.}}}\vspace{-0.0in}
    \label{fig:KVC}
\end{figure}
}

Figure~\ref{fig:KVC-wasted} shows
the percentages of the over-provisioning and under-provisioning KVC space over the allocated KVC space per request after adding the sweetspot padding in \emph{SyncDecoupled}. 
The over-provisioning percentages of Alpaca, ShareGPT, and BookCorpus are 2.62\%, 5.62\% and 5.12\%, respectively, whereas the under-provisioning percentages are 9.30\%, 13.42\%, and 21.92\%, respectively. In vLLM, when a request encounters a KVC allocation failure, other requests are likely to face similar issues. In contrast, \emph{SyncDecoupled} allocates KVC to each request in the batch based on its predicted RL, which may make KV data swapping unnecessity. Thus, in \emph{SyncDecoupled}, we used three solutions when a request does not receive sufficient KVC due to under-prediction. First, it is preempted, and its KV data is offloaded to CPU memory using the strategy in vLLM~\cite{vllm}. 
Second, it is preempted and its KV data won't be offloaded to the CPU memory. Third, it uses the KVC originally reserved for PTs. 
Figure~\ref{fig:memory-percentage} shows 
that the percentages of preemption time over the JCT of the preempted requests for these three solutions are 12\%, 4\%, and 0.5\%, respectively. 
\DEL{It means that an under-provisioned request is likely to receive KV space soon when it waits for another request to complete.} 

\DEL{When a request does not have enough KVC space, if we preempt it to the queue but do not move its KVC data, it will cause 
0.94\% of the JCT delay on average for this request for the three datasets. If the KV-cache is transferred to memory, it will cause 
a 2.08\% of the JCT.  Therefore, if we do not move KVC contents to CPU memory, it generates low preemption time overhead. {\color{red} However, if the reserved memory from the prompt is used instead of doing the movement, only 0.54\% of the JCT delay will be caused on average for the three datasets.}
}

\DEL{First, the request waits for released cache from other completed requests. Second, the request is preempted and runs again when further cache space becomes available to fulfill its required demand. In this scenario, a context switch is conduct, which takes a certain time. Third, we move the request's cache contents to the main memory, wait until other requests complete, and then fetch the contents back from the main memory.}

\DEL{\begin{figure}[t]
    \centering
\begin{minipage}[t]{0.48\linewidth}
\includegraphics[width=\linewidth,height=0.112\textheight]{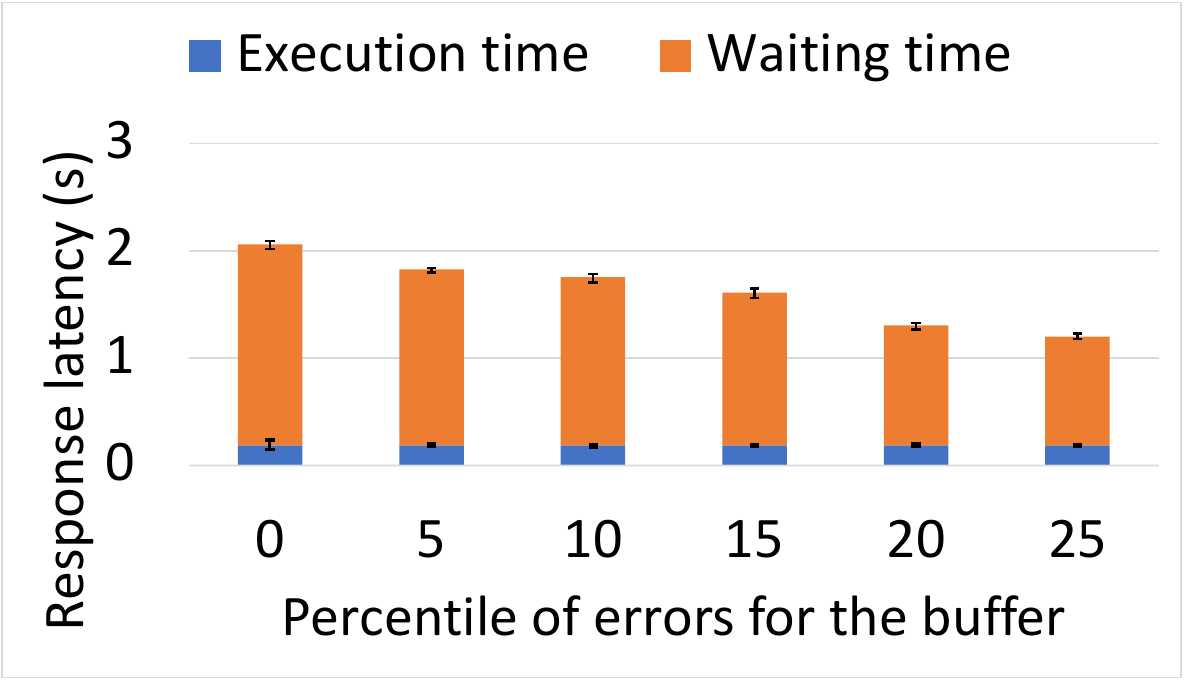} 
\vspace{-0.0in}    \caption{Response latency \sh{chagne Y to "Execution time"}.\vspace{-0.0in}}
    \label{fig:response-latency-buffer}
\end{minipage} 
\DEL{\begin{minipage}[t]{0.48\linewidth}
\includegraphics[width=\linewidth,height=0.112\textheight]{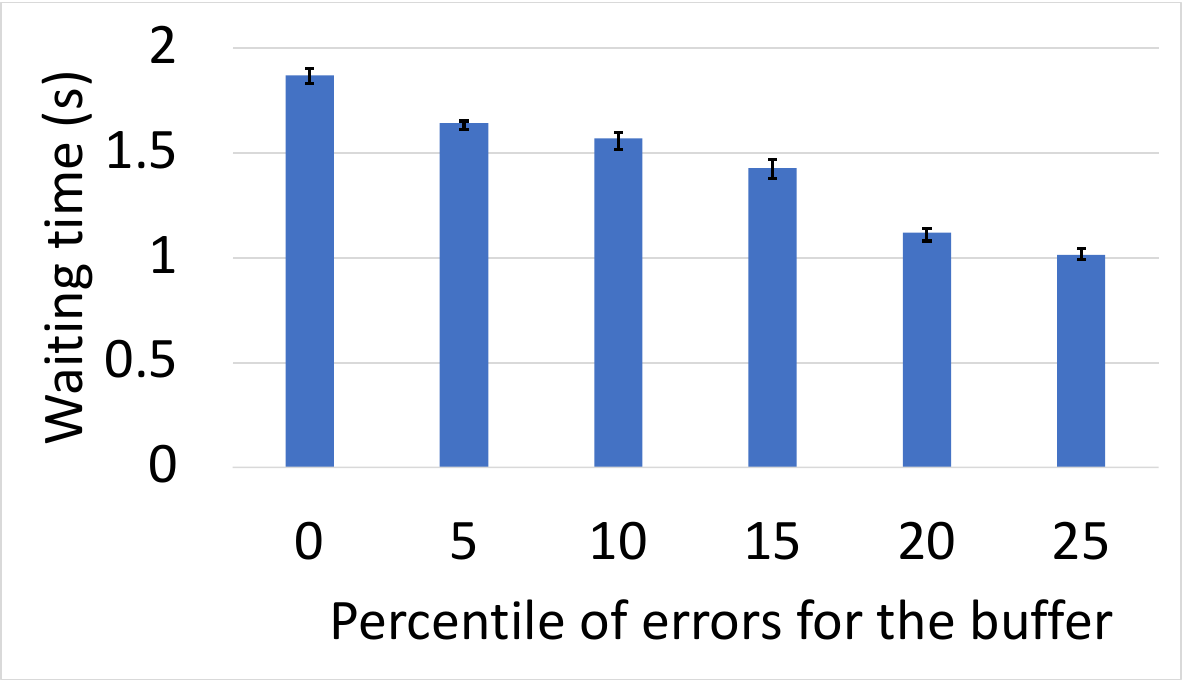}
 \vspace{-0.0in}   \caption{Waiting time. \vspace{-0.0in}}
    \label{fig:waiting-buffer}
\end{minipage} }
\end{figure}}

\DEL{\begin{figure*}[htb]
\centering
    \subfloat[Response latency.\vspace{-0.0in}\label{fig:response-latency-buffer}]{{\includegraphics[width=0.32\linewidth,height=0.112\textheight]{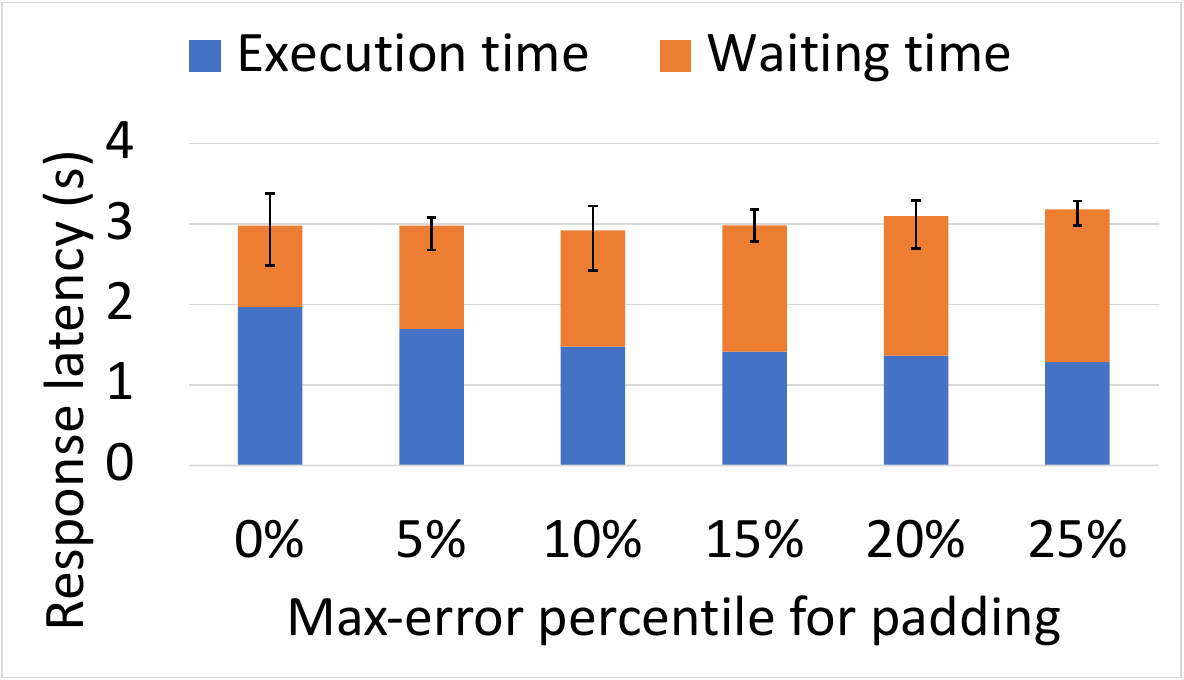}  }}
    \hfill
    \subfloat[Unused KVC percentage.\vspace{-0.0in}\label{fig:KVC-wsate-buffer}]{{\includegraphics[width=0.32\linewidth,height=0.112\textheight]{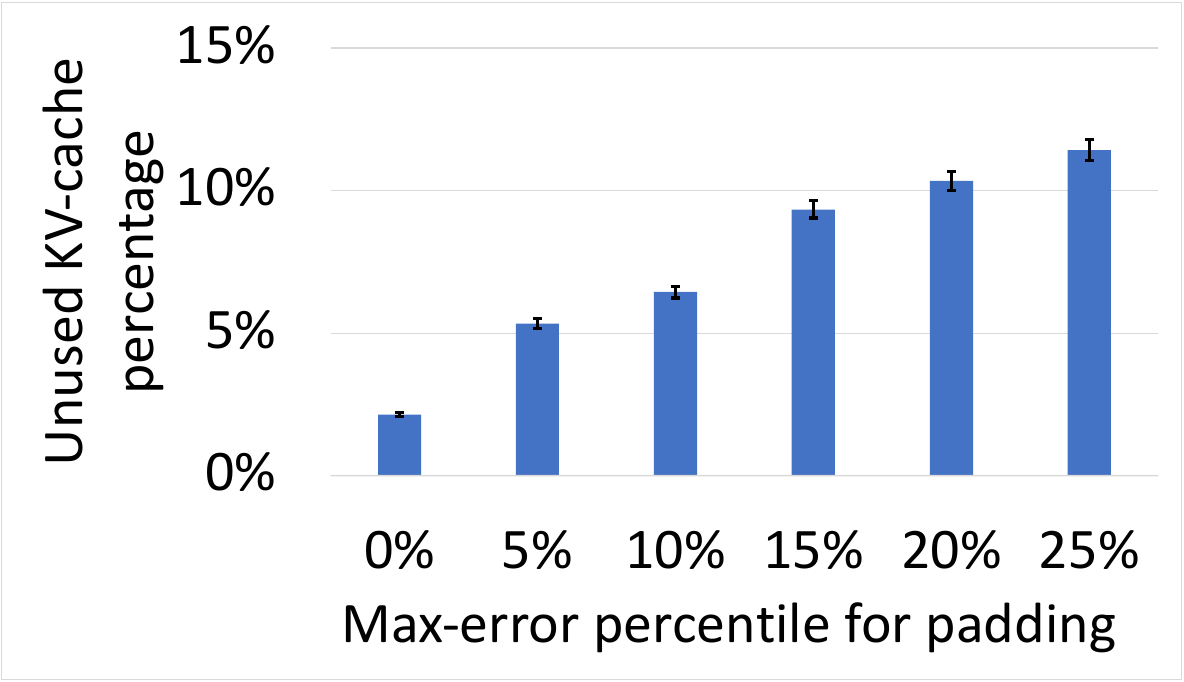} }}
    \hfill
    \subfloat[Under-provisioned requests.\vspace{-0.0in}\label{fig:under-provision-buffer}]{{\includegraphics[width=0.32\linewidth,height=0.112\textheight]{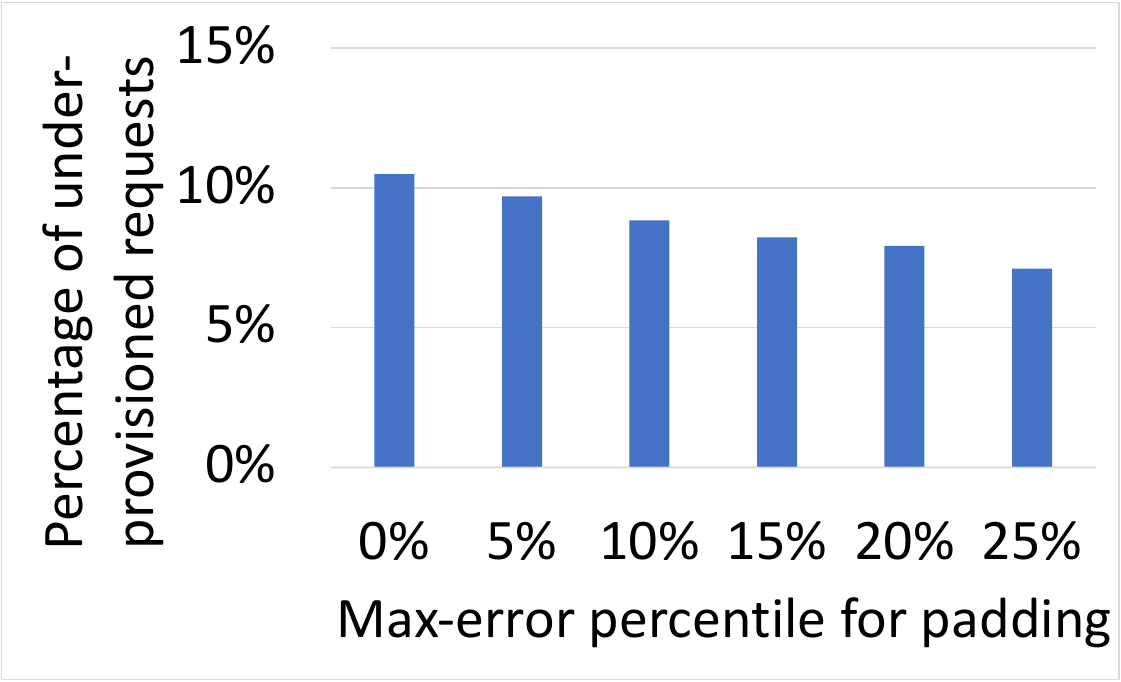} }}
    \hfill
    \vspace{-0.0in}
   \caption{\small{Impact of adding padding to the predicted response length.
\vspace{-0.0in}}}%
    \label{fig:buffer-measurement}
\end{figure*}}

\DEL{As the padding amount increases, the execution time decreases because under-predicted requests are less likely to experience under-provisioning, but the waiting time increases because increasing padding means less remaining KVC, and hence smaller batch size and lower throughput. }

\DEL{\vspace{-0.0in}
\begin{thm}\label{RLprediction1}
Adding padding to the predicted value can reduce the under-predicted requests, but it increases response time and the KVC waste.
\end{thm}\vspace{-0.0in}
}

\begin{thm}\label{RLprediction}
There exists a sweetspot padding ratio for the predicted output length that minimizes the average JCT. Upon a KVC allocation failure, using the reserved KVC and offload-free preemption may be more efficient than relying on offload-based preemption.\looseness=-1


\end{thm}
\vspace{-0.05in}

\vspace{-0.15in}
\subsection{Occupied KVC of Waiting Requests}
A waiting GT occupies a certain KVC space for the tokens of its prompt and also for its previously generated tokens if it is preempted. If a prompt is chunked, it also occupies
\begin{figure}
\centering
{{\includegraphics[width=0.96\linewidth,height=0.112\textheight]{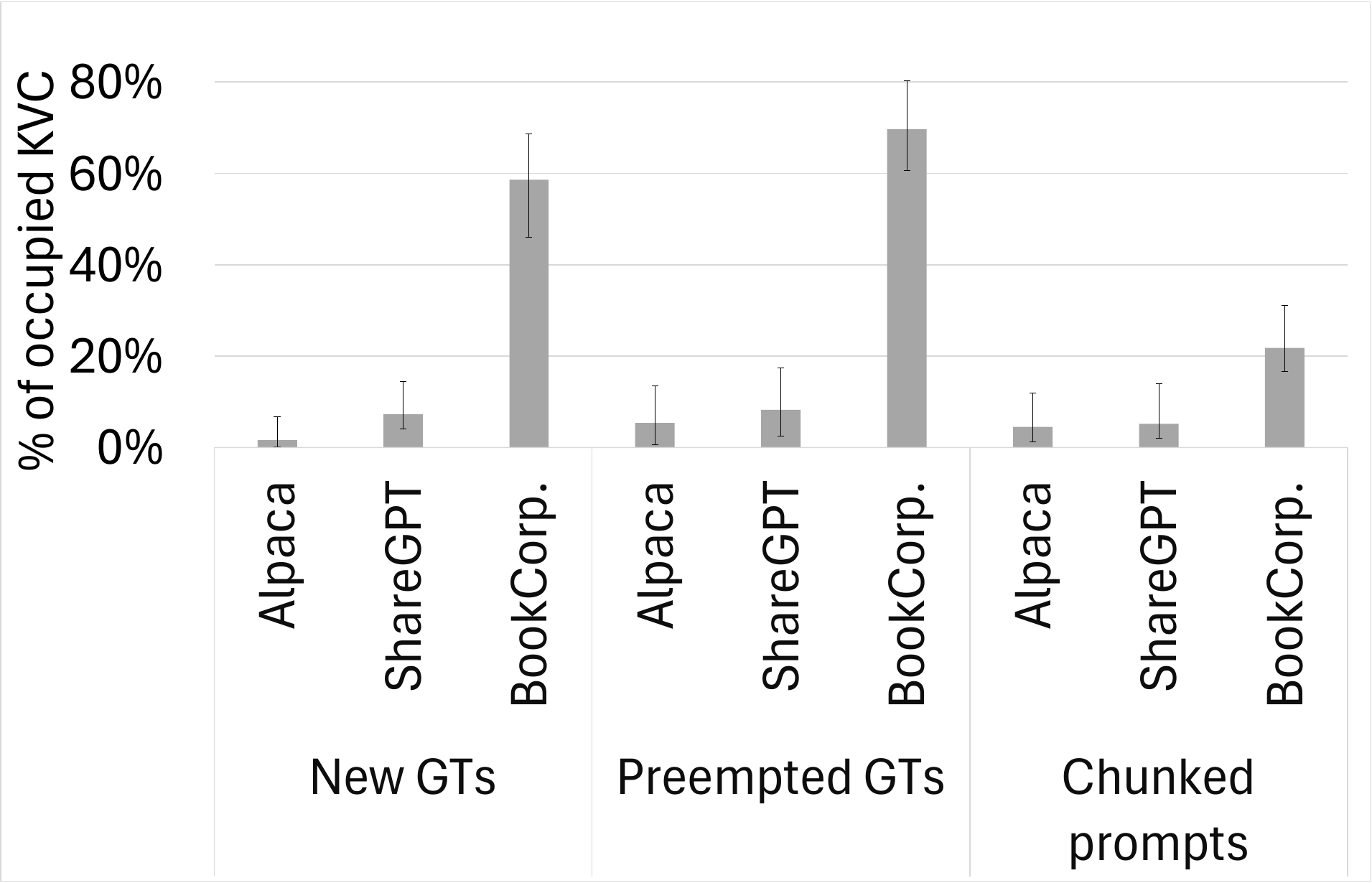}  }}
    \hfill
    \vspace{-0.17in}
   \caption{{Occupied KVC for each queued request.}}
\vspace{-0.2in}%
    \label{fig:memory-measurement-2}
\end{figure} 
a certain KVC. In the above experiment, we also measured the occupied KVC for the queued GTs and chunked prompts. Figure~\ref{fig:memory-measurement-2} shows the occupied KVC space for each new GT (just transitioned from a prompt), each preempted GT and each chunked prompt. The percentages of occupied KVC in each of these categories for each dataset vary greatly. Further, those of preempted GTs are higher than new GTs.  

\DEL{used by new GT ranges for the Alpaca from 1.41\%--6.70\%, and 1.60\% on average, the values of a preempted GT are for the Alpaca 4.81\%-8.11\%, and 5.42\% on average, and the values of the chunked prompt for the Alpaca are 3.2-7.3\%, and 4.53\% on average. For the ShareGPT the average values of new GT, and preempted GT and chunked prompts are 7.35\%, 8.25\%, and 5.18\%. For the BookCorpus the average values of new GT, and preempted GT and chunked prompts are 58.61\%, 69.71\%, and 21.86\%.}

\DEL{Figure~\ref{fig:8-b} shows the occupied KVC for all active requests after each iteration, for all PTs and for all GTs, respectively. 
The percentage of KVC used by all the requests, all PTs and all GTs after one iteration is 77.20\%, 13.91\% and 63.29\% of the total KVC on average.}




\DEL{\begin{figure}[htb]
\begin{minipage}[b]{0.25\textwidth}
    \centering    \includegraphics[width=0.5\columnwidth,height=0.112\textheight]{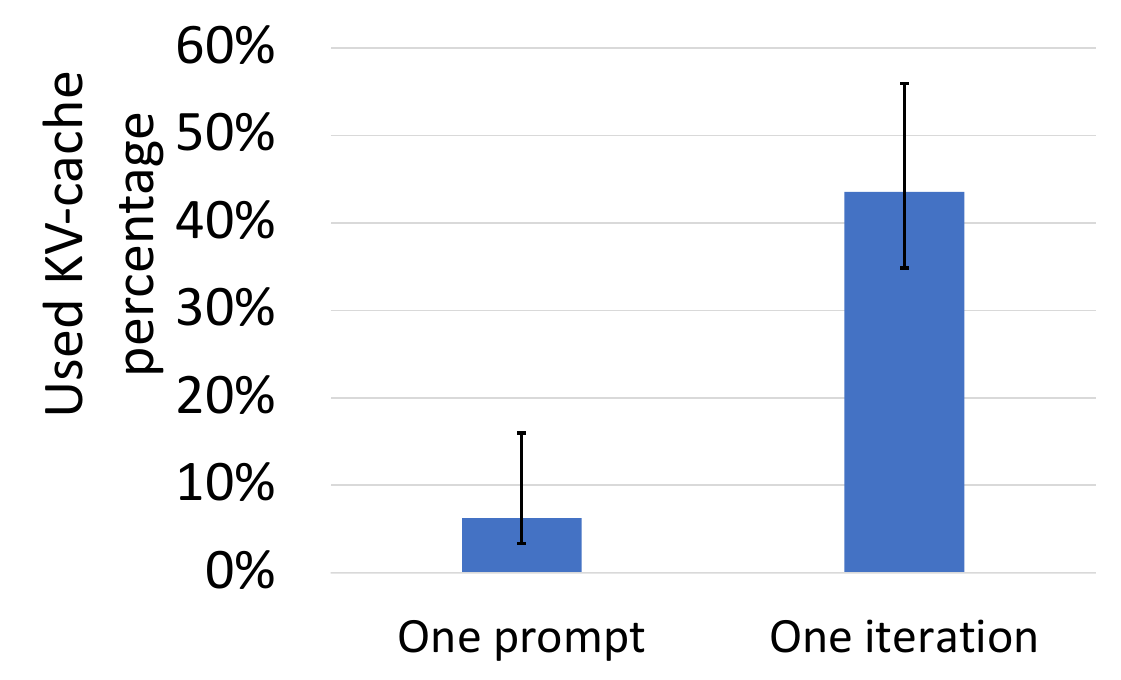}
    \caption{Occupied KVC.}
    \label{fig:KVC}
 \end{minipage}%
 \begin{minipage}[b]{0.25\textwidth}
  \centering
\includegraphics[width=0.5\linewidth,height=0.112\textheight]{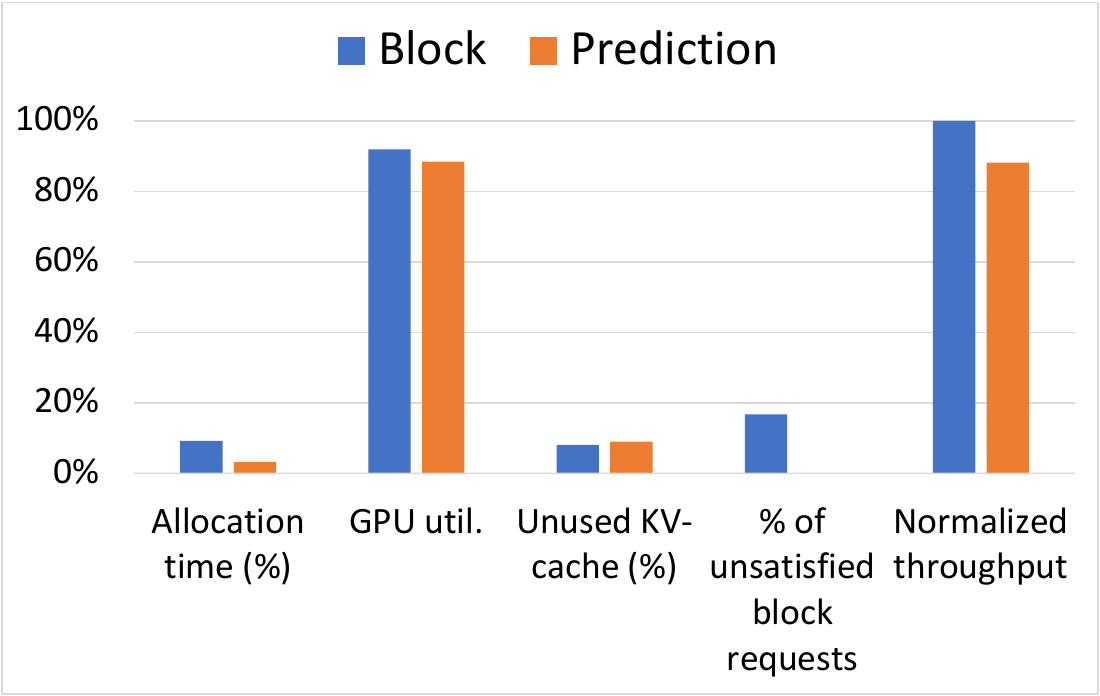}
    \caption{KVC allocation methods.}
    \label{fig:memory-fact}
 \end{minipage}
\end{figure}}


\begin{thm}\label{thm2-sort}
Waiting PTs and GTs occupy varying amounts of KVC space, so we should prioritize running those that occupy more KVC to release their KVC earlier.
\end{thm}
\vspace{-0.0in}

Finally, we ran DistServe on two A100 machines connected via Ethernet with a switch capacity of 100 Gb/s 
and added its results into Figure~\ref{fig:schedulers-measurement}. \emph{SyncDecoupled} has 9\% lower JCT than Distserve. 7\% of the JCT in DistServe is for the KVC value transfer.  \emph{SyncDecoupled} has 7\% higher throughput than the DistServe. DistServe has 8\% lower KVC utilization, 12\% lower forward size for prefilling and 82\% lower forward size for the decoding compared to \emph{SyncDecoupled}.


\begin{thm}\label{distServe}
Splitting the prefill and decode phases across two machines, as done in DistServe, can result in low GPU utilization in the decode machine and low KVC utilization in the prefil machine. This approach may also introduce latency due to the KVC value transfer, especially if the network bandwidth between the machines is limited.
\end{thm}
\vspace{-0.05in}

\DEL{\vspace{-0.0in}
\begin{thm}\label{thm2}
By allocating KVC to a request based on its predicted need, on average, {\color{red}29.80\%} \sh{this value should be close to 50\%, the KVC utilization of multiRes should be low}of the allocated cache is not used, providing an opportunity for optimization.??
\end{thm}
\vspace{-0.0in}
}

\DEL{\vspace{-0.0in}
\subsection{KVC Assignment}\vspace{-0.0in}
\DEL{??need to measure block-based
methodís drawback. whatís the overhead for the block
assignment? how to connect the blocks. ñif do not know}}
\DEL{\begin{figure}
    \centering
    \includegraphics[width=0.48\linewidth,height=0.112\textheight]{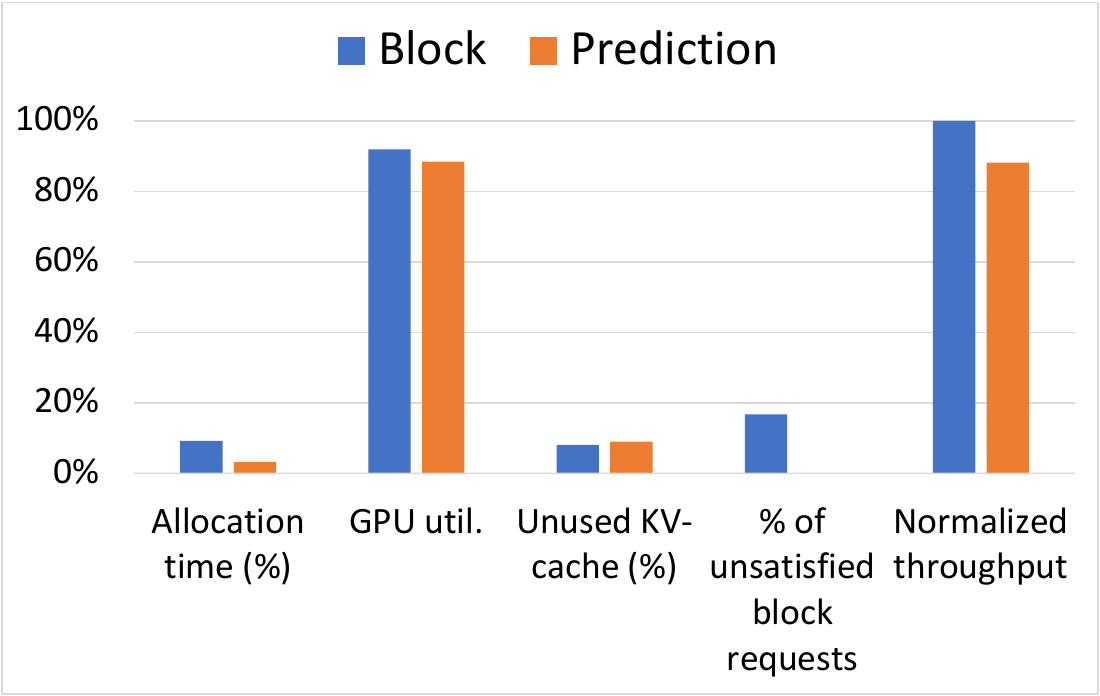}
    \caption{Performance of KVC allocation methods.}
    \label{fig:memory-fact}
\end{figure}}
\DEL{\begin{figure}[t]
\centering
    \subfloat[Cache allocation time.\vspace{-0.0in}\label{fig:allocation-time}]{{\includegraphics[width=0.48\linewidth,height=0.112\textheight]{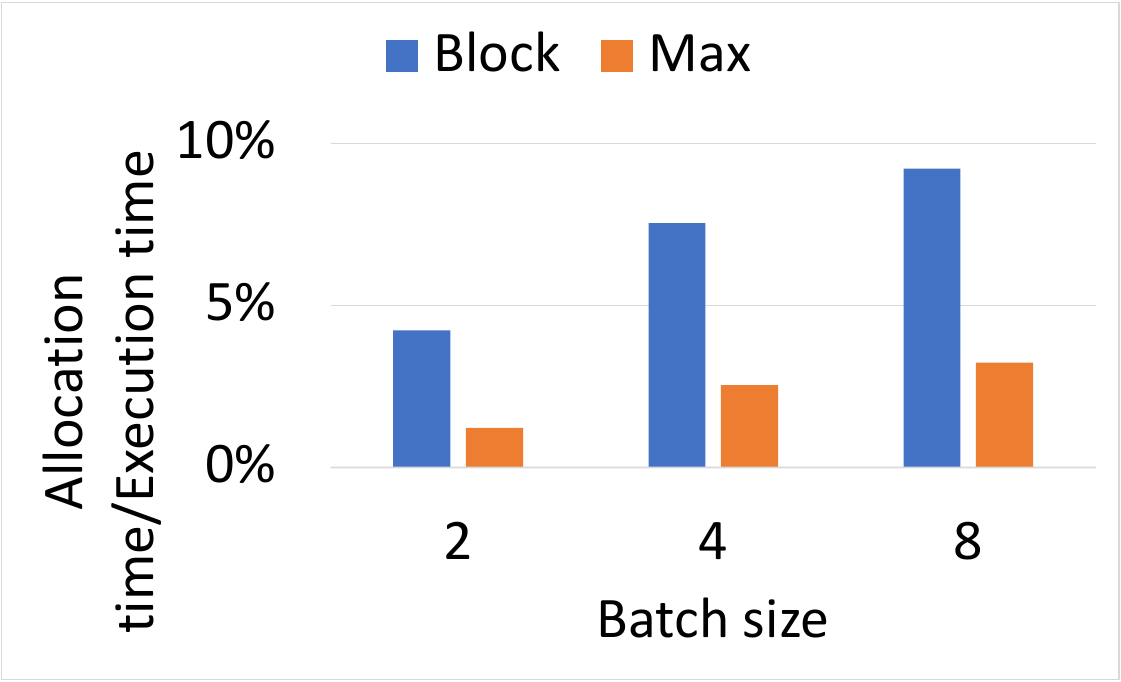} }}
    \hfill
    \subfloat[GPU\sh{block should be much higher-done}.\vspace{-0.0in}\label{fig:allocation-gpu}]{{\includegraphics[width=0.48\linewidth,height=0.112\textheight]{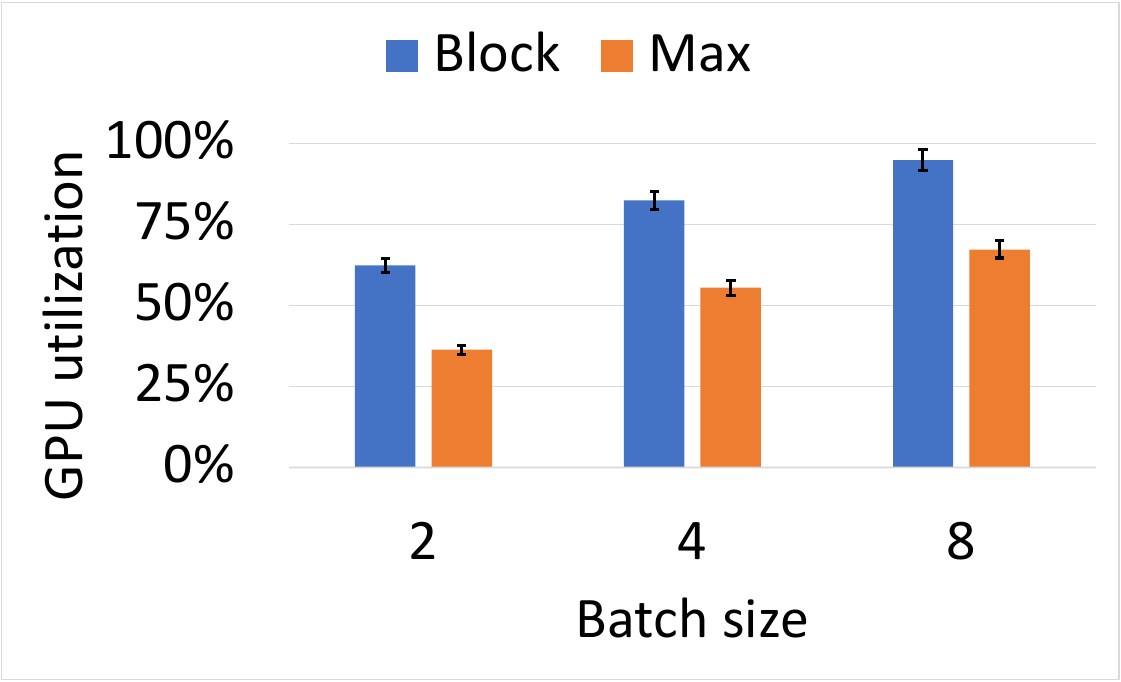} }}
    \hfill
    \subfloat[Throughput \sh{block should be much higher-done}.\vspace{-0.0in}\label{fig:allocation-throughput}]{{\includegraphics[width=0.48\linewidth,height=0.112\textheight]{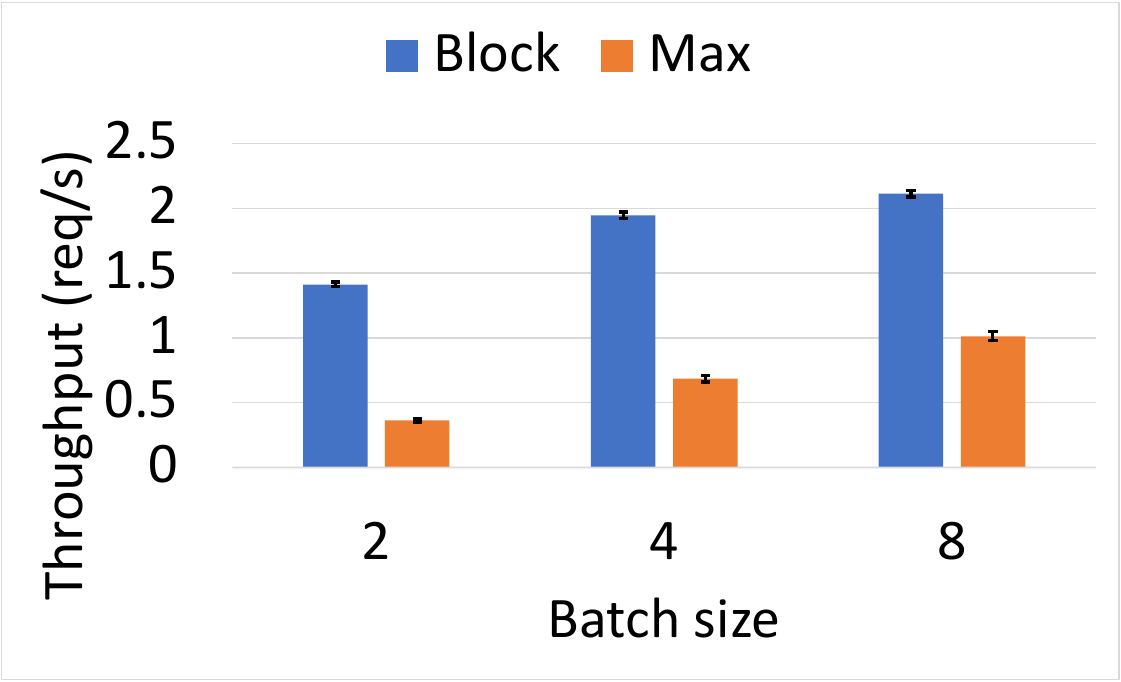} }}
    \hfill
    \subfloat[Memory.\vspace{-0.0in}\label{fig:factors4gt}]{{\includegraphics[width=0.48\linewidth,height=0.112\textheight]{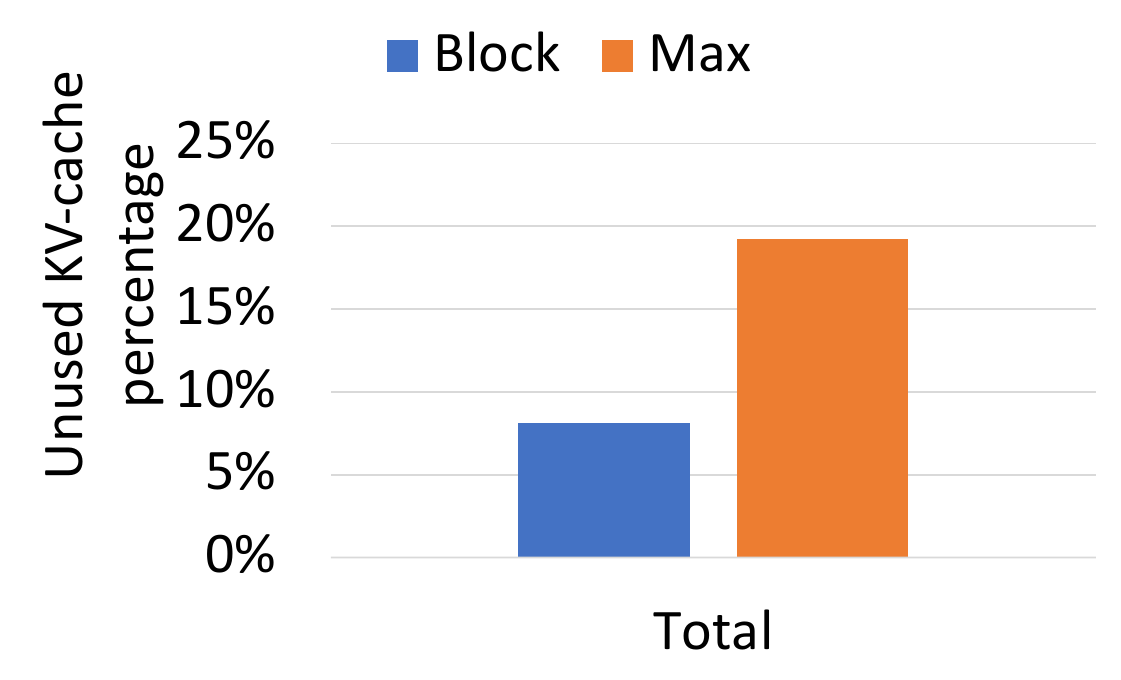} }}
    \hfill
    \subfloat[Iterations for block allocation.\vspace{-0.0in}\label{fig:factors5gt}]{{\includegraphics[width=0.48\linewidth,height=0.112\textheight]{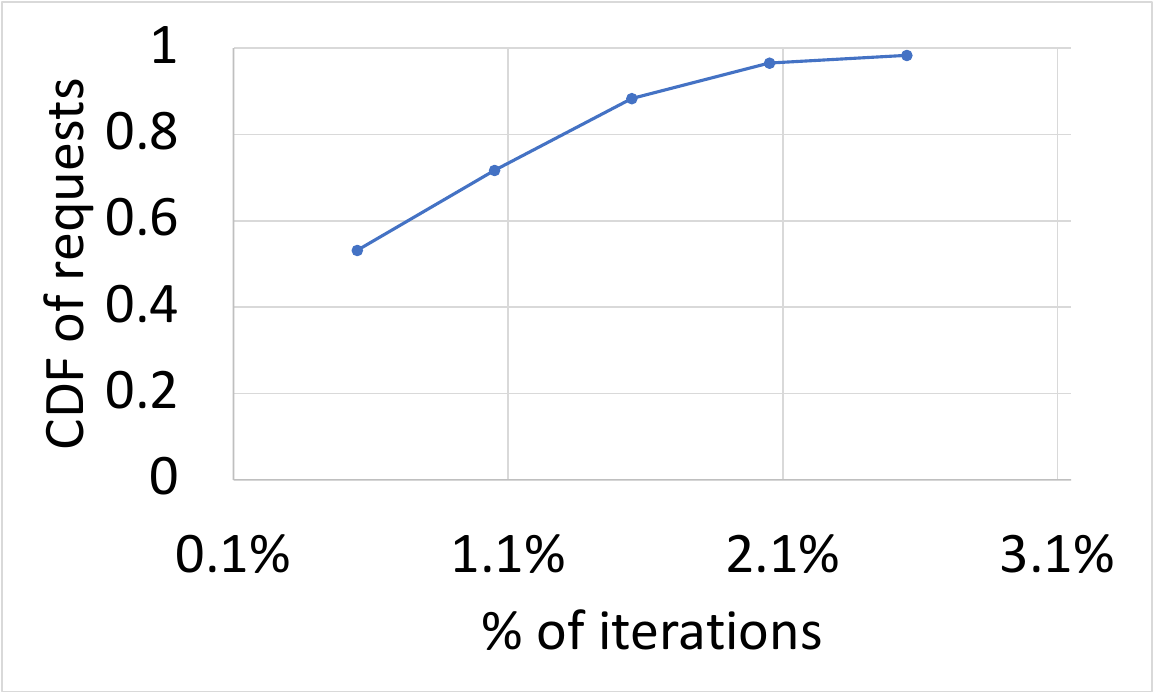} }}
    \hfill
    \vspace{-0.0in}
   \caption{\small{Performance of KVC allocation methods.?? These results are wrong, need to double check-done, updated\vspace{-0.0in}}}%
    \label{fig:memory-fact}
\end{figure}}

\DEL{Based on Observation \ref{thm2} (O\ref{thm2}), we want to see if the block-based method can improve the KVC utilization in \emph{SyncDecoupled} {\sh{instead of allocating all space equalling to the RL, it equals to a block??}}. We set the block size to 32{\sh{any reason for 80? what's the block size in vLLM?}}. Figure~\ref{fig:memory-fact} shows comparison results. The block-based method generates 4\%-9\% allocation time in execution time {\sh{after I redefine execution time, this term becomes response time, right?}}while the prediction-based method only generates 1\%-3\% allocation time. The \begin{wrapfigure}{c}{4cm}\vspace{-0.0in}
  \centering
\includegraphics[width=1\linewidth,height=0.112\textheight]{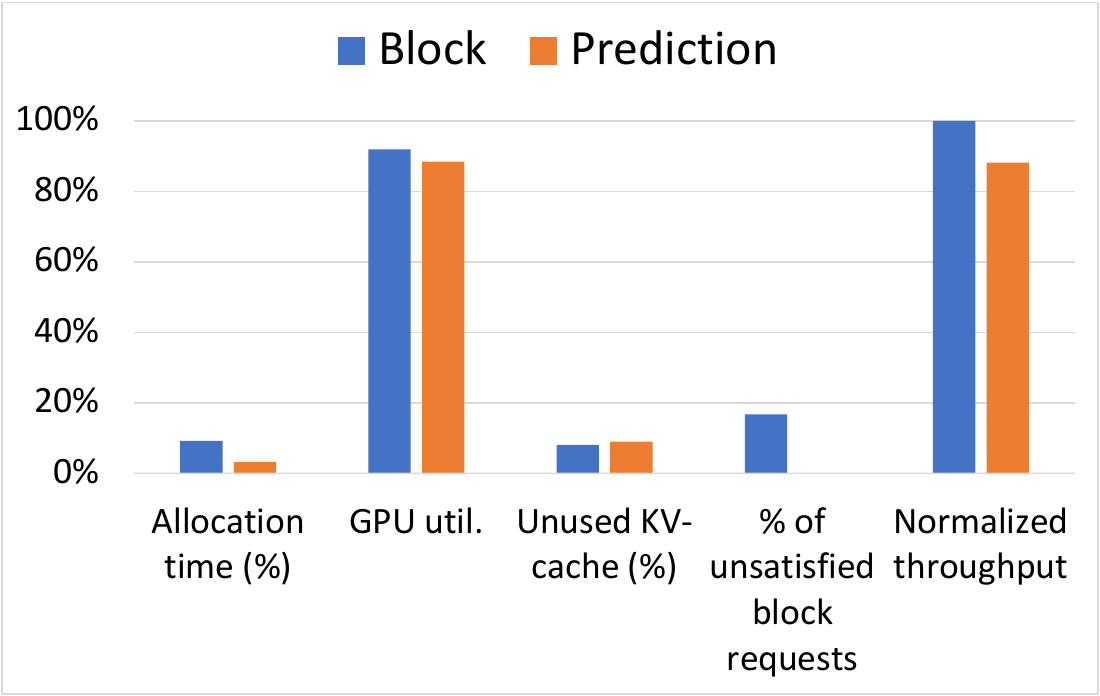}
  \vspace{-0.0in}  \caption{KVC allocation methods.}
    \label{fig:memory-fact}\vspace{-0.0in}
\end{wrapfigure}block-based method improves GPU utilization and throughput of the prediction-based method by 4.04\%  and 13\%, respectively and reduces the unused KVC percentage by 9\%. 
Finally, the block-based method experiences 16.74\% unsatisfied block requests {\sh{add this value for the prediction-based method}}, indicating a failure to allocate a block due to insufficient spare KVC space. Due to the \emph{SyncCoupled} method, simultaneous block requests within a GT group raise the likelihood of an unsatisfied block request. {\sh{this block-based is built upon our method. add vLLM that is built upon baseline. should be added in the figs at the beginning}}}


 

\DEL{\vspace{-0.0in}
\begin{thm}\label{thmKV}

In addition to higher KVC allocation time, the block-based method results in 16.74\% unsatisfied block requests in the Decouple+\emph{SyncCoupled} approach. Therefore, a new KVC allocation method is essential to address these limitations.
\end{thm}
\vspace{-0.0in}}
%

\vspace{-0.1in}
\section{System Design of \sys} \vspace{-0.1in}
\label{sec:design}


\begin{figure}[t]
    \centering
\includegraphics[width=1\columnwidth,height=0.135\textheight]{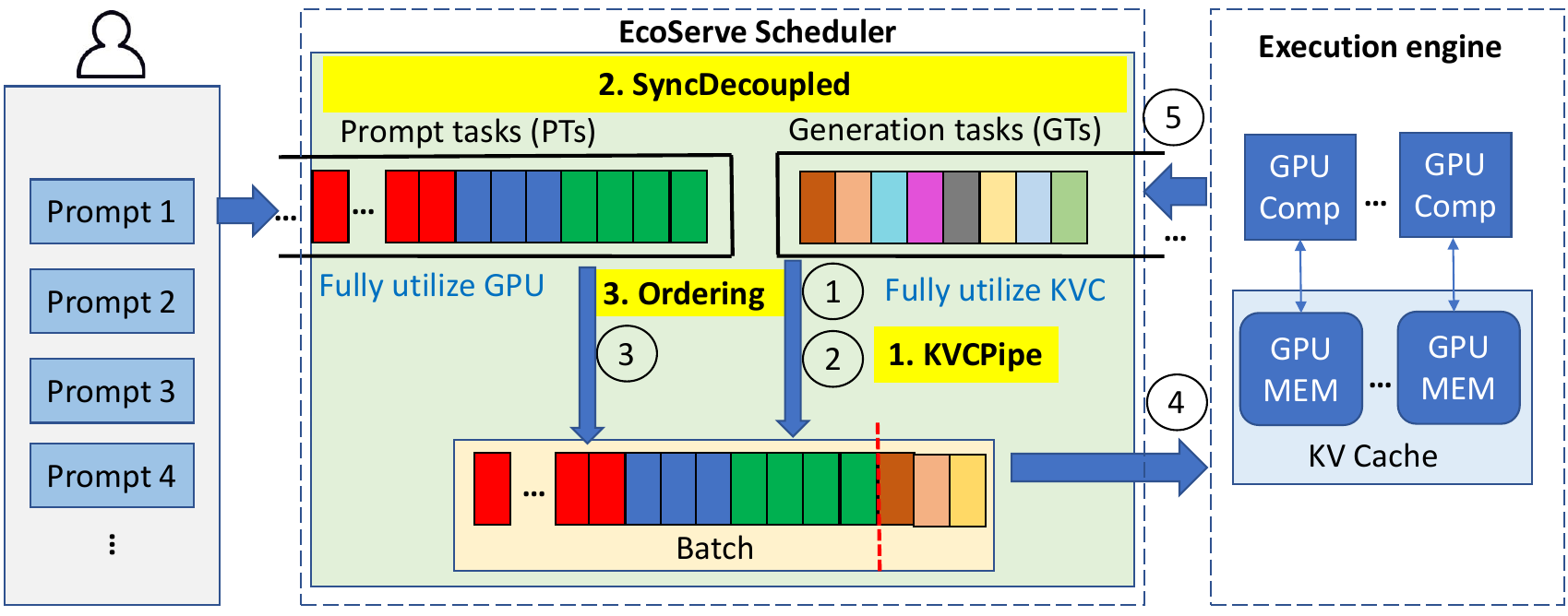}
\label{fig:component1}\vspace{-0.2in}
\end{figure}

\subsection{Overview}

Based on our observations, we propose \sys as shown in  Figure~\ref{fig:component1}. It consists of the following components marked in yellow in the figure.





\squishlist

\item[1.] {\textbf{KVC Pipelining (\emph{KVCPipe}).}}
This method solves the remaining issue of \emph{SyncDecoupled} indicated in O\ref{thm2}, in which each GT enables other requests to use its allocated but unused KVC space to improve KVC utilization (\cref{sec:comp4}).

\item[2.]{\textbf{Resource Responsibility Decoupling with Time-Synced Batching  (\emph{SyncDecoupled}).}} Guided by O\ref{thm-decoupling}, \sys incorporates \emph{SyncDecoupled}, which consists of the decoupling method (\cref{sec:comp1}) and the time-synced batching method, and also handles KVC under-provisioning by using reserved KVC, padding and offload-free preemption based on O\ref{RLprediction} (\cref{sec:comp2}).


\DEL{\item[(2)] {\textbf{Decoupling Prompt and GT Processing (Decouple).}} Method 1 decouples the prompt processing
and GT processing of a request by setting two queues for them (\cref{sec:comp1}). 

\item  {\textbf{Completion Time Synchronized Batching (SyncCoupled).}} Method 2 groups GTs with the same predicted RL and tries to batch them together. Based on O\ref{misprediction}, O\ref{RLprediction} and O\ref{RLprediction1}, it handles KVC under-provisioning by preemption instead of relying on adding a padding to the prediction value (\cref{sec:comp2}).
}


\item[3.] {\textbf{Prompt and Generation Task Ordering (\emph{Ordering}).}} It orders waiting requests, aiming to satisfy SLOs and prioritizing requests with higher occupied KVC in order to release it earlier (based on O\ref{thm2-sort}). It also facilitates quickly selecting requests from each queue to fully utilize each resource (\cref{sec:comp3}).



\DEL{\item[(3)] Based on O\ref{thm-decouple} and O\ref{thm2-sort}, Method 3 groups and orders the prompts based on prompt length, SLO and occupied KVC, and groups and orders GTs based on predicted response length, SLO and occupied KVC for task selection from the queues (\cref{sec:comp3}). 
}

\squishend

In Figure~\ref{fig:component1}, the blocks of the same color represent the tokens from the same request. Users' requests are entered to the prompt waiting queue and ordered by \emph{Ordering}. After each iteration, \emph{SyncDecouple} is executed. Specifically, if a GT group in the batch completes, other same-RL GT groups from the GT queue are fetched to fully allocate the available KVC (\circled{1}). Then, \emph{KVCPipe} is executed to select more GTs to use allocated but unused KVC of these selected GTs (\circled{2}). Next, PTs are fetched from the prompt queue to reach the TFS to fully utilize the GPU (\circled{3}). The formed batch is sent to the execution engine to be executed (\circled{4}). After the execution, the PTs become GTs and are entered to the GT waiting queue and ordered by \emph{Ordering} (\circled{5}). Then, the process repeats again.


\IdeatoRead{?analysis section, need to add results for the deadlock for block-based KVC allocation}

\vspace{-0.0in}


\vspace{-0.1in}\subsection{KVC Pipelining\label{sec:comp4}\vspace{-0.05in}}
\IdeatoRead{(??analsysis: need to show the CDF distribution of response lengths, need to show the length varies)}

\DEL{In LLM inference, the KVC or GPU memory is a bottleneck. \Orca allocates the KVC space to each request based on the maximum total sequence length, thus limiting the batch size and the GPU utilization. Alternatively, the Transformers library by Huggingface \cite{Face3} allocates cache space to each token once a new token is generated but it generates high memory overhead. The block-based allocation method~\cite{vllm} assumes an always available cache block when needed, which may not hold true in practice. (??analysis to show it). }

\DEL{{\sh{Need to add "The prefilling requests’ KV-cache values are stored in the KV-cache. The techniques in vLLM are adopted for memory access, with a table holding the memory locations. A GT with under-predicted RL will be preempted, and its occupied cache will be moved to CPU memory (2nd paragraph in page 8) and will be moved back when the GT is selected to run." somewhere}}

??I wonder how to move the KV cache of the next decode batch into the GPU. The GPU memory is already occupied by the previous batch, so it seems hard to asynchronously move the KV cache of the next batch to the GPU. If this happens synchronously, what is the overhead of this movement?

{\sh{We used the swapping strategy with a copy-on-write mechanism in vLLM [8]. If a selected GT group has a request with data in CPU memory, the data is moved to GPU synchronously within the GT group (instead of a batch). The overhead is around 0.088s per request. We will add it to Figure 16."}}

{\sh{"It is not clear how \sys handles the generated KVC of prefilling stages. Does the KVC move back to CPU memory after each layer or after completing the prefilling? What is the size of GPU memory reserved for the prefilling phase? What is the CPU memory usage? Assuming in an industry setting, for DGX boxes, eight GPUs share the CPU memory. Can we accumulate enough similar decode length requests for good throughput?" -- C4. See Q4 above. \sys does not reserve GPU memory. The CPU memory usage is ~63\%. Currently, we consider scheduling for one machine with multiple GPUs, which cooperate in handling each request. Your idea is fantastic for the scenario where multiple GPUs handle multiple requests. We will study it in our future work.}}
}
Motivated by O\ref{thm2} that the exact-allocation cannot fully utilize KVC though it prevents KVC allocation failures, we propose this \emph{KVCPipe} method, which allows GTs to share allocated but unused cache space. In this method, the second half of the allocated  KVC part of one GT request (denoted as $r_1$) is reallocated to another GT (denoted as $r_2$). A request whose RL is no more than but closest to half of $r_1$'s RL is identified as $r_2$. Then, when $r_1$'s KVC space usage reaches the middle of its allocated KVC space, i.e., the starting point of $r_2$'s KVC space, $r_2$ completes and releases its occupied KVC. 
Here, we call $r_1$ the hosting GT of $r_2$ and call $r_2$ the hosted GT of $r_1$. Figure~\ref{fig:KVPipeline} shows an example in which different colors represent different GTs. GT $r_1$'s predicted RL (RL for simplicity) is 32 tokens. We allocate 32-token KVC to $r_1$ and reallocate the 16-token KVC from the middle of the allocated space to $r_2$, which has an RL of 16 tokens. Then,  Both GTs start running simultaneously, and when $r_1$'s occupied KVC reaches the middle, $r_2$ completes and releases its cache space. Then, $r_1$ can continue using the cache until it completes. 

We generalize this method to multiple GTs as shown in Figure~\ref{fig:KVPipeline1}. Within the left and right half of the allocated space for $r_1$, in the second half of the space, we can further embed another GT, i.e., $r_3$ and $r_4$ in the figure. $r_3$ and $r_4$ will complete when their hosting GTs' KVC usages reach their memory starting points. Similarly, each of these GTs, $r_1$-$r_4$, can host another GT in itself. This recursive process continues, akin to Russian nesting dolls, until no more GTs can be accommodated. 
\begin{figure}[t]
\centering
    \subfloat[Example for two GTs.\vspace{-0.0in} \label{fig:KVPipeline}]{{\includegraphics[width=0.48\linewidth,height=0.08\textheight]{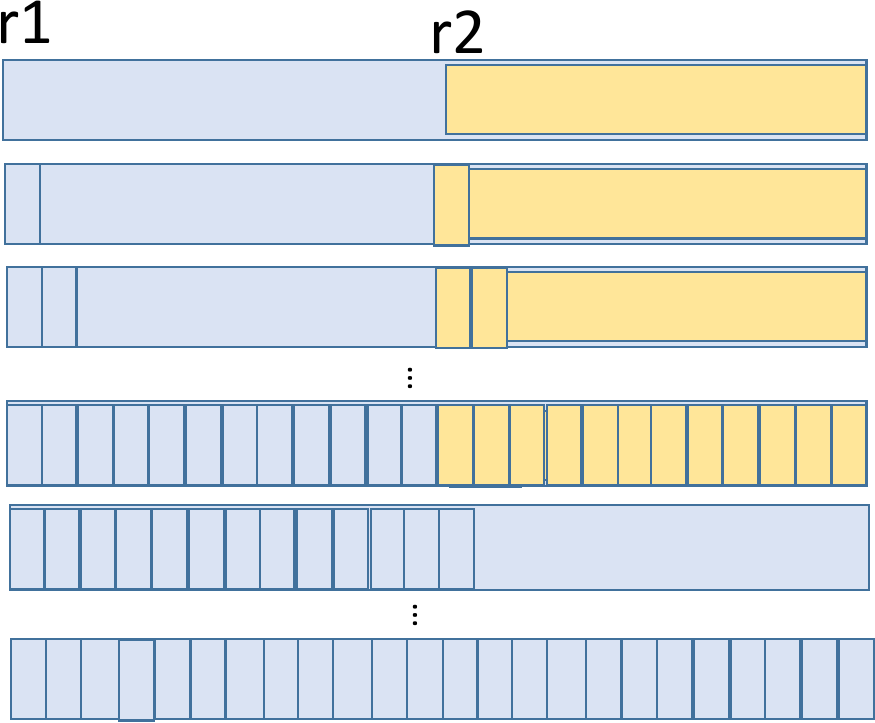}  }}
    \hfill
    \subfloat[Example for four GTs. \vspace{-0.0in} \label{fig:KVPipeline1}]{{\includegraphics[width=0.48\linewidth,height=0.08\textheight]{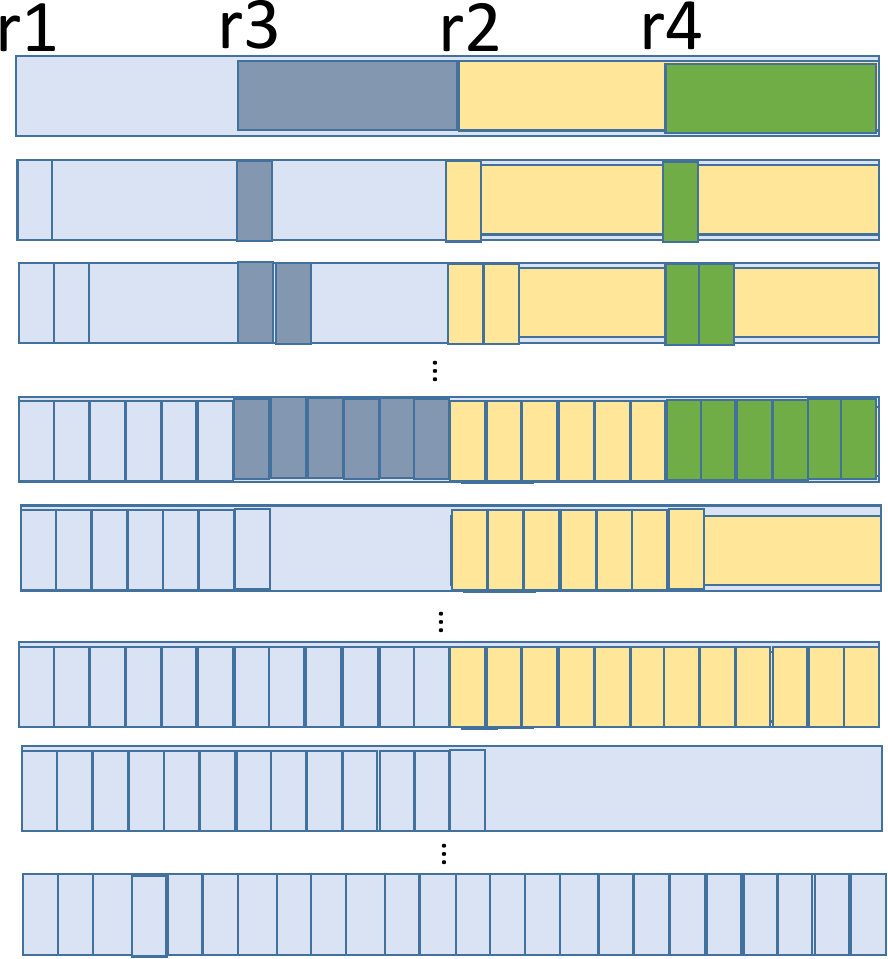} }}
    \hfill
    \vspace{-0.15in}
   \caption{Examples of KVC pipelining.
\vspace{-0.1in}}
    \label{fig:memory-measurement}\vspace{-0.15in}
\end{figure}
\DEL{To coordinate this method with other methods, after selecting GTs (with $x$ predicted response length) to maximize KVC utilization, we find GT groups with response lengths $L_r=x/2, ~x/4, ~x/8, ...$ until the half of the shortest $L_r$ is reached. Subsequently, the sum of tokens in the forward sequence is calculated, and prompts are selected to reach the TFS. This integrated approach enhances efficiency in both KVC utilization and GPU processing.}

Recall that the RL may be under-estimated, and then a hosted GT may not complete by the predicted time.
To avoid this case, we use a buffer of $b$ tokens here; that is, \sys finds a hosted GT that has RL no more than but closest to half of the hosing GT's RL subtracting $b$ tokens. In \sys, after the GT groups are selected in batching, for each of the GT groups with $m$ GTs and RL $L_r=l$ in the GT waiting queue, \sys finds GT groups with RLs no more than but closest to $l/2-b,~l/4-b, ~l/8-b...$, respectively. \sys needs to find $m\cdot 2^1$ GTs, $m\cdot 2^2$ GTs, $m\cdot 2^3$ GTs and so on for each of the above RL values, respectively. The GTs in the hosted GT group and in the hosting GT group are randomly mapped. With the buffer strategy, if a hosted GT $r_2$ still does not complete when its hosting GTs need the KVC space to be returned, $r_2$ is preempted and its occupied cache will be moved to the main memory temporarily using the copy-on-write mechanism in~\cite{vllm}. 
\looseness=-1 




\DEL{Note that the hosted GTs do not need to be allocated with additional KVC space and they only increase the KVC utilization. Subsequently, the sum of tokens in the forward sequence is calculated, and PTs are selected from the PT queue to reach the TFS. }

\DEL{{\sh{The paper proposes pipelining KVC. However, if paged attention is used, the KVC can intrinsically be pipelined even in a more fine-grained way. For example, if request 3 is reusing request 1’s KVC space, and request 4 is reusing request 2’s cache space. Using paged attention allows leaving all memory space available at the beginning of the processing and all requests consume the space at the same pace. When 3 and 4 finish, they free all the space, and the KVC of requests 1 and 2 can continue to grow. Moreover, the paged attention solution is more flexible. Assuming request 3 is longer than expected and request 4 is shorter, it is possible for request 3 to use the KVC space originally for request 4 to avoid an out-of-memory issue, but in the pipelining setup, request 3 can at most utilize the space for request 1.}}
}

\vspace{-0.1in}
\subsection{Resource Responsibility Decoupling with Time-Synced
Batching} \label{sec:comp2}

\DEL{Based on O\ref{thm-decoupleMotivation}, sophisticated iteration-level schedulers that select prompts to fully utilize resources have two shortcomings: long scheduling time and being unable to fully utilize the resources. To address these shortcomings, }

\DEL{In \cref{sec:analysis}, we proposed the \emph{SyncDecouple} method to achieve our goal. 
}

\DEL{\emph{SyncDecouple} employs exact-allocation and comprises two components detailed in the following two subsections.}


\DEL{simultaneously tackle the triple challenges of i) fully utilizing both GPU and KVC, ii) preventing KVC allocation failures in execution, and iii) minimizing scheduling time. \DEL{iteration-level scheduling becomes imperative. Though request-level scheduling does not need scheduling after each iteration, it introduces challenges such as the delay of completed requests waiting for unfinished ones, and the postponement of new requests. To address this dilemma, based on O\ref{thm-groupGT}, we propose grouping and batching same-RL GTs.} 
}

\vspace{-0.05in}
\subsubsection{Resource Responsibility Decoupling}\label{sec:comp1}\vspace{-0.0in}

\DELMayNeed{\bl{(??Analysis section: need to show the advantage of decoupling and the pre-sorting, instead of selecting prompt one by one to reach TFS. e.g., we need to fill up 1000 prompt tokens, checking requests in the queue sequentially takes time. if you group them based on prompt length, then you can quickly find the group with 1000 tokens. the same for the GT group below--need analysis figs to show this)}}

\IdeatoRead{?idea: balance between the two processors or queues
}

\DEL{The max-allocation method results in 19\% of allocated but unused KVC (O\ref{thm2}), while the block-based allocation method, which avoids this issue, cannot ensure the provisioning of the required KVC space and also leads to high memory allocation overhead (O\ref{thmKV}). To tackle these challenges, \sys predicts a request's response length and allocates KVC space equivalent to its total sequence length. The response length prediction also enables \sys to group same-RL GTs together in a batch to avoid iteration-level scheduling during the group running time. }

\DEL{The GPU utilization depends on the forward size, while the KVC utilization depends on the total sequence lengths.}

\DEL{(?may remove)The max-allocation method results in low batch size and throughput, while the block-based allocation method, which avoids this issue, cannot ensure the provisioning of the required KVC space and also leads to high memory allocation overhead (O\ref{thmKV}). To address the issues, aiming to batch same-RL GTs to save scheduling time, \sys predicts a GT's RL and allocates KVC space equivalent to the predicted RL.\looseness=-1 
}

\DEL{(?this paragraph may be removed)The PTs in a batch will be transformed
into one-token-input GTs after the first iteration, diminishing
GPU utilization based on Equ.\eqref{eq:totop}. A GPU's computation and memory can be fully utilized by batching compute-intensive prompts with memory-intensive GTs~\cite{Agrawal2023SARATHIEL,MicrosoftRef}. However, it is feasible only when there are available KVC space for the prompts. Therefore, it is not feasible when a batch is formed to fully utilize the KVC and no requests complete to release KVC. That means only when a request completes, then prompts can be added to the batch {\sh{need to measure and draw a fig, X=different percent values of requests in a batch completed after an iteration, Y=the percent of such iterations over all iterations}}. O\ref{release} shows that the percentage of such occasions is low. In addition, in \sys, the added prompts are not guaranteed to complete simultaneously with existing GTs (Figure~\ref{fig:decoupling}), which is a contradiction to the design principle of SyncCoupled, which is to synchronize the completion time of the requests in a batch. To address this, we propose deferring the scheduling of these newly generated GTs for later execution.}

\DEL{Prompt processing tasks are computation intensive while GTs are memory intensive. To optimize throughput, we designate prompts for fully utilizing GPU resources and designate GTs for fully utilizing GPU memory resources when creating a batch. As a result, as shown in Figure~\ref{fig:combine},  However, the prompt tasks in a batch will be transformed into one-token-input GTs after the first iteration, diminishing GPU utilization while escalating demands on KVC. To ensure that TFS is achieved at each iteration, prompts must be added to the batch after each iteration. These prompts' newly generated GTs may not receive the required KVC, and also may not complete simultaneously with existing GTs (Figure~\ref{fig:decoupling}). This is against \sys's design principle that tries to let GTs in a batch start and complete at the same time by grouping same-RL GTs in a batch. To address this, we propose deferring the scheduling of these newly generated GTs for later execution. 
}

\DEL{In the previous LLM inference serving systems, prompts wait in a waiting queue and they will be processed to become GTs and then complete continuously. Though Coupled tries to fully utilize the GPU and memory resources in this scenario,  MultiRes introduces insufficient effectiveness in finding prompts with specific prompt lengths and specific total sequence lengths to fully utilize GPU and KVC, as indicated in O\ref{thm-decoupleMotivation}.}


This method aims to facilitate fully utilizing the compute and memory resources of a GPU. 
Instead of letting each request to contribute to the utilizations of the dual-resources (see Figure~\ref{fig:DecoupleDemo11}), it assigns PTs and GTs the responsibility of fully utilizing the GPU and KVC, respectively (see Figure~\ref{fig:DecoupleDemo12}).
Users' prompts are placed in the PT waiting queue, while the GTs resulting from the prompts enter the GT waiting queue. The KVC size allocated to a PT equals its prompt length, and that allocated to a GT equals to its predicted RL. For easy memory management, the KVC allocation is still in the unit of a block~\cite{vllm}. In the GT queue, GTs with the same predicted RL are grouped (details are in \cref{sec:comp2}). 
We will present how to order the PTs and GT groups in their queues in \cref{sec:comp3}. 
A small amount KVC is reserved for PTs.\looseness=-1 

In the system, at the initial iteration, prompts are selected sequentially to form a batch until the available KVC space is fully allocated or the TFS is reached (i.e., GPU will be fully utilized). 
For subsequent iterations, the procedure is as follows. After an iteration, the PTs become GTs, which are entered into the GT queue and grouped based on predicted RL. The KV values of the prompt tokens are stored in KVC. 
Then, a new batch is formed for the next iteration. That is, if there is available KVC (in the second iteration or when a GT group completes), the queued GT groups are selected sequentially until the KVC is fully allocated. If the KVC's available space cannot accommodate an entire group, the group is split to fit within the KVC. Next, the PTs from the PT queue are selected sequentially until the TFS is reached.
As a result, 
\sys can fully utilize the compute and memory resources of the GPU in each iteration, while reducing scheduling time. 

\begin{figure}[t]
\centering
    \subfloat[W/o decoupling: cannot add PTs until a GT completes.\vspace{-0.0in}\label{fig:DecoupleDemo11}]{{\includegraphics[width=0.45\linewidth,height=0.08\textheight]{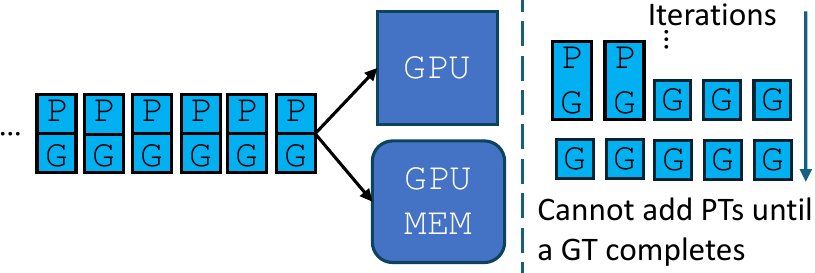}  }}
    \hfill
    \subfloat[W/ decoupling: can add PTs in each iteration. \vspace{-0.0in}\label{fig:DecoupleDemo12}]{{\includegraphics[width=0.45\linewidth,height=0.08\textheight]{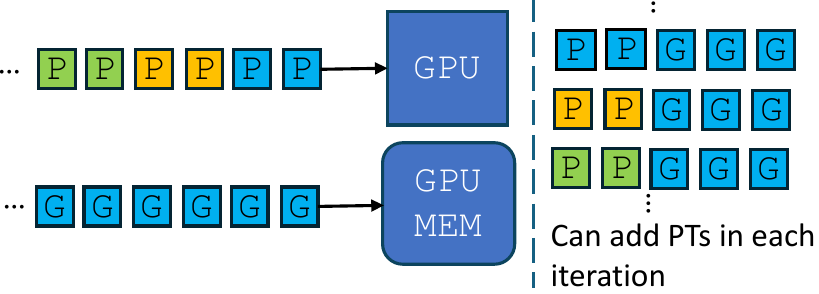} }}
    \hfill
    \vspace{-0.15in}
   \caption{{Decoupling enables adding PTs to the batch in each iteration with reserved KVC.
\vspace{-0.2in}}}
    \label{fig:DecoupleDemo}
\end{figure}

\DEL{By grouping same-RL GTs, we eliminate the need of iteration-level scheduling for GTs and reduce the time for scheduling. When the ongoing GTs complete simultaneously, GTs with the same predicted response length from the GT queue are added into the batch, aiming to fully use the KVC.} 

\DEL{After each iteration, if a GT group completes, queuing GT groups are added to the batch sequentially, aiming to fully use the KVC. If the KVC's spare space cannot accommodate an entire group, the group needs to split to be accommodated into the KVC. Then, PTs from the PT queue are added to the batch to reach the TFS. The process repeats. As a result, as shown in Figure~\ref{fig:combine}, \sys can fully utilize GPU and memory in each iteration, while reducing scheduling time.
}


\DEL{For easy memory management, the KVC allocation is still in the unit of a block~\cite{vllm}. That is, \sys allocates the minimum number of blocks that can host the GT or PT. Since a GT may receive excess KVC beyond its demand, its PT will continue to use its excess KVC. Then, a GT's KVC demand is its predicted RL subtracting the excess KVC.}

\DEL{Now, some prompt tasks become token GTs, and we need to add prompt tasks to reach the TFS. these newly produced token GTs and the token GTs of the added prompts may not complete with the existing token GTs. Therefore, we propose returning the GTs to be scheduled later. Recall that we let token same-RL GTs run together, so we do not need to do iteration-level scheduling for the GTs. Therefore, we need to gather GTs first before we can group the GTs with the same RL. If we put prompts and GTs in the same waiting queue, by checking each task, this makes finding prompts with a certain prompt length more time-consuming (??need analysis results). In addition, finding the same-RL GTs also is time-consuming (??need analysis results). Due to these two reasons, we propose decoupling the prompt processing and GT processing. Then, after each iteration, only prompts (selected from the prompt queue) need to be added to the batch to reach the TFS, i.e., fully utilize the GPU capacity. After the current GTs complete at the same time, GTs from the GT queue are added to the batch to fully utilize the KVC.}



\DEL{??if under-prediction, how many more iterations are needed -- draw a CDF-done
??if over-prediction, how many are overpredicted—get total and CDF-done
}
\IdeatoRead{?Now, some prompt tasks become token GTs, and we need to add prompt tasks to reach the TFS, then we need to make sure these newly produced token GTs and the token GTs of the added prompts will have complete with the existing token GTs, i.e., their response length is one token and two tokens less than that of the existing token GTs. It might not be easy to find such prompts (?this can be another paper)}

\DEL{If we just put the arrival
e.g., we need to fill up 1000 prompt tokens, checking requests in the queue sequentially takes time. if you group them based on prompt length, then you can quickly find the group with 1000 tokens. the same for the GT group below--need analysis figs to show this}

\vspace{-0.1in}
\subsubsection{Time-Synced Batching} \label{sec:comp2}\vspace{-0.0in}

This method enables the GTs in a group in a batch to commence and conclude execution simultaneously, eliminating the need for iteration-level scheduling. 
\sys groups the GTs with the same predicted RL in the GT queue. \DEL{When selecting GTs to be included in a batch, \sys selects GTs sequentially until their KVC demand reaches the available KVC space. }
During execution, a GT group may complete earlier than other GT groups. Then, the responses of the GTs in the group are returned to the users, and other GT groups in the queue are selected to add to the batch to fully allocate the KVC. 

\DEL{\begin{wrapfigure}{c}{4.5cm}\vspace{-0.0in}
    \centering
\includegraphics[width=0.52\columnwidth,height=0.08\textheight]{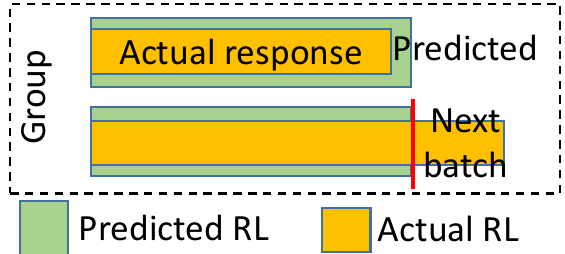}\vspace{-0.0in}
    \caption{Batch same-RL GT.}
    \label{fig:Prediction}\vspace{-0.0in}
\end{wrapfigure}}

Given the autoregressive nature of LLMs, predicting the RL of a request poses a significant challenge. 
Given the high perception ability of LLMs to understand input prompts and generate outputs, we use an LLM to predict the RL~\cite{Zheng2023ResponseLP} based on the input prompt. When a request arrives at the queue, \sys uses the RL predictor (e.g., built upon the OPT-13B model and explained in \cref{sec:analysis}) running in another server to predict its RL~\cite{Zheng2023ResponseLP}. This RL predictor can be used for many inference servers, and it undergoes continual retraining in the system. The LLM predicts a request's RL concurrently during its waiting and PT processing. Therefore, \sys is suitable for the scenario where the sum of these two time latencies is no less than the RL prediction latency, which is true in our scenario with a high arrival rate as indicated in the Table~\ref{tab:trace-table}. In this case, the prediction latency cost ($\approx0.921s$ in our analysis) is unlikely to affect JCT. \looseness=-1

\DEL{Predicting response lengths enables us to allocate KVC to a GT based on its anticipated RL.} 

\DEL{Conducting scheduling after each iteration takes certain time (e.g., 8\% of the iteration time). As indicated in ??, because of the scheduling, ??\% of GPU resources will be wasted. Therefore, it is necessary to avoid the iteration-level scheduling. However, on the other hand, request-level scheduling has the problem of delaying the completed requests, waiting for uncompleted requests, and delayed waiting for new requests. To tackle this challenge, we propose a method that batches GTs with similar response lengths. Therefore, these GTs can start at the same time, and complete the execution at approximately the same time, eliminating the need for returning after each iteration to execute iteration-level scheduling.

However, due to the autoregressive feature of LLM, a formidable challenge here is how to predict the response length of a request. For this purpose, we reply on LLM. The inputs include...?? We classify the response lengths to a certain range levels. For example, ...??

Currently, using \Orca's KVC allocation mechanism, the batch size is limited and much memory is wasted as many requests won't use up the allocated KVC. Predicting the response lengths also enable us to allocate KVC based on the predicted response length, thus avoiding the problems. Therefore, when we select GTs to be input to a batch, we aim to let the GTs' responses to fill up the KVC.

It is possible that we cannot find GTs with similar response lengths to fill up the KVC. Then, we need to find another group of GTs with another response length. In this case, during the execution, a GT group may complete earlier than other groups, and then this group returns back and another group with similar response length is selected and entered to the batch to fully utilize the KVC.
}



\noindent\textbf{Misprediction.}
Underestimation and overestimation can occur, with underestimation being more critical as it leads to request preemptions, while overestimation only wastes some KVC space. O\ref{RLprediction} suggests padding, reserved KVC, and offload-free preemption as effective solutions. Based on this, we propose a method to address underestimation effectively.\looseness=-1
Based on this, we propose an effective method to handle underestimation.\looseness=-1

\DEL{There could be underestimation and overestimation. Underestimation is more critical than overestimation since the request do not have enough KVC space and may have to be offloaded to the main memory, while overestimation just wastes some KVC space. O\ref?? and O\ref?? show that adding buffer and preempting the request are not effective regarding response latency and KVC use. We then propose another method.
}


In the method, the predicted RL is increased by a certain percent. 
In addition, when a request encounters an under-provisioning, it first tries to use the reserved KVC. If the reserved KVC is not enough, this request stops processing along with other completed requests in the batch. By subtracting its currently generated response length from its predicted RL, we obtain its new predicted RL $L_{new}$. It will be grouped with other GTs with the predicted RLs equal to $L_{new}$. \DEL{to expedite the completion of this request. This allows the request to release its KVC earlier.}For instance, 
when the first GT ($r_1$) and the second GT ($r_2$) reach their predicted RL, $r_1$ completes but $r_2$ does not, and then both GTs are returned. 
Subsequently, $r_2$ will be grouped with other GTs that share the same predicted RL as its $L_{new}$ to be scheduled. This request won't wait long in the queue because it occupies a certain KVC and will be given a higher priority in the queue ordering method, as presented in the next subsection.\looseness=-1

\DEL{Next, additional GT groups will be selected using the previous method to fully utilize the KVC space. The last GT group may split, and only a portion of the GTs is added to the batch if the KVC's spare space cannot accommodate the entire group.}

\DEL{The request that needs more KVC stops running along with other requests. Based on its currently generated response length, we predict its subsequent response length $L_{new}$ (??you need to use "currently generated response length" as an input, in order to use the same LLM for the prediction, you had better keep this input and set it 0 for initial requests). Then, in the next GT task scheduling, we choose GTs that have predicted response length equalling to $L_{new}$ in order to complete this request soon so that it can release its KVC earlier.}

\IdeatoRead{(??-you need to show measurement figs for "if not considering one factor, what would be the result" in order to show the factor needs to be considered.).??--by this testing, you also need to determine the order of the importance or effectiveness of the 3 factors.
The GPU computation workload is determined by the prompt length. 
}

\vspace{-0.1in}
\subsection{Prompt and Generation Task Ordering}\label{sec:comp3}\vspace{-0.0in}
\DEL{As illustrated in Figure~\ref{fig:component1}, \sys maintains two waiting queues: one for prompt requests and another for GT requests. Processed prompt requests move to the GT waiting queue.
The responsibility of prompts is to fully utilize the GPU resources. }

\DEL{Recall that in each iteration, prompts are picked from the prompt waiting queue to be entered to the batch. The GTs are picked from the GT waiting queue to be entered to the batch after a GT group in the batch is processed.}

This approach determines how to order the tasks in the PT waiting queue and GT waiting queue, respectively. O\ref{thm2-sort} indicates that waiting PTs and GTs occupy varying amounts of KVC space, so we should prioritize running those that occupy more KVC in order to release their KVC earlier.


\DEL{O\ref{thm2-sort} also indicates that chunked PTs may have different occupied KVC sizes too. A long prompt may be chunked in processing~\cite{deepspeed-fastgen,Agrawal2023SARATHIEL}. After a prompt's chunks are processed, the KV values of the processed tokens are stored in the KVC until this job completes. In order to reduce the time duration of such occupied KVC, the prompts with more processed prompt tokens should be processed earlier. 
}

To order the tasks in each queue, we consider three factors in order: 1) JCT SLO, 2) occupied KVC space, and 3) predicted RL for GTs and prompt length for PTs. We set this order because the SLO of a request must be satisfied, and the KVC is a bottleneck compared to GPU. In addition, in order to quickly select tasks to fully allocate KVC or utilize GPU to save scheduling time, GTs with longer predicted RLs or PTs with longer prompts should be selected earlier.

\DEL{We classify, order and select the GTs in the same manner as for PTs explained above. 

Therefore, when choosing the prompts to be processed, we consider three factors in order of their importance: 1) SLO, 2) occupied KVC space, and 3) prompt length. }


Each task has a deadline to satisfy its SLO. 
For each factor, we set up certain magnitude ranges and order tasks accordingly (e.g., 0.2-0.5s, 0.5-2s, $>$2s for the deadline, and 0-128, 128-256, 256-384, 384-512 for the predicted RL). 
First, we order the tasks in ascending order of the deadline. Then, within each deadline range, we further order the tasks in descending order of their occupied KVC sizes.  Next, within each occupied KVC size range, we order the tasks in descending order of the predicted RL for GTs or prompt length for PTs. Consequently, when selecting GTs or PTs, \sys picks up the tasks in sequence and uses the binary search to find a task with the predicted RL or prompt length close to the required length to fully allocate the KVC or fully utilize the GPU.\looseness=-1 


\DEL{Concerning GTs, the objective is to execute same-RL GTs, reducing the need for scheduling during execution. }

\DEL{\DEL{Based on O\ref{thm2-sort}, in order to let the job occupying more cache space to complete and release its KVC earlier, we give a higher priority to the waiting GTs that occupy more KVC space. We classify response lengths into different ranges (e.g., 32-64, 64-128, 128-256), and also classify GTs' SLOs and occupied KVC to different ranges.} GTs are first ordered in ascending SLO ranges. Within the same SLO range, GTs are ordered in the descending order of their occupied KVC space. Further, within the same occupied KVC space range, the same-RL GTs are grouped.}

\DEL{When a GT group completes, \sys selects GT groups in order from the ordered list. Among the GT groups, \sys uses the binary search to find a group to fully utilize the KVC. That is, the sum of the GTs' predicted RLs in the group equals to the available KVC space. 
}

\DEL{We pick tasks from the tightest SLO to the loosest SLO, select tasks within each SLO range based on occupied KVC range, and then select same-RL GTs range. (?the following may already in the previous section)In the ideal case, all GTs in a batch have the same response length, so the GTs in one batch complete at the same time. In scenarios with different GT groups having distinct response lengths, when a group of same-RL GTs range complete, the batch is returned and new GTs are selected to fill the batch using the aforementioned method. 
}



\DEL{When a batch completes, we need to pick up GTs. Among the GTs with the tightest SLO range, the GTs are sequentially selected. In the ideal case, we select the tasks from the tightest SLO to the loosest SLO. In each SLO range, we select the tasks with the same occupied KVC range, and then select the tasks with the same response length. In the ideal case, all GTs have the same response length, so the requests in one batch complete at the same time. Sometimes, the requests in the batch may not have the same response length. Then, when a group of same-RL GTs complete, the batch is returned and new GTs are selected to be put into the batch using the above method.}

\DEL{In a more complex situation, some GTs must start running to satisfy their SLOs. To handle this case, we classify the GTs into two categories: urgent and non-urgent. GTs falling under the urgent category require immediate execution to fulfill their SLO requirements, while those in the non-urgent category can wait. When a batch completes, we prioritize picking up GTs from the urgent category and employ the previously outlined method to fill up the batch efficiently. This approach ensures timely processing of urgent GTs while maintaining an optimal scheduling strategy.}


\IdeatoRead{(??for GTs, you need to measure the effectiveness of SLO, predicted response length, and the occupied KVC too, and order the effectiveness of considering them )}

\DEL{\begin{algorithm}[t]
    \SetAlgoLined
    \LinesNumbered
    \SetCommentSty{small}
    \SetKwInOut{Input}{Input}
    \SetKwInOut{Output}{Output}
    \Input{A newly received prompt task $p$; Sorted prompt waiting queue ($Q_P$).} 
\Output{Sorted prompt waiting queue}
Use binary search to find SLO-prompt-group for $p$\\
In $p$'s SLO-prompt-group, use binary search to find Cache-prompt-group for $p$\\
In $p$'s Cache-prompt-group, use binary search to find Length-prompt-group for $p$\\
    \caption{Pseudocode of $\mathrm{Insert(p,$Q_P$)}$.}
    \label{alg-checkCache}
    \end{algorithm}
}




\vspace{-0.1in}
\subsection{Putting All Together}\vspace{-0.0in}
Algorithms~\ref{alg-checkCache} presents the pseudocode of \sys, integrating the three methods and mapping its steps to those in Figure~\ref{fig:component1}. Upon returning a batch, if a GT group completes, \sys fetches GT groups from the GT queue to replenish the batch. First, GT groups are selected to fully utilize the KVC (line 2), followed by the iterative selection of hosted GTs for the chosen GTs using the KVC pipelining method (line 3). Next, PTs are chosen from the PT queue to meet the TFS (line 5). The batch is then executed (line 6), and after execution, GTs generated from prompts and preempted GTs are reinserted into the GT queue (lines 7-9).

\DEL{by calling Algorithm~\ref{alg-checkCache1}.  Algorithm~\ref{alg-checkCache1} 
uses binary search to find the GT's location in the queue based on its SLO, occupied KVC and predicted RL.}  

\begin{algorithm}[t]
    \SetAlgoLined
    \LinesNumbered
    \SetCommentSty{small}
    \SetKwInOut{Input}{Input}
    \SetKwInOut{Output}{Output}
\footnotesize
    \Input{Returned batch; Sorted PT queue ($Q_P$); Sorted GT queue ($Q_G$); available KVC ($A_{KVC}$)}
    \Output{Form and execute a new batch}
\If{a GT group completes}{

 Select GT groups from $Q_G$ until $\sum L_{r}=A_{KVC}$ \circled{1} (\emph{SyncDecouple});\\
Iteratively select hosted GT groups from $Q_G$ for each hosting GT group \circled{2} (\emph{KVCPipe});\\ 
}
Select PTs from $Q_P$ until $\sum L_{p}=TFS$ \circled{3} (\emph{SyncDecouple});\\
Execute the batch \circled{4}\\
\tcc{Enter the GTs transformed from PTs to GT queue}
        \For{each newly generated GT and preempted GT $t$ }{
    Ordering($t$, $Q_G$); \circled{5} (\emph{Ordering});
        } 
 Return tokens generated from GTs to users;\\
Release KVC space of completed requests;

    \caption{Pseudocode of an iteration's execution.}
    \label{alg-checkCache}
    \end{algorithm}

\DEL{\begin{algorithm}[t]
    \SetAlgoLined
    \LinesNumbered
    \SetCommentSty{small}
    \SetKwInOut{Input}{Input}
    \SetKwInOut{Output}{Output}
    \Input{A newly received GT $t$; Sorted GT waiting queue ($Q_G$).} 
\Output{Sorted GT waiting queue ($Q_G$)}
Get the top SLO-GT-group for $t$\\
Get the Cache-GT-group for $t$\\
In the Cache-GT-group, use binary search to find Length-GT-group for $t$ based on $t$'s response length\\
    \caption{Pseudocode of $\mathrm{Insert(t,$Q_G$)}$.}
    \label{alg-checkCache1}
    \end{algorithm}
}

\DEL{
\begin{algorithm}[t]
    \SetAlgoLined
    \LinesNumbered
    \SetCommentSty{small}
    \SetKwInOut{Input}{Input}
    \SetKwInOut{Output}{Output}
    \footnotesize
    \Input{A newly received GT $t$; Ordered GT waiting queue ($Q_G$).} 
\Output{Ordered GT waiting queue ($Q_G$)}
\tcc{Use binary search to find $t$'s location}
Find the SLO-GT-group of $t$ based on its SLO\\
In $t$'s SLO-GT-group, find the KVC-GT-group for $t$ based on its occupied KVC\\
In $t$'s KVT-GT-group, find the RL-GT-group for $t$ based on its predicted RL\\
    \caption{Pseudocode of $\mathrm{Ordering(t, $Q_G$)}$.}
    \label{alg-checkCache1}
    \end{algorithm}
}

\DEL{\begin{figure}
    \centering
\includegraphics[width=1\columnwidth,height=0.145\textheight]{fig/KVPipeline}
    \caption{An example of KVC pipelining for two GTs.}
    \label{fig:KVPipeline}
\end{figure}

\begin{figure}
    \centering
\includegraphics[width=1\columnwidth,height=0.145\textheight]{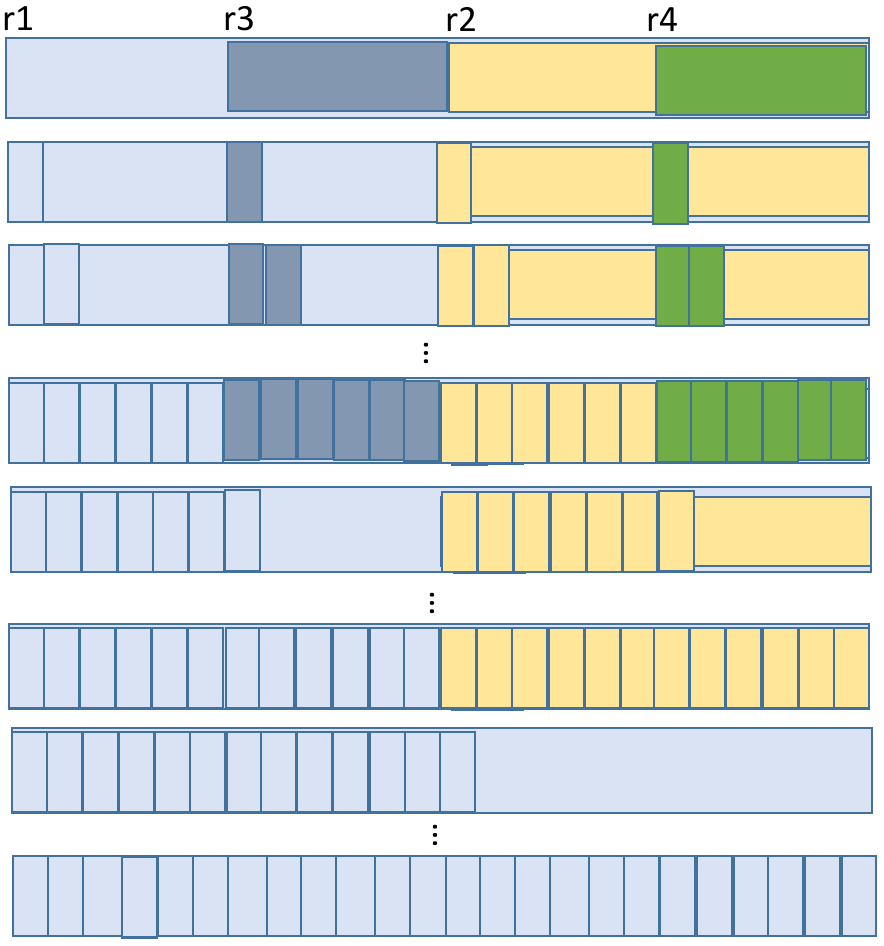}
    \caption{An example of KVC pipelining for four GTs.}
    \label{fig:KVPipeline1}
\end{figure}
?need to cite algorithm
}

\DEL{\newpage
\begin{figure*}[t]
\centering
\subfloat[Response latency for Alpaca(i.e., JCT).\vspace{-0.0in}\label{fig:exp-1-a}]{{\includegraphics[width=0.23\linewidth,height=0.112\textheight]{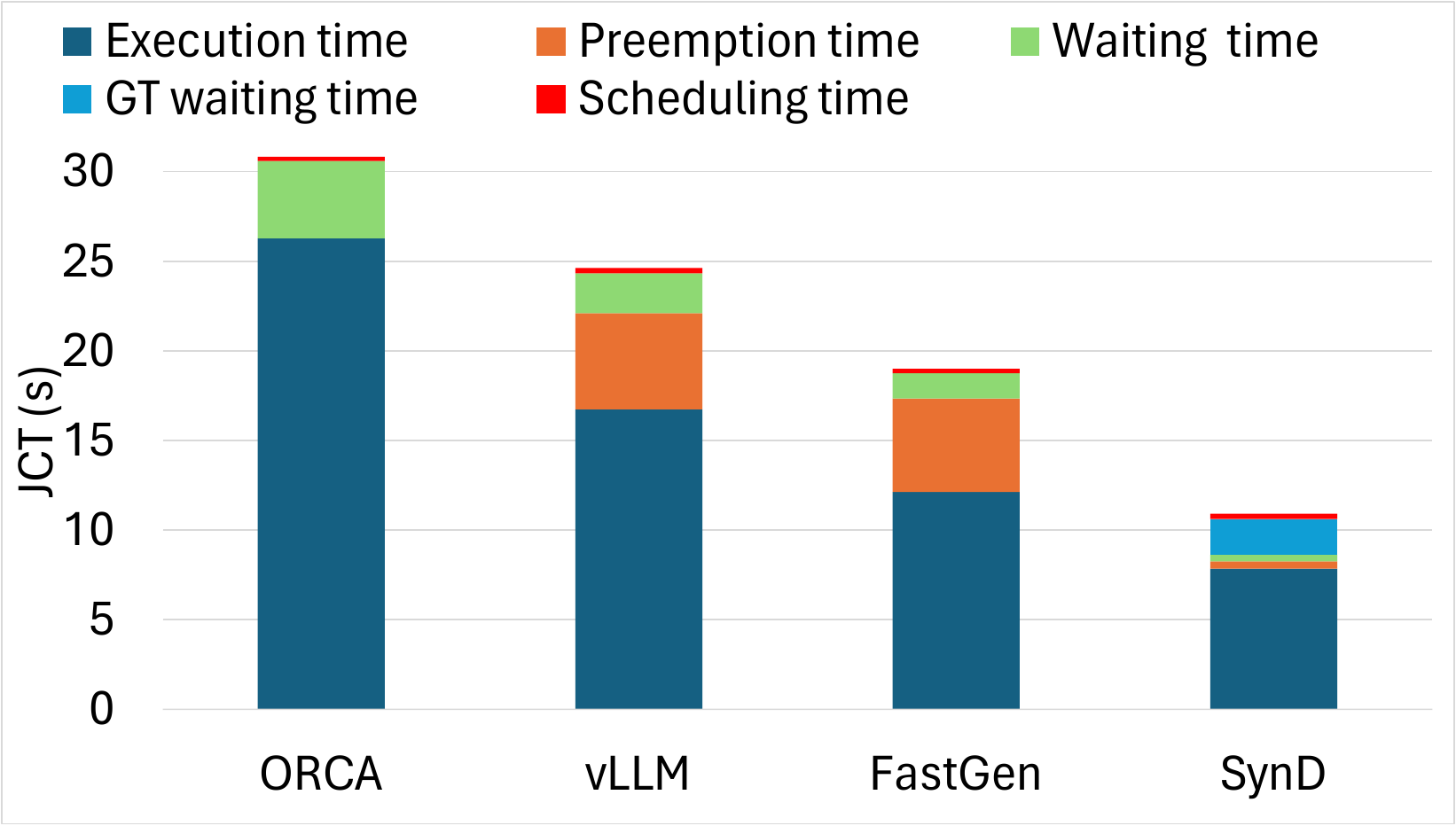} }}
\hfill
\subfloat[Response latency for ShareGPT(i.e., JCT).\vspace{-0.0in}\label{fig:exp-1-s}]{{\includegraphics[width=0.23\linewidth,height=0.112\textheight]{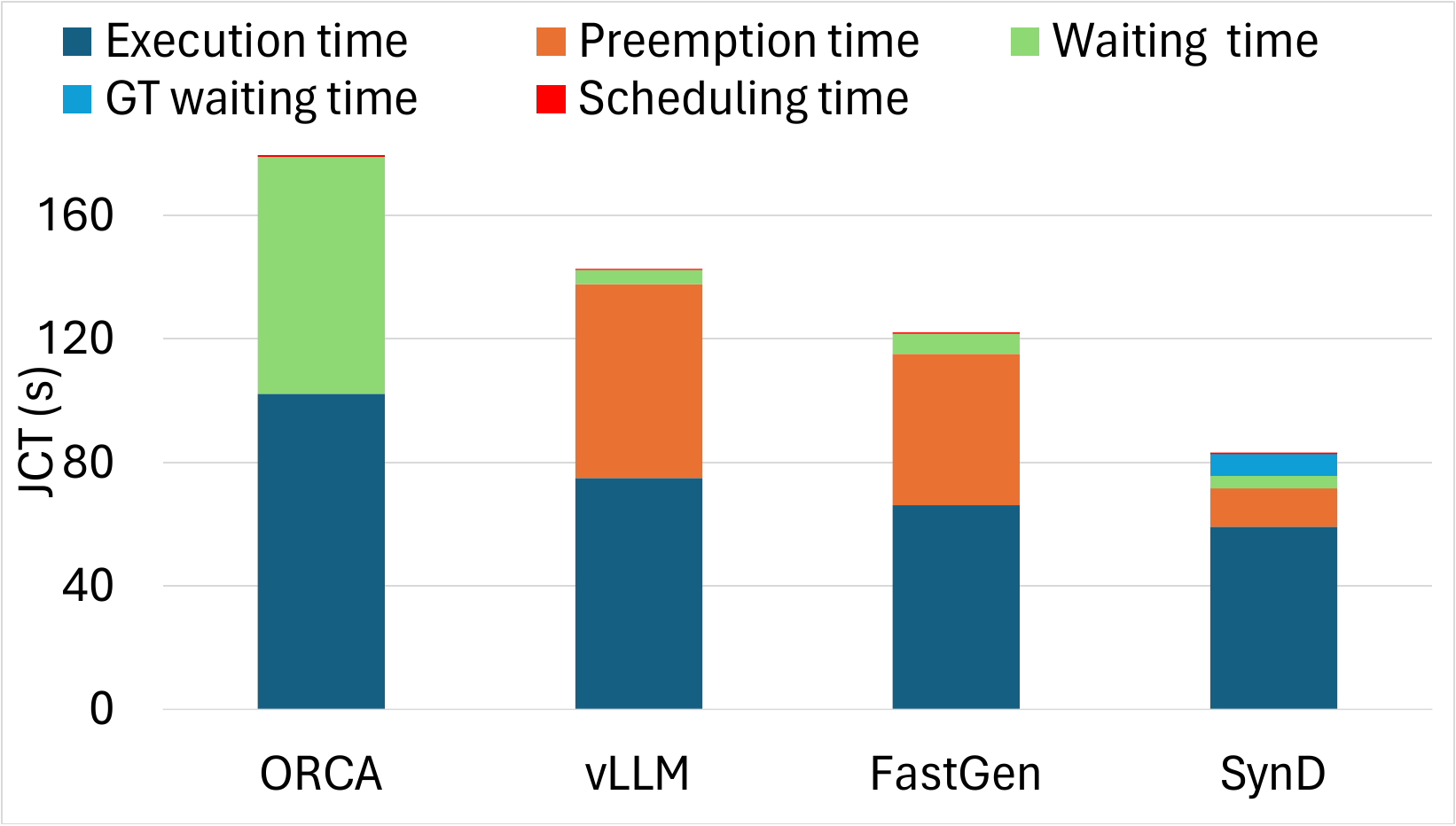} }}
\hfill
\subfloat[Response latency for BookCorpus(i.e., JCT).\vspace{-0.0in}\label{fig:exp-1-b}]{{\includegraphics[width=0.23\linewidth,height=0.112\textheight]{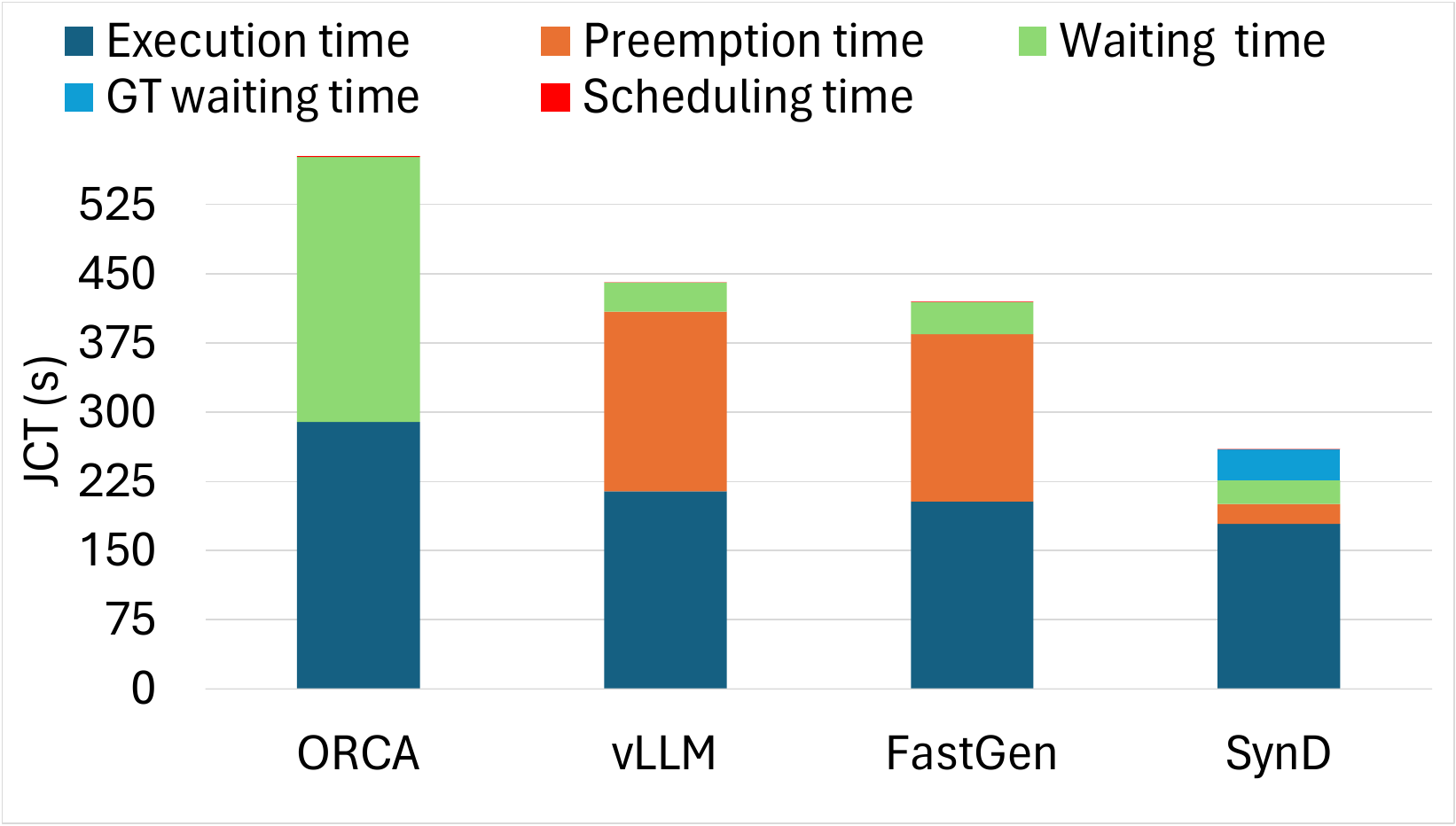} }}
\hfill
\subfloat[TBT.\vspace{-0.0in}\label{fig:tbt-1}]{{\includegraphics[width=0.23\linewidth,height=0.112\textheight]{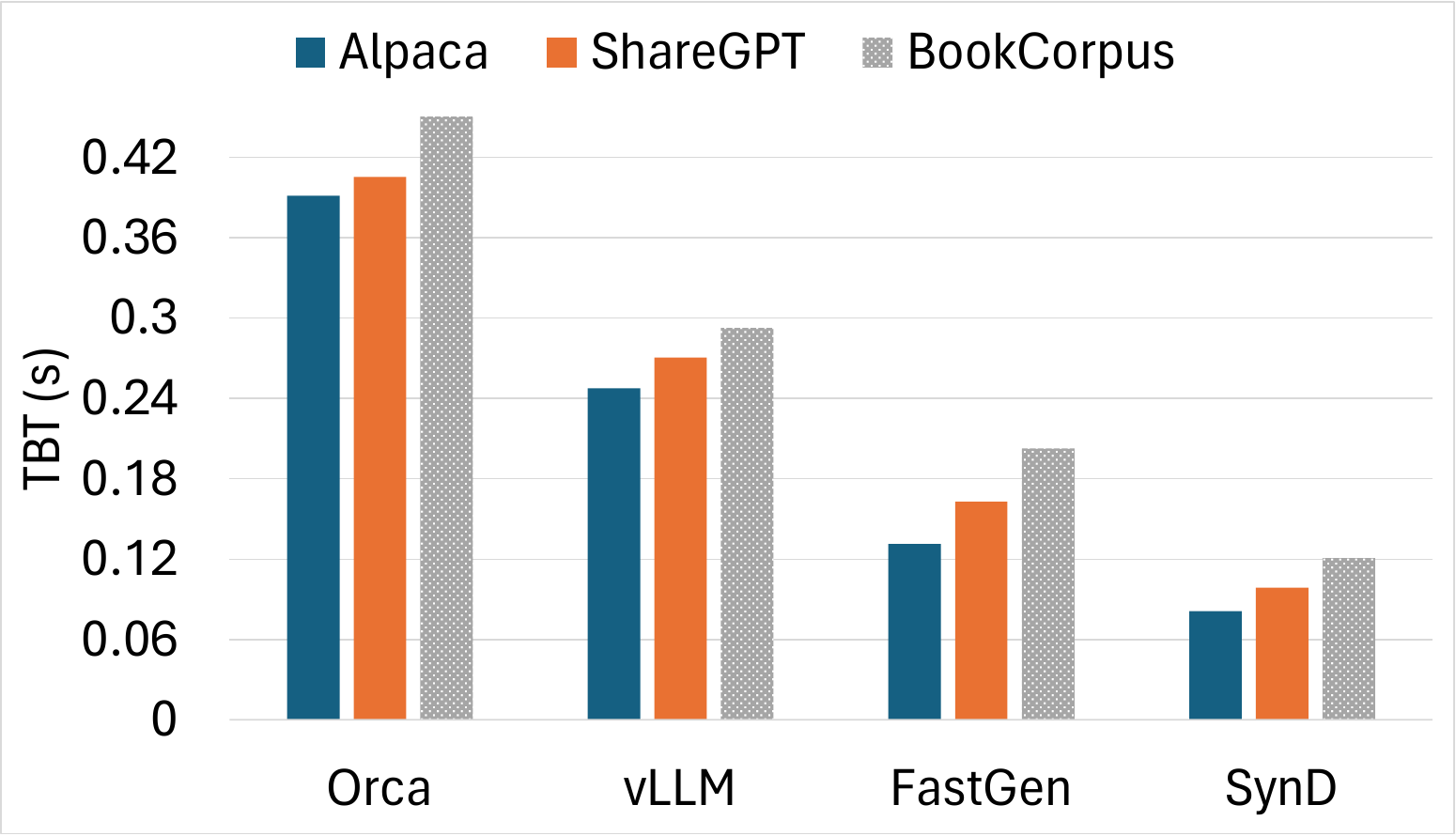} }}
\hfill
\DEL{\subfloat[Normalized latency.\vspace{-0.0in}\label{fig:exp-2}]{{\includegraphics[width=0.23\linewidth,height=0.112\textheight]{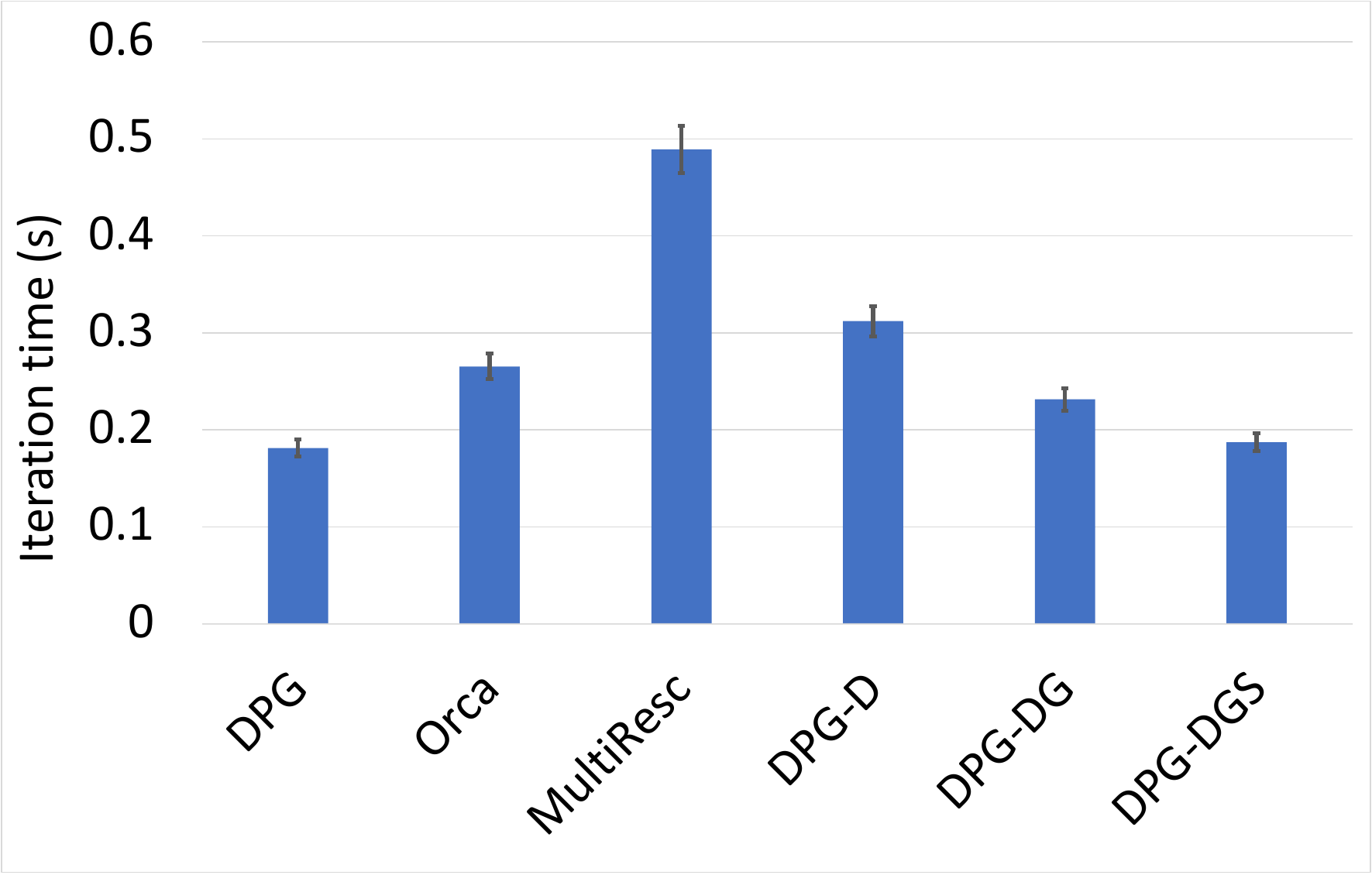} }}
\hfill}
\subfloat[SLO.\vspace{-0.0in}\label{fig:exp-slo}]{{\includegraphics[width=0.23\linewidth,height=0.112\textheight]{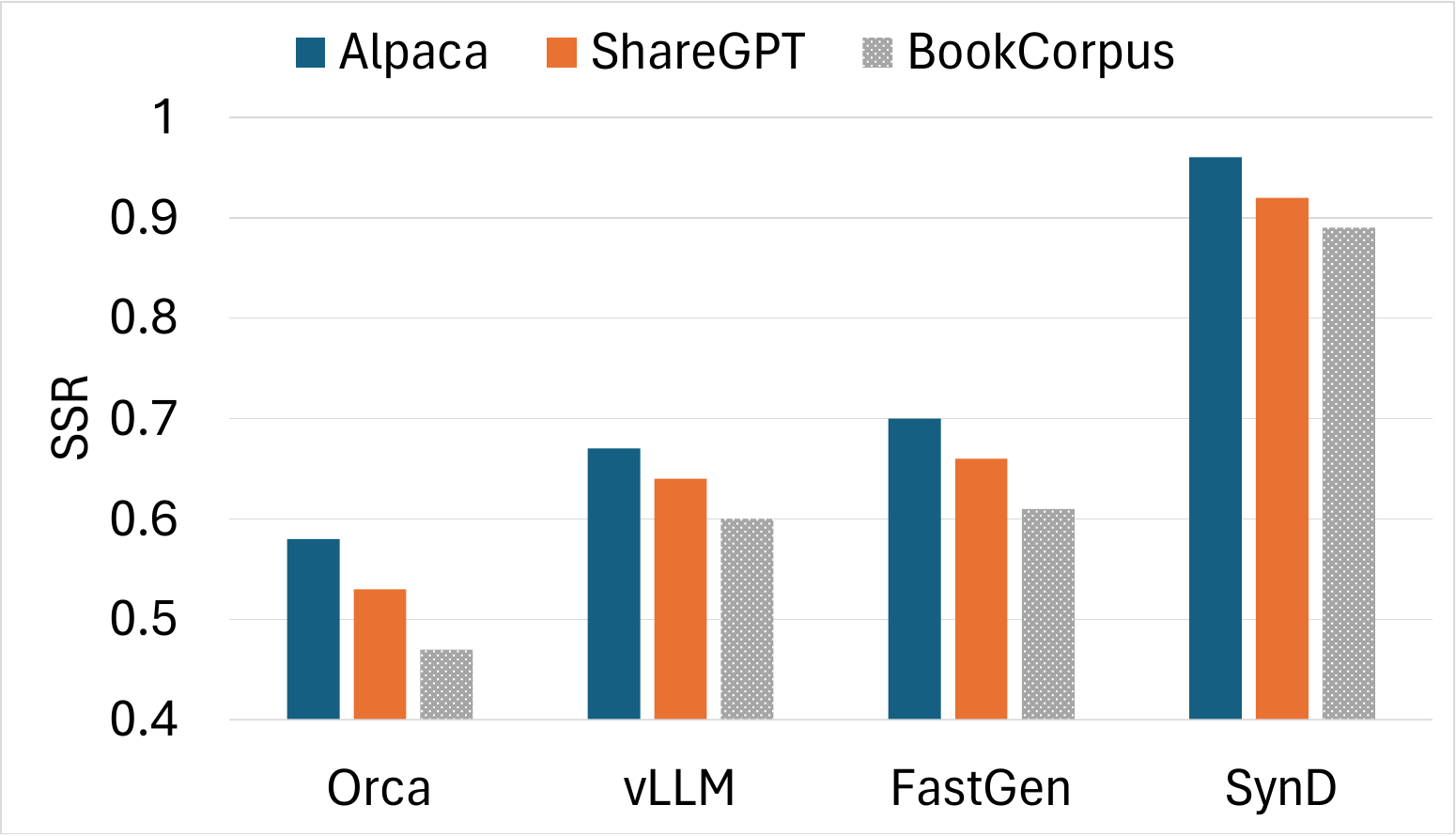} }}
    \hfill
\subfloat[Throughput.\vspace{-0.0in}\label{fig:exp-req}]{{\includegraphics[width=0.23\linewidth,height=0.112\textheight]{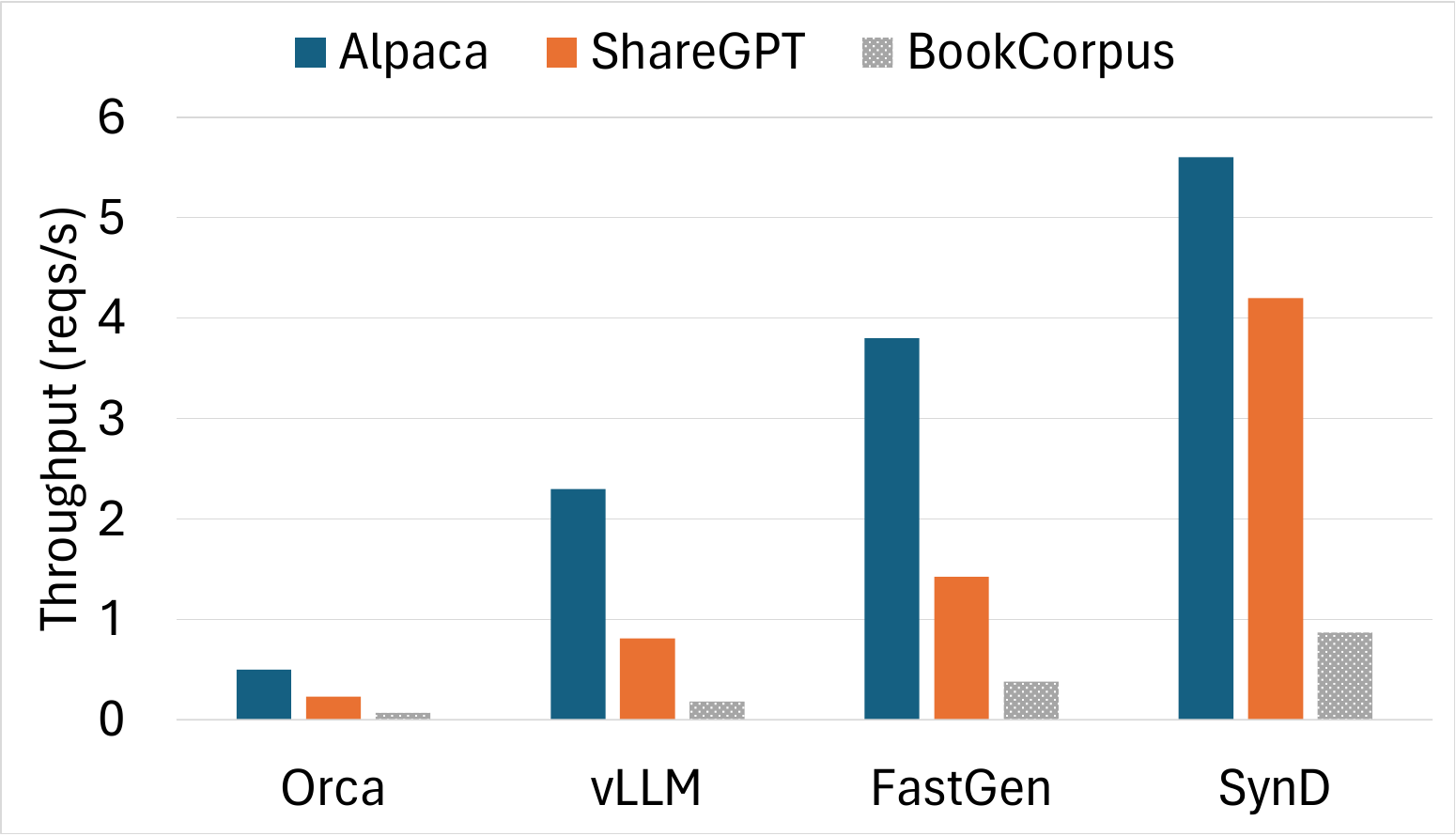} }}
    \hfill
\subfloat[Throughput (token/s).\vspace{-0.0in}\label{fig:exp-tokens}]{{\includegraphics[width=0.23\linewidth,height=0.112\textheight]{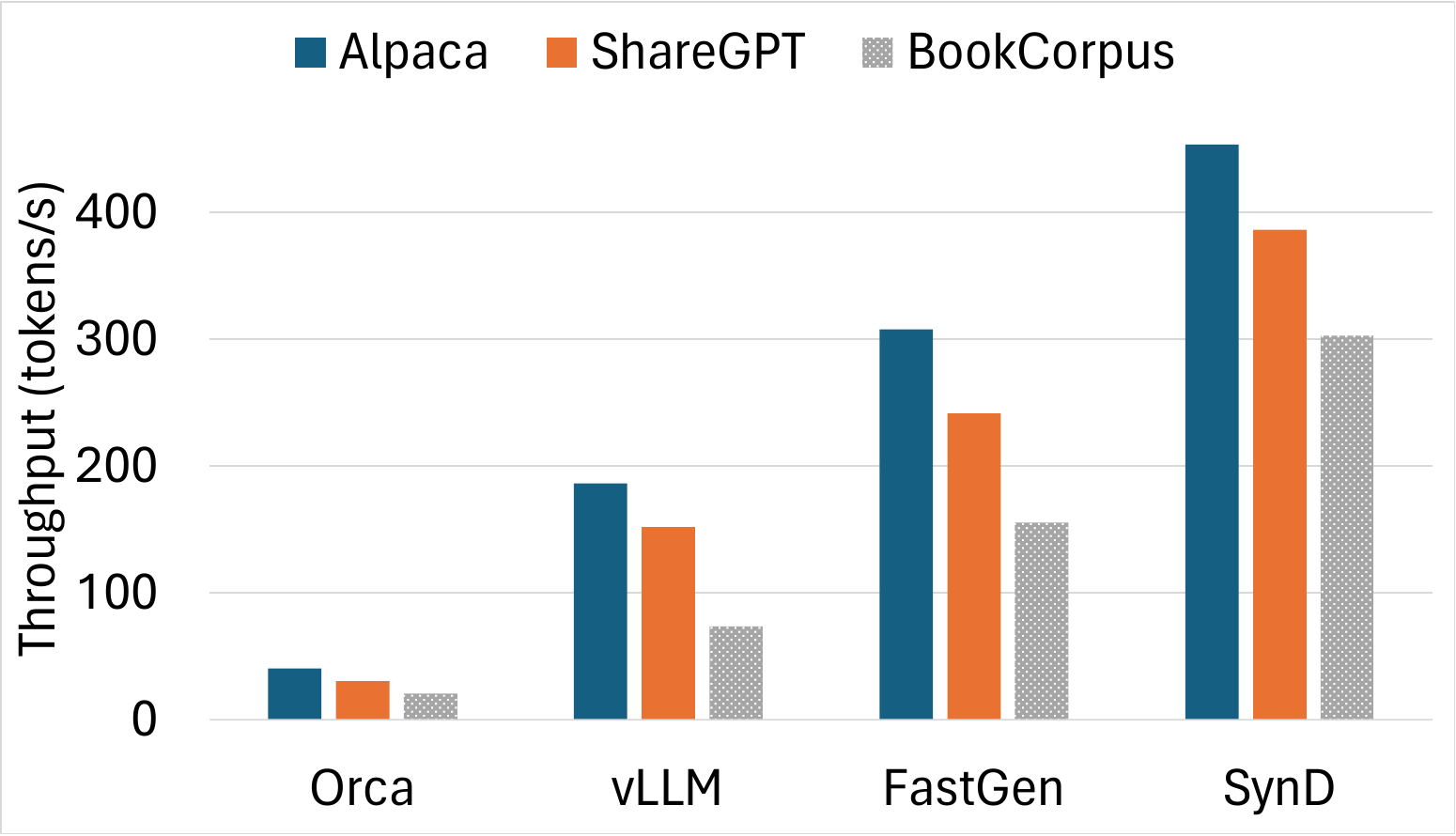} }}
\hfill
\subfloat[KVC utilization.\vspace{-0.0in}\label{fig:exp-kvc}]{{\includegraphics[width=0.23\linewidth,height=0.112\textheight]{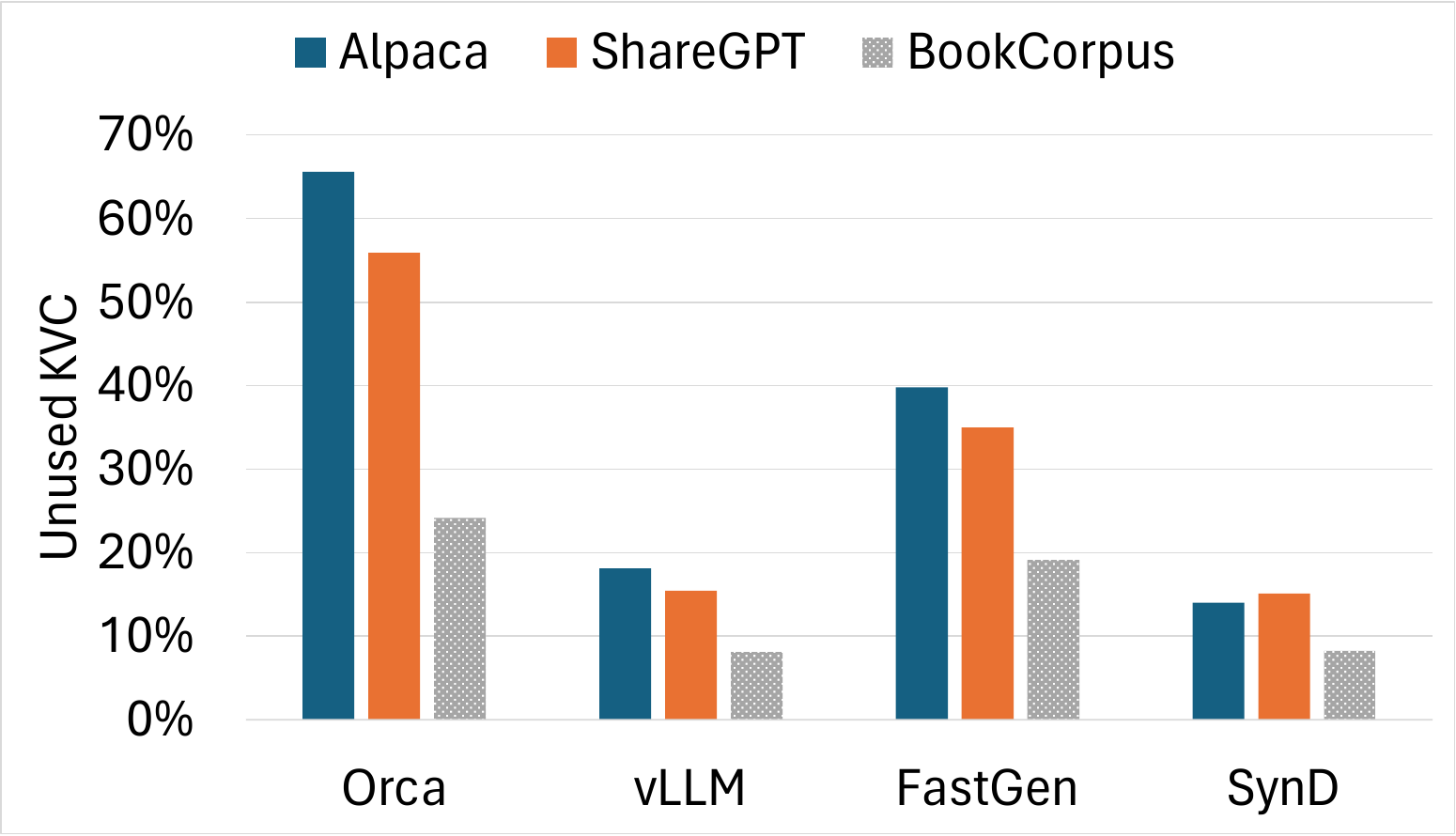} }}
    \hfill
\subfloat[KVC allocation failures.\vspace{-0.0in}\label{fig:exp-kvc-alloc}]{{\includegraphics[width=0.23\linewidth,height=0.112\textheight]{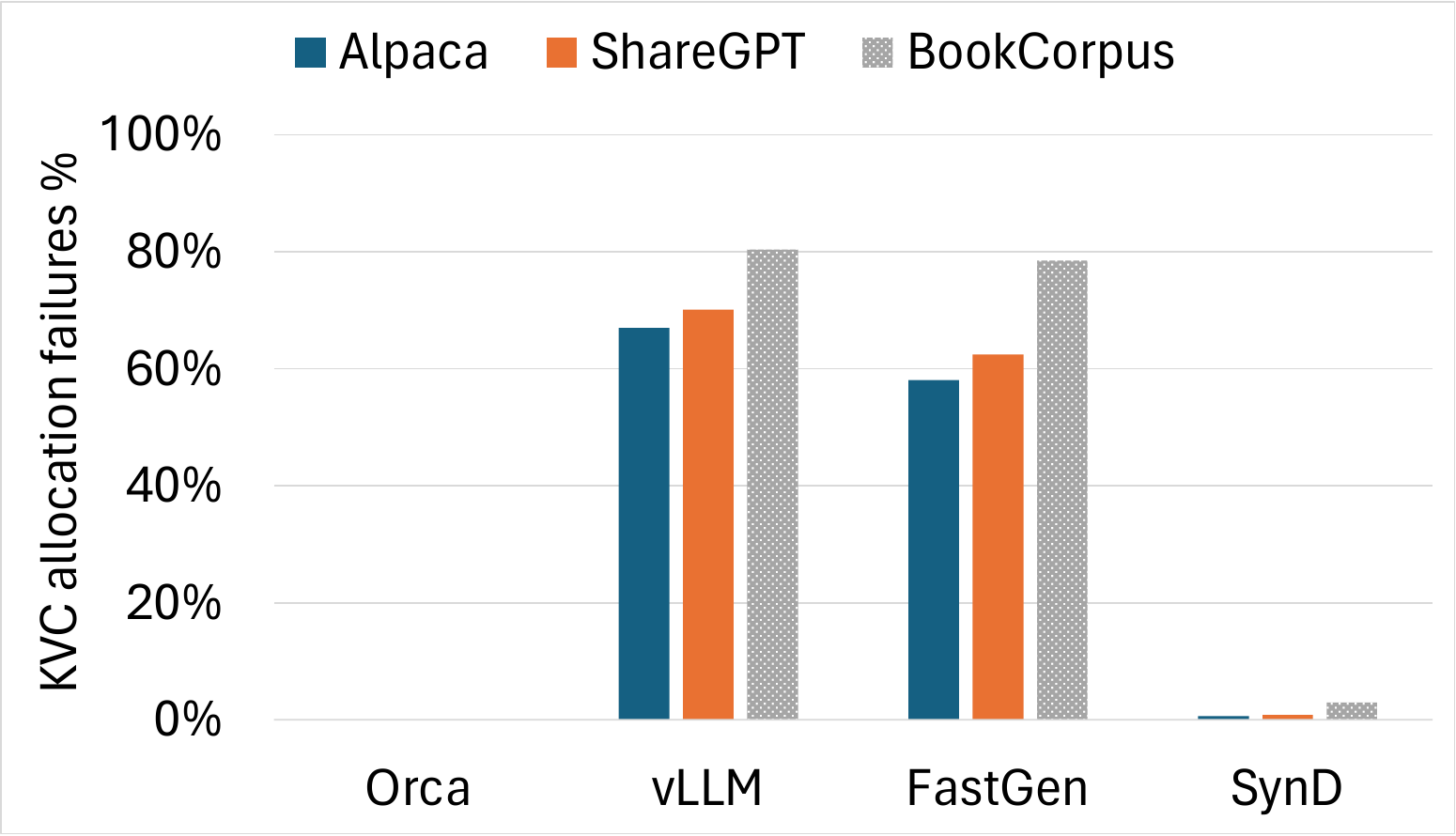}}}
    \hfill
    \subfloat[Scheduling time overhead.\vspace{-0.0in}\label{fig:exp-7}]{{\includegraphics[width=0.23\linewidth,height=0.112\textheight]{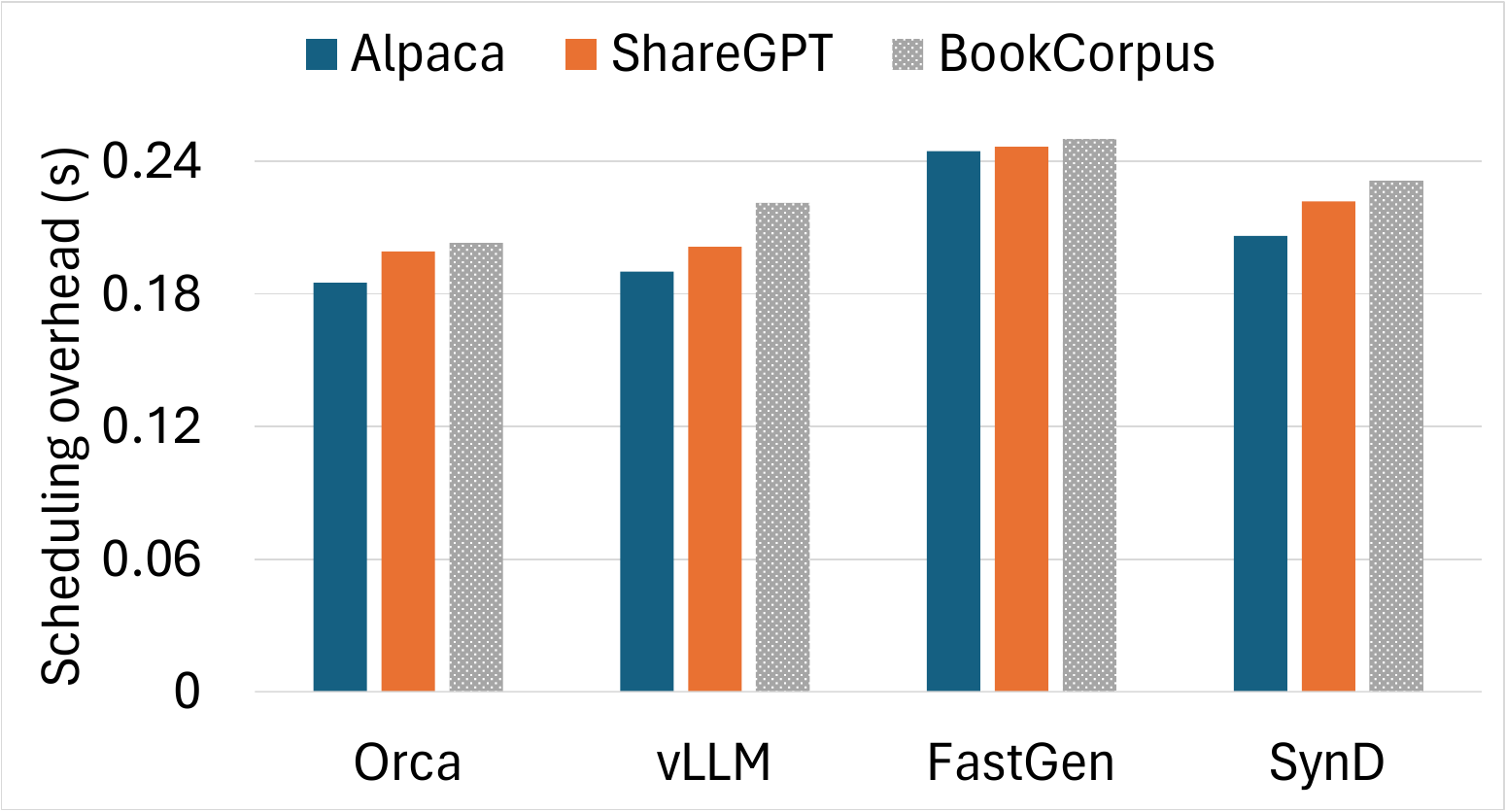} }}
    \hfill
    \DEL{\subfloat[Normalized latency vs. request rates.\vspace{-0.0in}\label{fig:exp-7}]{{\includegraphics[width=0.23\linewidth,height=0.112\textheight]{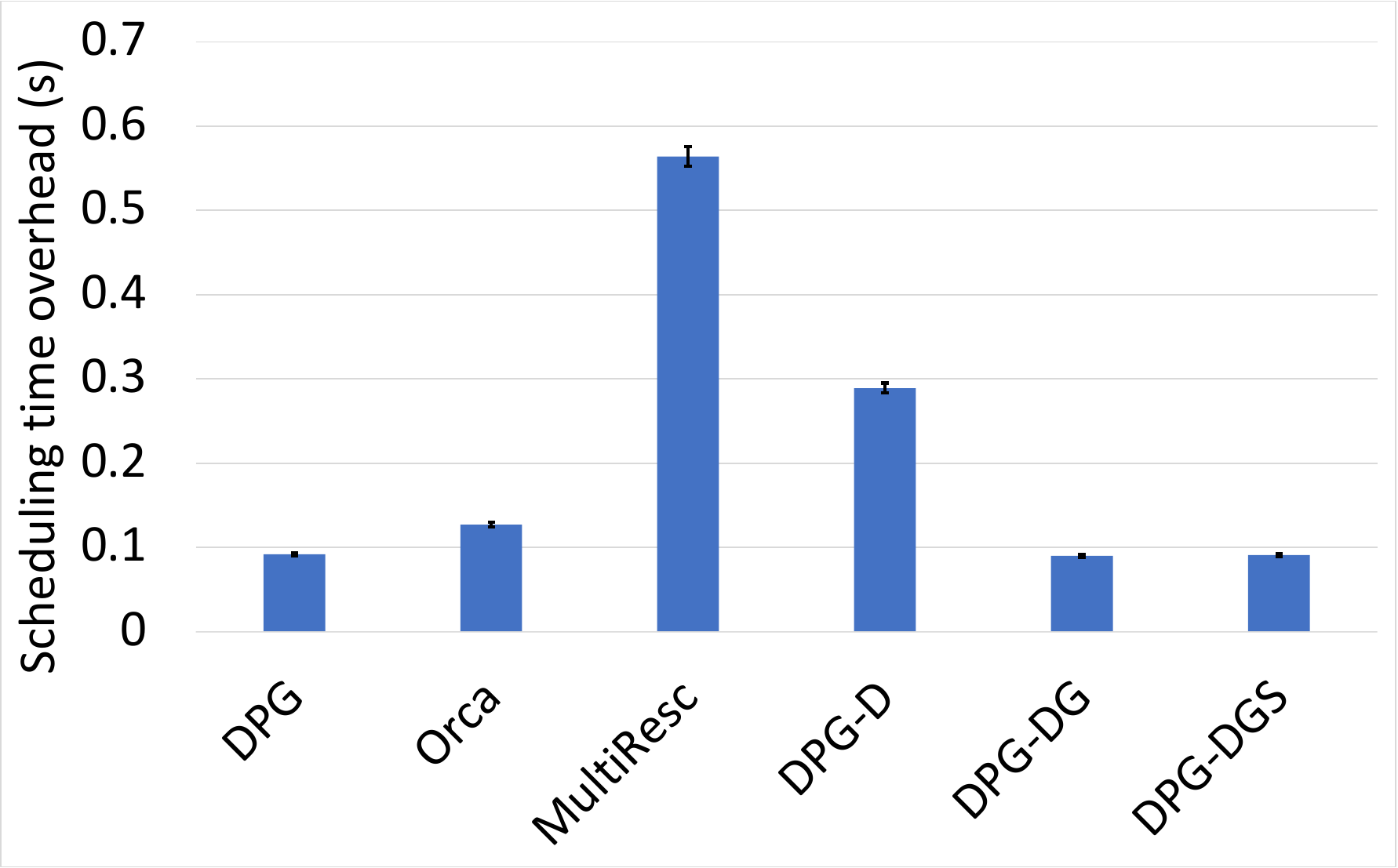} }}
    \hfill}
    \hfill
    \vspace{-0.0in}
   \caption{\small{Performance of different methods for the OPT-13B model for all traces (??each fig shows 3 traces, the same for the following).\vspace{-0.0in}}}%
    \label{fig:prompt-methods}
\end{figure*}}

\DEL{\begin{figure*}[t]
\centering
\subfloat[Response latency for Alpaca(i.e., JCT).\vspace{-0.0in}\label{fig:exp-1-a-175}]{{\includegraphics[width=0.23\linewidth,height=0.112\textheight]{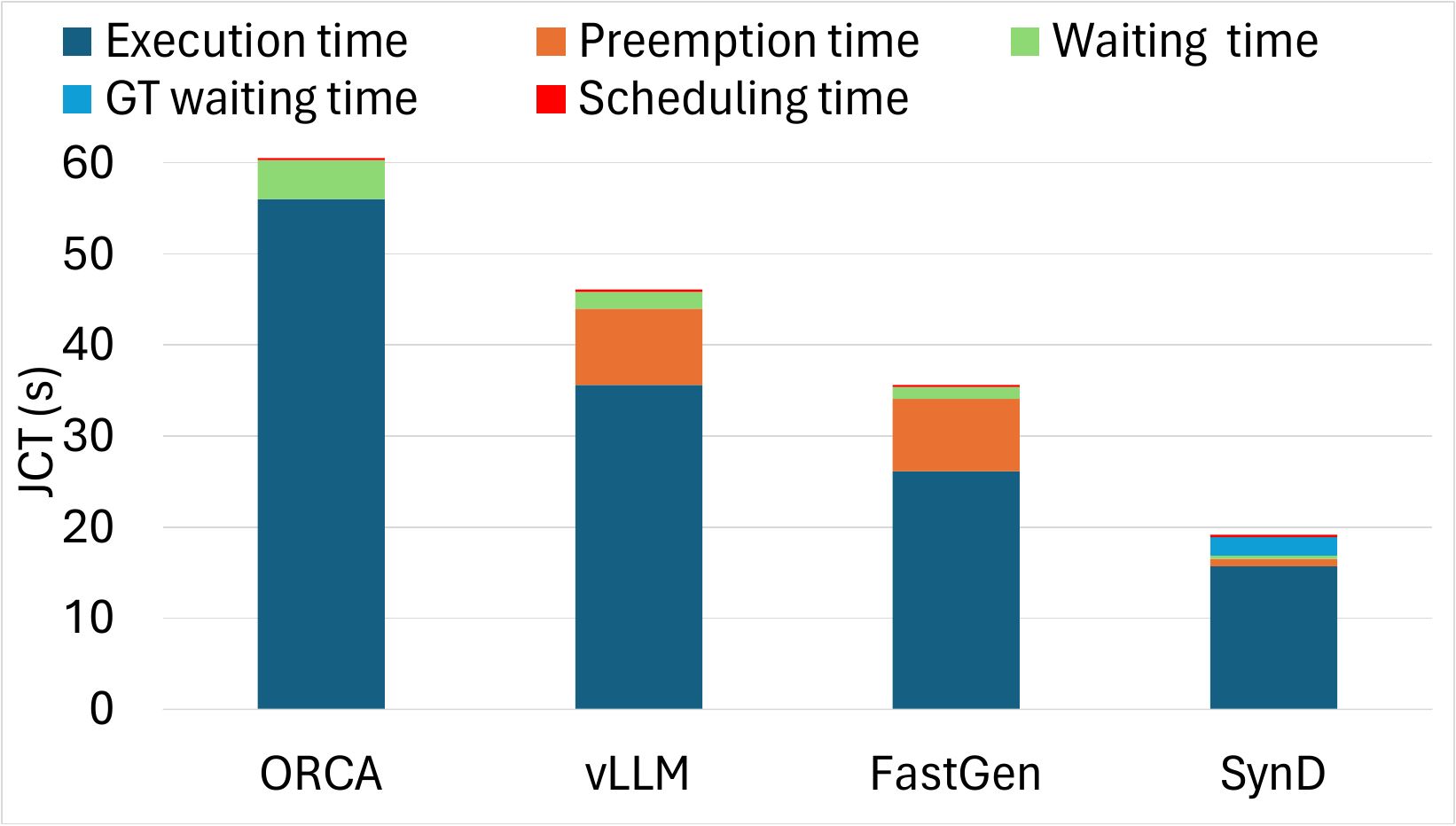} }}
\hfill
\subfloat[Response latency for ShareGPT(i.e., JCT).\vspace{-0.0in}\label{fig:exp-1-s-175}]{{\includegraphics[width=0.23\linewidth,height=0.112\textheight]{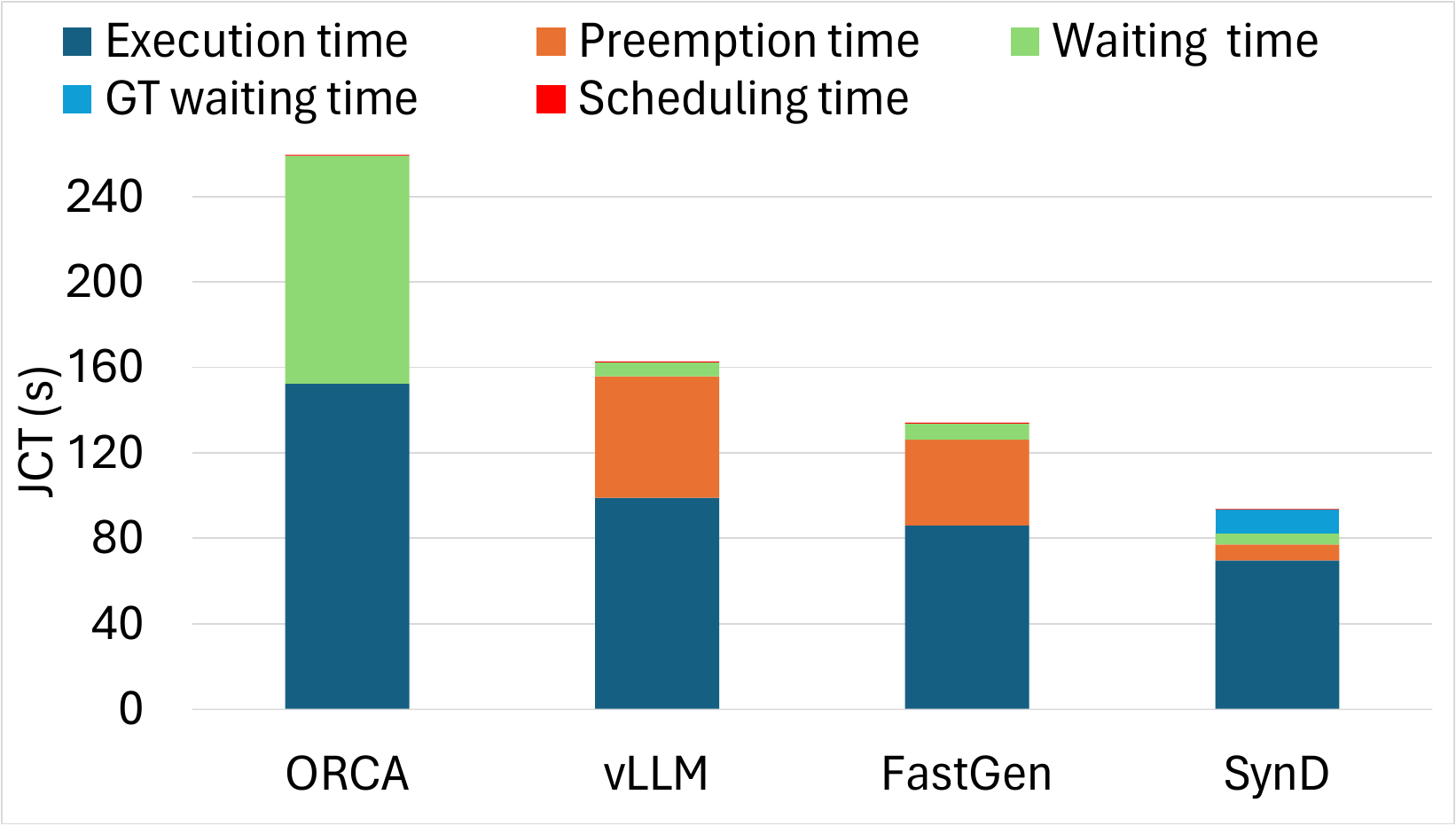} }}
\hfill
\subfloat[Response latency for BookCorpus(i.e., JCT).\vspace{-0.0in}\label{fig:exp-1-b-175}]{{\includegraphics[width=0.23\linewidth,height=0.112\textheight]{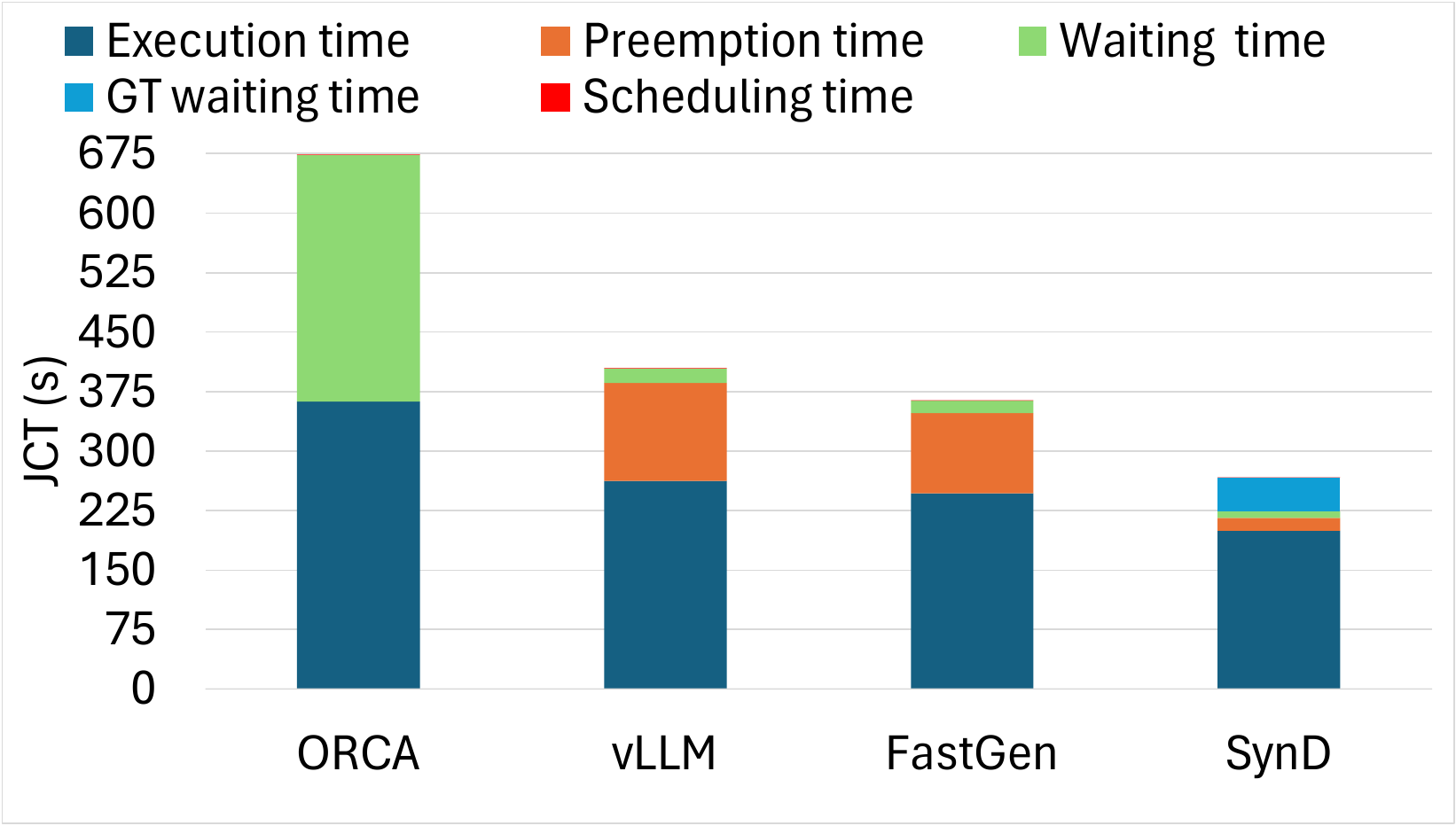} }}
\hfill
\subfloat[TBT.\vspace{-0.0in}\label{fig:tbt-1-175}]{{\includegraphics[width=0.23\linewidth,height=0.112\textheight]{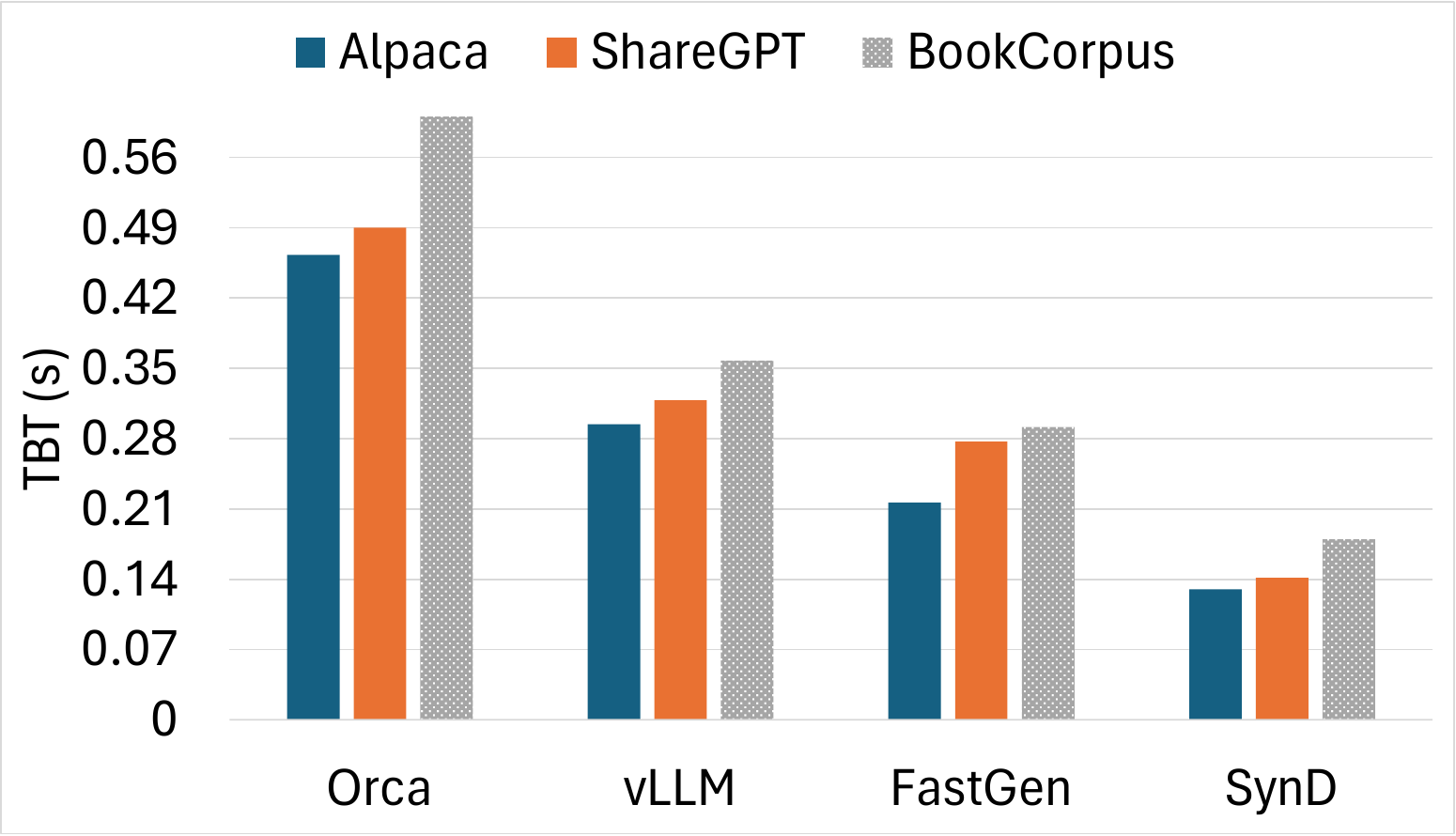} }}
\hfill
\DEL{\subfloat[Normalized latency.\vspace{-0.0in}\label{fig:exp-2}]{{\includegraphics[width=0.23\linewidth,height=0.112\textheight]{Fig/iteration-time-exp-up.pdf} }}
\hfill}
\subfloat[SLO.\vspace{-0.0in}\label{fig:exp-slo-175}]{{\includegraphics[width=0.23\linewidth,height=0.112\textheight]{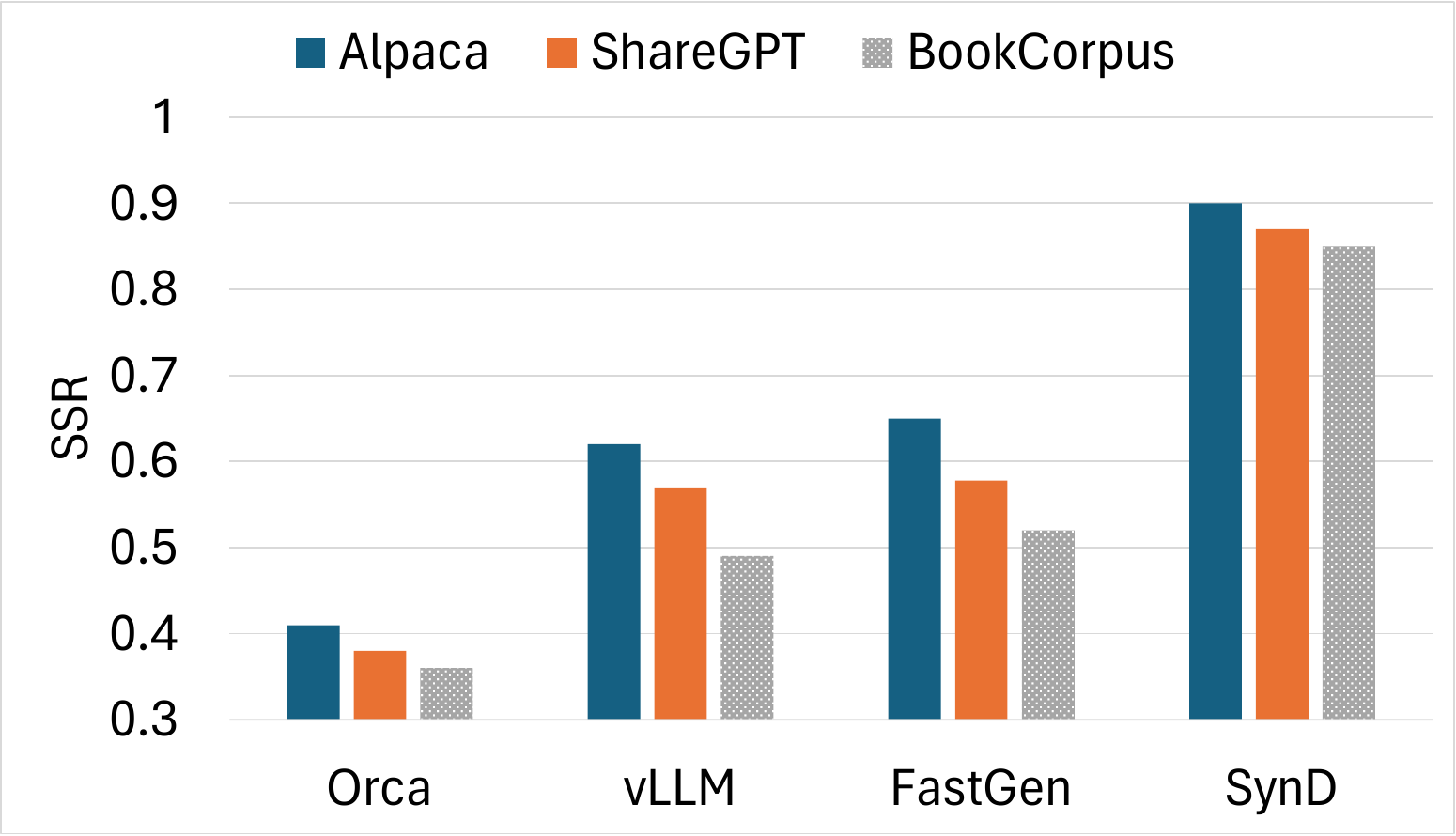} }}
    \hfill
\subfloat[Throughput.\vspace{-0.0in}\label{fig:exp-req-175}]{{\includegraphics[width=0.23\linewidth,height=0.112\textheight]{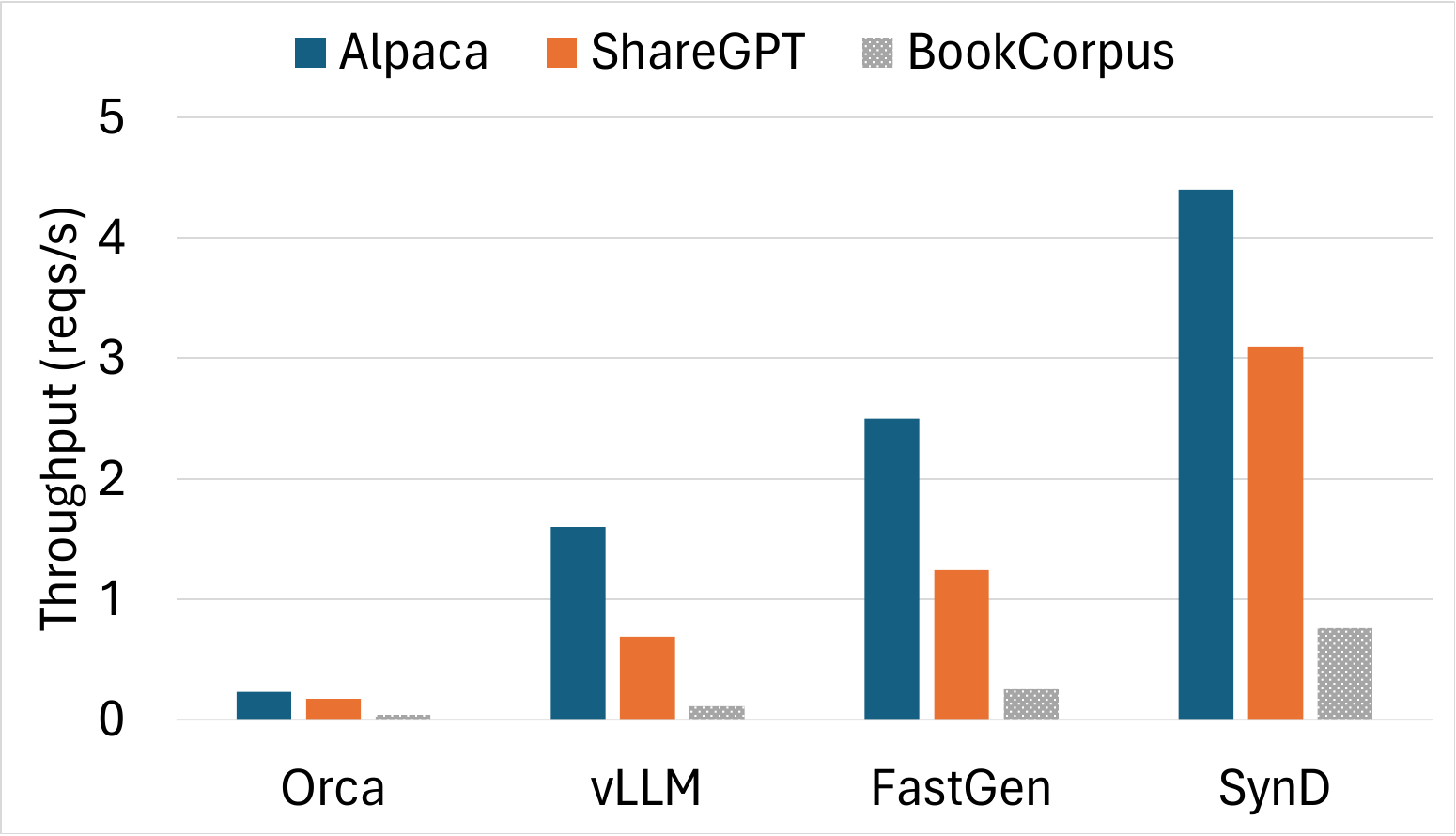} }}
    \hfill
\subfloat[Throughput (token/s).\vspace{-0.0in}\label{fig:exp-tokens-175}]{{\includegraphics[width=0.23\linewidth,height=0.112\textheight]{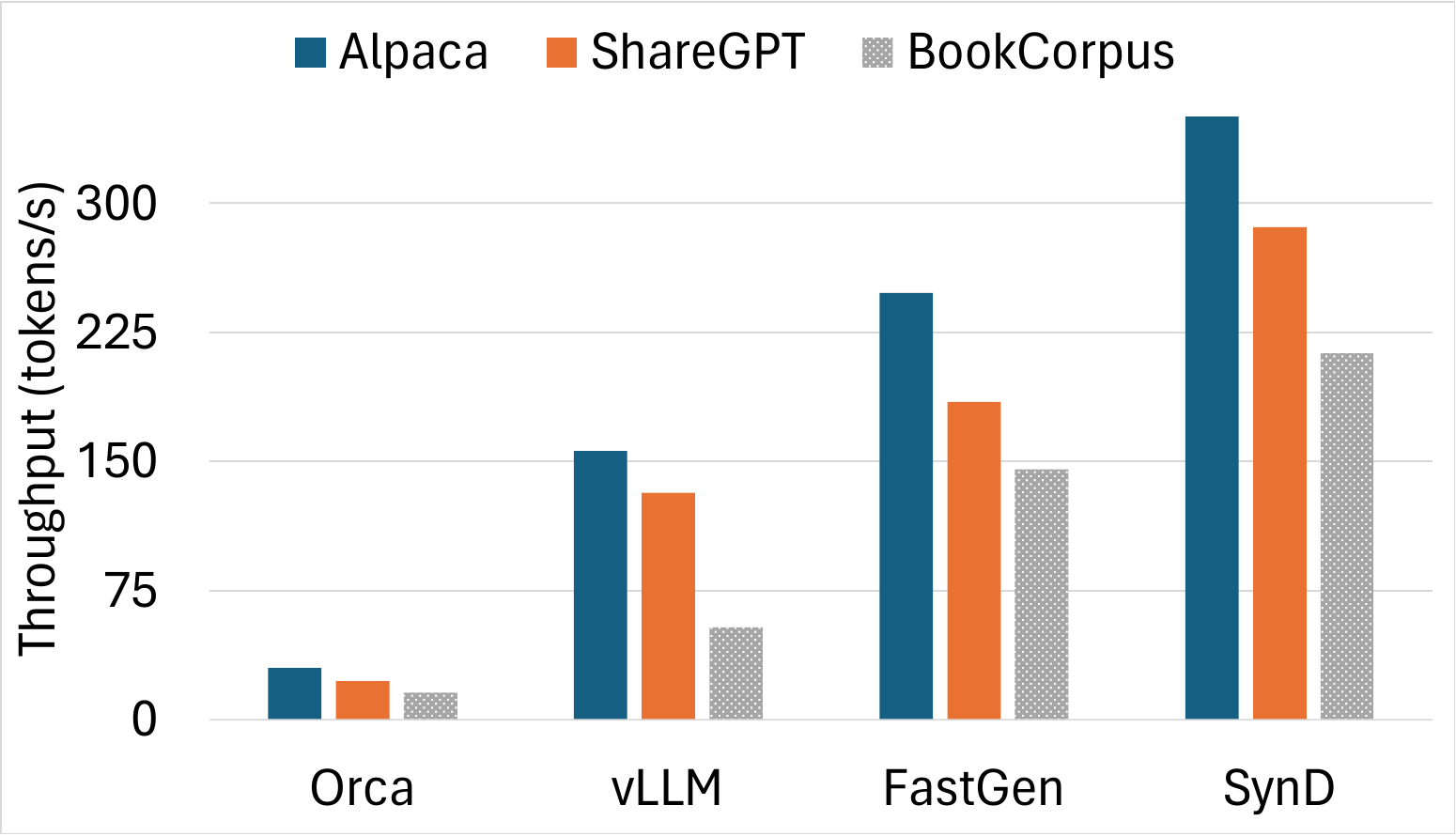} }}
\hfill
\subfloat[KVC utilization.\vspace{-0.0in}\label{fig:exp-kvc-175}]{{\includegraphics[width=0.23\linewidth,height=0.112\textheight]{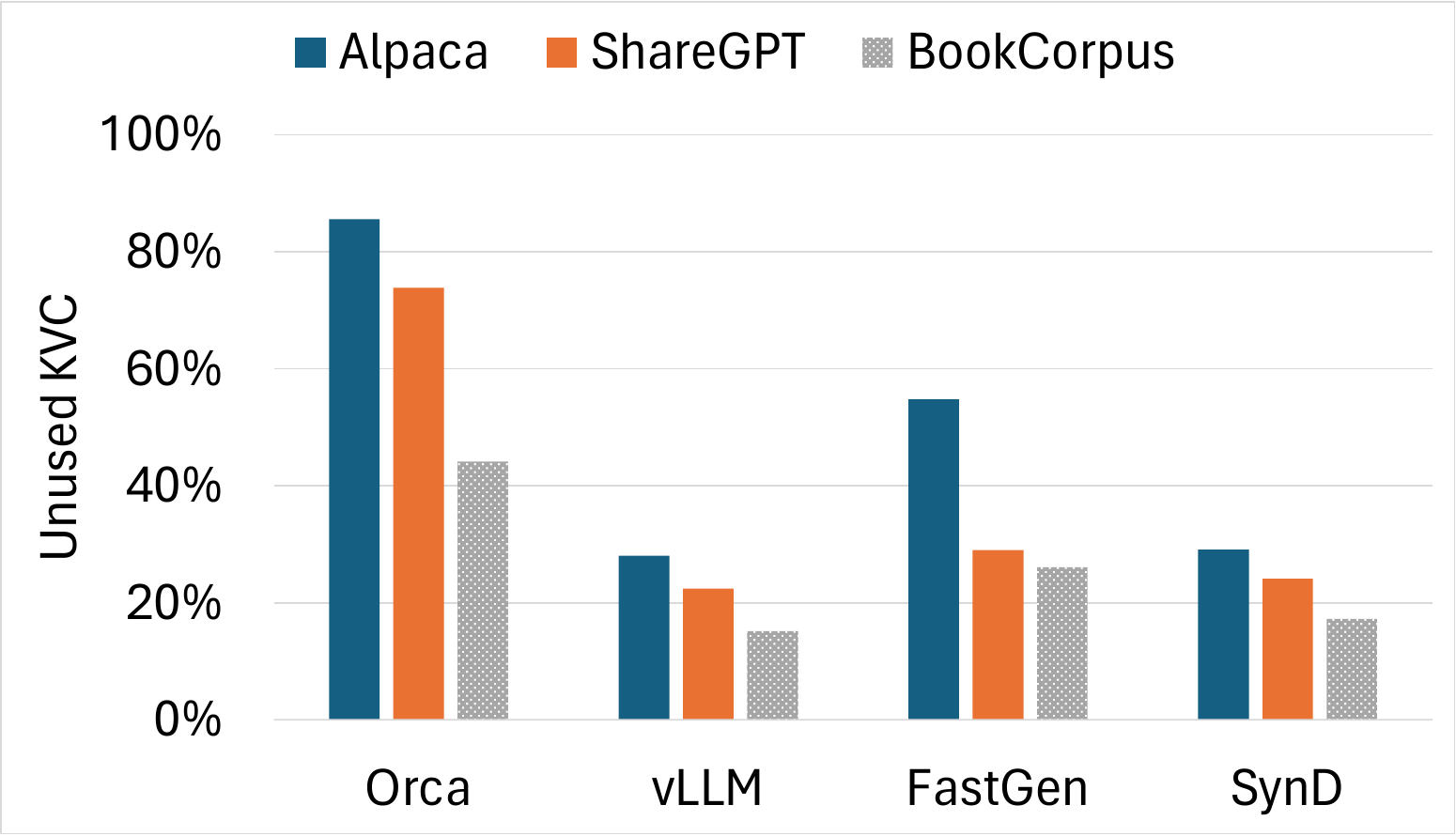} }}
    \hfill
\subfloat[KVC allocation failures.\vspace{-0.0in}\label{fig:exp-kvc-alloc-175}]{{\includegraphics[width=0.23\linewidth,height=0.112\textheight]{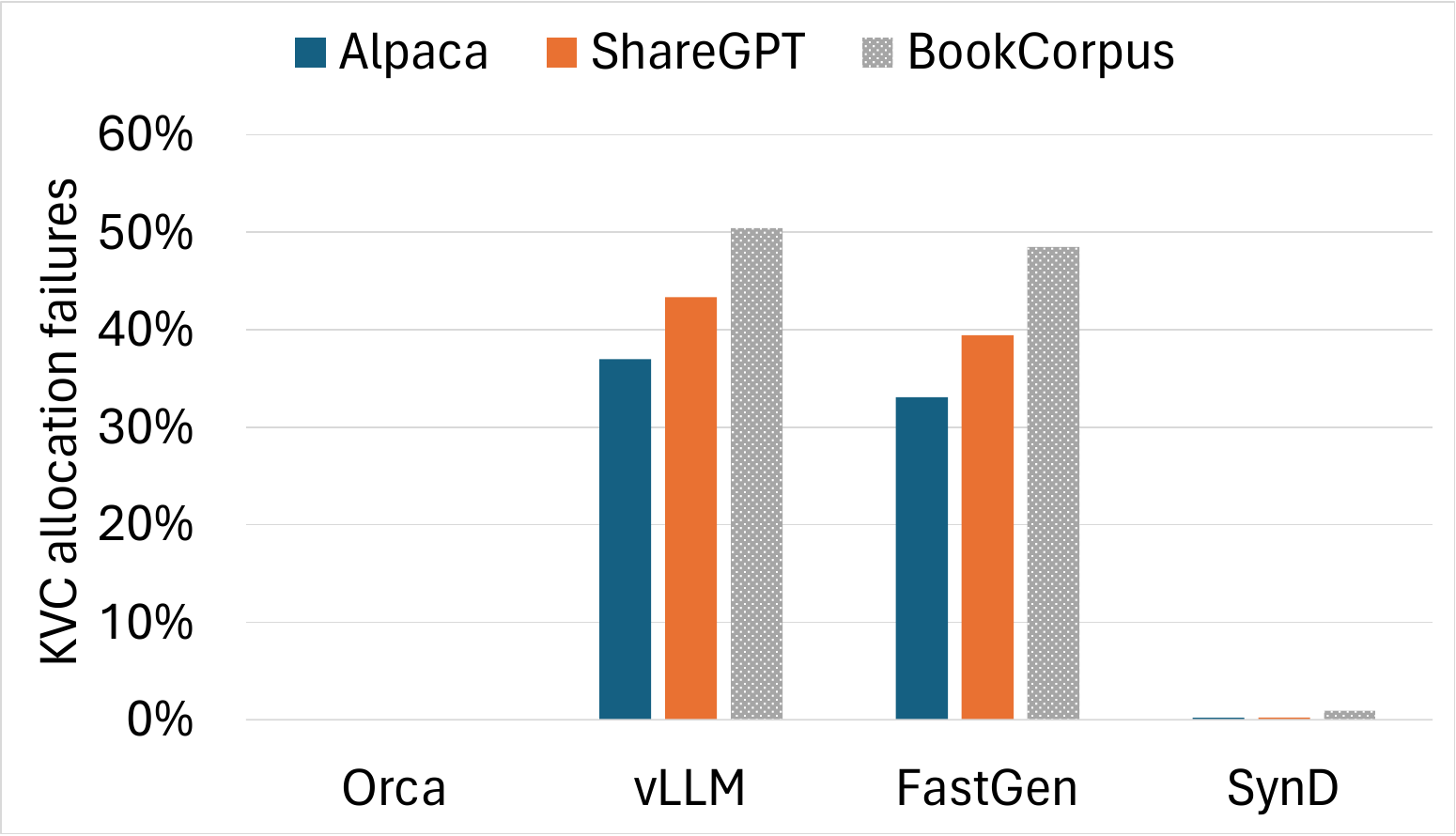}}}
    \hfill
    \subfloat[Scheduling time overhead.\vspace{-0.0in}\label{fig:exp-sch-175}]{{\includegraphics[width=0.23\linewidth,height=0.112\textheight]{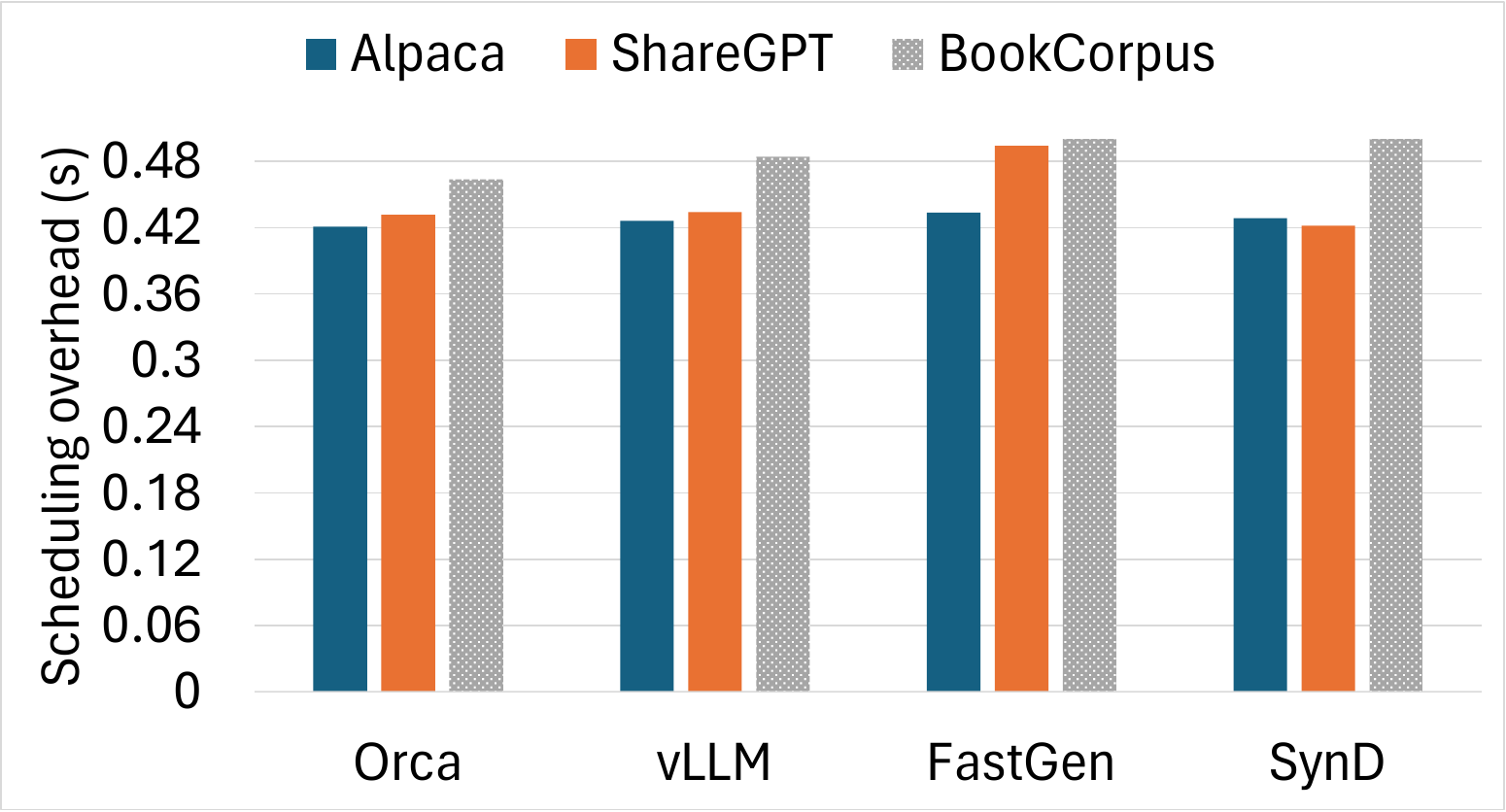} }}
    \hfill
    \DEL{\subfloat[Normalized latency vs. request rates.\vspace{-0.0in}\label{fig:exp-7}]{{\includegraphics[width=0.23\linewidth,height=0.112\textheight]{Fig/scheduling-overhead-exp-error.pdf} }}
    \hfill}
    \hfill
    \vspace{-0.0in}
   \caption{\small{Performance of different methods for the OPT-175B model for all traces.\vspace{-0.0in}}}%
    \label{fig:exp-175}
\end{figure*}}

\DEL{\begin{figure}[t]
\centering
    \subfloat[Example for two GTs.\vspace{-0.0in} \label{fig:KVPipeline}]{{\includegraphics[width=0.48\linewidth,height=0.112\textheight]{fig/KVPipeline}  }}
    \hfill
    \subfloat[Example for four GTs. \vspace{-0.0in} \label{fig:KVPipeline1}]{{\includegraphics[width=0.48\linewidth,height=0.112\textheight]{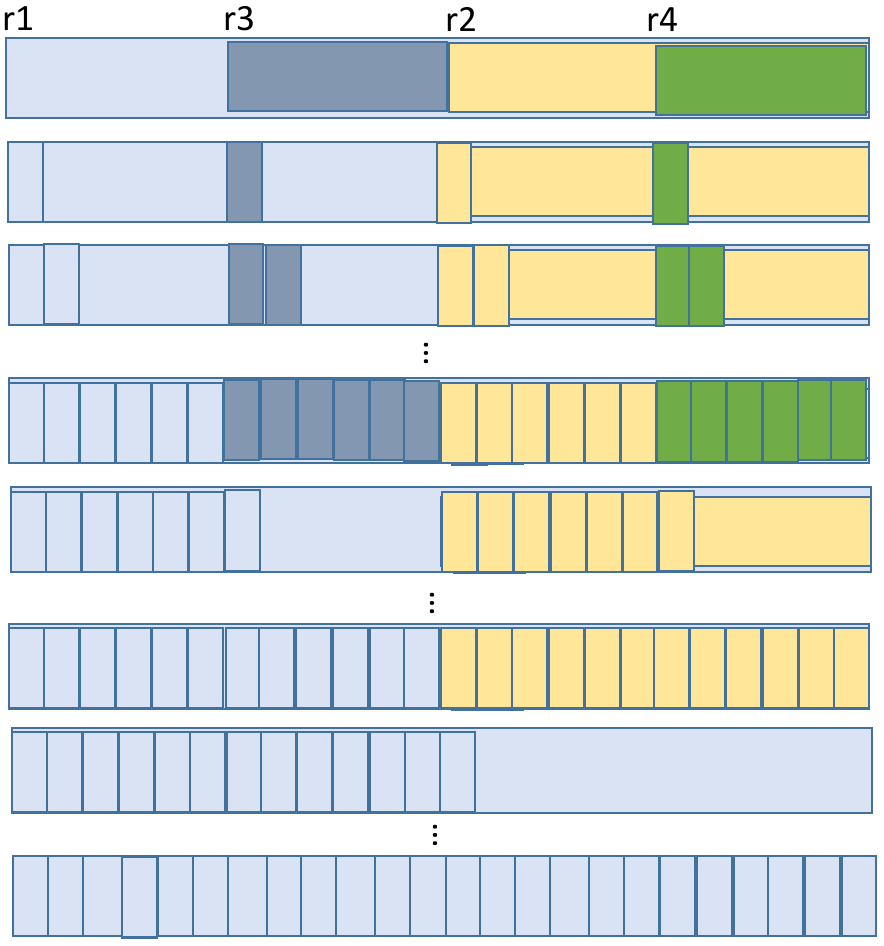} }}
    \hfill
    \vspace{-0.0in}
   \caption{Examples of KVC pipelining.
\vspace{-0.0in}}
    \label{fig:memory-measurement}\vspace{-0.0in}
\end{figure}}
\DEL{\begin{figure*}[t]
\centering
       \subfloat[Response latency (i.e., JCT).\vspace{-0.0in}\label{fig:exp-1}]{{\includegraphics[width=0.23\linewidth,height=0.112\textheight]{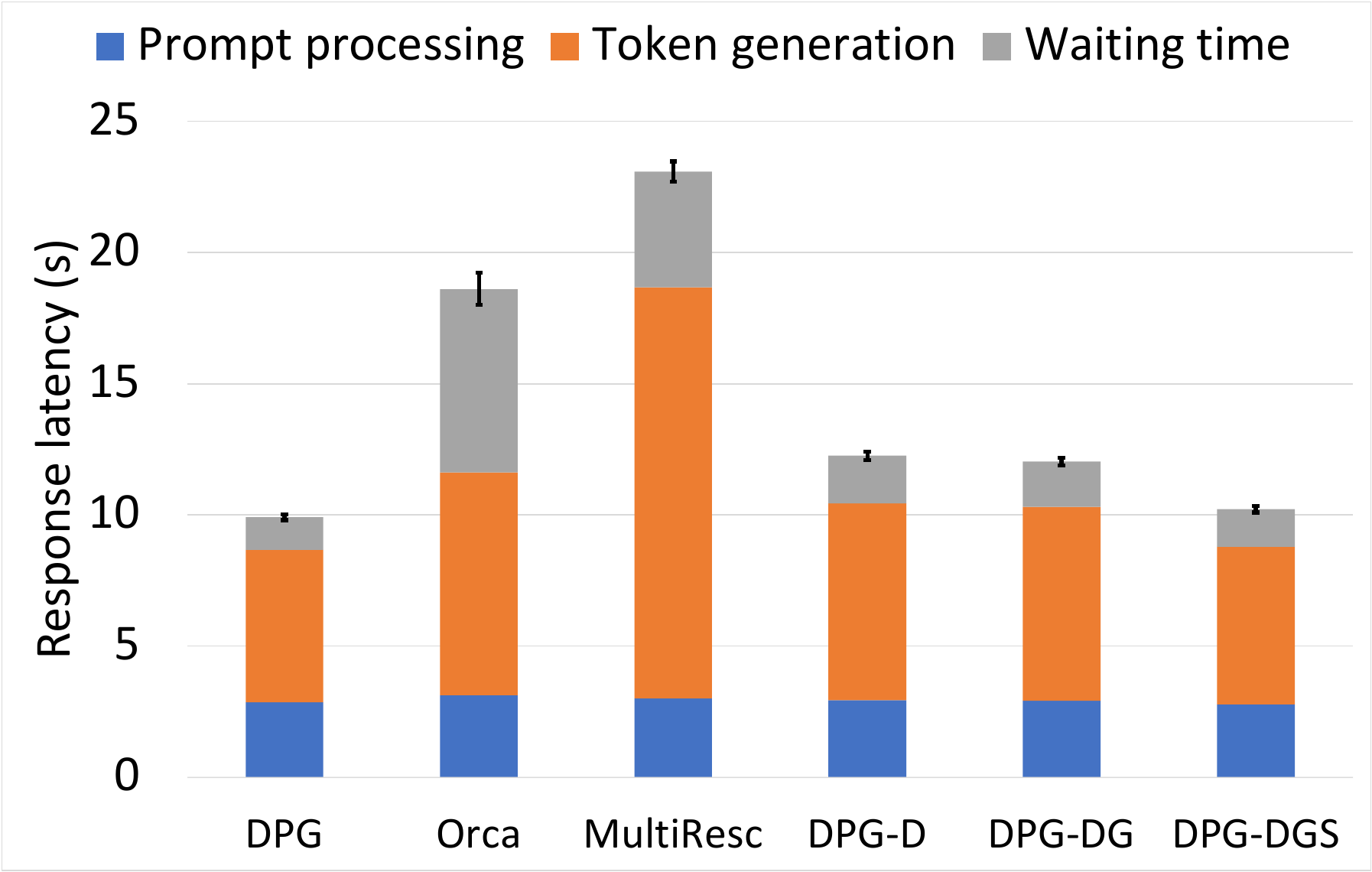} }}
    \hfill
    \DEL{\subfloat[Normalized latency.\vspace{-0.0in}\label{fig:exp-2}]{{\includegraphics[width=0.23\linewidth,height=0.112\textheight]{Fig/iteration-time-exp-up.pdf} }}
    \hfill}
    \subfloat[SLO.\vspace{-0.0in}\label{fig:exp-3}]{{\includegraphics[width=0.23\linewidth,height=0.112\textheight]{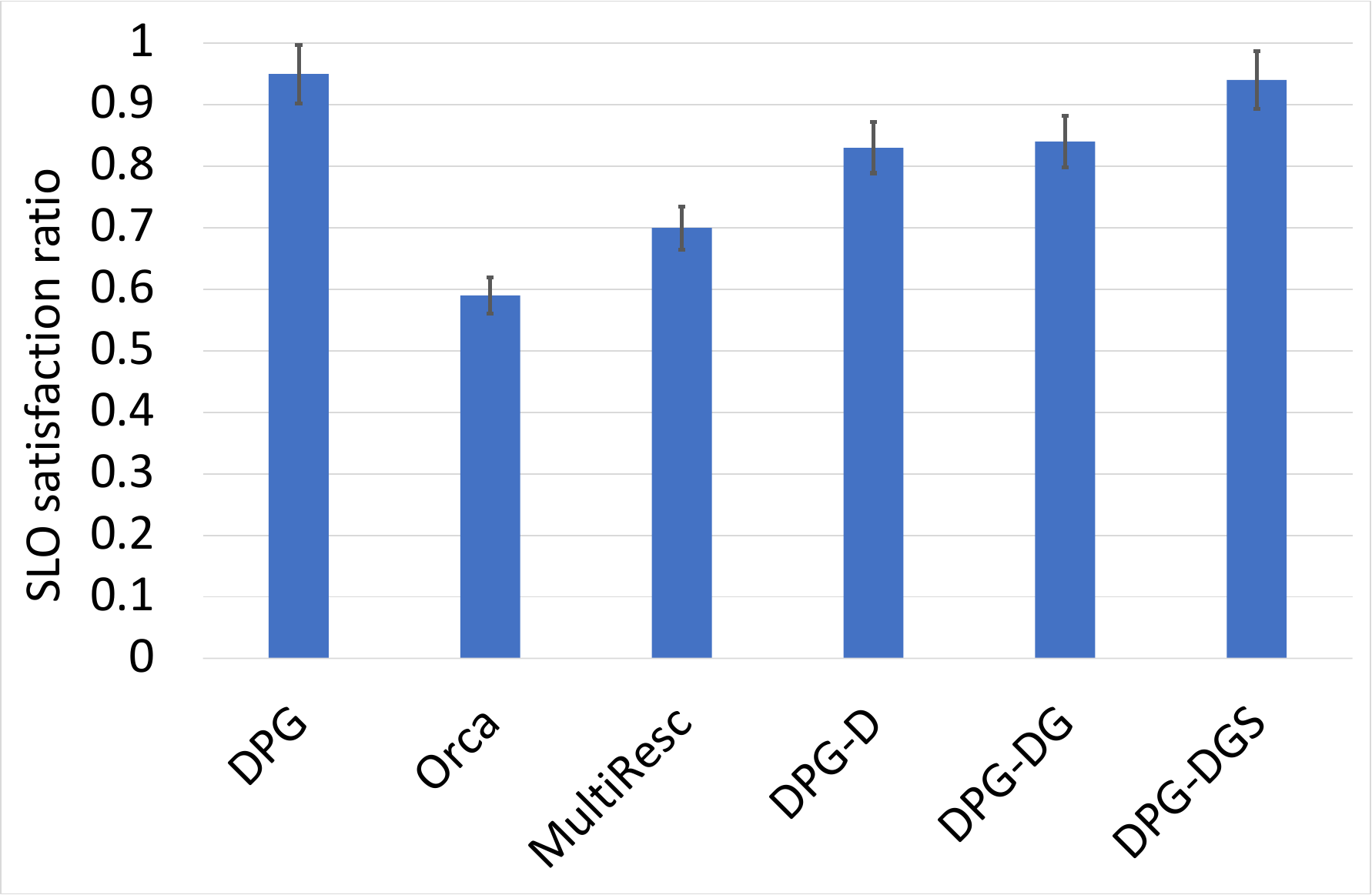} }}
    \hfill
\subfloat[Throughput.\vspace{-0.0in}\label{fig:exp-5}]{{\includegraphics[width=0.23\linewidth,height=0.112\textheight]{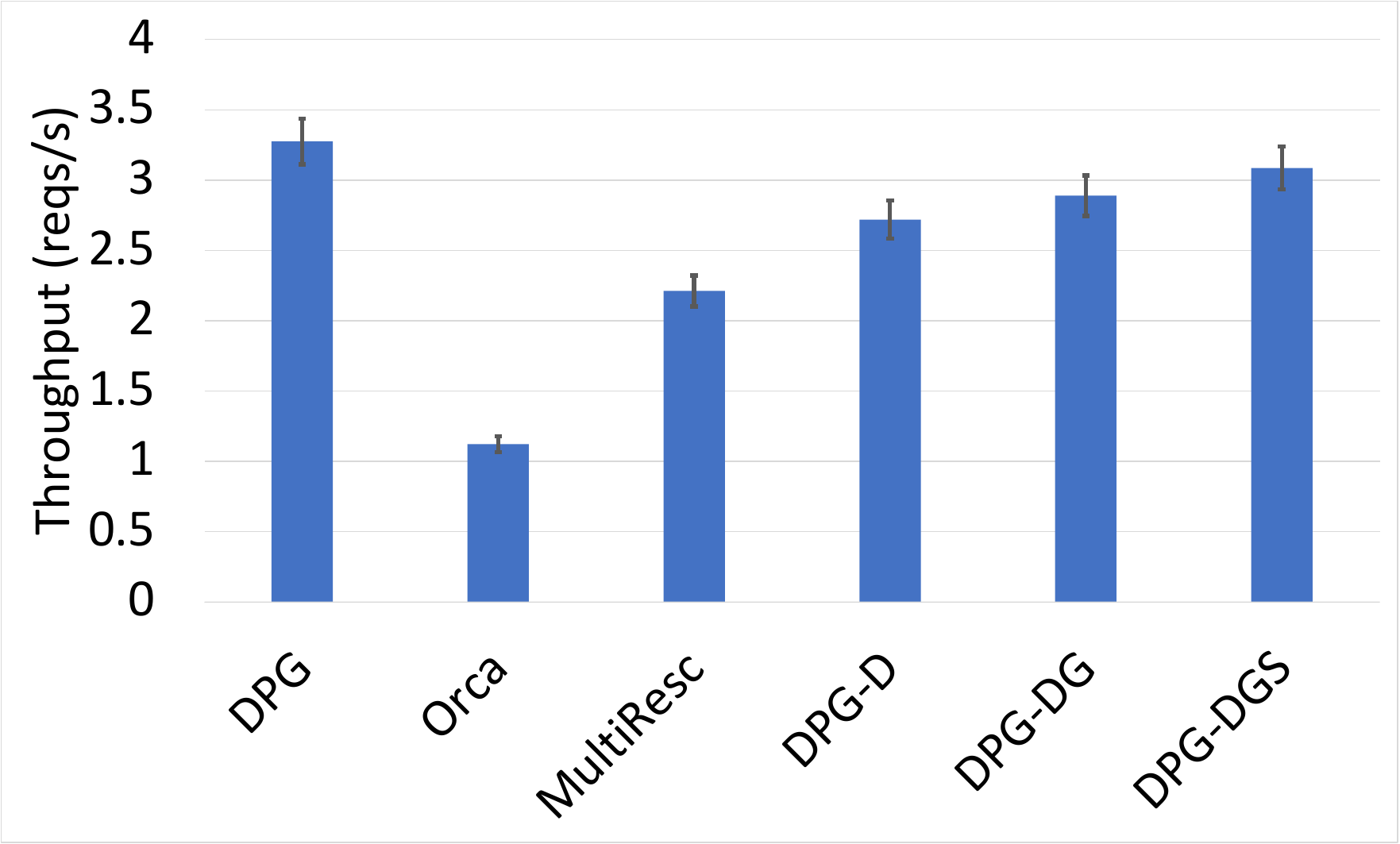} }}
    \hfill
    \subfloat[GPU utilization (token/s).\vspace{-0.0in}\label{fig:exp-4}]{{\includegraphics[width=0.23\linewidth,height=0.112\textheight]{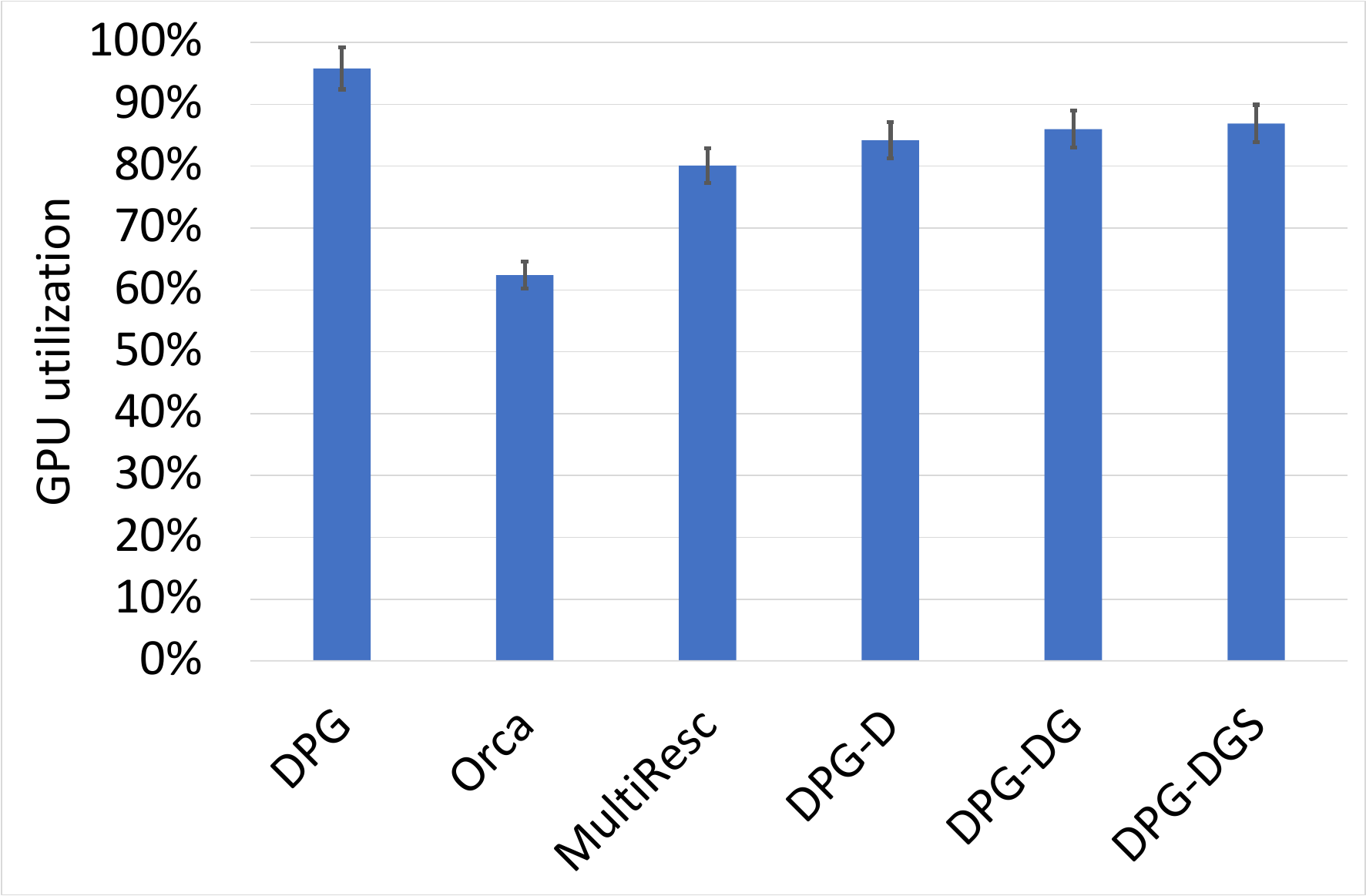} }}
    \hfill
    \subfloat[KVC utilization.\vspace{-0.0in}\label{fig:exp-6}]{{\includegraphics[width=0.23\linewidth,height=0.112\textheight]{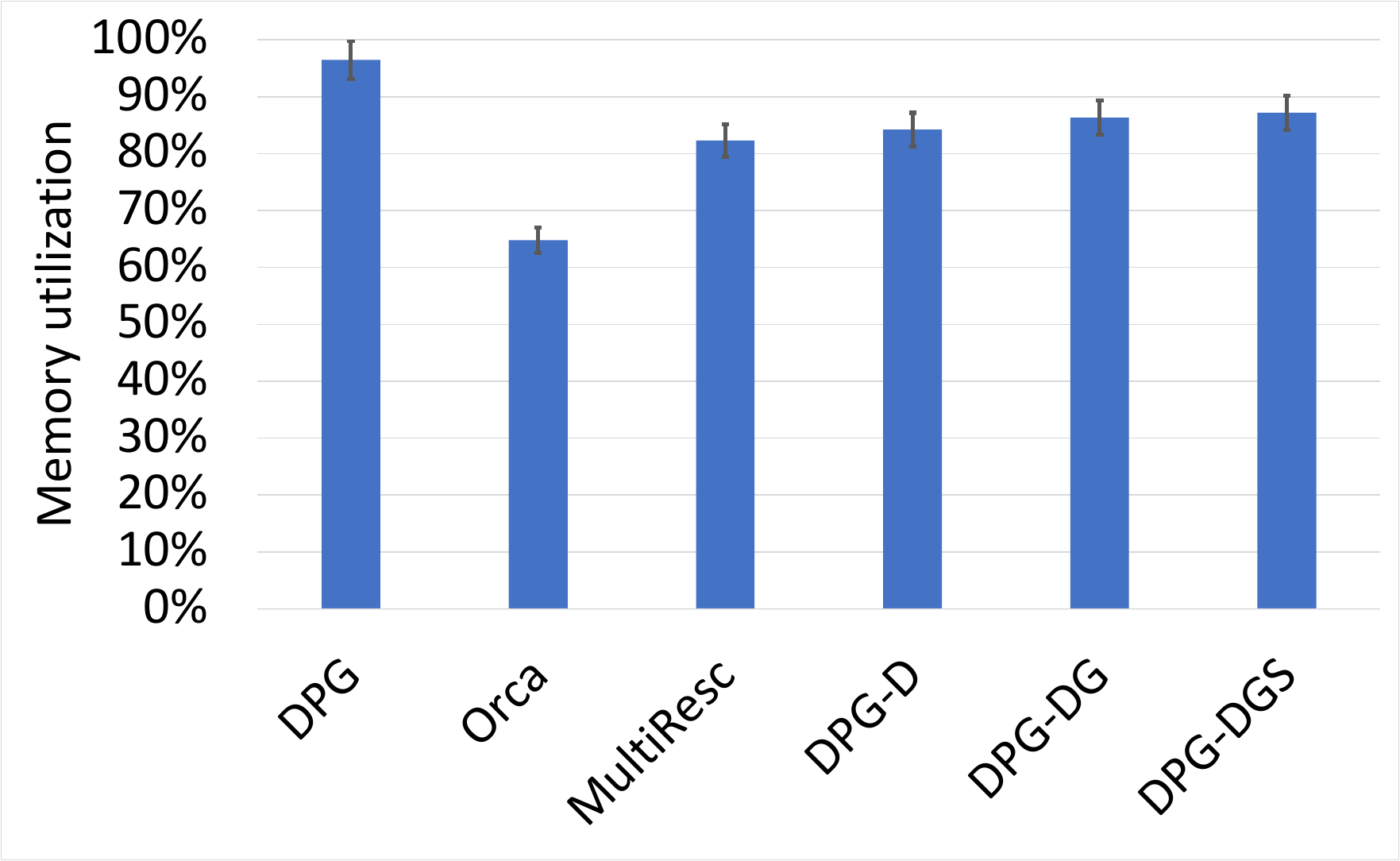} }}
    \hfill
    \subfloat[KVC allocation failures.\vspace{-0.0in}\label{fig:exp-6}]{{\includegraphics[width=0.23\linewidth,height=0.112\textheight]{Fig/Memory-exp.pdf} }}
    \hfill
    \subfloat[Scheduling time overhead.\vspace{-0.0in}\label{fig:exp-7}]{{\includegraphics[width=0.23\linewidth,height=0.112\textheight]{Fig/scheduling-overhead-exp-error.pdf} }}
    \hfill
    \hfill
    \vspace{-0.0in}
   \caption{\small{Performance of different methods for the OPT-175B model for all traces.\vspace{-0.0in}}}%
    \label{fig:prompt-methods}
\end{figure*}}

\DEL{\begin{figure}[t]
\centering
    \subfloat[Example for two GTs.\vspace{-0.0in} \label{fig:KVPipeline}]{{\includegraphics[width=0.48\linewidth,height=0.112\textheight]{fig/KVPipeline}  }}
    \hfill
    \subfloat[Example for four GTs. \vspace{-0.0in} \label{fig:KVPipeline1}]{{\includegraphics[width=0.48\linewidth,height=0.112\textheight]{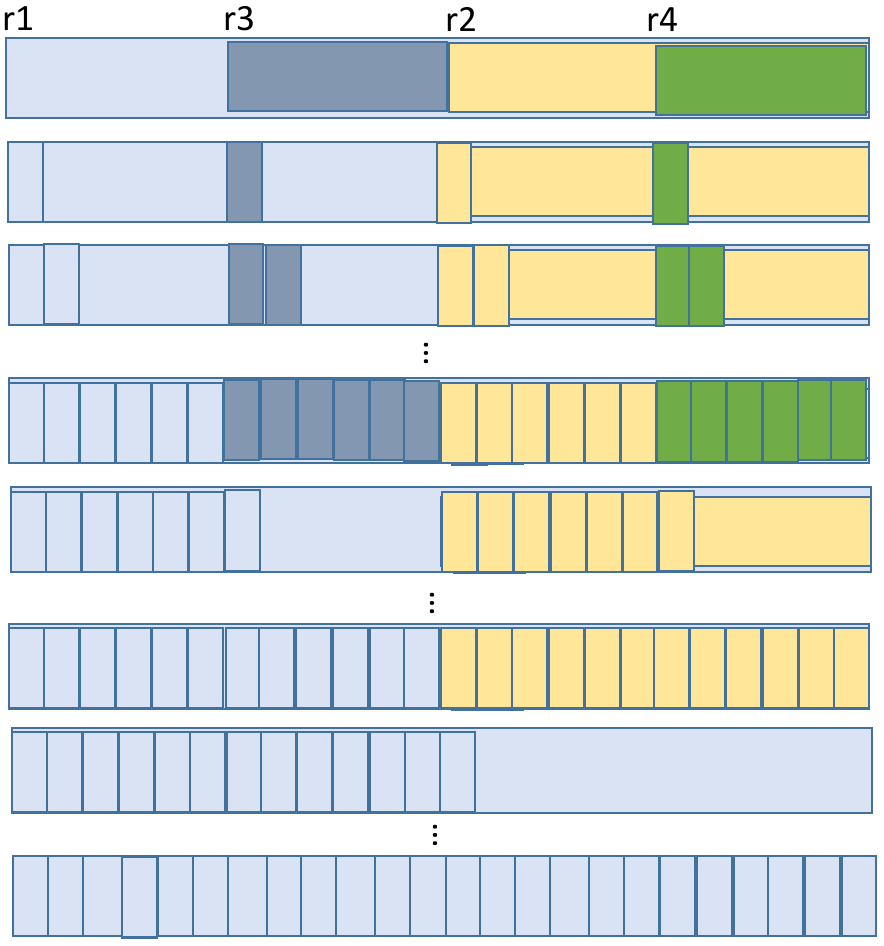} }}
    \hfill
    \vspace{-0.0in}
   \caption{Examples of KVC pipelining.
\vspace{-0.0in}}
    \label{fig:memory-measurement}\vspace{-0.0in}
\end{figure}}
\DEL{\begin{figure*}[t]
\centering
       \subfloat[Response latency (i.e., JCT (Waiting time+ Scheduling time (picking up the prompt and GT) + Execution time +  Preemption/Swapping)). Record the prompt and GT inserting time and write in words no figure\vspace{-0.0in}\label{fig:exp-1}]{{\includegraphics[width=0.23\linewidth,height=0.112\textheight]{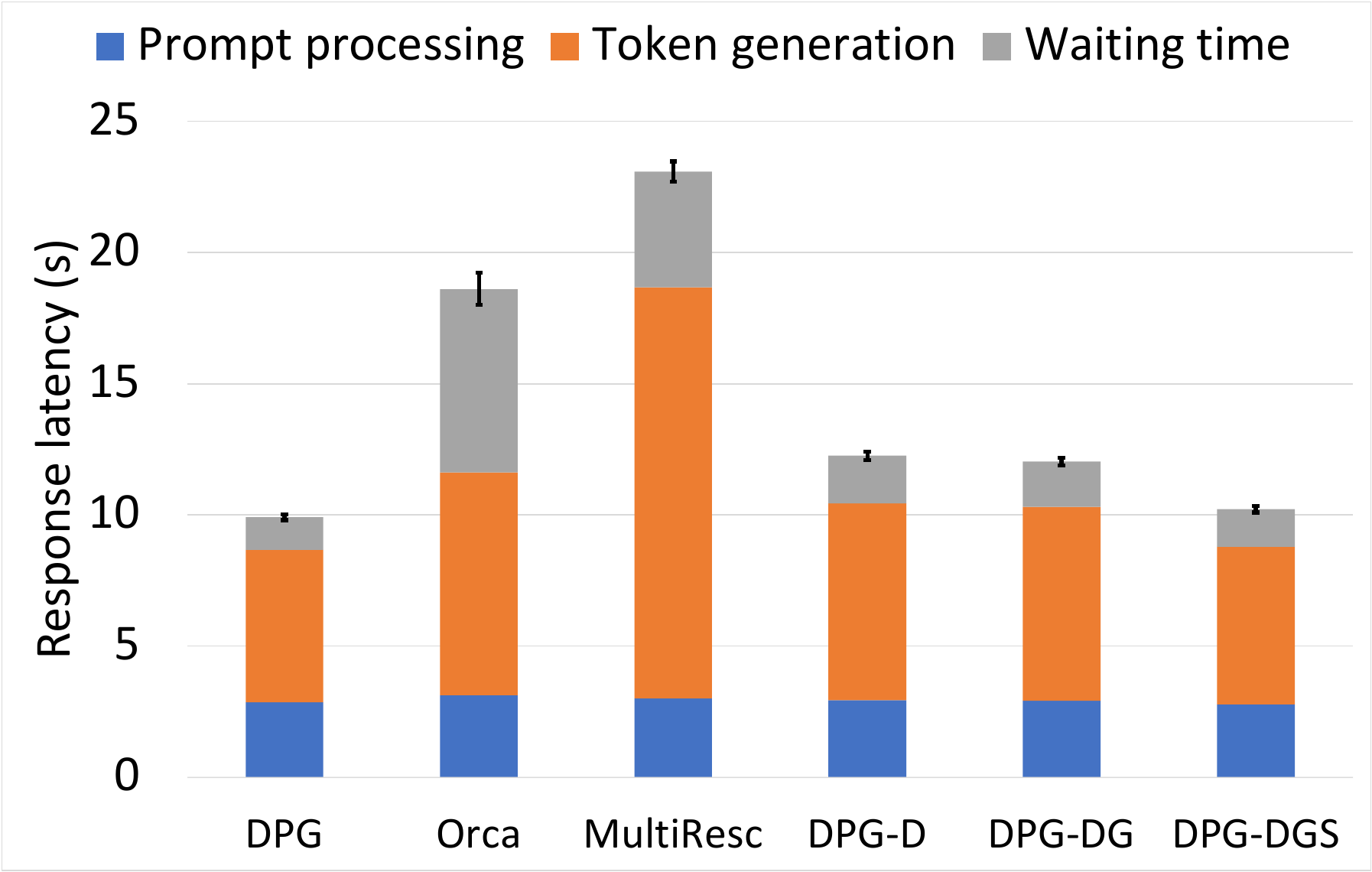} }}
    \hfill
\DEL{    \subfloat[Normalized latency.\vspace{-0.0in}\label{fig:exp-2}]{{\includegraphics[width=0.23\linewidth,height=0.112\textheight]{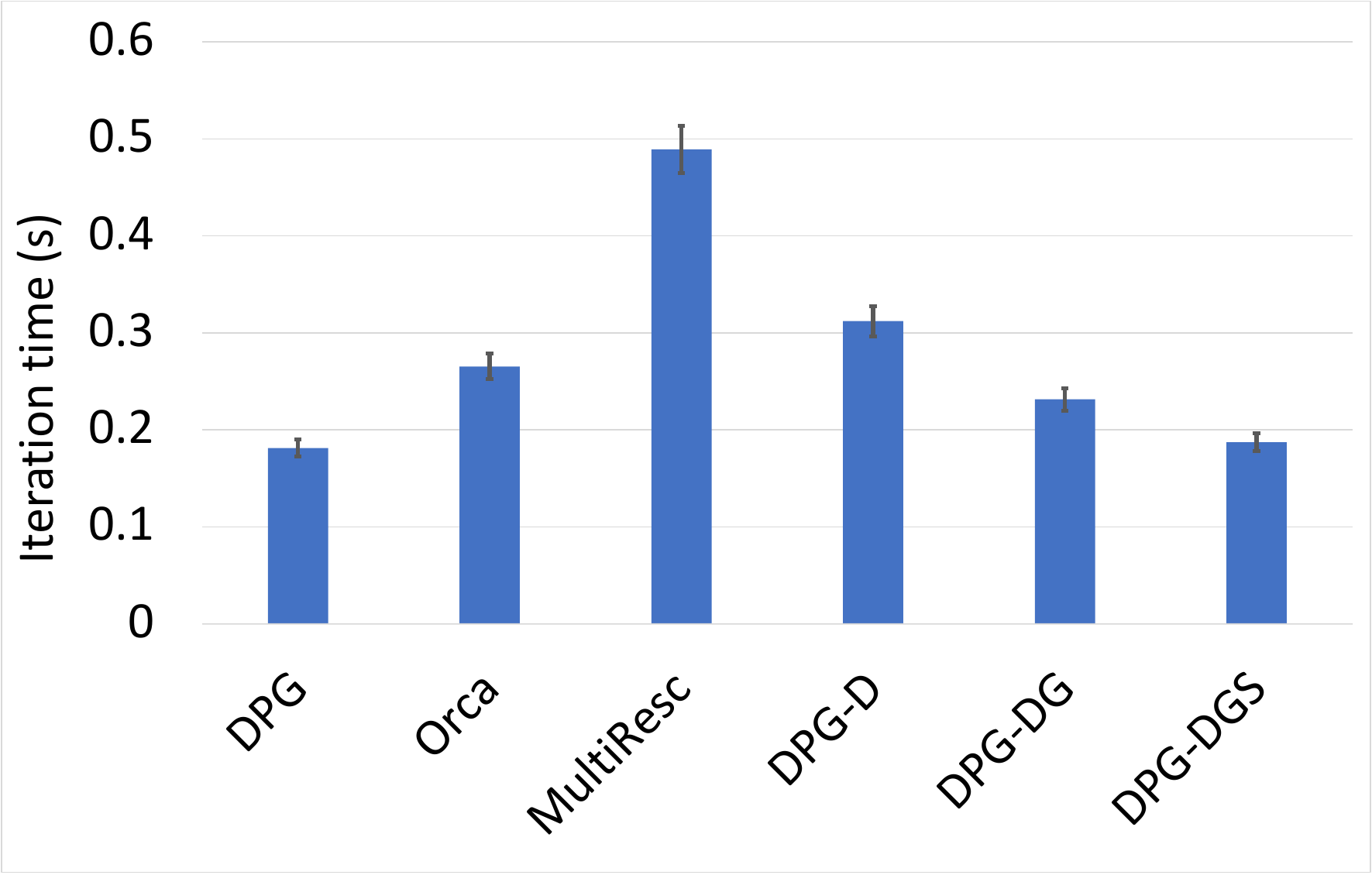} }}
    \hfill}
\DEL{   \subfloat[TBT.\vspace{-0.0in}\label{fig:exp-3}]{{\includegraphics[width=0.23\linewidth,height=0.112\textheight]{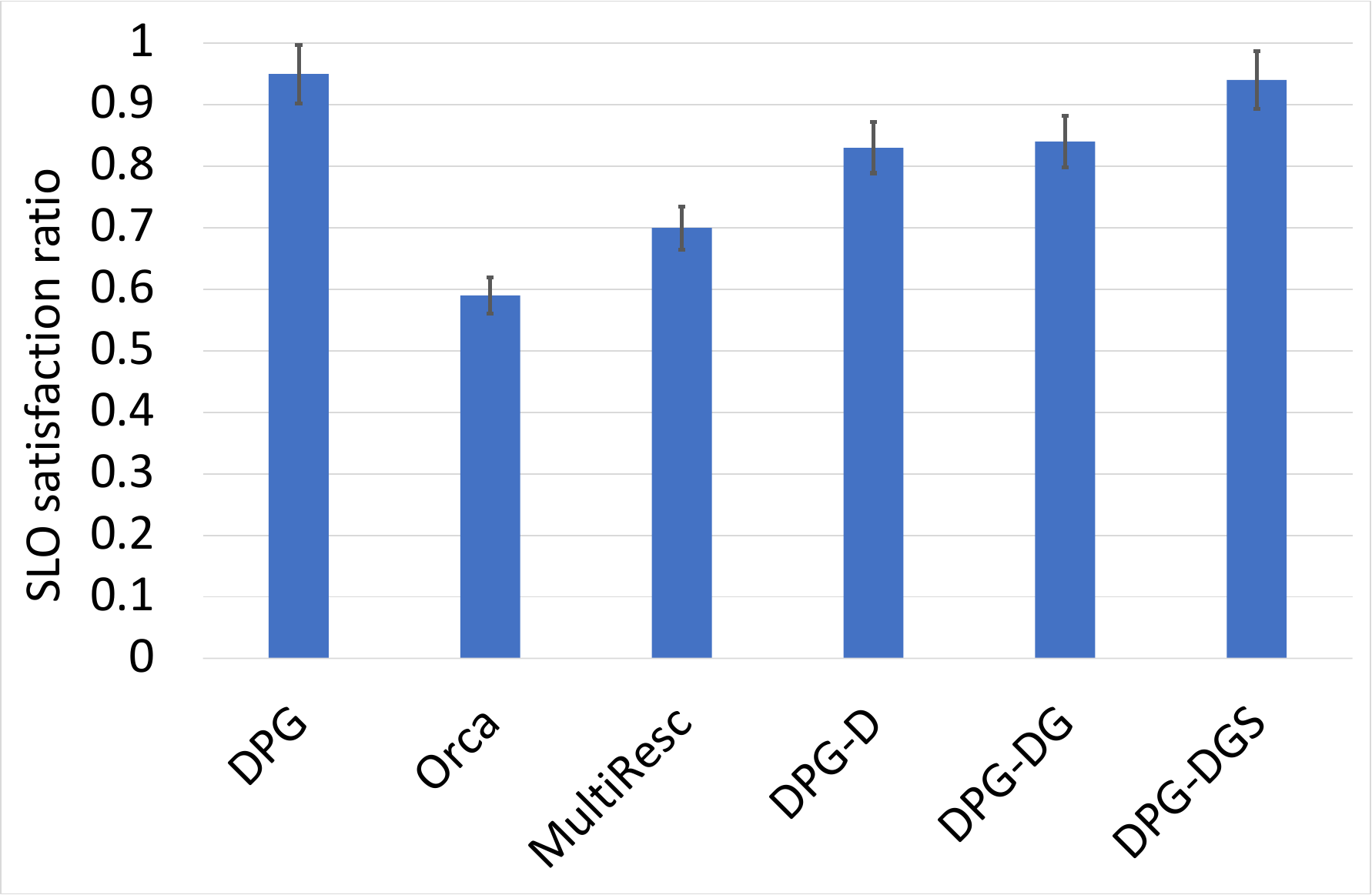} }}
    \hfill}\subfloat[SLO.\vspace{-0.0in}\label{fig:exp-3}]{{\includegraphics[width=0.23\linewidth,height=0.112\textheight]{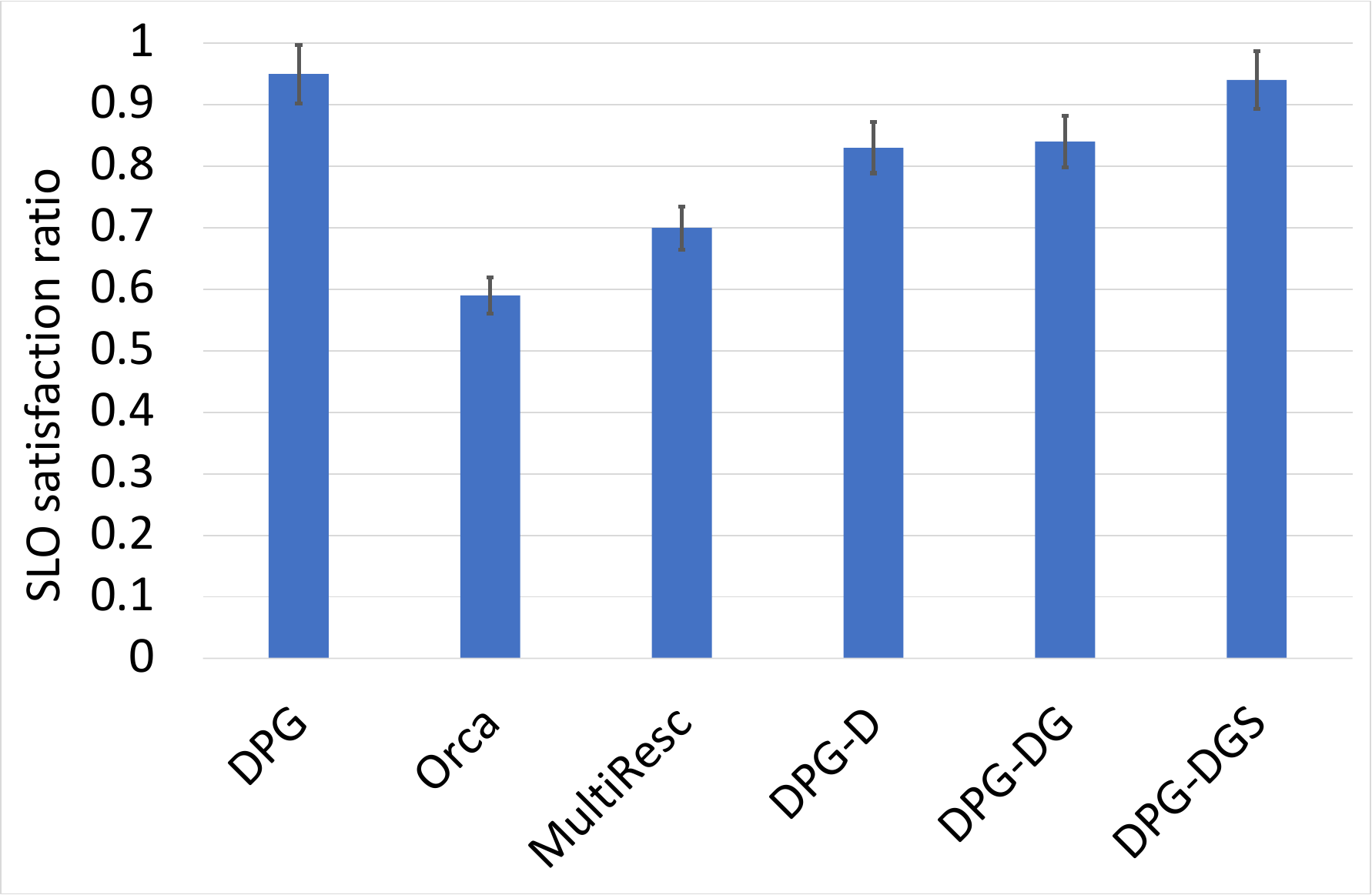} }}
    \hfill
\subfloat[Throughput.\vspace{-0.0in}\label{fig:exp-5}]{{\includegraphics[width=0.23\linewidth,height=0.112\textheight]{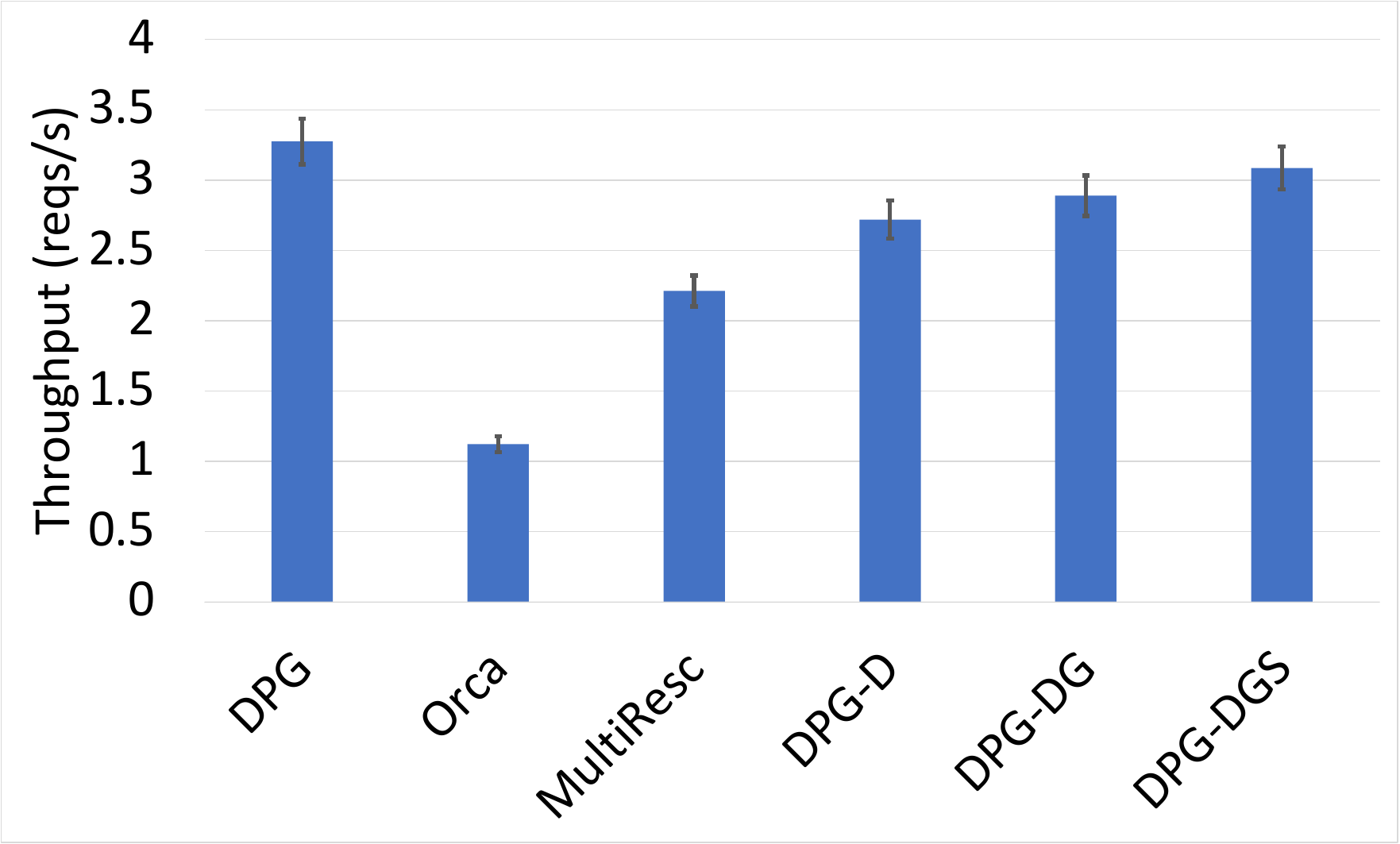} }}
    \hfill
    \subfloat[GPU utilization (token/s).\vspace{-0.0in}\label{fig:exp-4}]{{\includegraphics[width=0.23\linewidth,height=0.112\textheight]{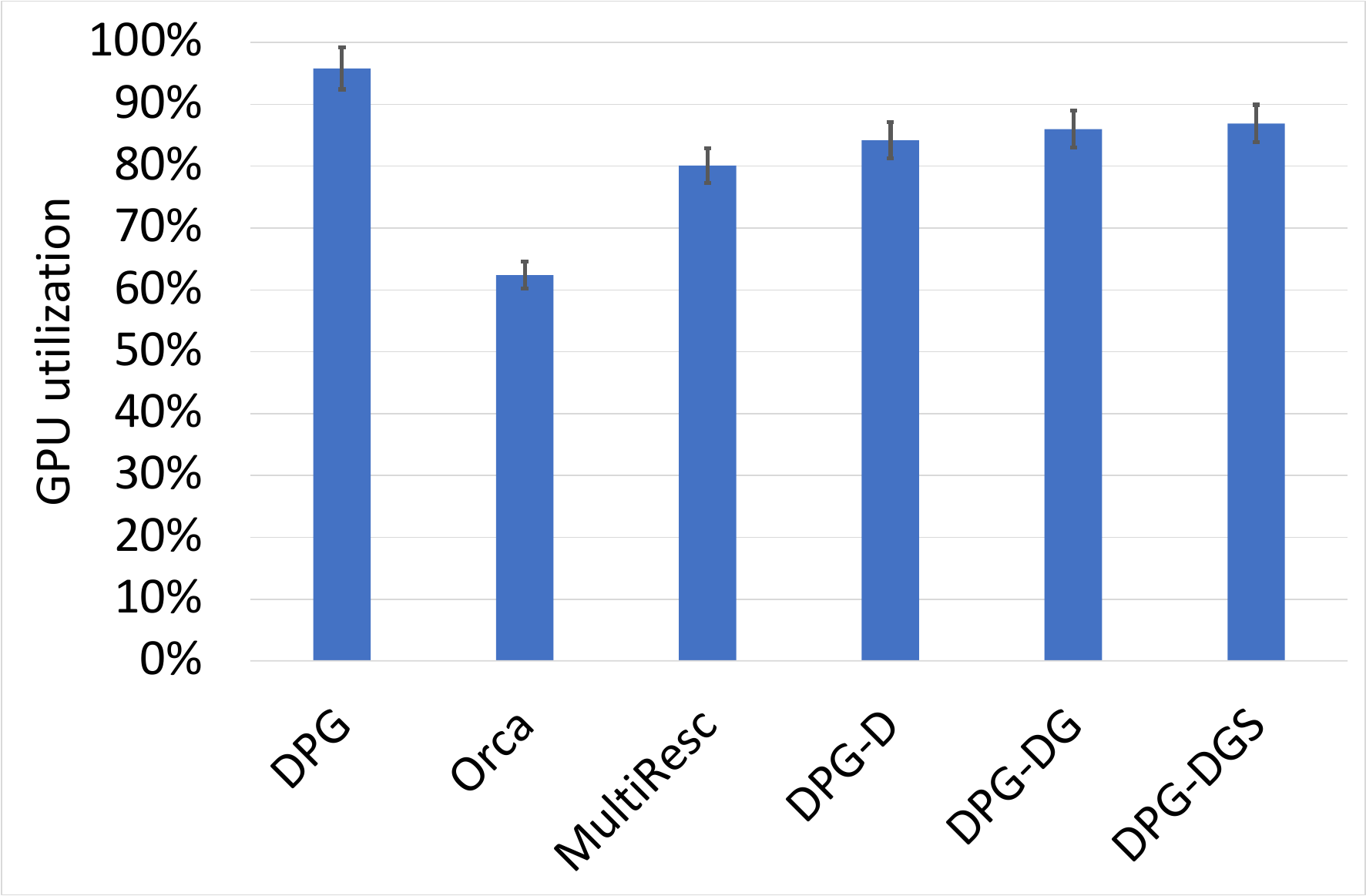} }}
    \hfill
    \subfloat[KVC utilization.\vspace{-0.0in}\label{fig:exp-6}]{{\includegraphics[width=0.23\linewidth,height=0.112\textheight]{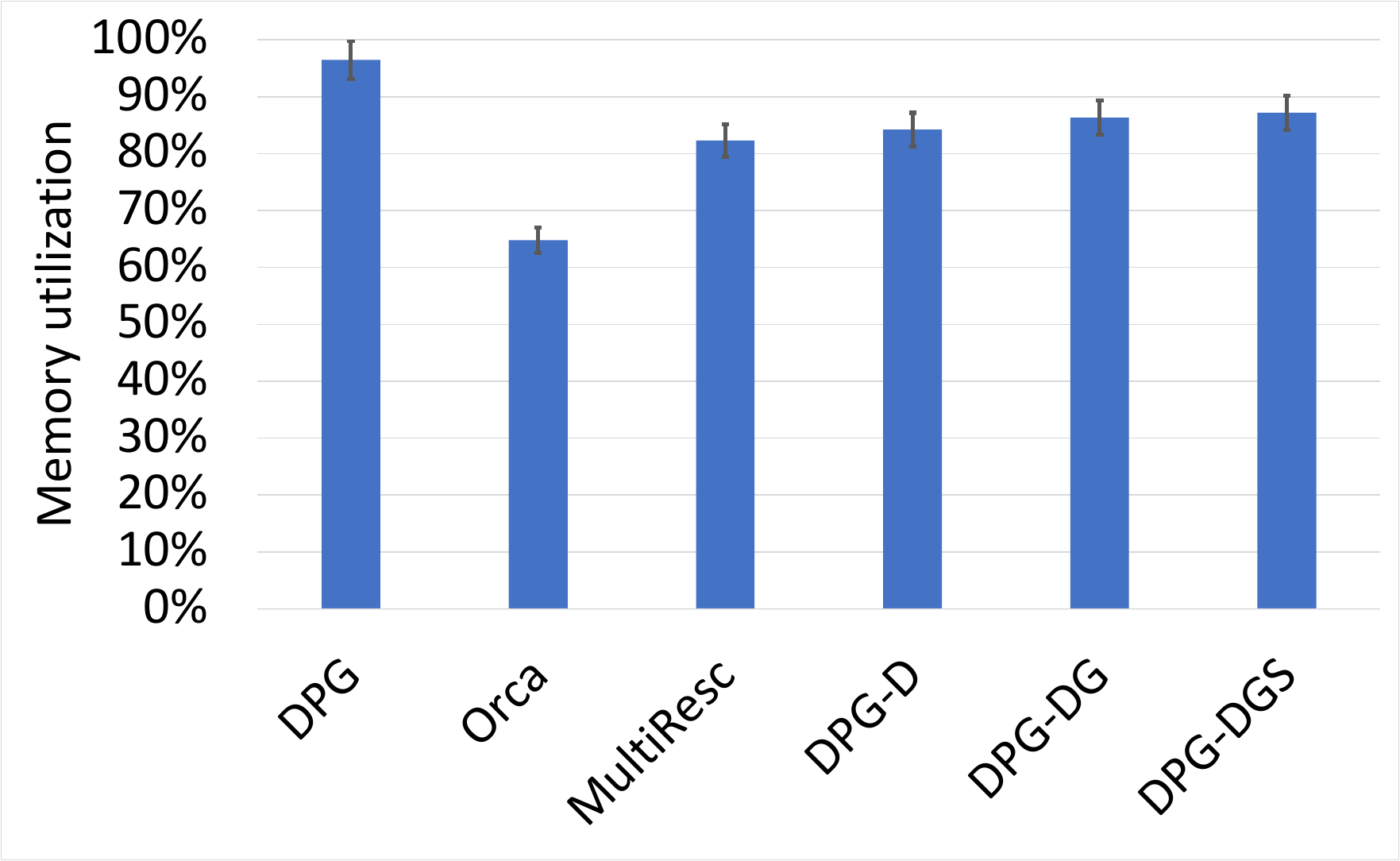} }}
    \hfill
    \subfloat[KVC allocation failures.\vspace{-0.0in}\label{fig:exp-6}]{{\includegraphics[width=0.23\linewidth,height=0.112\textheight]{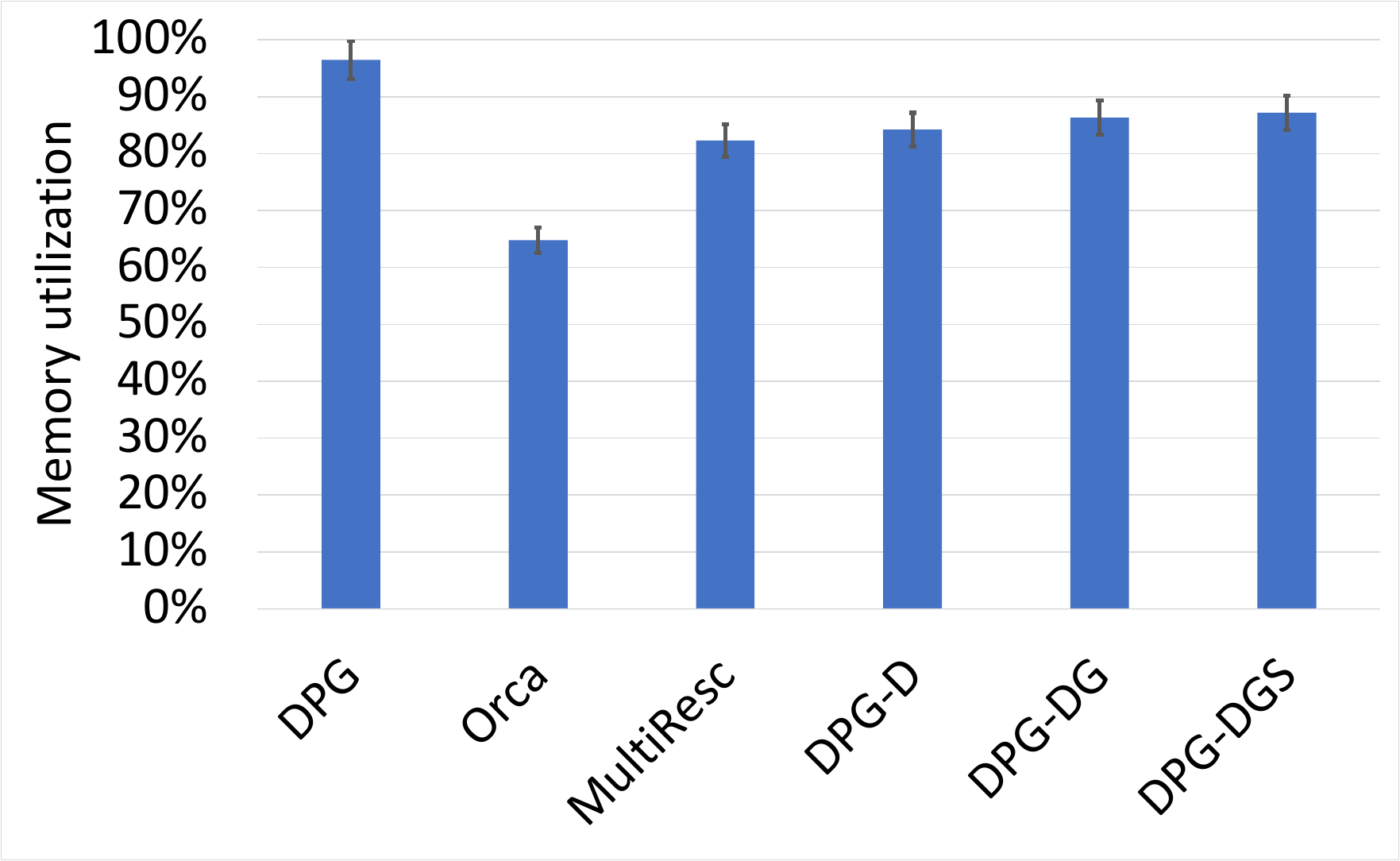} }}
    \hfill
    \subfloat[Scheduling time overhead.\vspace{-0.0in}\label{fig:exp-7}]{{\includegraphics[width=0.23\linewidth,height=0.112\textheight]{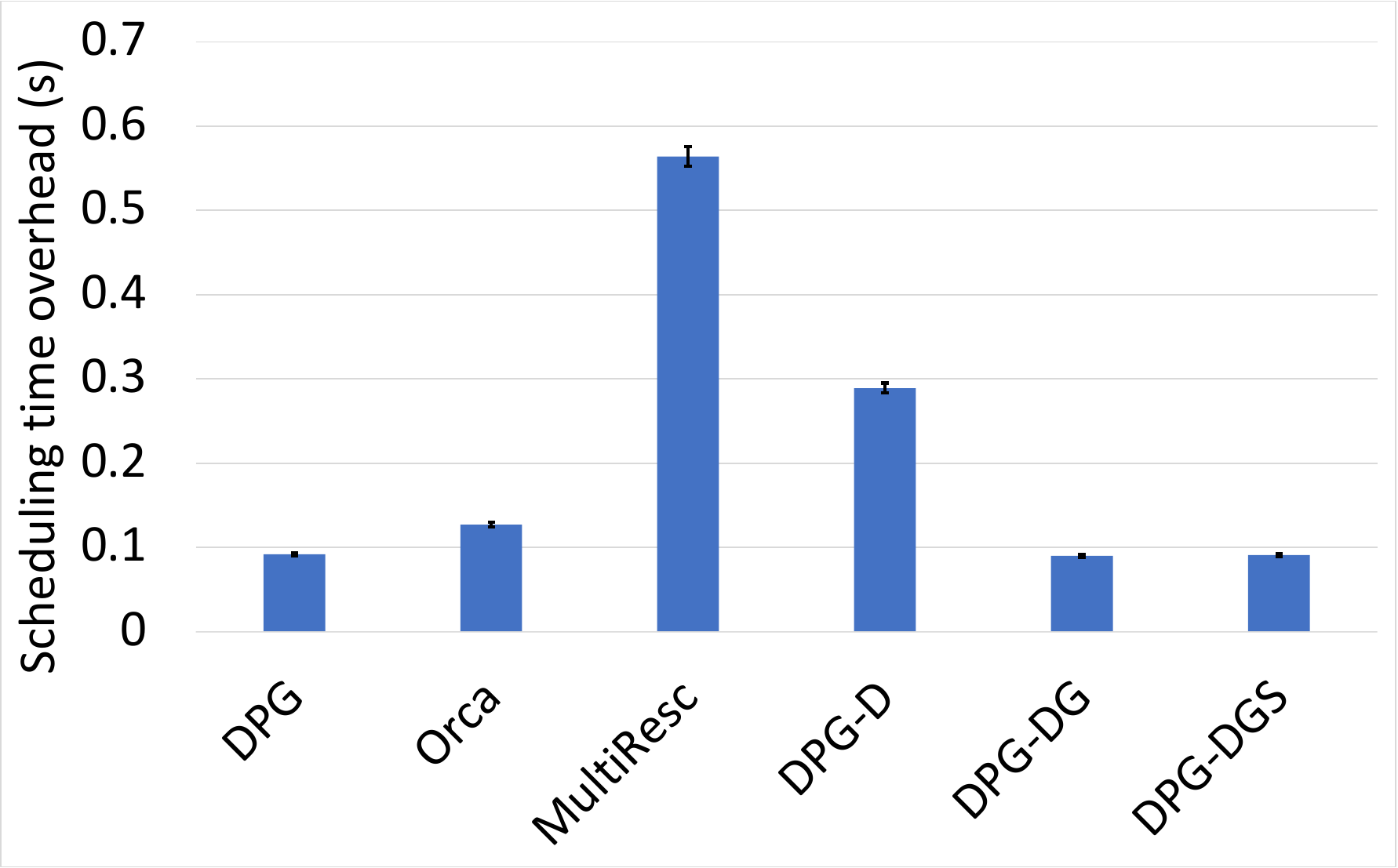} }}
    \hfill
    \hfill
    \vspace{-0.0in}
   \caption{\small{Performance of different methods for the Llama-13B model for all three {\sh{I changed to all three traces, we cannot miss any trace}} traces.\vspace{-0.0in}}}%
    \label{fig:prompt-methods}
\end{figure*}}

\DEL{\begin{figure*}[h]

    \centering

    \subfigure[Llama-2 7B (The Piles).\label{expfig:llama2_7b_piles}]

    {\includegraphics[width=0.32\linewidth,height=0.1\textheight]{fig_nsdi25/exp/llama2_7b_piles.pdf}}

    \subfigure[Llama-2 13B (The Piles).\label{expfig:llama2_13b_piles}]

    {\includegraphics[width=0.32\linewidth,height=0.1\textheight]{fig_nsdi25/exp/llama2_13b-piles.pdf}}

    \subfigure[Llama-2 70B (The Piles).\label{expfig:llama2_70b_piles}]

    {\includegraphics[width=0.32\linewidth,height=0.1\textheight]{fig_nsdi25/exp/llama2_70b_piles.pdf}}

    \subfigure[OPT 13B (The Piles).\label{expfig:opt_13b_piles}]

    {\includegraphics[width=0.32\linewidth,height=0.1\textheight]{fig_nsdi25/exp/opt_13b_piles.pdf}}

    \subfigure[OPT 66B (The Piles).\label{expfig:opt_66b_piles}]

    {\includegraphics[width=0.32\linewidth,height=0.1\textheight]{fig_nsdi25/exp/opt_66b_piles.pdf}}

    \subfigure[OPT 175B (The Piles).\label{expfig:opt_175b_piles}]

    {\includegraphics[width=0.32\linewidth,height=0.1\textheight]{fig_nsdi25/exp/opt_175b_piles.pdf}}

    \subfigure[Llama-2 7B (ShareGPT).\label{expfig:llama2_7b_sharegpt}]

    {\includegraphics[width=0.32\linewidth,height=0.1\textheight]{fig_nsdi25/exp/llama2_7b_sharegpt.pdf}}

    \subfigure[Llama-2 13B (ShareGPT).\label{expfig:llama2_13b_sharegpt}]

    {\includegraphics[width=0.32\linewidth,height=0.1\textheight]{fig_nsdi25/exp/llama2_13b_sharegpt.pdf}}

    \subfigure[Llama-2 70B (ShareGPT).\label{expfig:llama2_70b_sharegpt}]

    {\includegraphics[width=0.32\linewidth,height=0.1\textheight]{fig_nsdi25/exp/llama2_70b_sharegpt.pdf}}

    \subfigure[OPT 13B (ShareGPT).\label{expfig:opt_13b_sharegpt}]

    {\includegraphics[width=0.32\linewidth,height=0.1\textheight]{fig_nsdi25/exp/opt_13b_sharegpt.pdf}}

    \subfigure[OPT 66B (ShareGPT).\label{expfig:opt_66b_sharegpt}]

    {\includegraphics[width=0.32\linewidth,height=0.1\textheight]{fig_nsdi25/exp/opt_66b_sharegpt.pdf}}

    \subfigure[OPT 175B (ShareGPT).\label{expfig:opt_175b_sharegpt}]

    {\includegraphics[width=0.32\linewidth,height=0.1\textheight]{fig_nsdi25/exp/opt_175b_sharegpt.pdf}}

    \caption{Normalized latency v.s. request rate.}

    \label{expfig:normalized_latency}

\end{figure*}}

\DEL{\begin{figure}
    \centering
    \includegraphics[width=\columnwidth,height=0.24\textheight]{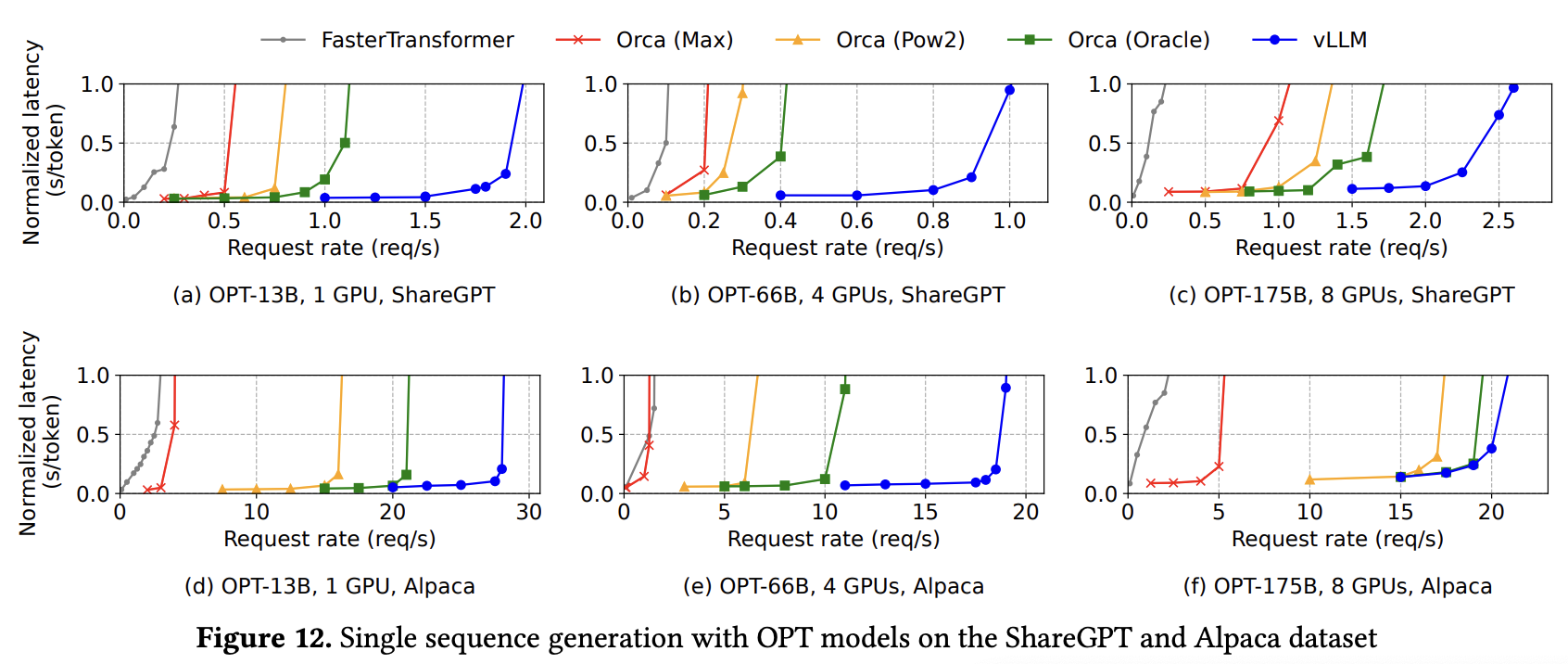}
    \caption{ Sequence generation with varied request rates OPT models on the ShareGPT and Alpaca dataset}
    \label{fig:enter-label}
\end{figure}}

\vspace{-0.0in}

\vspace{-0.2in}
\section{Performance Evaluation} 
\label{sec:evaluation}\vspace{-0.1in}


\begin{figure*}[t]
\centering
     \subfloat[OPT-13B on ShareGPT.\vspace{-0.0in}\label{fig:exp-13-s}]{{\includegraphics[width=0.32\linewidth,height=0.112\textheight]{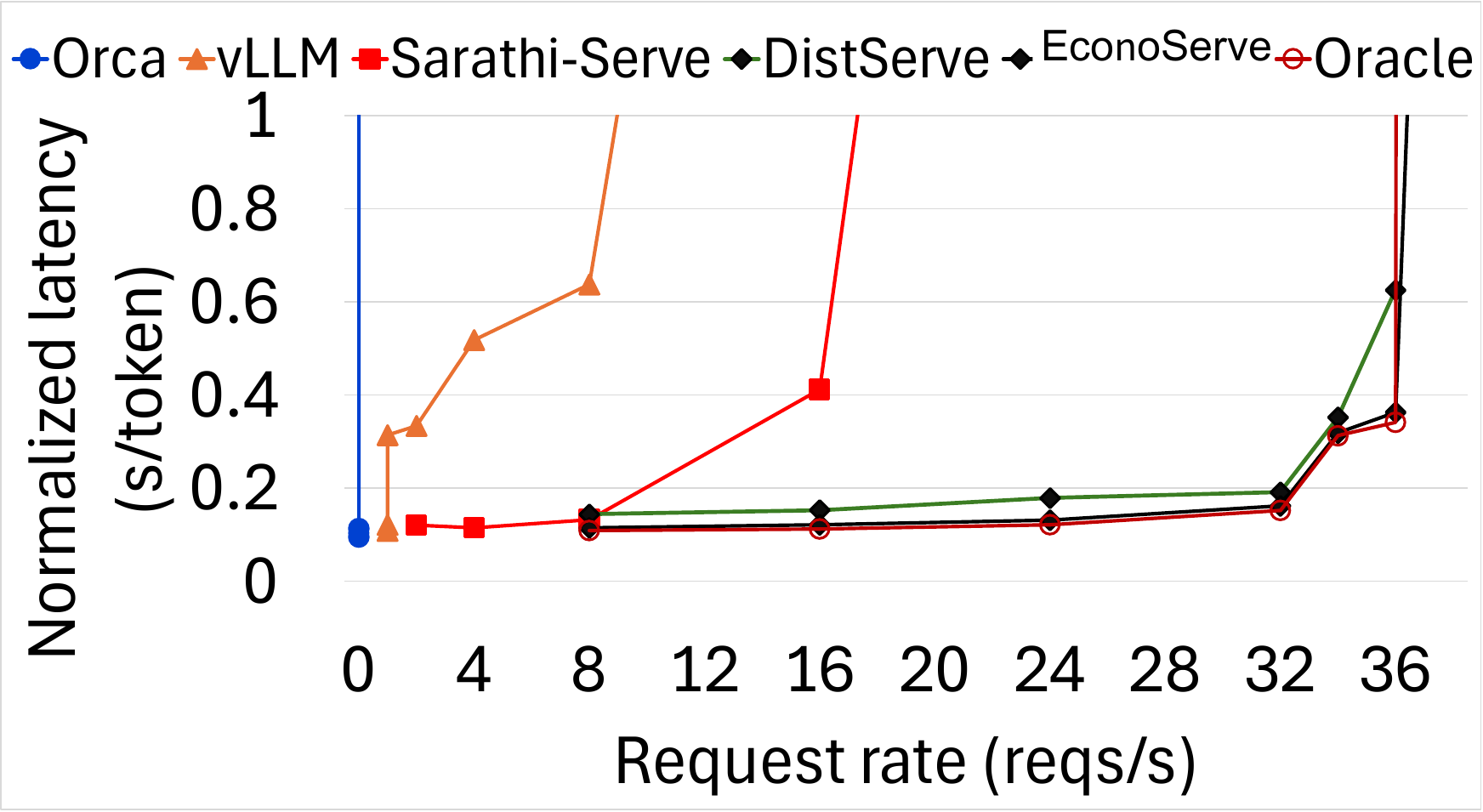} }}
    \hfill
    \subfloat[OPT-175B on ShareGPT.\vspace{-0.0in}\label{fig:exp-175-s}]{{\includegraphics[width=0.32\linewidth,height=0.112\textheight]{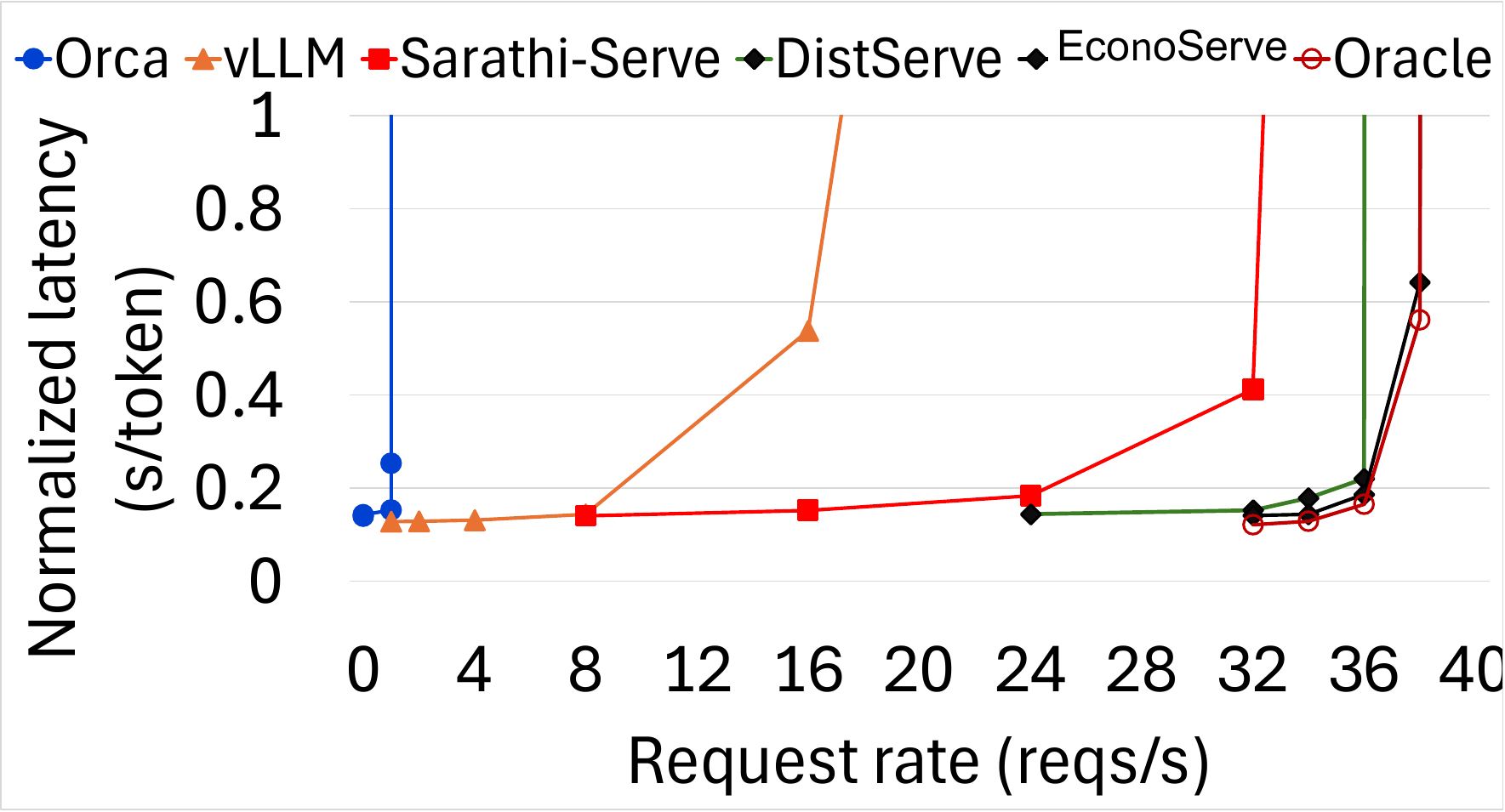} }}
    \hfill
    \subfloat[Llama-33B on ShareGPT.\vspace{-0.0in}\label{fig:exp-l3-s}]{{\includegraphics[width=0.32\linewidth,height=0.112\textheight]{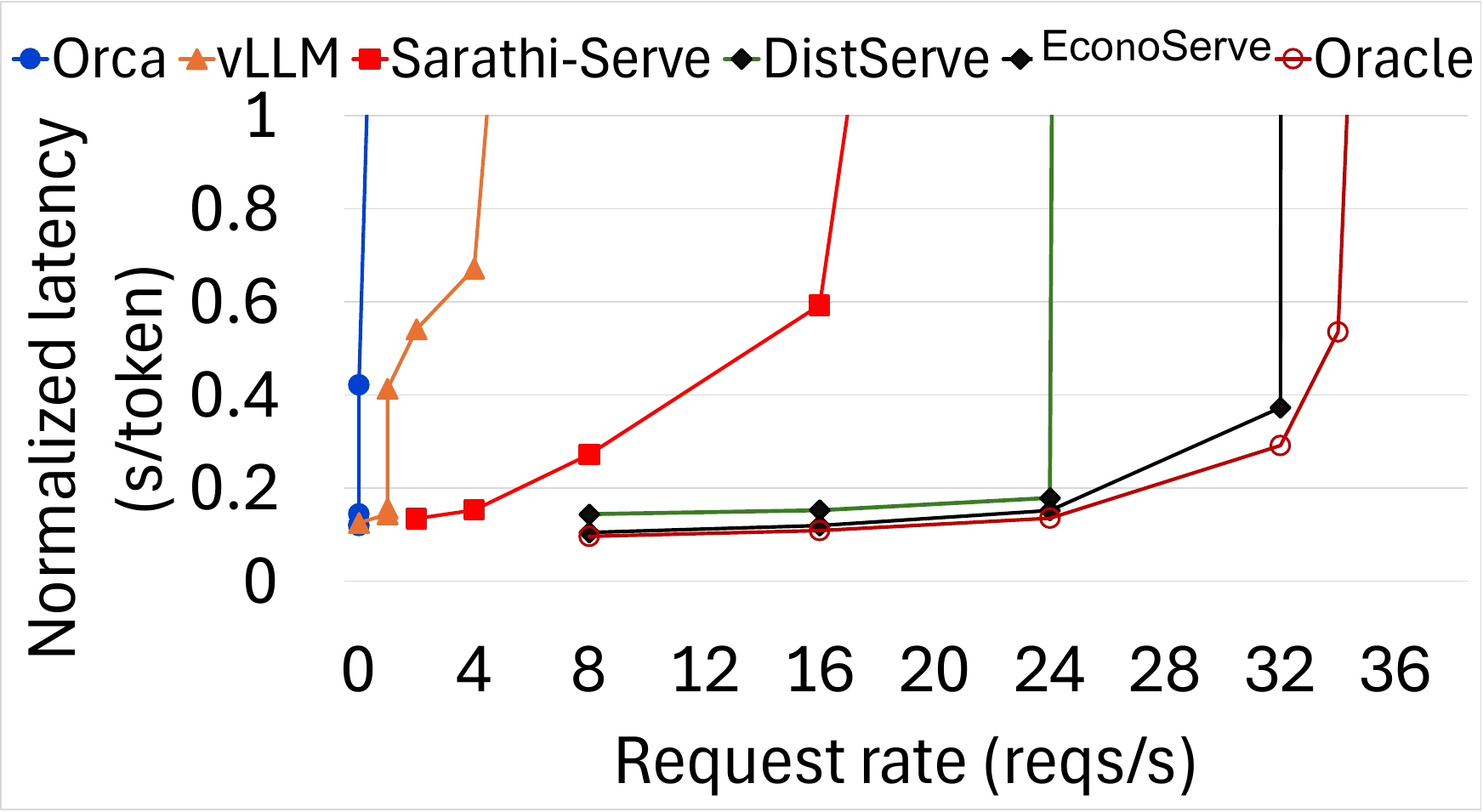} }}
    \hfill
     \subfloat[OPT-13B with BookCorpus.\vspace{-0.0in}\label{fig:exp-13-b}]{{\includegraphics[width=0.32\linewidth,height=0.112\textheight]{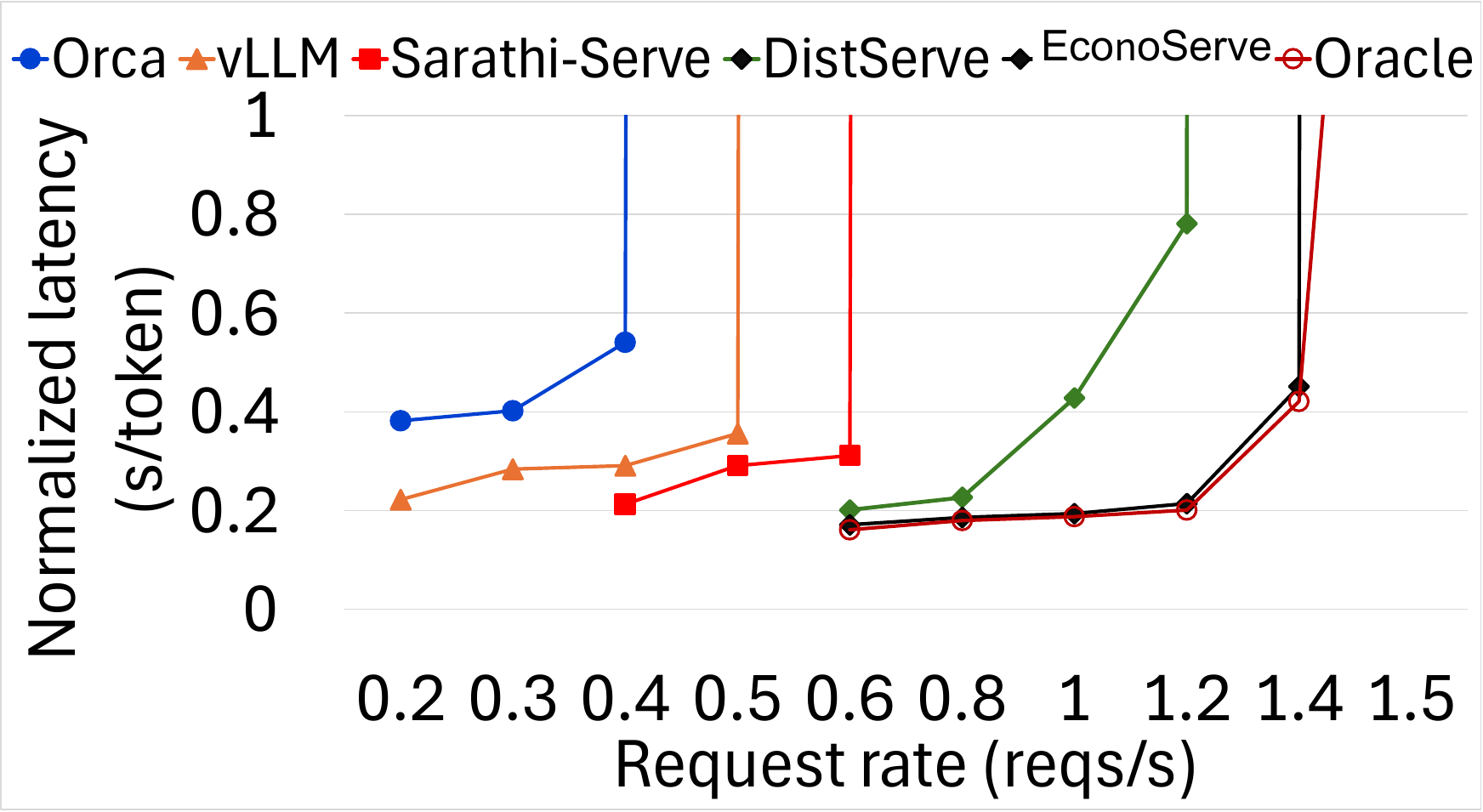} }}
    \hfill
    \subfloat[OPT-175B with BookCorpus.\vspace{-0.0in}\label{fig:exp-175-b}]{{\includegraphics[width=0.32\linewidth,height=0.112\textheight]{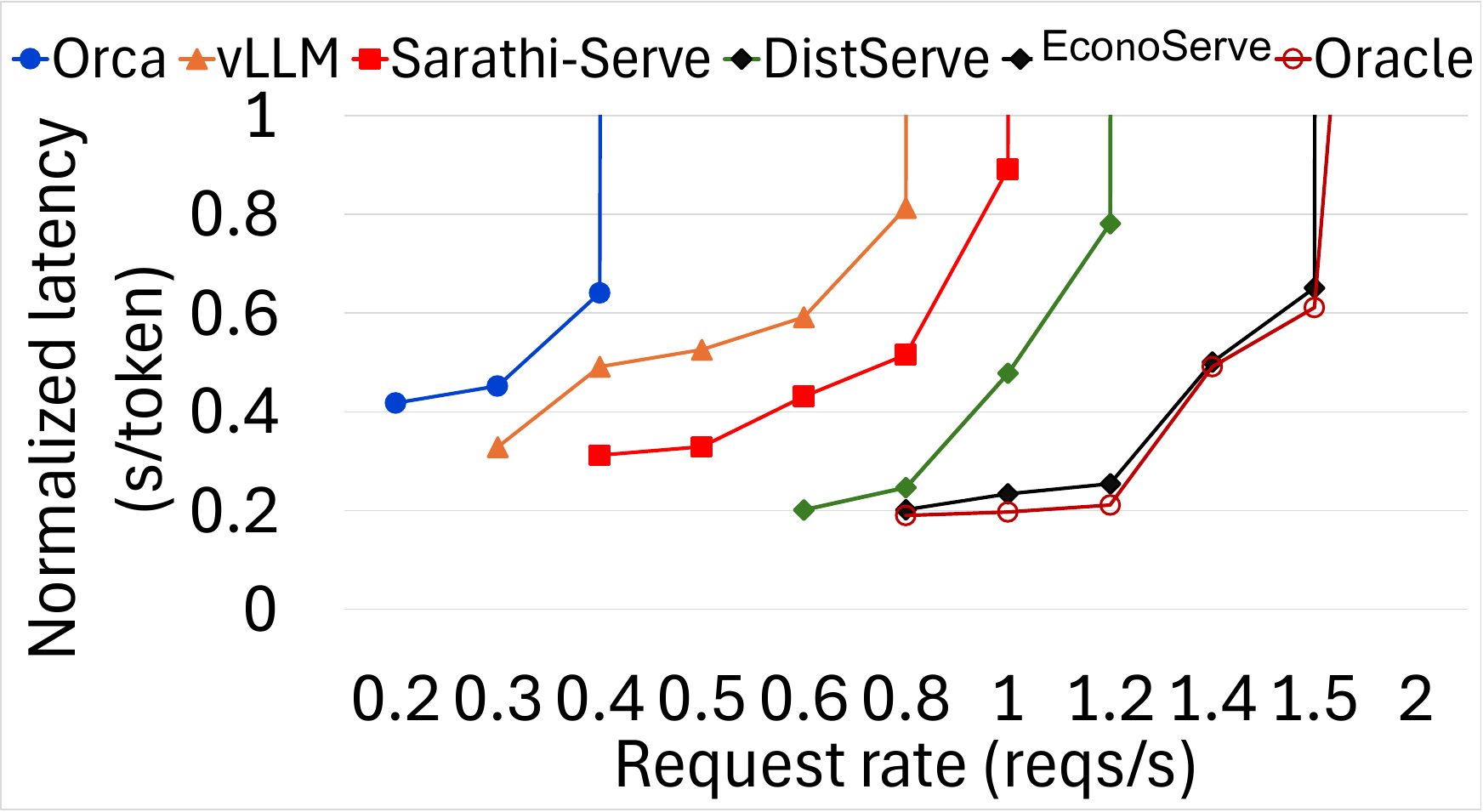} }}
    \hfill
    \subfloat[Llama-33B with BookCorpus.\vspace{-0.0in}\label{fig:exp-3}]{{\includegraphics[width=0.32\linewidth,height=0.112\textheight]{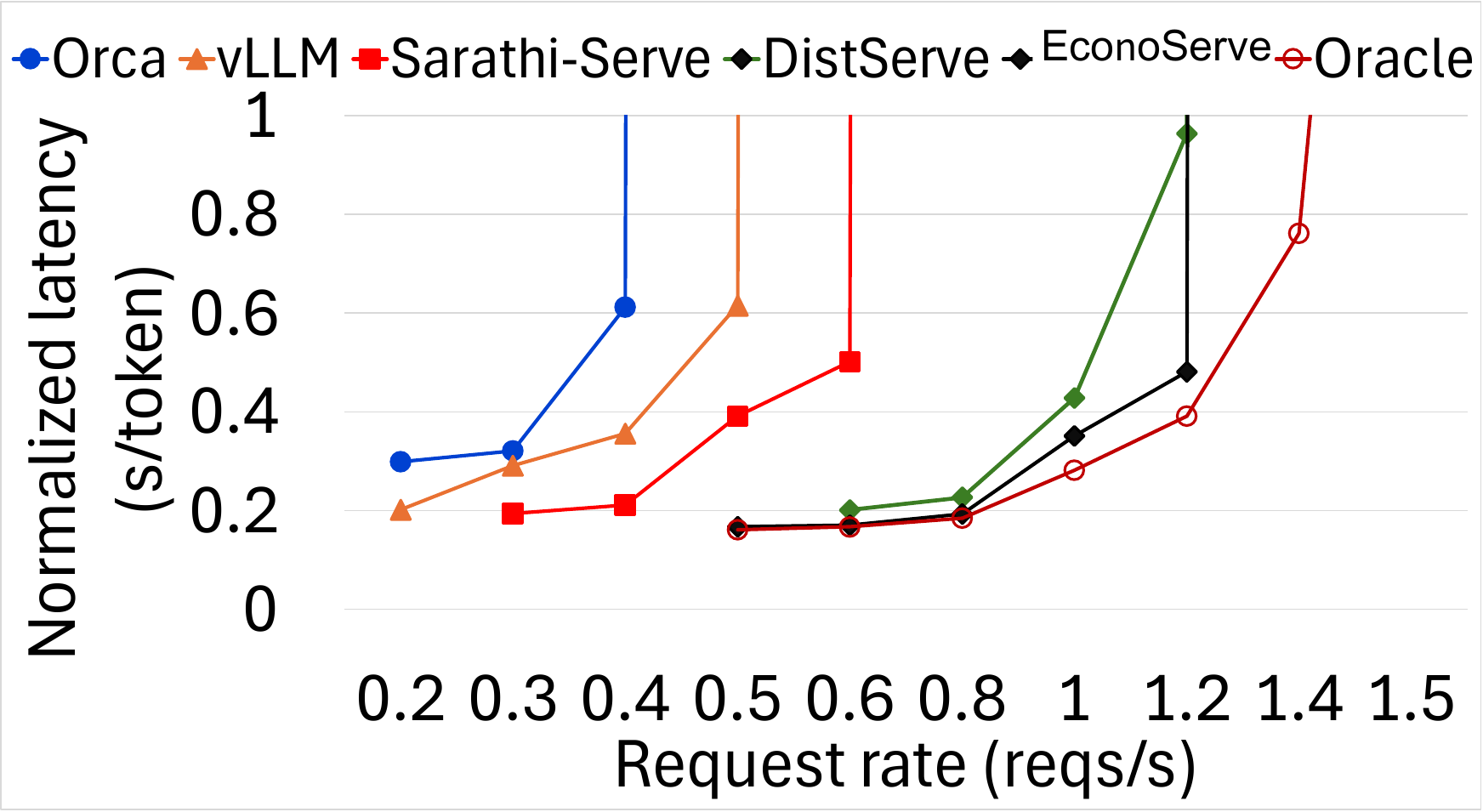} }}
    \hfill
    \subfloat[OPT-13B on Alpaca.\vspace{-0.0in}\label{fig:exp-13-a}]{{\includegraphics[width=0.32\linewidth,height=0.112\textheight]{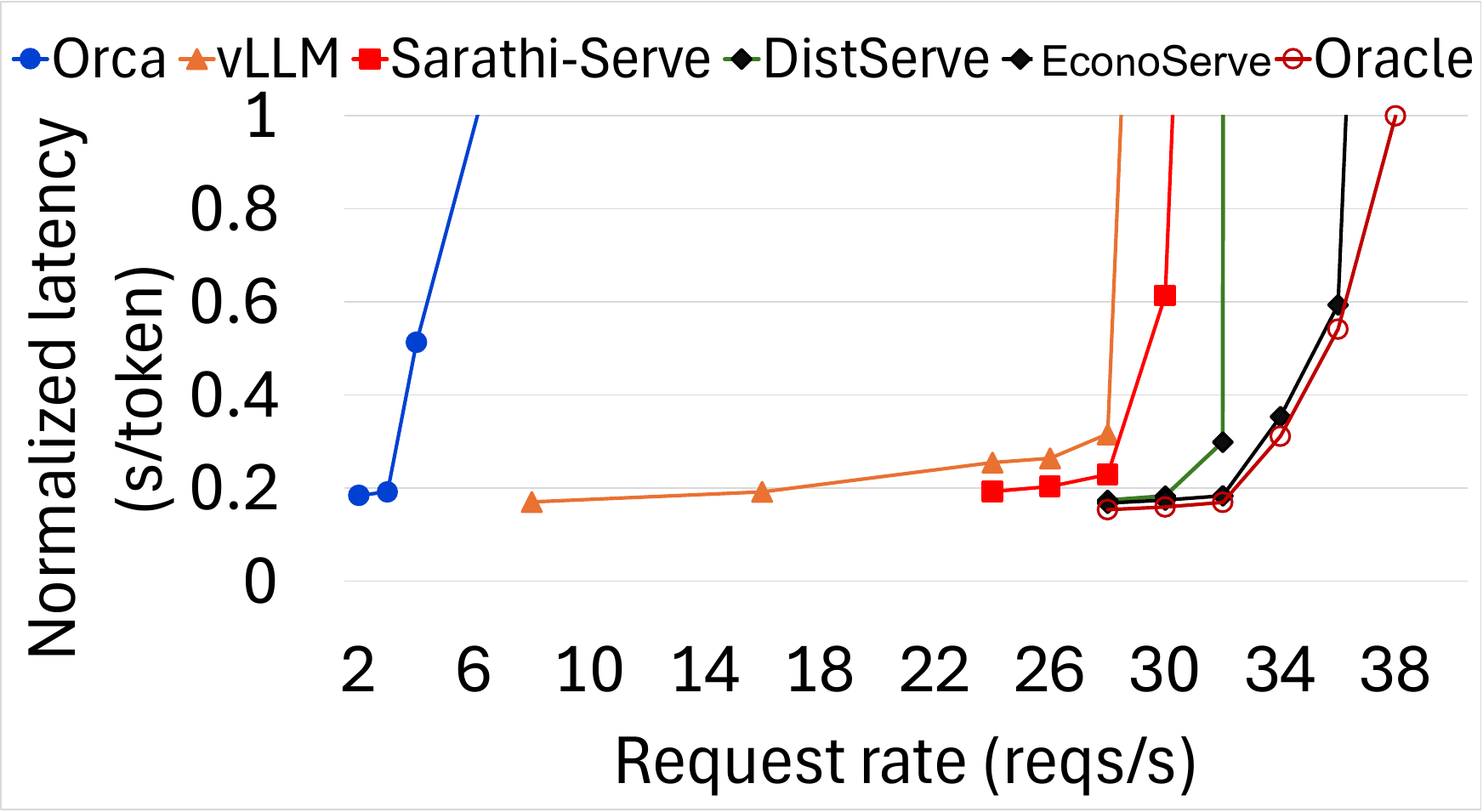} }}
    \hfill
    \subfloat[OPT-175B on Alpaca.\vspace{-0.0in}\label{fig:exp-175-a}]{{\includegraphics[width=0.32\linewidth,height=0.112\textheight]{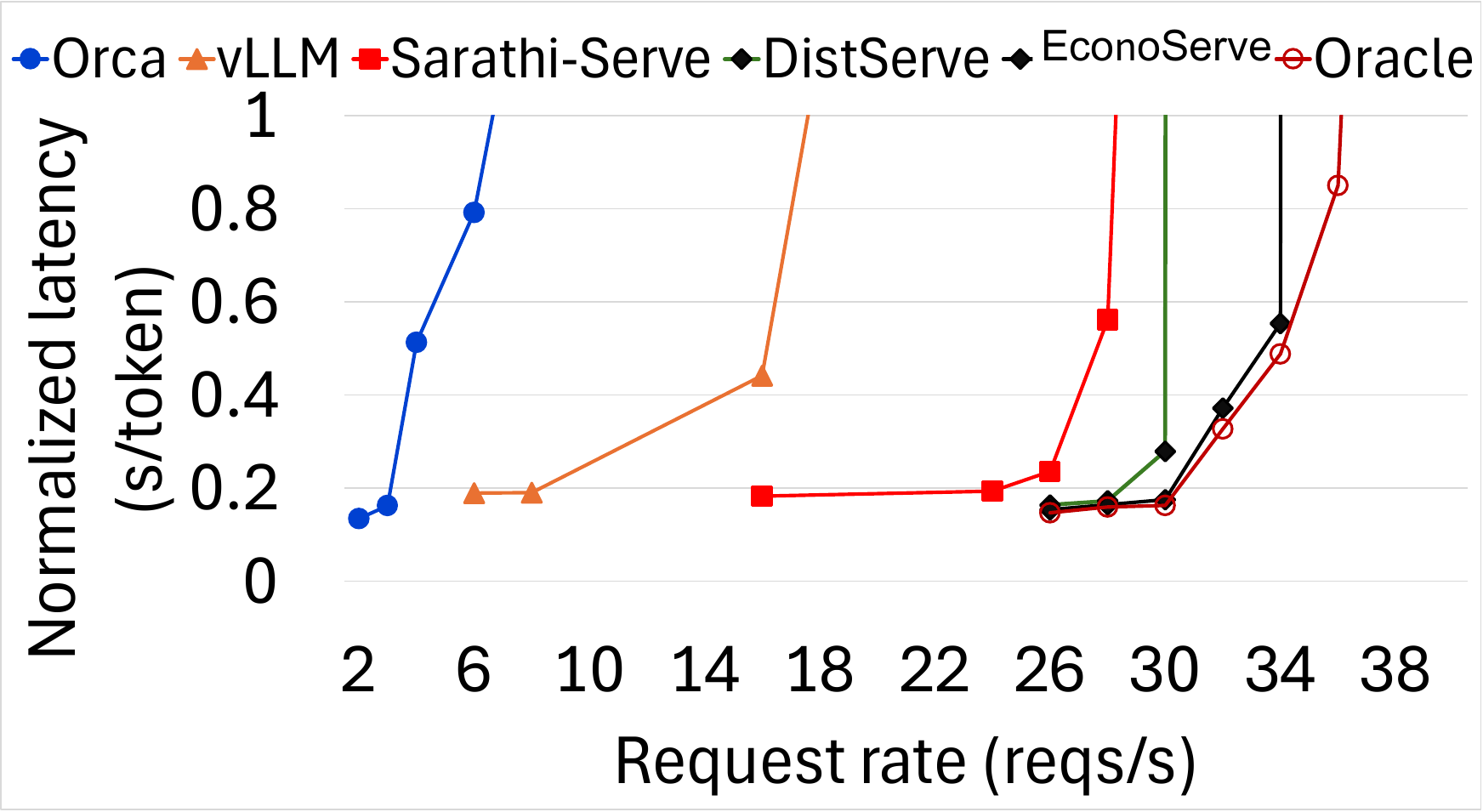} }}
    \hfill
    \subfloat[Llama-33B on Alpaca.\vspace{-0.0in}\label{fig:exp-l3-al}]{{\includegraphics[width=0.32\linewidth,height=0.112\textheight]{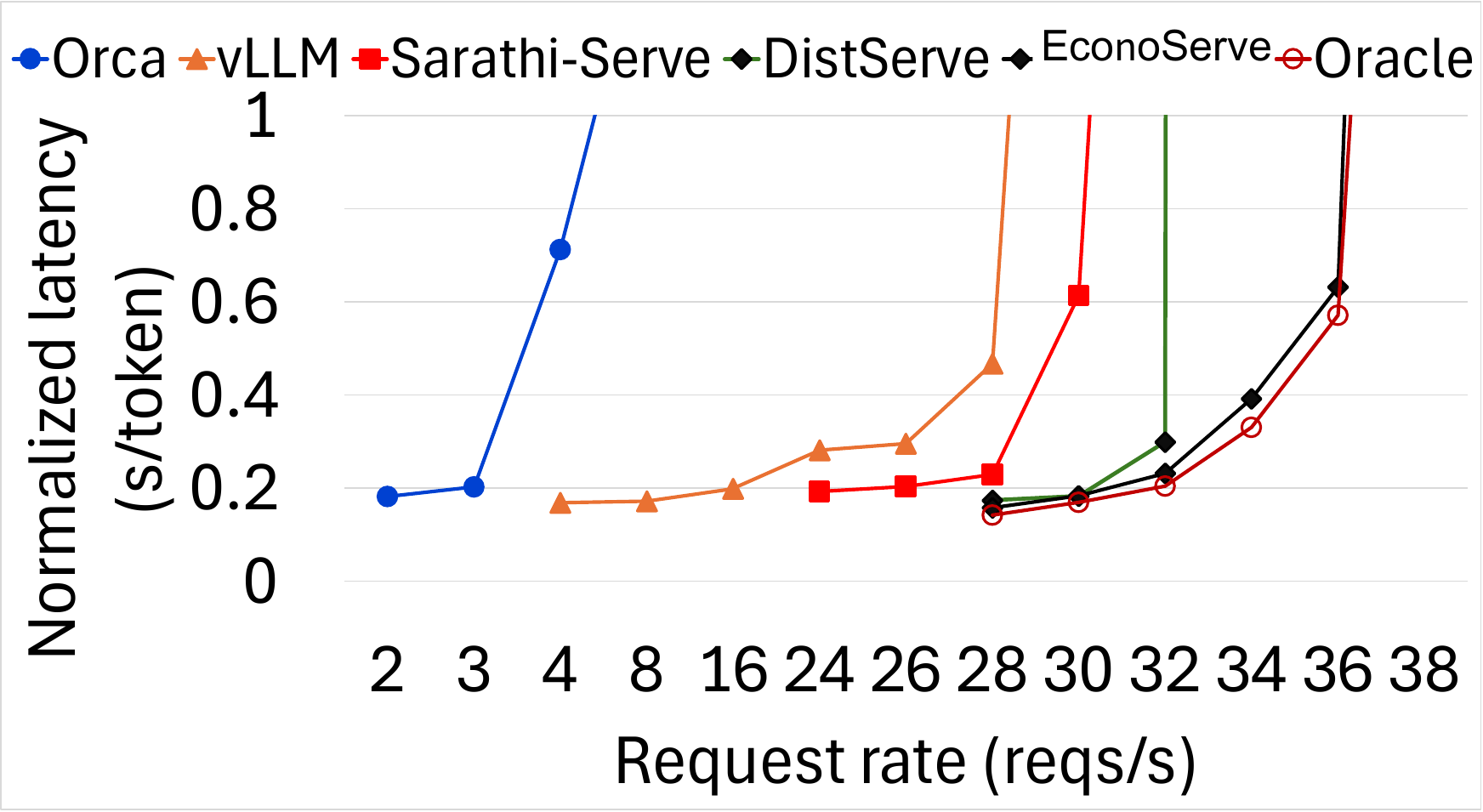} }}
    \hfill
    \vspace{-0.1in}
   \caption{\small{Performance comparison for all models on the three traces (DistServe utilizes twice as many GPUs compared to other methods).\vspace{-0.2in}}}%
    \label{fig:prompt-seq-vllm}
\end{figure*}

\begin{figure*}[t]
\centering
    \subfloat[SSR of OPT-13B.\vspace{-0.0in}\label{fig:ssr-13b}]{{\includegraphics[width=0.32\linewidth,height=0.112\textheight]{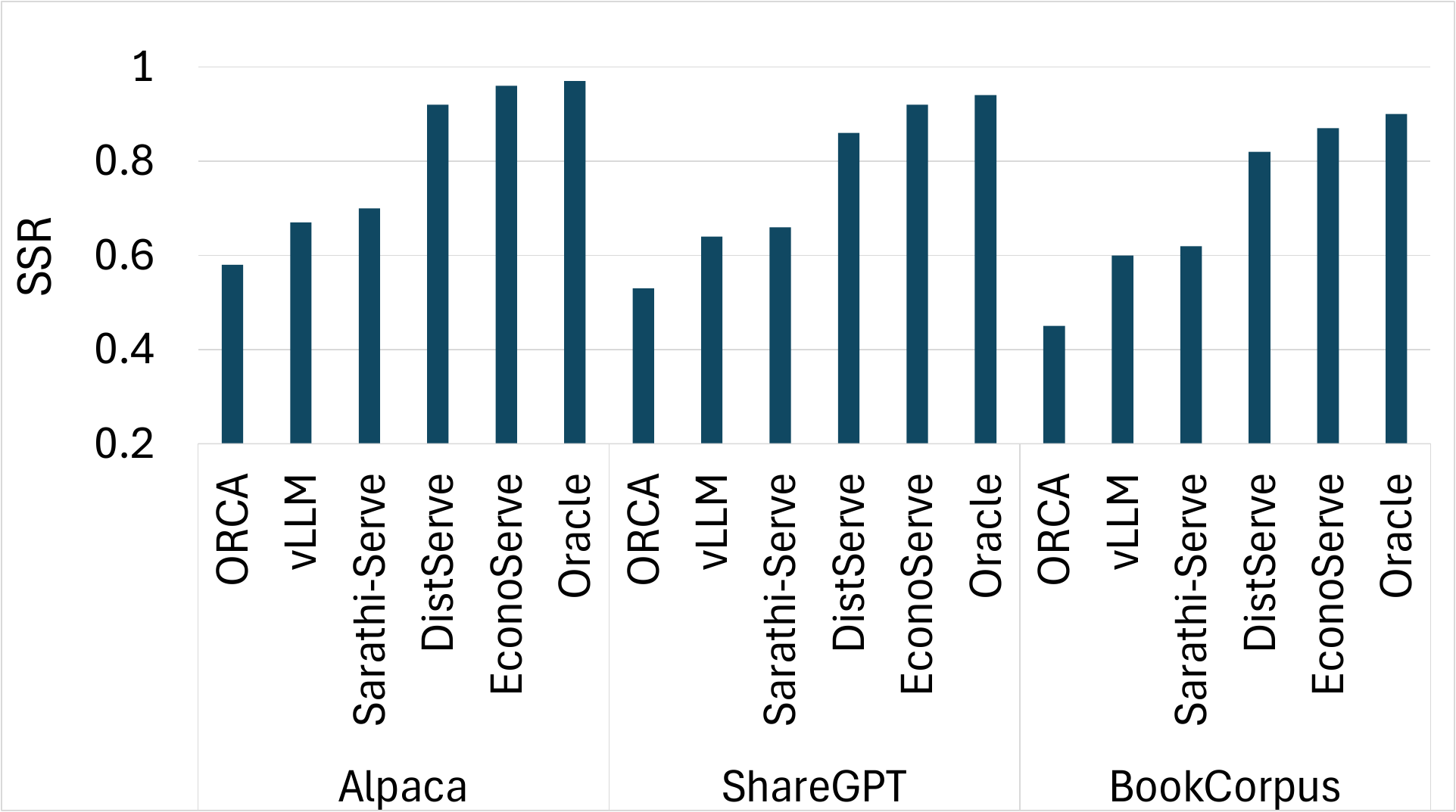} }}
    \hfill
    \subfloat[SSR of OPT-175B.\vspace{-0.0in}\label{fig:ssr-175b}]{{\includegraphics[width=0.32\linewidth,height=0.112\textheight]{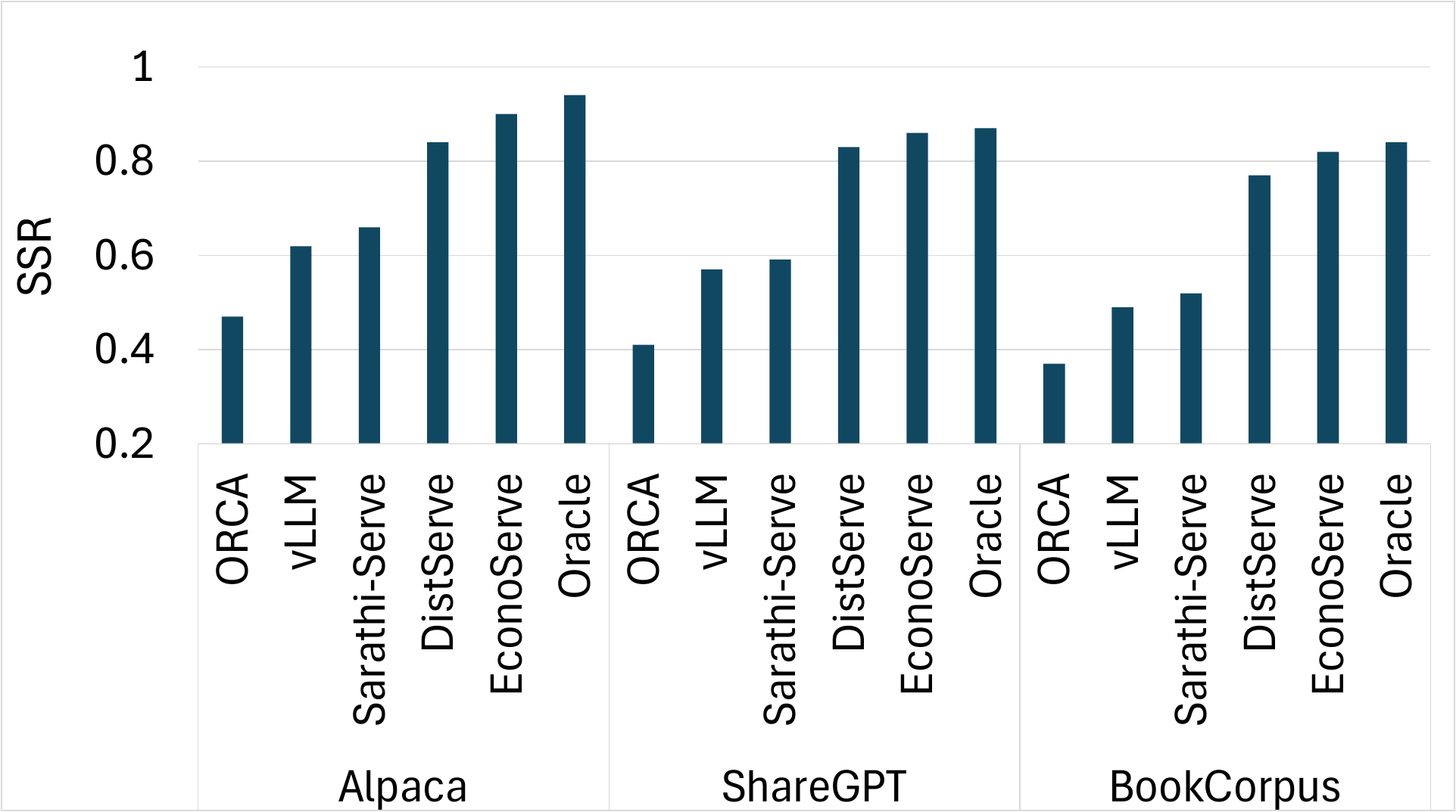} }}
    \hfill
    \subfloat[SSR of Llama-33B.\vspace{-0.0in}\label{fig:ssr-llama}]{{\includegraphics[width=0.32\linewidth,height=0.11\textheight]{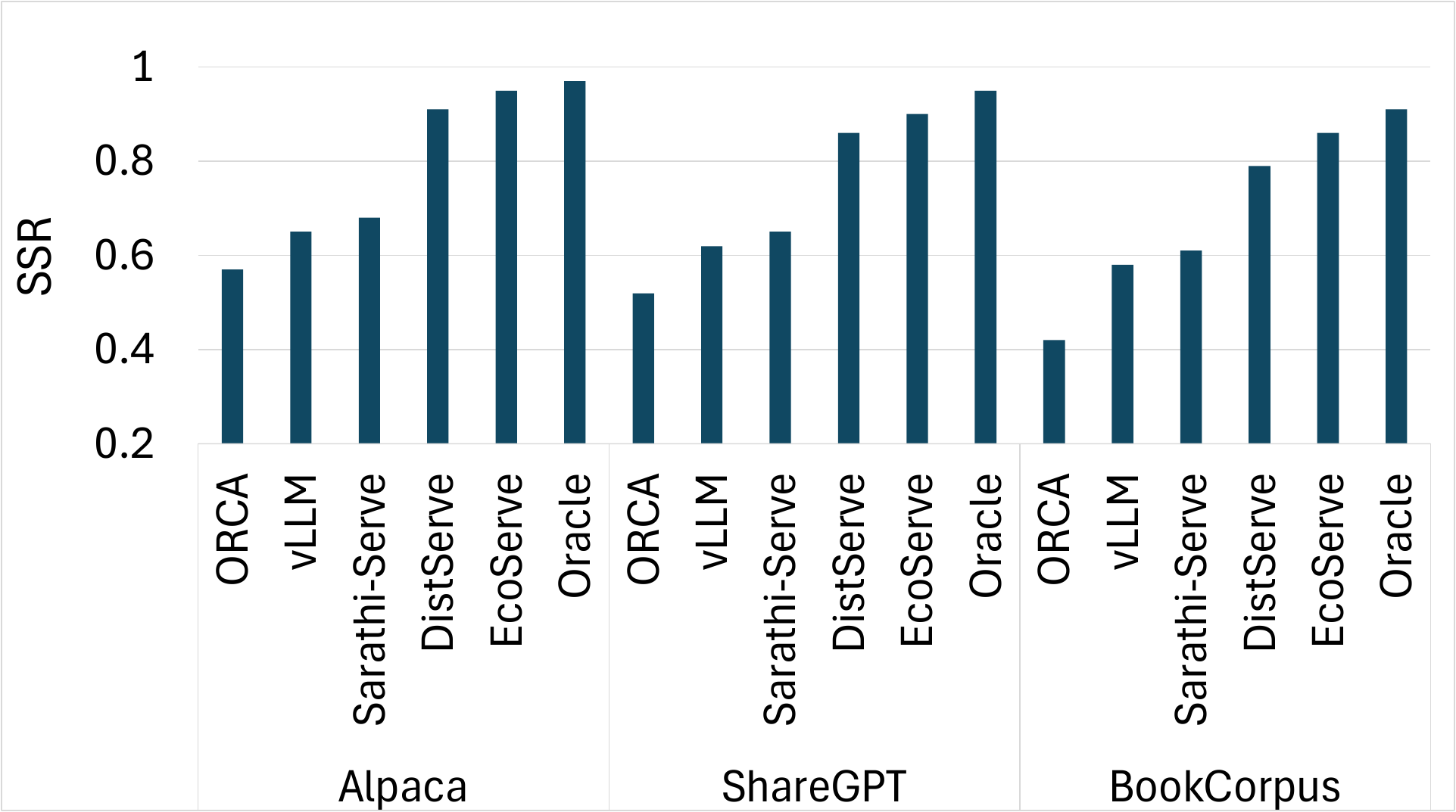} }}
    \hfill
    \vspace{-0.15in}
   \caption{\small{Overall SSR comparison for all models on the three traces (DistServe utilizes twice as many GPUs compared to other methods).\vspace{-0.15in}}}%
    \label{fig:ssr-seq-vllm}
\end{figure*}
\noindent{\textbf{Experiment Settings.}} The experiment settings are the same as in \cref{sec:analysis} unless otherwise specified.
We additionally tested Llama-33B and OPT-175B~\cite{Zhang2022OPTOP} and using model parallelism on two and eight GPUs of the AWS machine mentioned in~\cref{sec:analysis}, respectively, following strategies in~\cite{vllm}. The KVC memory for them is 19.2GB and 264GB, respectively. We set the batch size of \Orca to 16 for OPT-175B as in~\cite{280922}. 
\DEL{The default padding ratio is 10\%, 15\%, and 20\% and the reserved KVC is 1.2\%, 3\%, and 5\%, for the Alpaca, ShareGPT, and BookCorpus datasets, respectively.} We used the sweetspot padding ratios identified in \cref{sec:analysis} and empirically set the buffer  $b$ for \emph{KVCPipe} (\cref{sec:comp4}) to 15\%, 15\%, and 10\% of the predicted RL for Alpaca, ShareGPT, and BookCorpus, respectively. 
For each dataset, we measured the average prompt processing latency, $t_p$, and the average token generation latency, $t_g$. We then set the JCT SLO of a request to \emph{SLO-scale}$\times(t_p+t_g \times l_g)$, where $l_g$ denotes its RL, following~\cite{slo-ref}. As in~\cite{Li2023AlpaServeSM}, the default SLO-scale was set to 2. We used 3-hour trace for OPT-13B and Llama-33B and used 1-hour trace for OPT-175B. \looseness=-1




\DEL{{{correct. we only have JCT SLO now-done}} 
We reported the average with the 5th and 95th percentile values{\sh{make sure of it}}. }

\noindent\textbf{Implementation.} We implemented \sys using the vLLM source code~\cite{vllm}, comprising about 7K lines of Python and 2K lines of C++ code. The KVC pipelining method was written in C++, while the batching scheduler and queue management were implemented in Python. To order the prompts and same-RL GTs, we maintained multiple priority queues from the class \emph{queue.PriorityQueue}. The memory management process was built using the \emph{append\_token} function scheme in~\cite{vllm}. Required metadata, such as a table containing the start and end memory addresses of the requests, was stored in the host memory. The kernel in vLLM was adopted to read and write in the KVC. To ensure parallel memory access, a GPU thread block was assigned to read/write the memory. Allocation for the predicted RL for the request groups was handled using a different thread block. The \emph{cudaMemcpyAsync} API was used to manage data movements between various levels of memory. Prompt and token generation were batched in one iteration as in nanoGPT~\cite{nanogpt}. \sys also allows parallelization of the attention computation using the \emph{flash\_attn\_qkv\_packed} function from the FlashAttention~\cite{Dao2023FlashAttention2FA} API. 
\looseness=-1

\DEL{We maintained multiple priority queues from the class \emph{queue.PriorityQueue} to order the prompts and same-RL GTs. The memory management process was built using the scheme \emph{append\_token} function in~\cite{vllm}. We stored the required metadata, such as a table containing the start and end memory addresses of the requests, in the host memory. We adopted the kernel in vLLM to read and write in the  KVC. To ensure parallel memory access, we assigned a GPU thread block to read/write the memory. We handled the allocation for the predicted RL for the request groups using a different thread block. We used the \emph{cudaMemcpyAsync} API to handle the data movements between the various levels of memory. We batched the prompt and token generation in one iteration as in nanpGPT~\cite{nanogpt}. \sys also allows the parallelization of the attention computation using \emph{flash\_attn\_qkv\_packed} function from the FlashAttention~\cite{Dao2023FlashAttention2FA} API.?? }

\DEL{Upon the RL prediction on a request, the request is entered to the queue. The prompts and GTs are selected by a customizable scheduler implemented in Python to be added to the batch. }

\noindent{\textbf{Compared Methods.}} We compared \sys with \Orca, vLLM, Sarathi-Serve and DistServe. 
By default, DistServe runs on two GPUs distributed across two machines as in \cref{sec:analysis}.
We also tested the variants of \sys. \sys-D (i.e., \emph{UnsynedDecoupled})
\underline{d}ecouples prompt and GT processing, 
selects tasks sequentially from the two queues to fully utilize the GPU and KVC in each iteration, and uses the exact-allocation. \sys-SD is \emph{SyncDecouple}. \sys-SDO further incorporates the \emph{Ordering} method. 
Note that the performance difference between \sys and \sys-SDO indicates the effectiveness of \emph{KVCPipe}. We also tested \sys
with full knowledge of the RLs, denoted as \emph{Oracle}. Normalized latency of the system is the mean of every request's end-to-end latency divided by its output length~\cite{vllm}.\looseness=-1 


\begin{figure*}[!t]
\centering
    \subfloat[KVC of OPT-13B.\vspace{-0.0in}\label{fig:kvc-13b}]{{\includegraphics[width=0.32\linewidth,height=0.112\textheight]{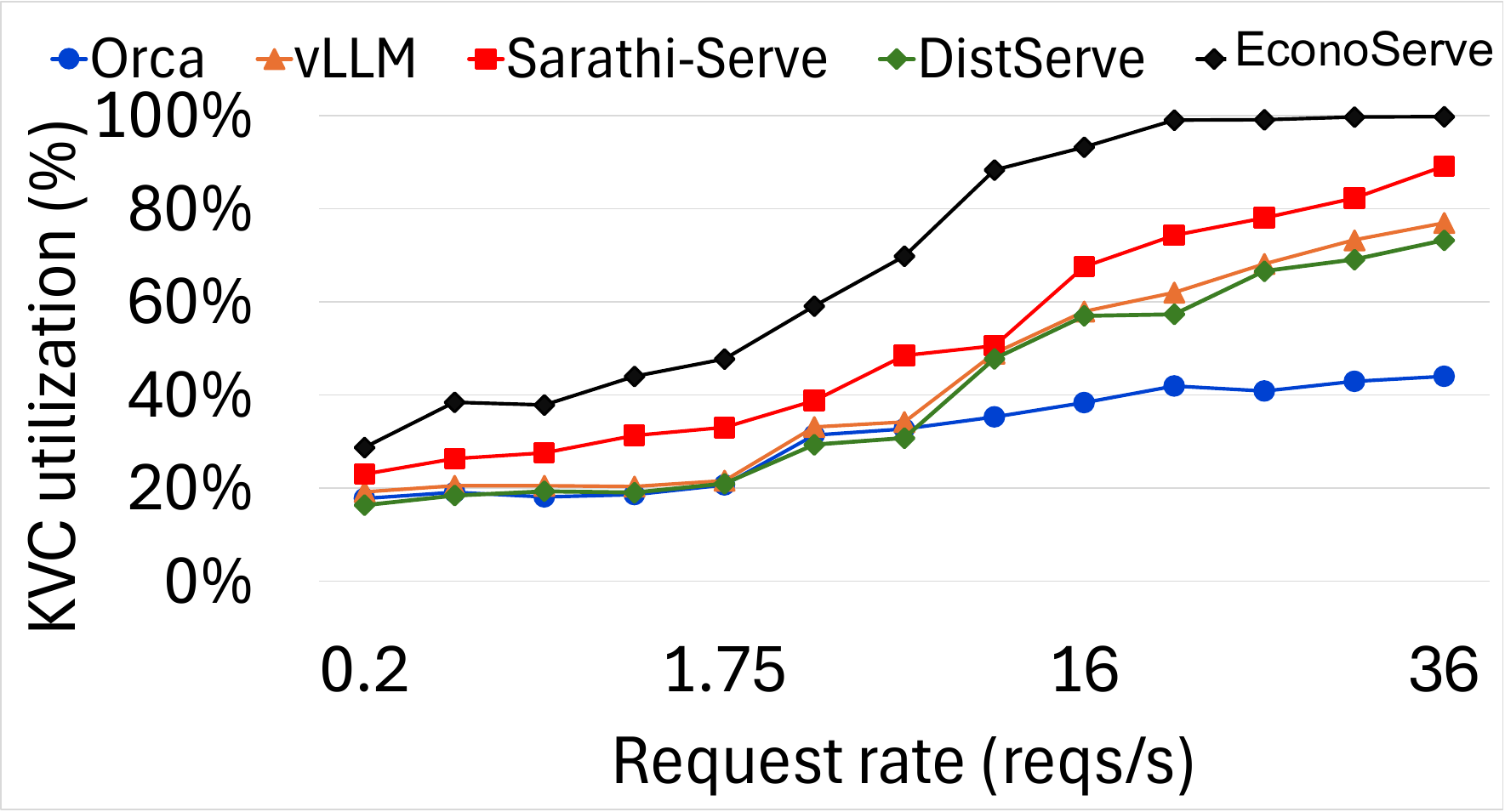} }}
    \hfill
    \subfloat[KVC of OPT-175B.\vspace{-0.0in}\label{fig:kvc-175b}]{{\includegraphics[width=0.32\linewidth,height=0.112\textheight]{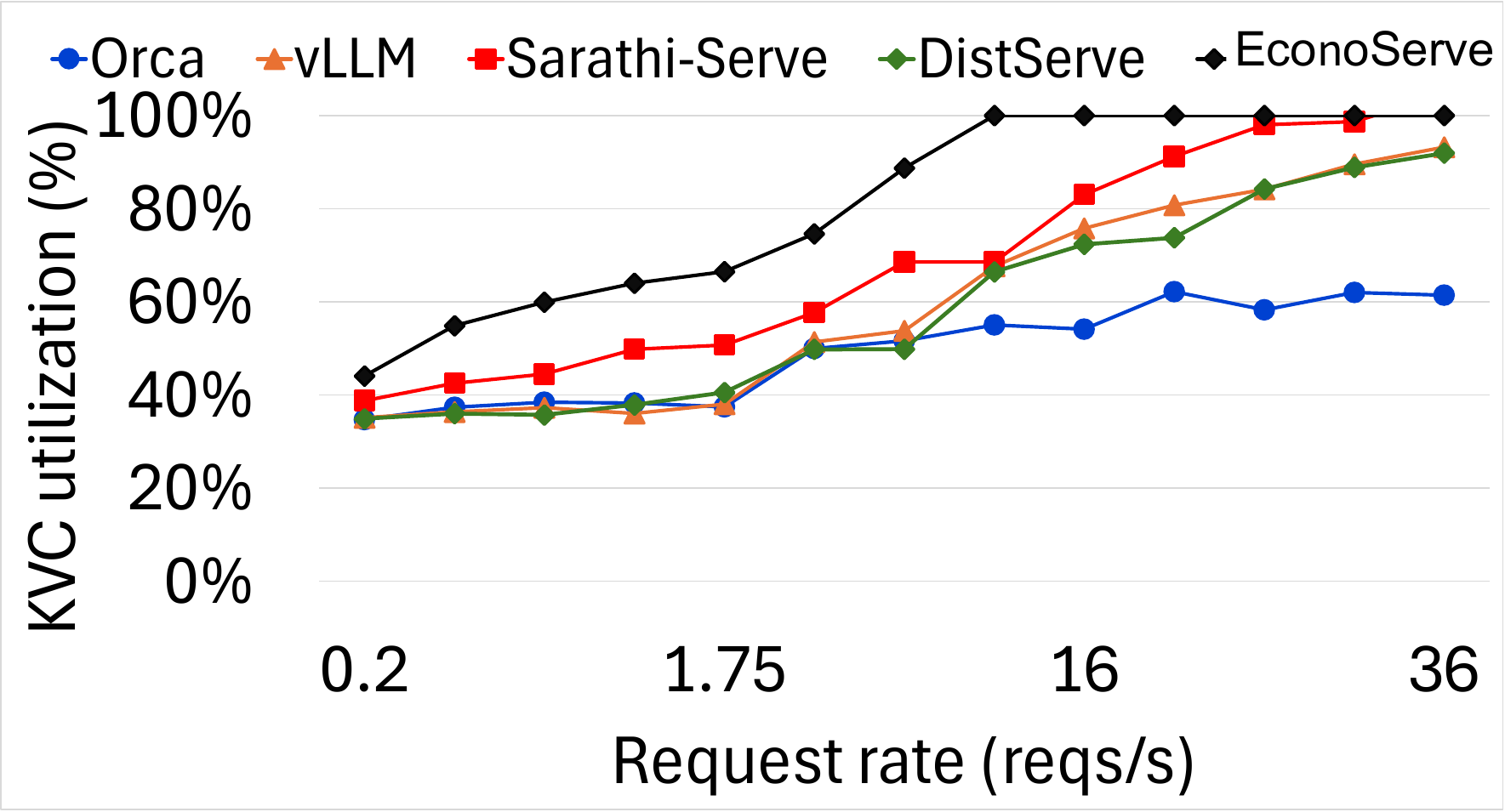} }}
    \hfill
    \subfloat[KVC of Llama-33B.\vspace{-0.0in}\label{fig:kvc-llama}]{{\includegraphics[width=0.32\linewidth,height=0.11\textheight]{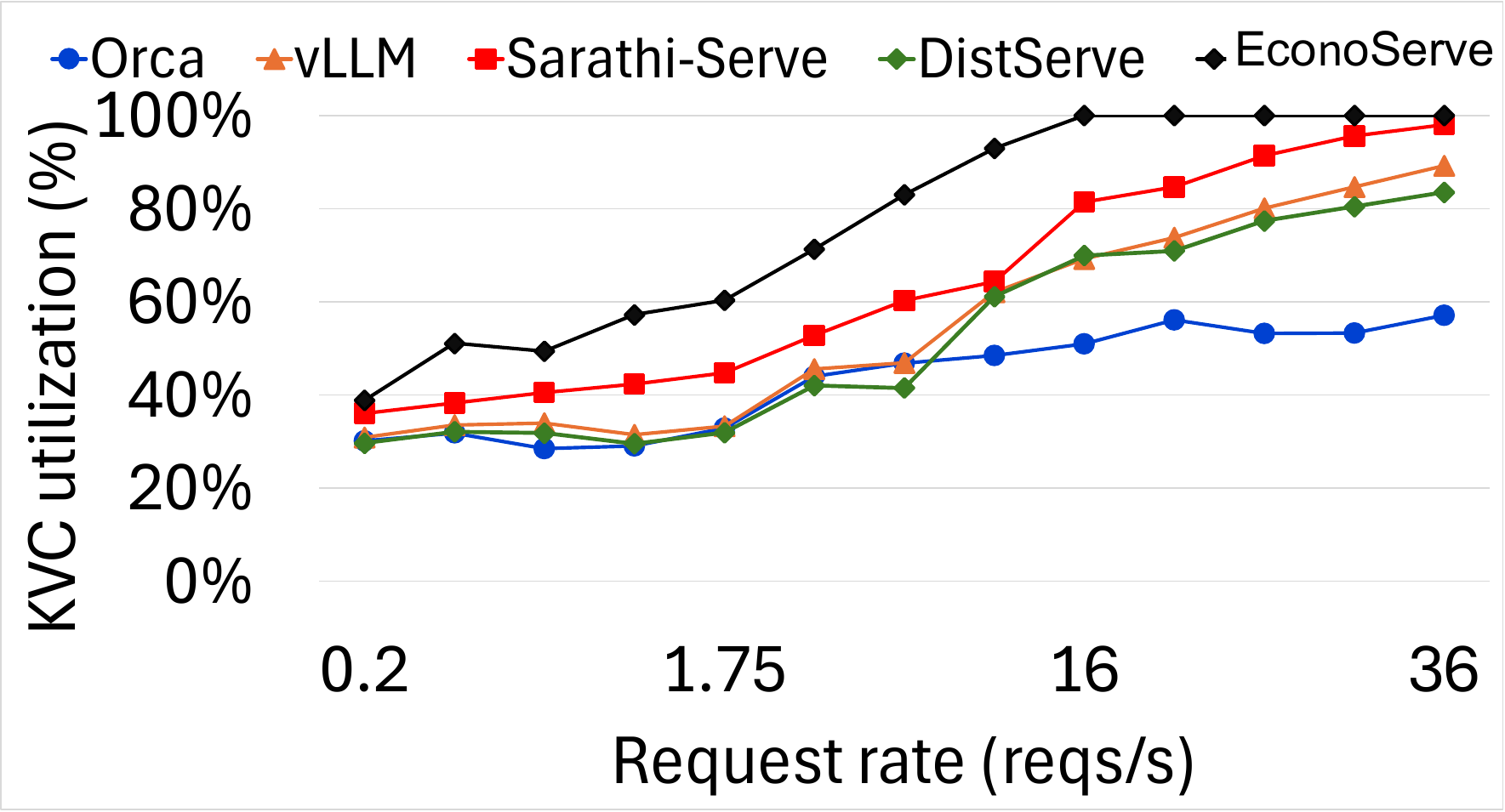} }}
    \hfill
    \subfloat[GPU utilization of OPT-13B.\vspace{-0.0in}\label{fig:gpu-13b}]{{\includegraphics[width=0.32\linewidth,height=0.112\textheight]{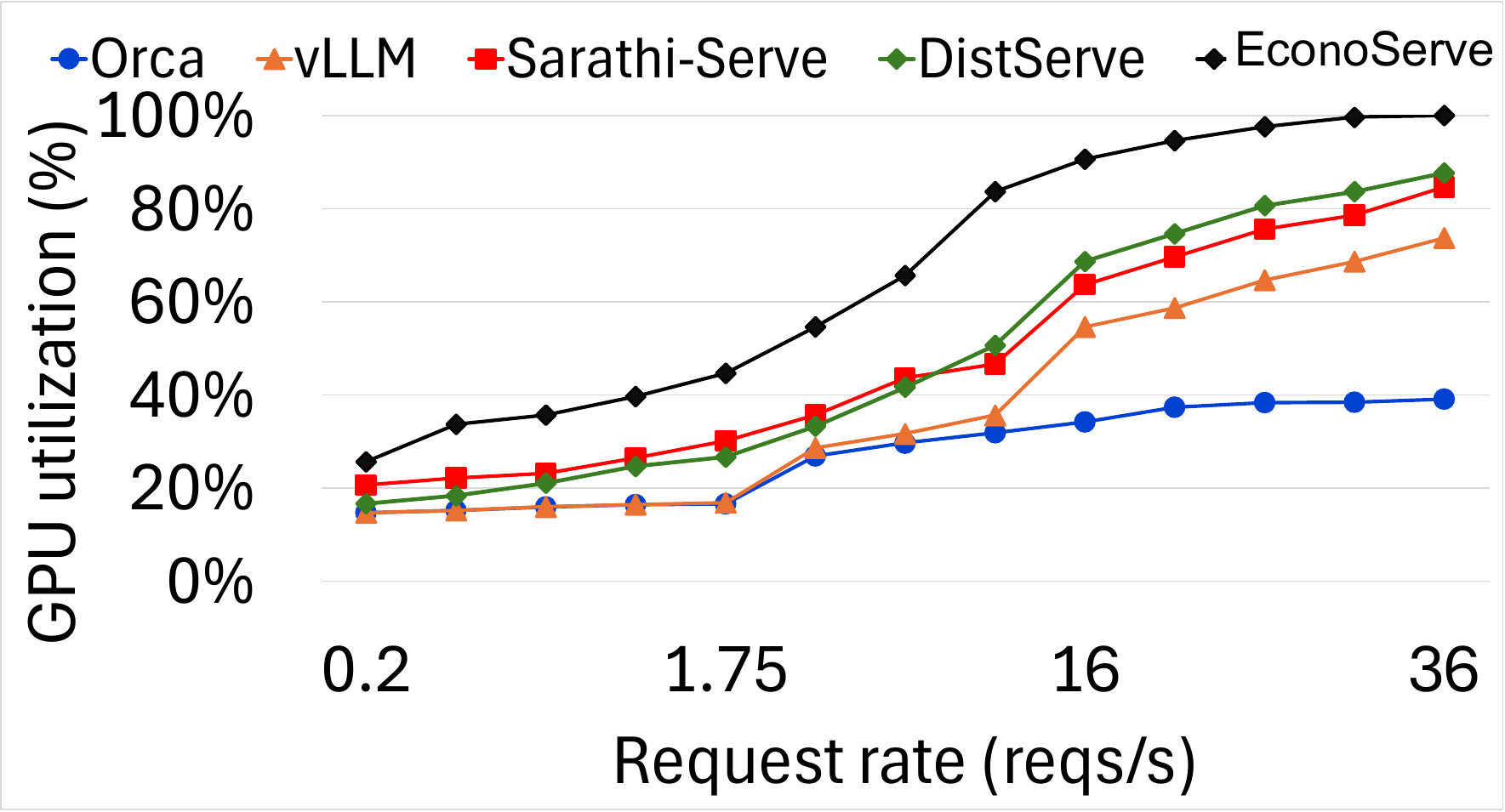} }}
    \hfill
    \subfloat[GPU utilization of OPT-175B.\vspace{-0.0in}\label{fig:gpu-175b}]{{\includegraphics[width=0.32\linewidth,height=0.112\textheight]{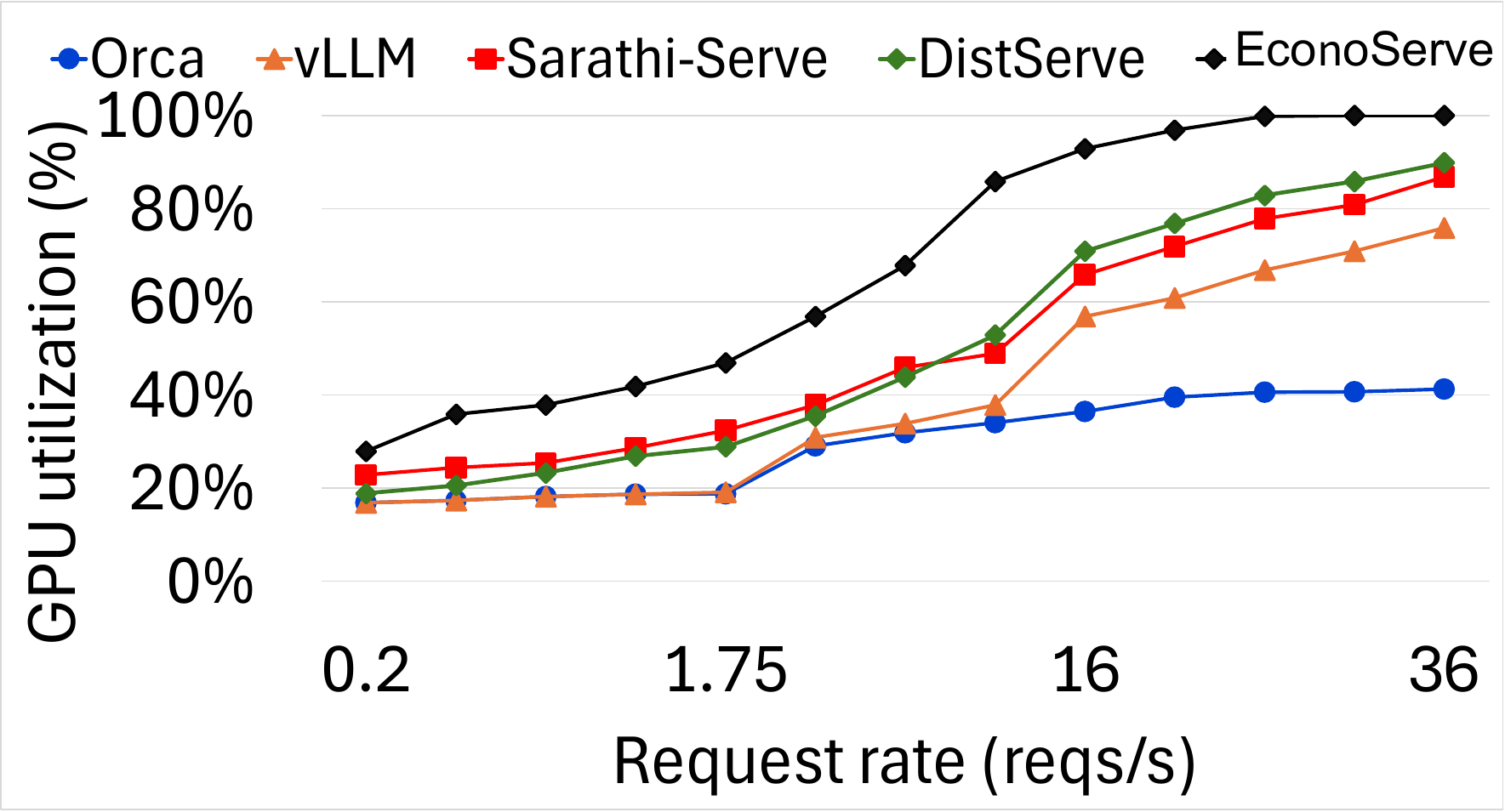} }}
    \hfill
    \subfloat[GPU utilization of Llama-33B.\vspace{-0.0in}\label{fig:gpu-llama}]{{\includegraphics[width=0.32\linewidth,height=0.11\textheight]{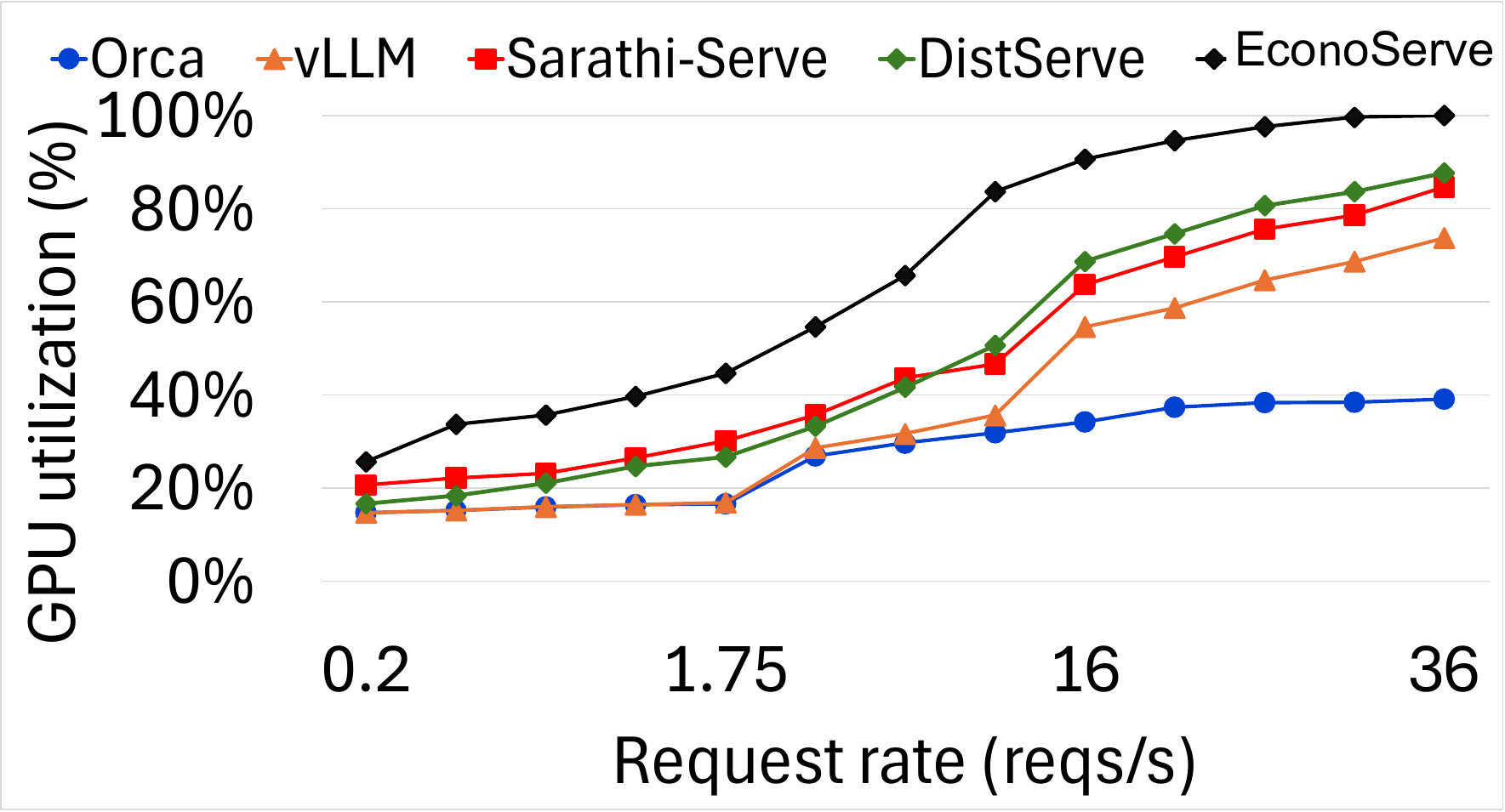} }}
    \hfill
    \vspace{-0.1in}
   \caption{\small{Overall utilization comparison for all models on ShareGPT (DistServe utilizes twice as many GPUs compared to other methods).\vspace{-0.15in}}}%
    \label{fig:utilization-seq-vllm}
\end{figure*}

\begin{figure*}[]
    \centering
  \subfloat[Homogeneous.\vspace{-0.0in}\label{fig:gpu-homo}]{{\includegraphics[width=0.325\linewidth,height=0.112\textheight]{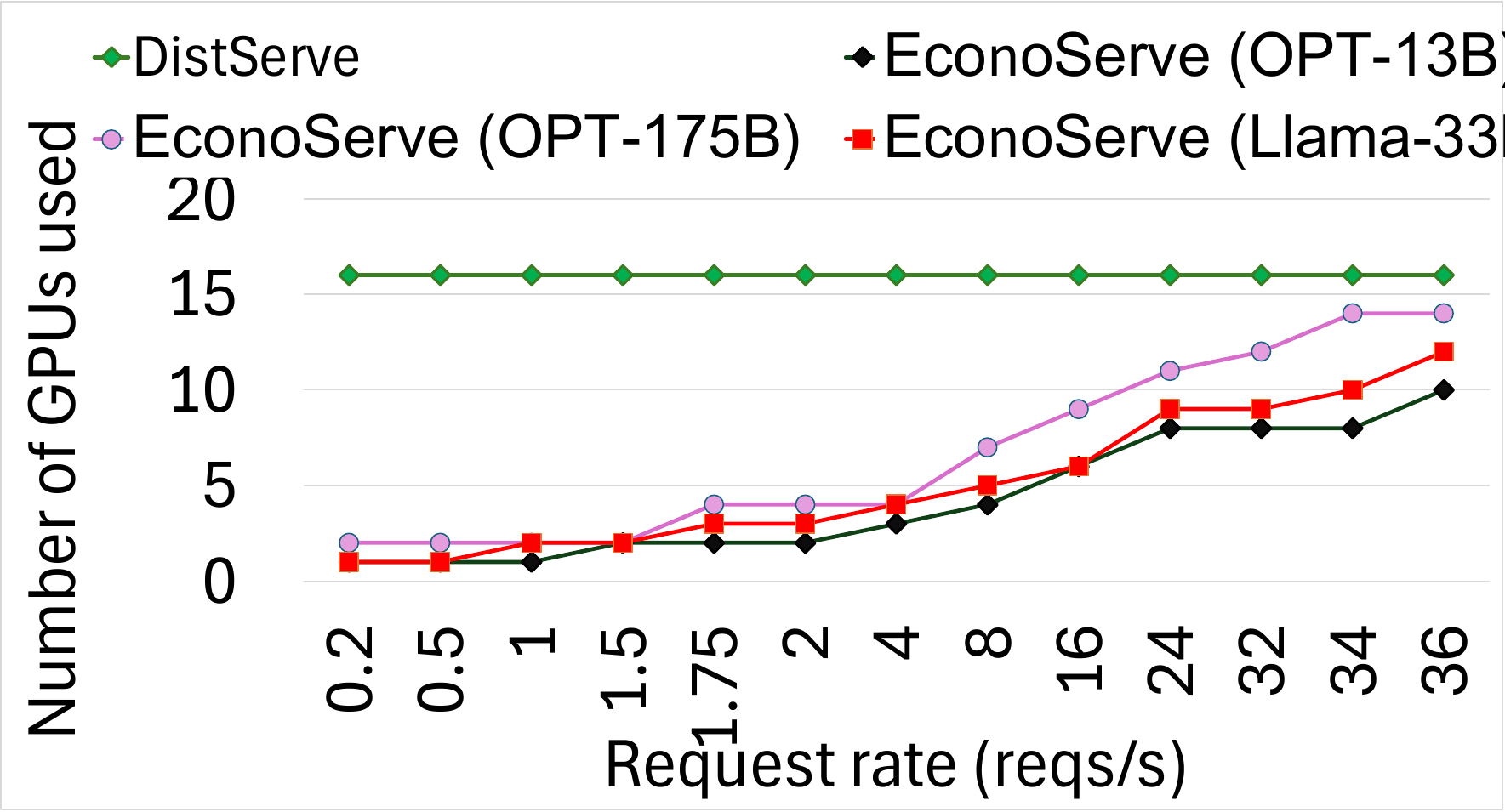}  }}
    \hfill
    \subfloat[Heterogeneous.\vspace{-0.0in}\label{fig:gpu-hetero}]{{\includegraphics[width=0.325\linewidth,height=0.112\textheight]{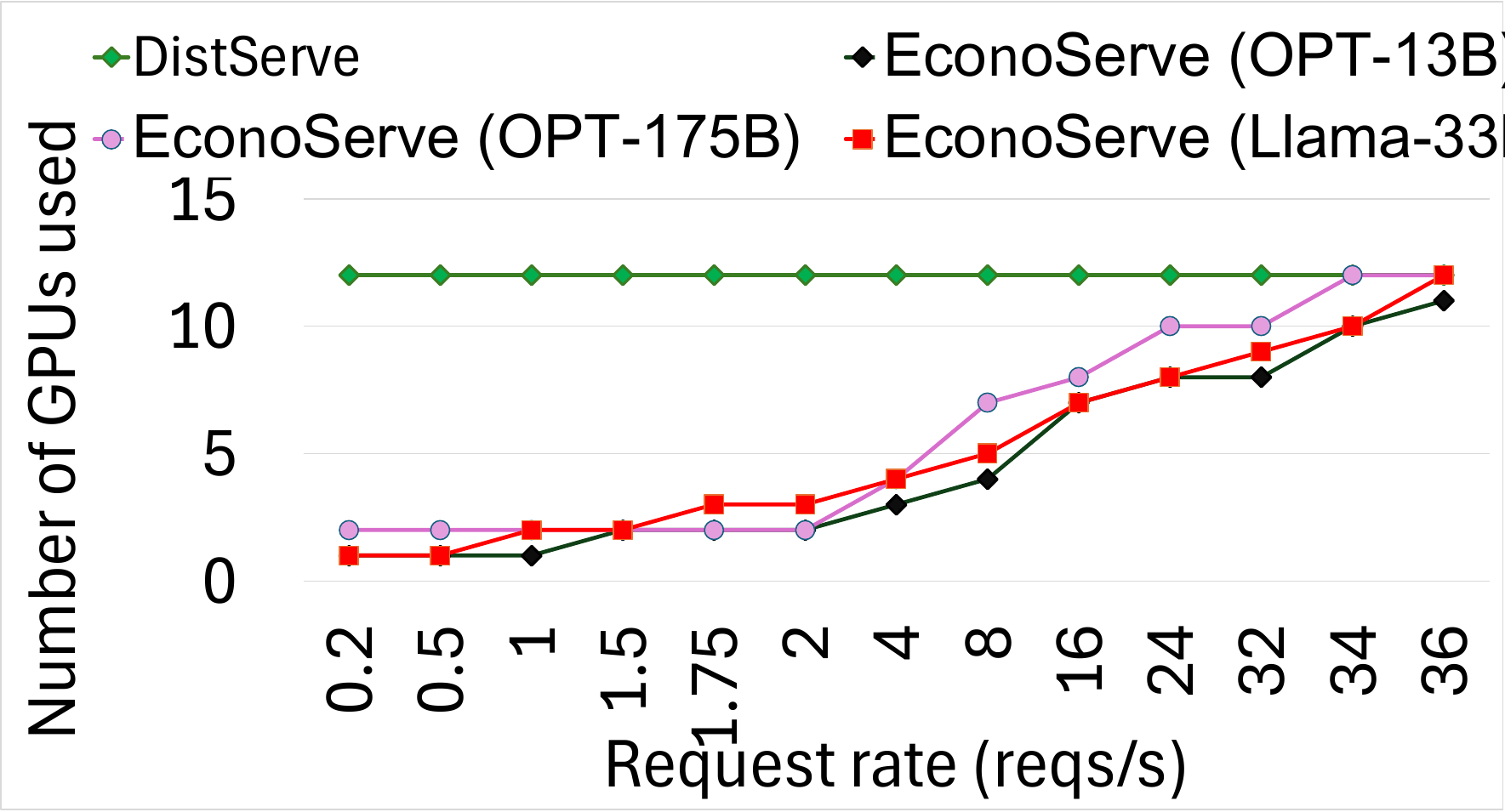} }}
    \hfill
    \subfloat[Simulator.\vspace{-0.0in}\label{fig:gpu-simulator}]{{\includegraphics[width=0.325\linewidth,height=0.112\textheight]{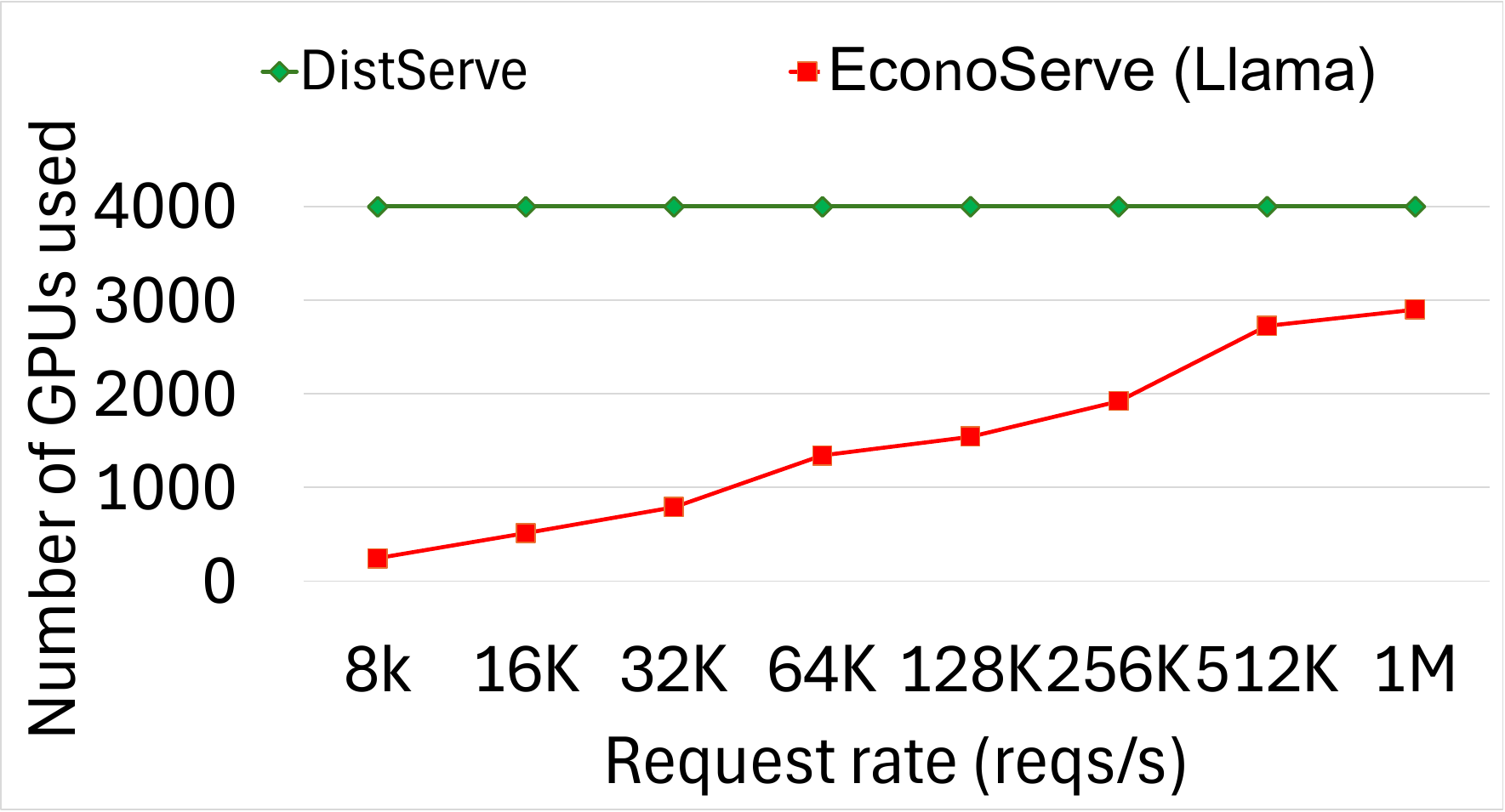} }}
    \hfill
\vspace{-0.01in}\caption{Comparison with DistServe regarding the number of GPUs used to achieve the same goodput.}\vspace{-0.15in}
\label{fig:num-of-gpus}
\end{figure*}

\DEL{\begin{table}[h]
\centering
\caption{Offline time for \sys and its variants.}
\label{tab:offline-time}\vspace{-0.0in}
\resizebox{\columnwidth}{!}{%
\begin{tabular}{|l|l|l|l|l|}
\hline
Method & \sys & \sys-D & \sys-DG & \sys-DGS \\ \hline
Offline time (s) & 0.421 & 0.244 & 0.301 & 0.408 \\ \hline
\% to JCT & 9.8\%  & 4.8\%  & 6.2\%  & 9.2\%  \\ \hline
\end{tabular}%
}\vspace{-0.0in}
\end{table}}

\DEL{\begin{figure*}[t]
\centering
     \subfloat[OPT-13B on ShareGPT.\vspace{-0.0in}\label{fig:exp-13-s}]{{\includegraphics[width=0.32\linewidth,height=0.112\textheight]{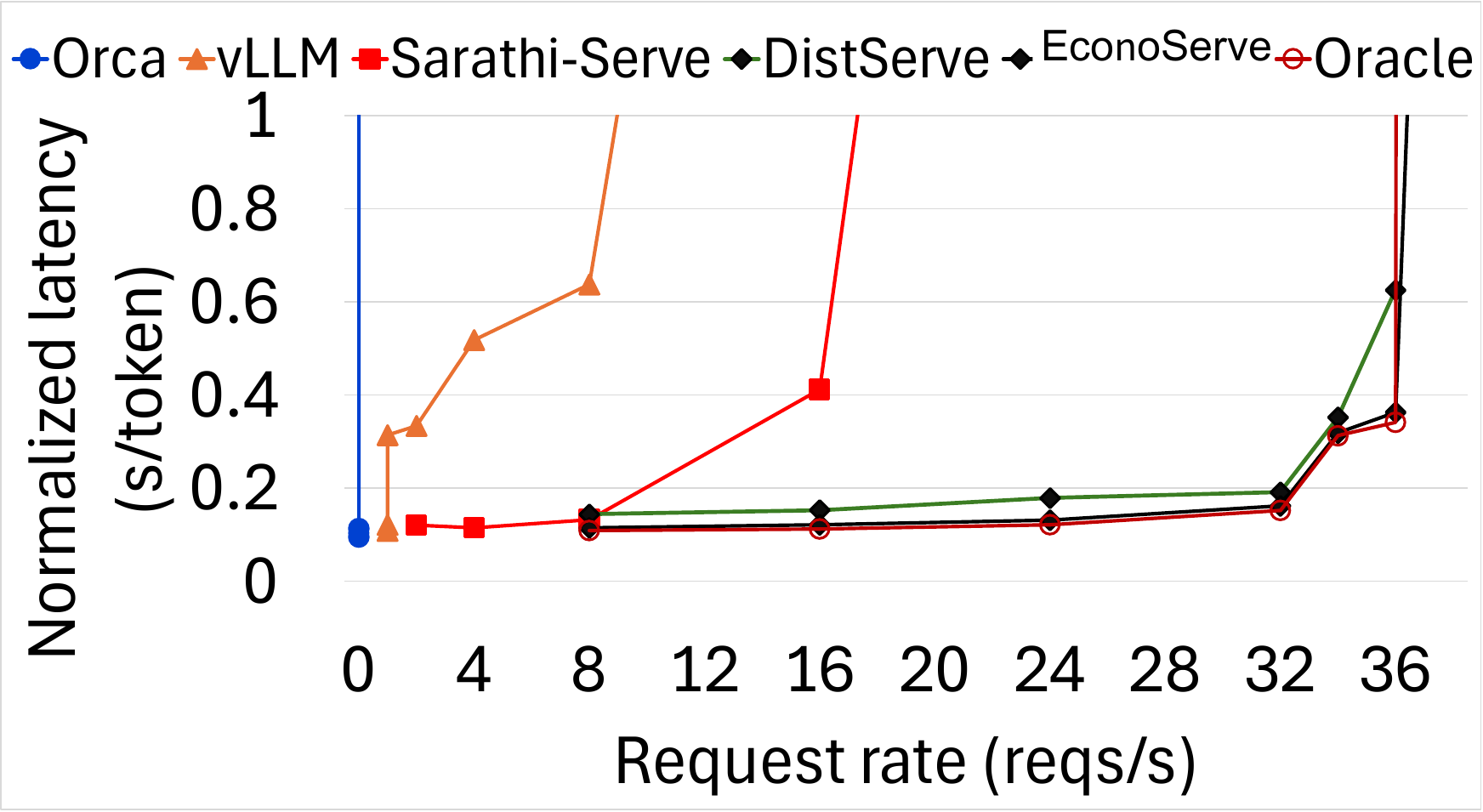} }}
    \hfill
    \subfloat[OPT-175B on ShareGPT.\vspace{-0.0in}\label{fig:exp-175-s}]{{\includegraphics[width=0.32\linewidth,height=0.112\textheight]{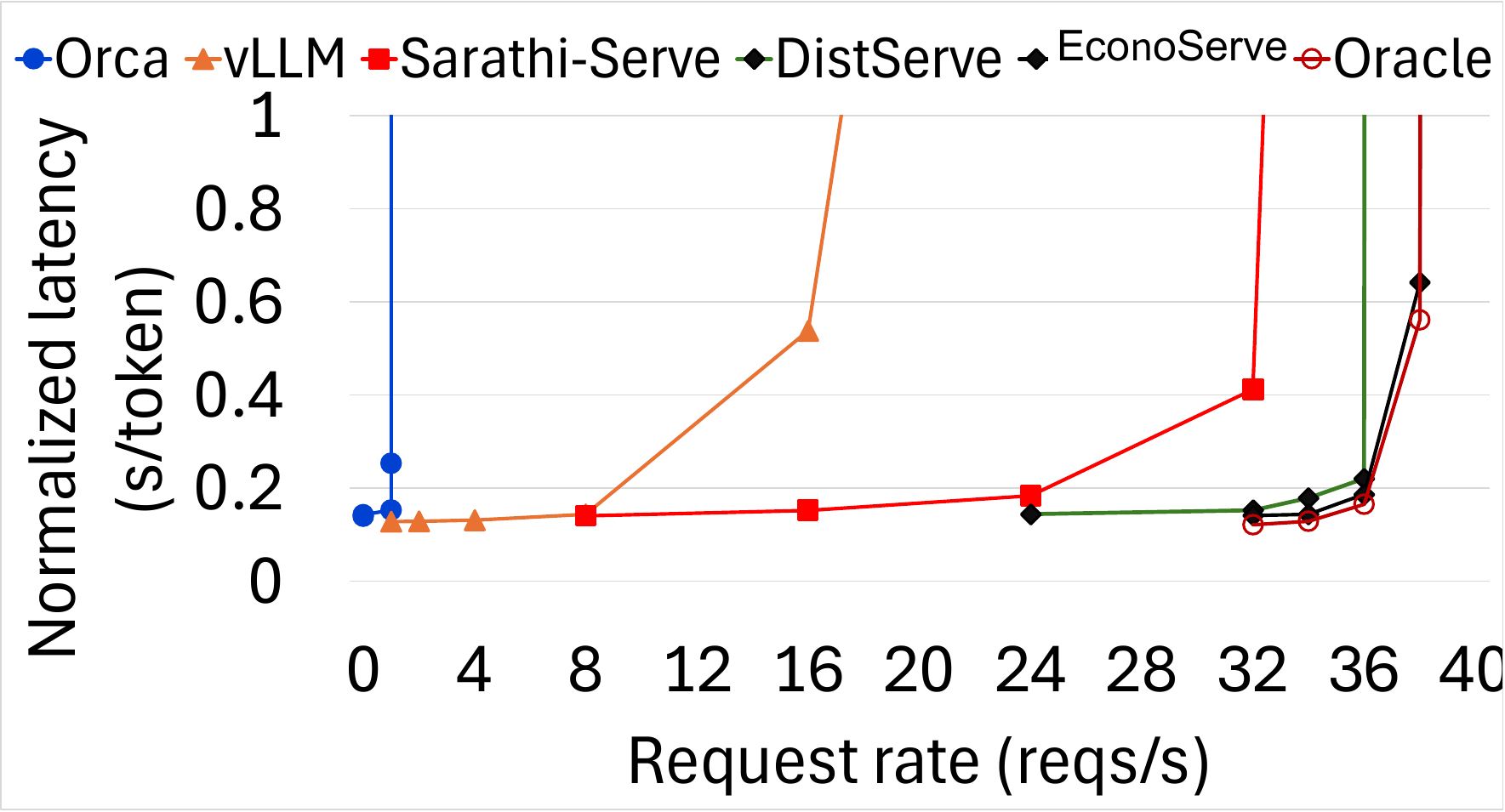} }}
    \hfill
    \subfloat[Llama-33B on ShareGPT.\vspace{-0.0in}\label{fig:exp-l3-s}]{{\includegraphics[width=0.32\linewidth,height=0.112\textheight]{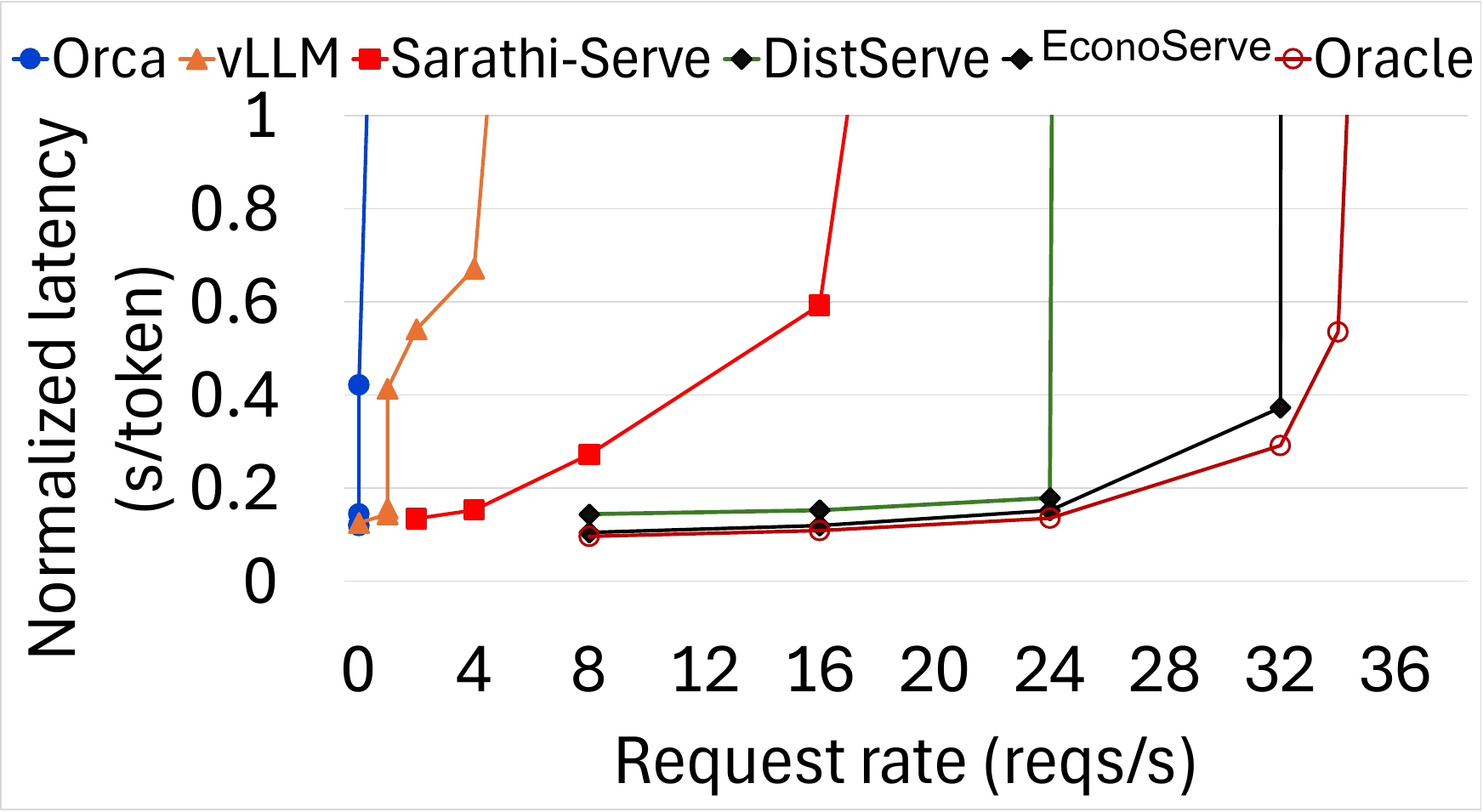} }}
    \hfill
     \subfloat[OPT-13B with BookCorpus.\vspace{-0.0in}\label{fig:exp-13-b}]{{\includegraphics[width=0.32\linewidth,height=0.112\textheight]{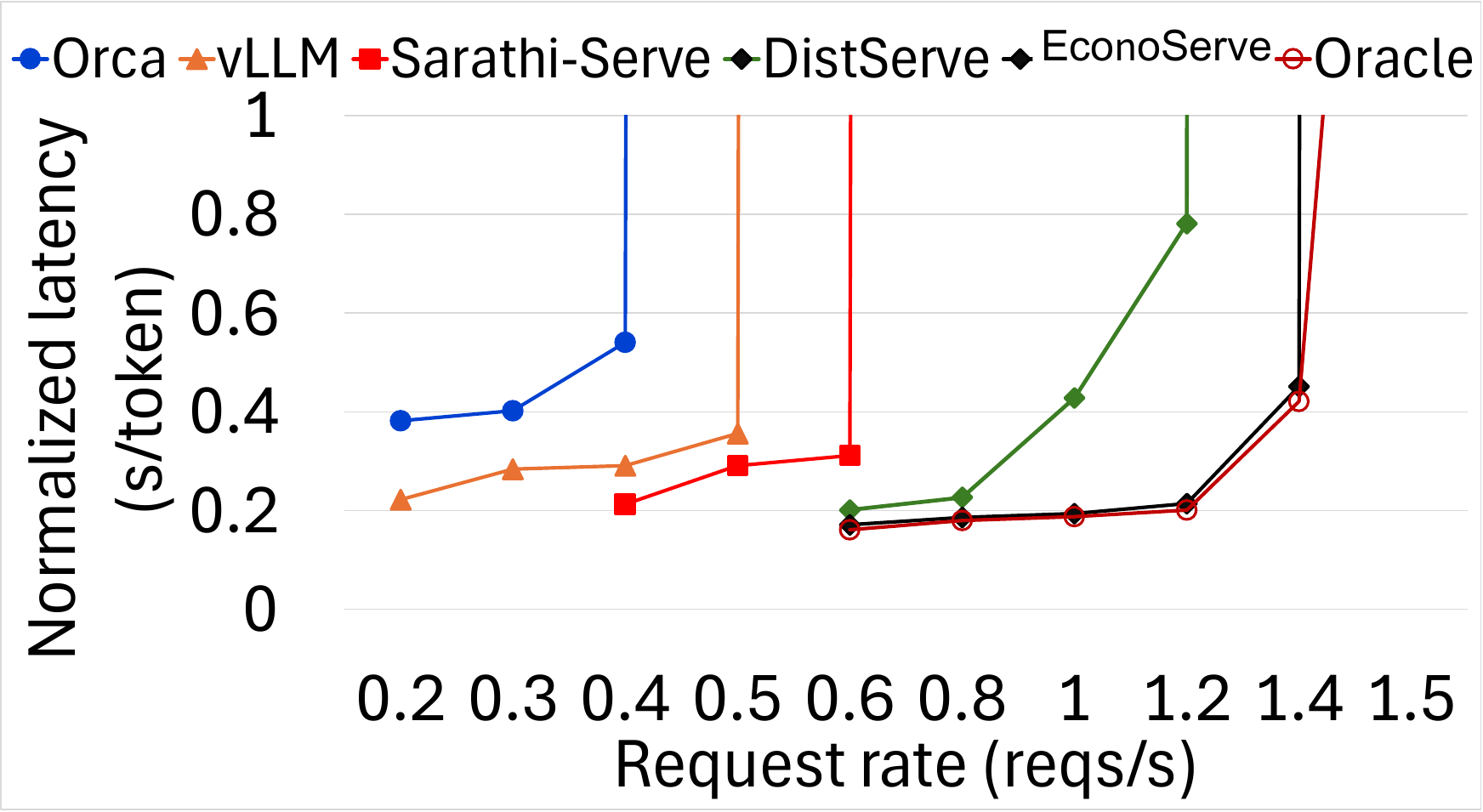} }}
    \hfill
    \subfloat[OPT-175B with BookCorpus.\vspace{-0.0in}\label{fig:exp-175-b}]{{\includegraphics[width=0.32\linewidth,height=0.112\textheight]{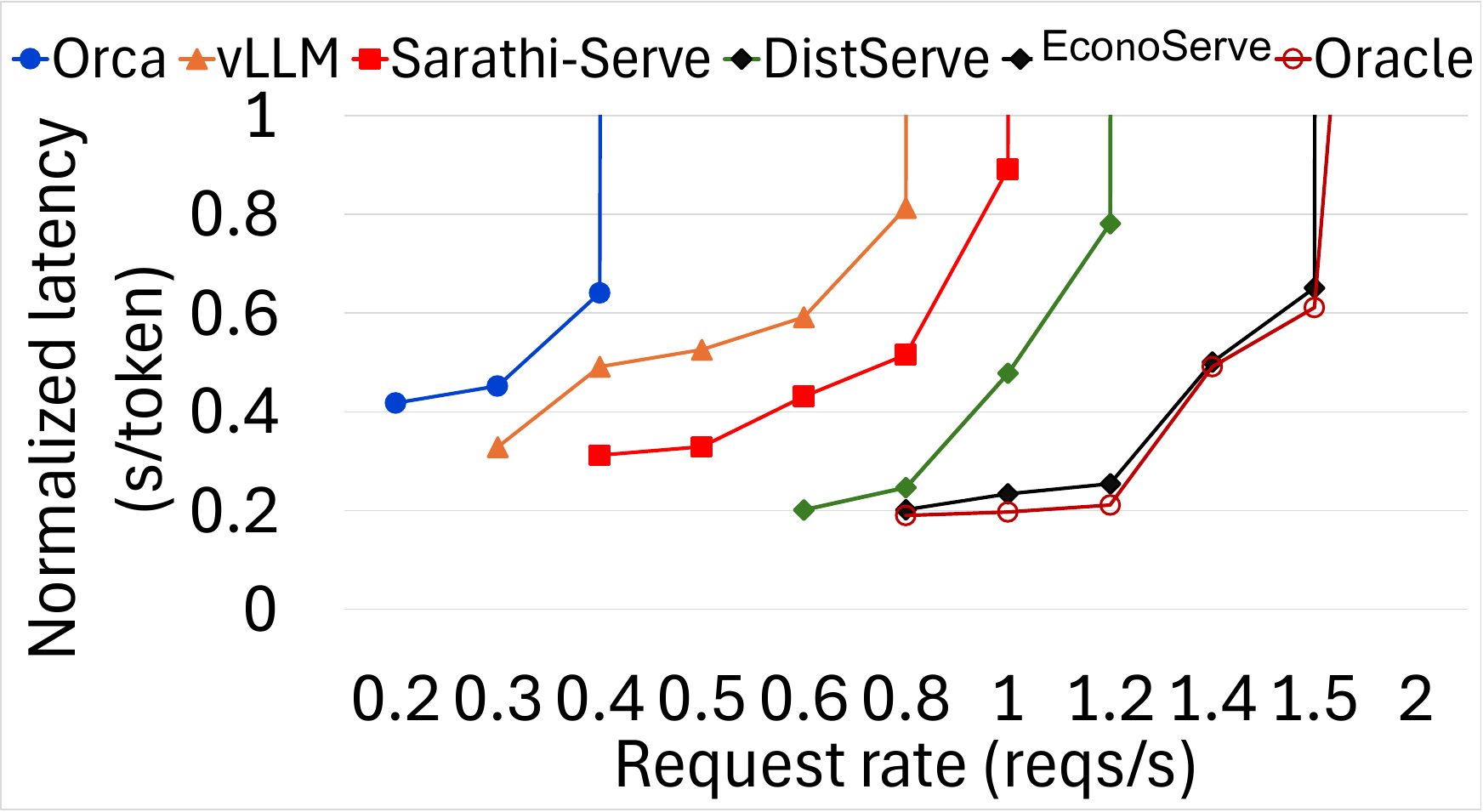} }}
    \hfill
    \subfloat[Llama-33B with BookCorpus.\vspace{-0.0in}\label{fig:exp-3}]{{\includegraphics[width=0.32\linewidth,height=0.112\textheight]{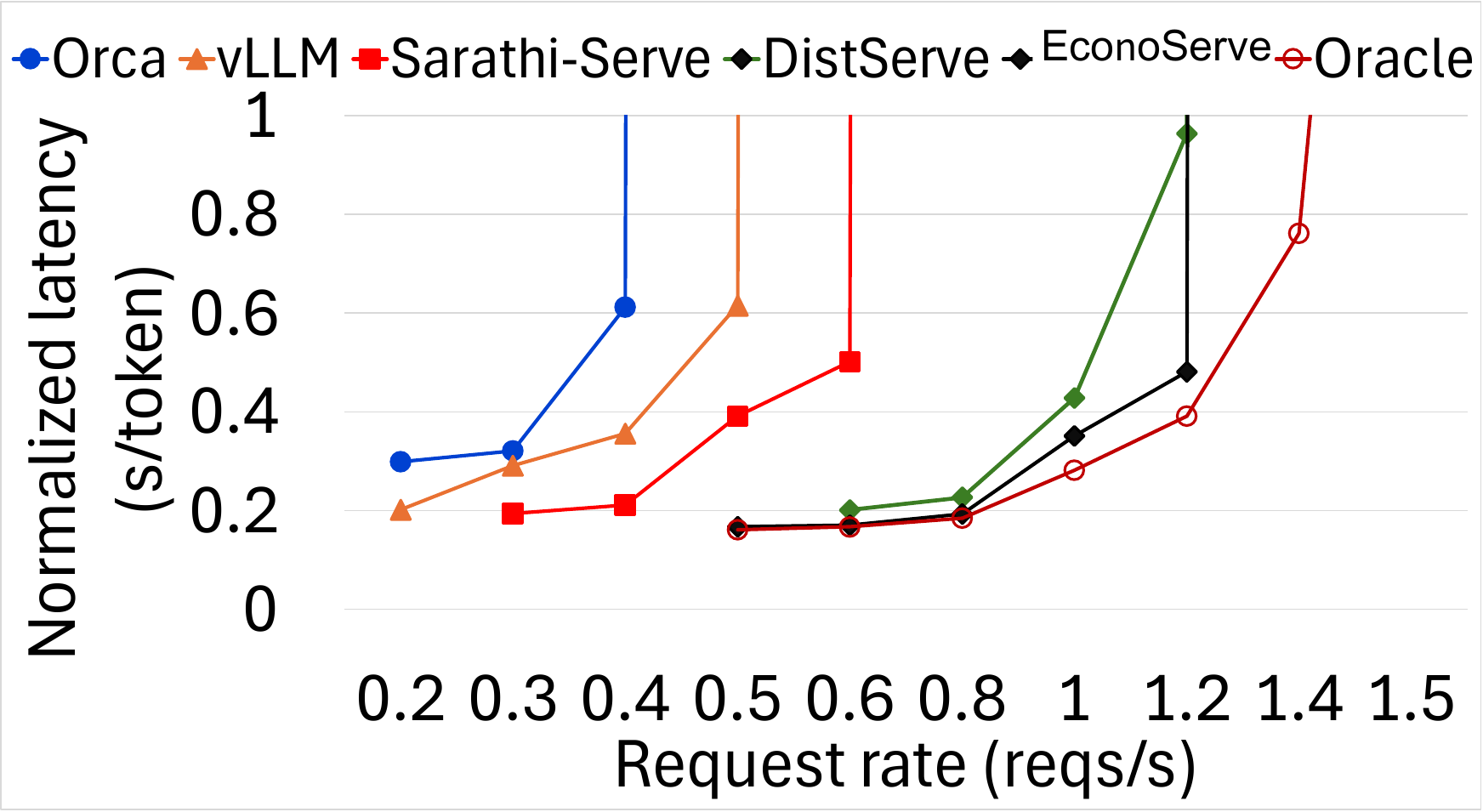} }}
    \hfill
    \subfloat[OPT-13B on Alpaca.\vspace{-0.0in}\label{fig:exp-13-a}]{{\includegraphics[width=0.32\linewidth,height=0.112\textheight]{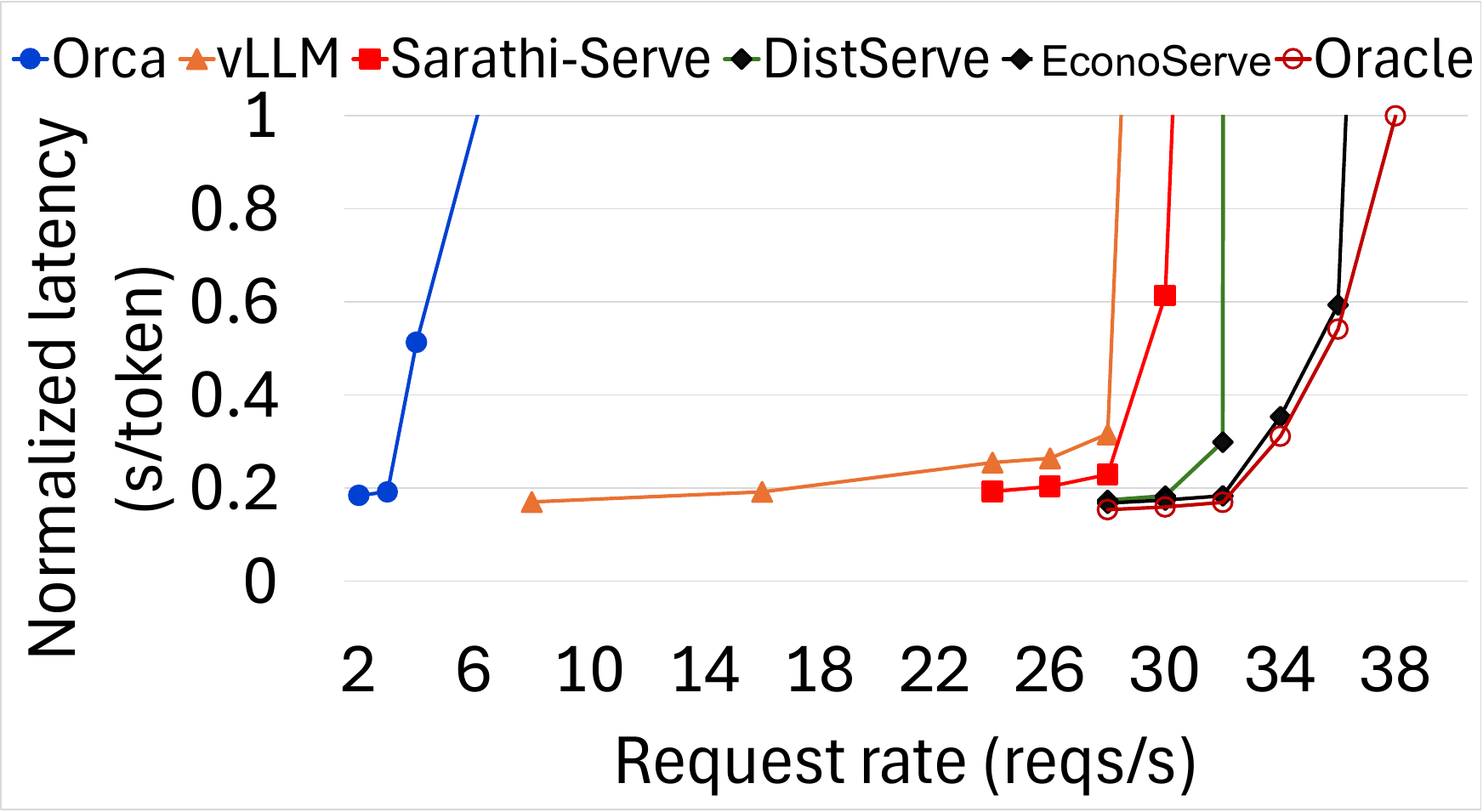} }}
    \hfill
    \subfloat[OPT-175B on Alpaca.\vspace{-0.0in}\label{fig:exp-175-a}]{{\includegraphics[width=0.32\linewidth,height=0.112\textheight]{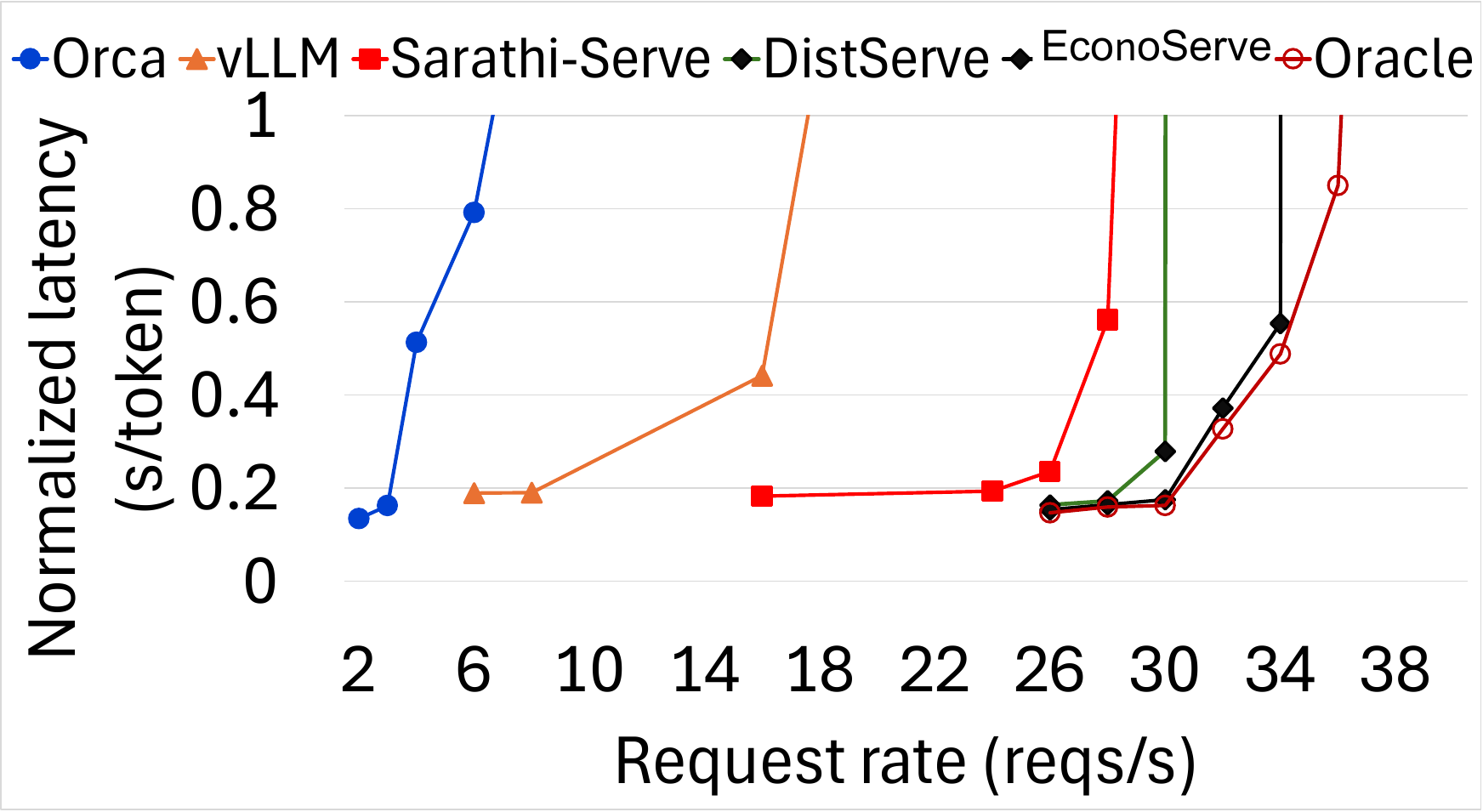} }}
    \hfill
    \subfloat[Llama-33B on Alpaca.\vspace{-0.0in}\label{fig:exp-l3-al}]{{\includegraphics[width=0.32\linewidth,height=0.112\textheight]{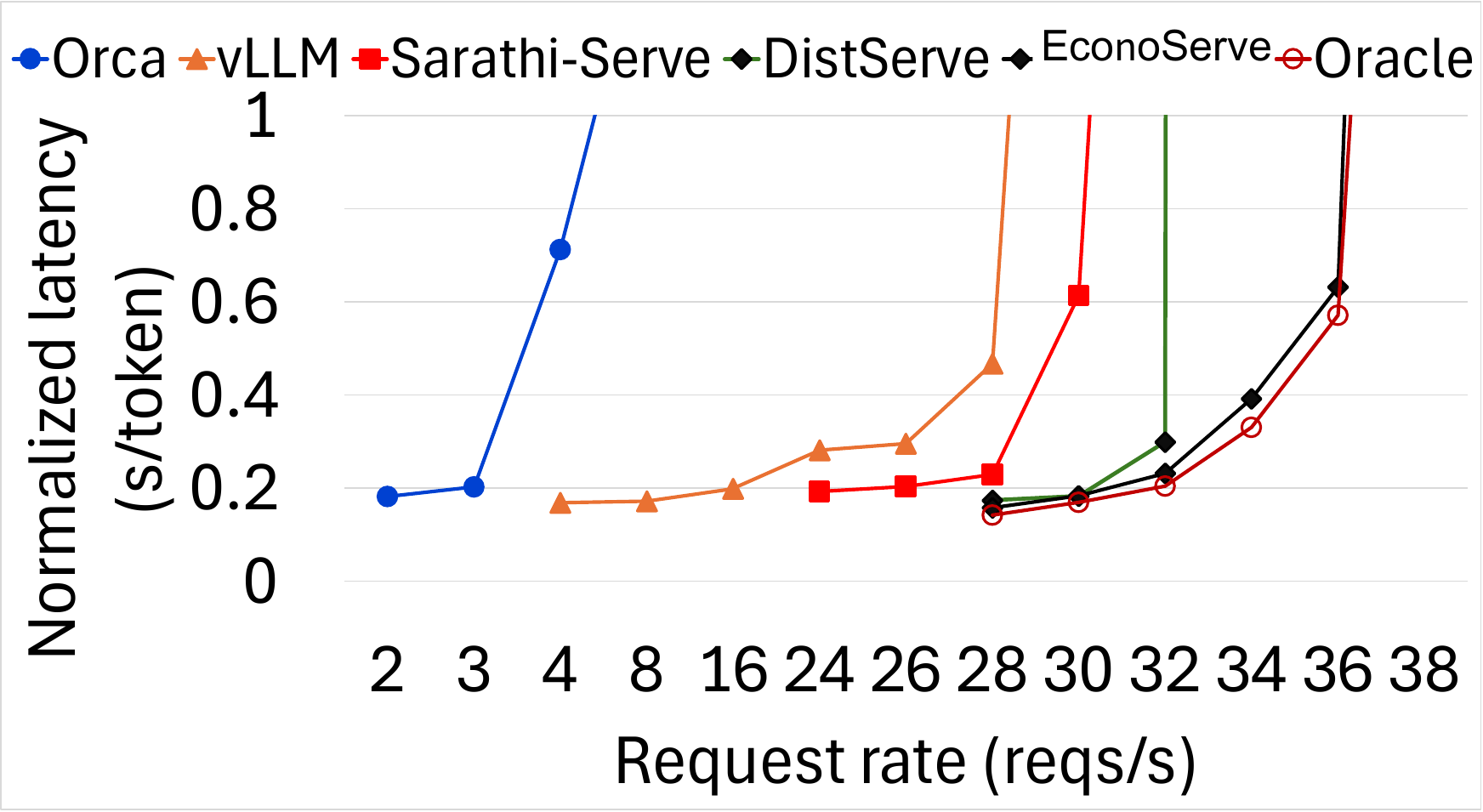} }}
    \hfill
    \subfloat[SSR of OPT-13B.\vspace{-0.0in}\label{fig:ssr-13b}]{{\includegraphics[width=0.32\linewidth,height=0.112\textheight]{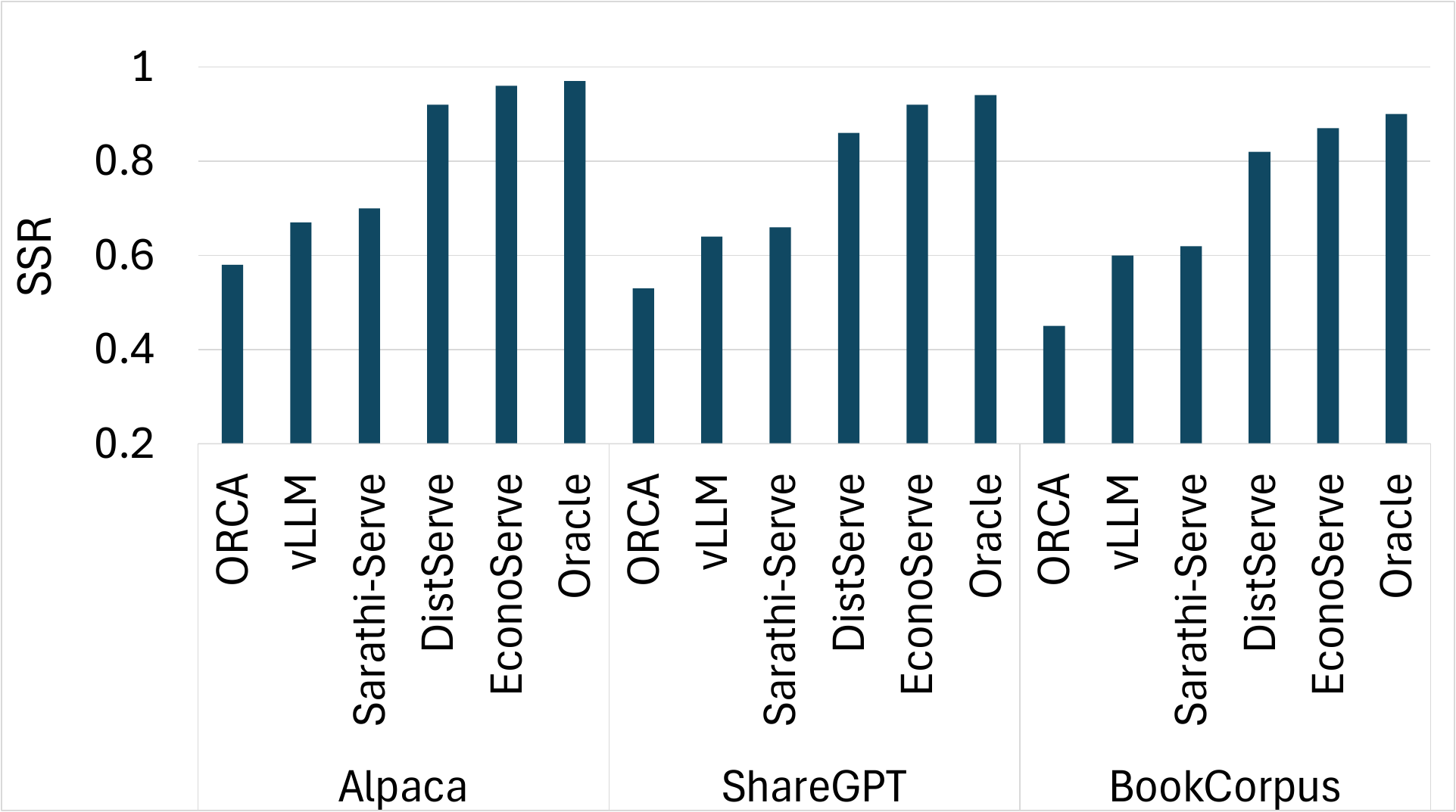} }}
    \hfill
    \subfloat[SSR of OPT-175B.\vspace{-0.0in}\label{fig:ssr-175b}]{{\includegraphics[width=0.32\linewidth,height=0.112\textheight]{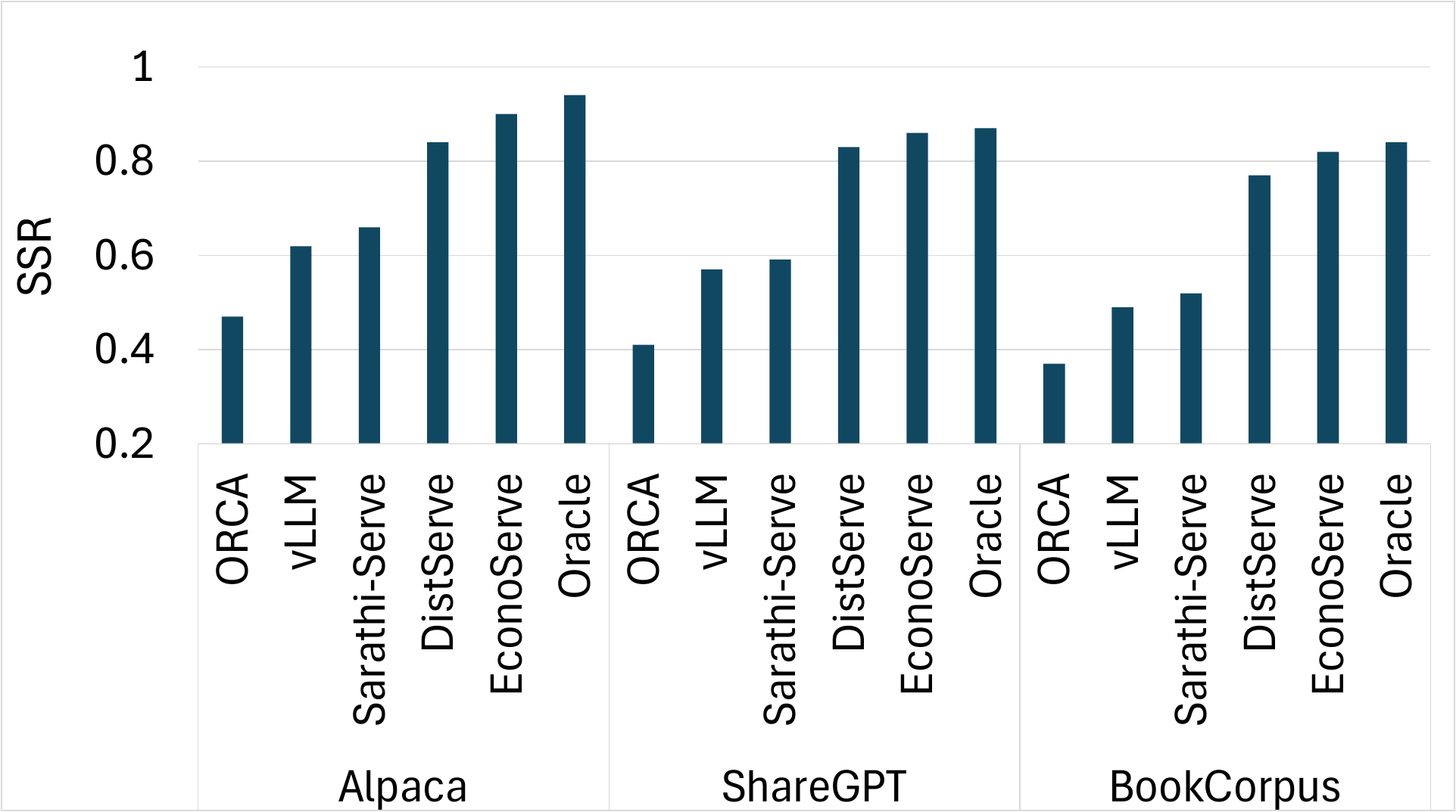} }}
    \hfill
    \subfloat[SSR of Llama-33B.\vspace{-0.0in}\label{fig:ssr-llama}]{{\includegraphics[width=0.32\linewidth,height=0.11\textheight]{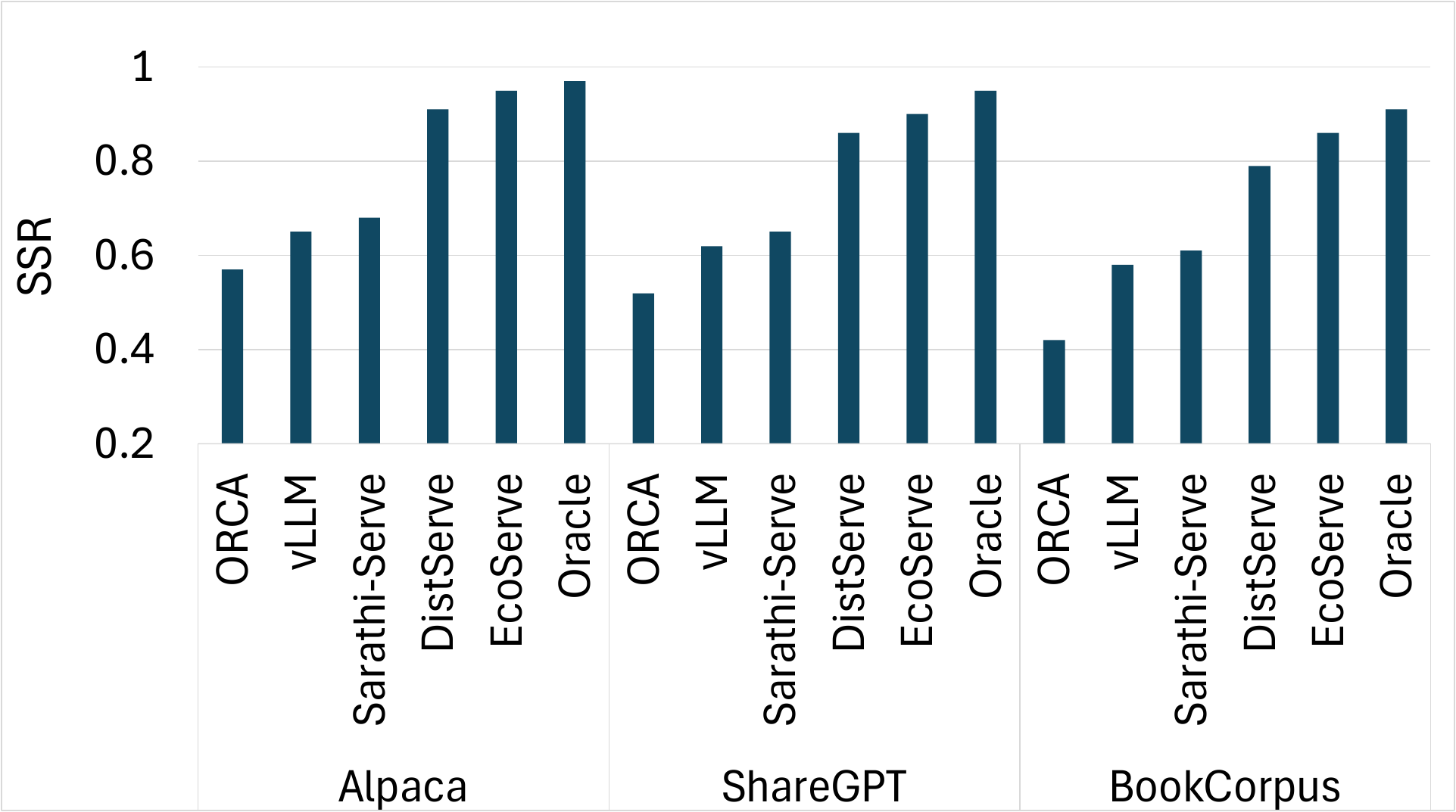} }}
    \hfill
    \subfloat[KVC of OPT-13B.\vspace{-0.0in}\label{fig:kvc-13b}]{{\includegraphics[width=0.32\linewidth,height=0.112\textheight]{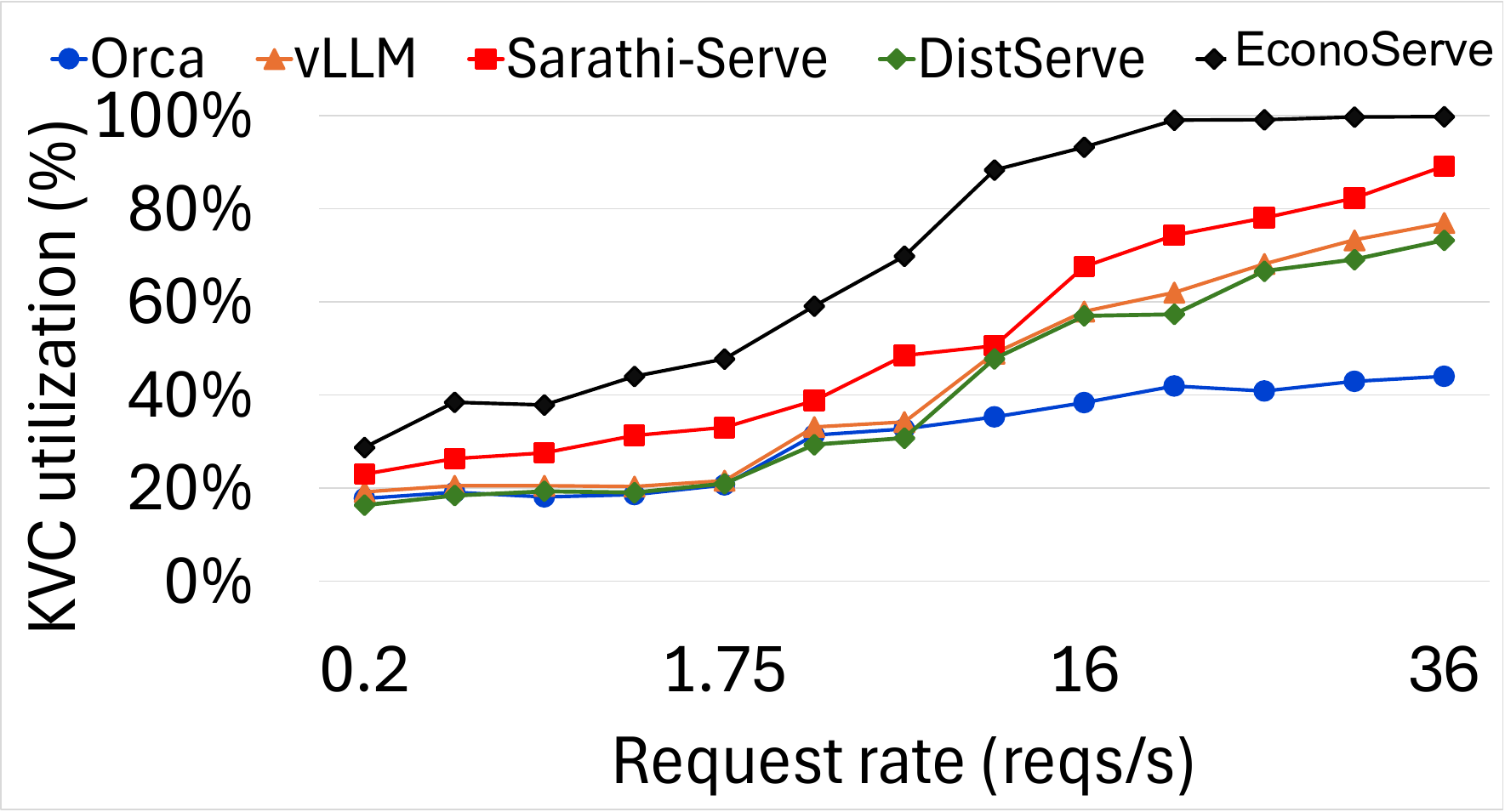} }}
    \hfill
    \subfloat[KVC of OPT-175B.\vspace{-0.0in}\label{fig:kvc-175b}]{{\includegraphics[width=0.32\linewidth,height=0.112\textheight]{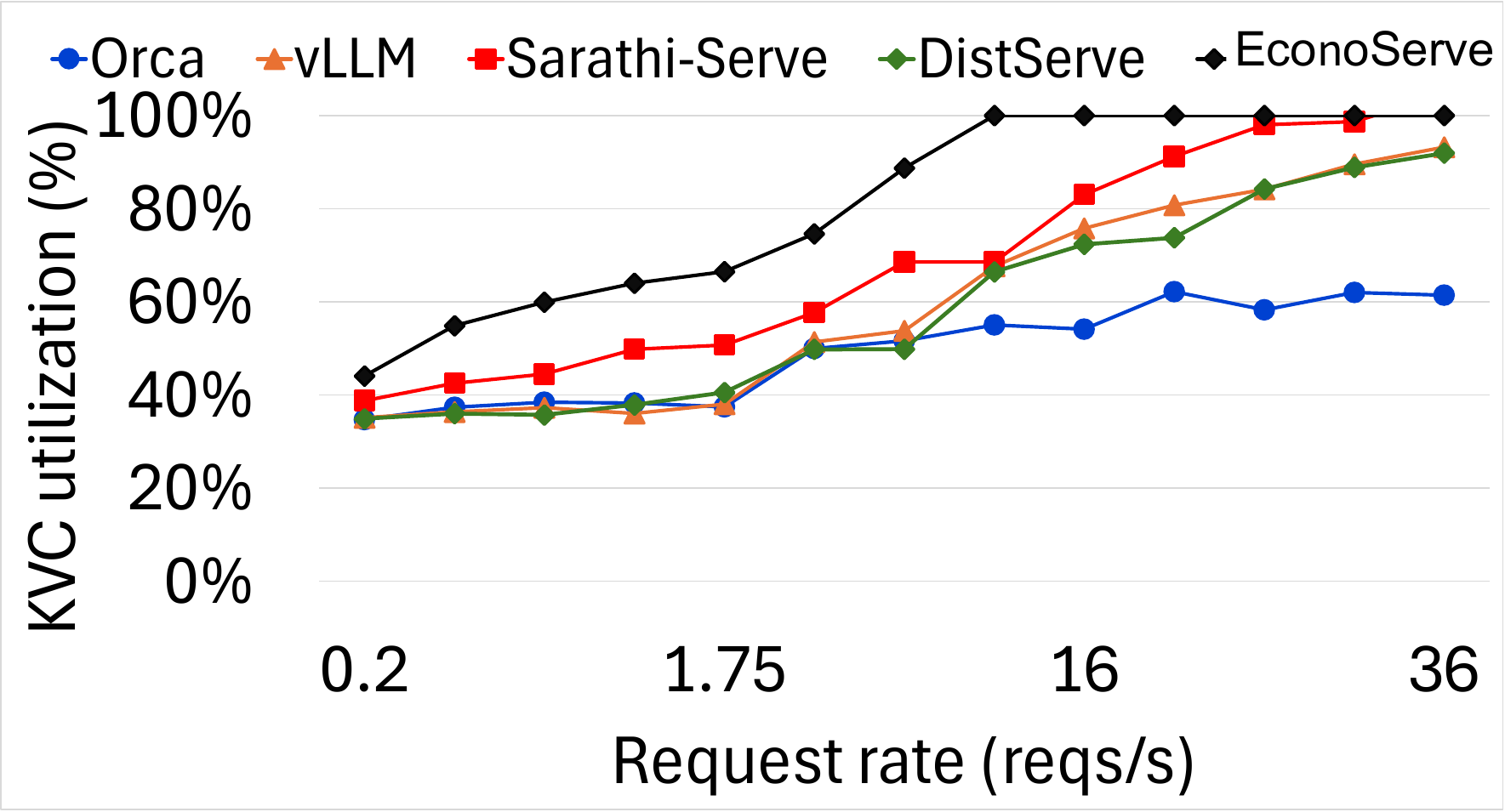} }}
    \hfill
    \subfloat[KVC of Llama-33B.\vspace{-0.0in}\label{fig:kvc-llama}]{{\includegraphics[width=0.32\linewidth,height=0.11\textheight]{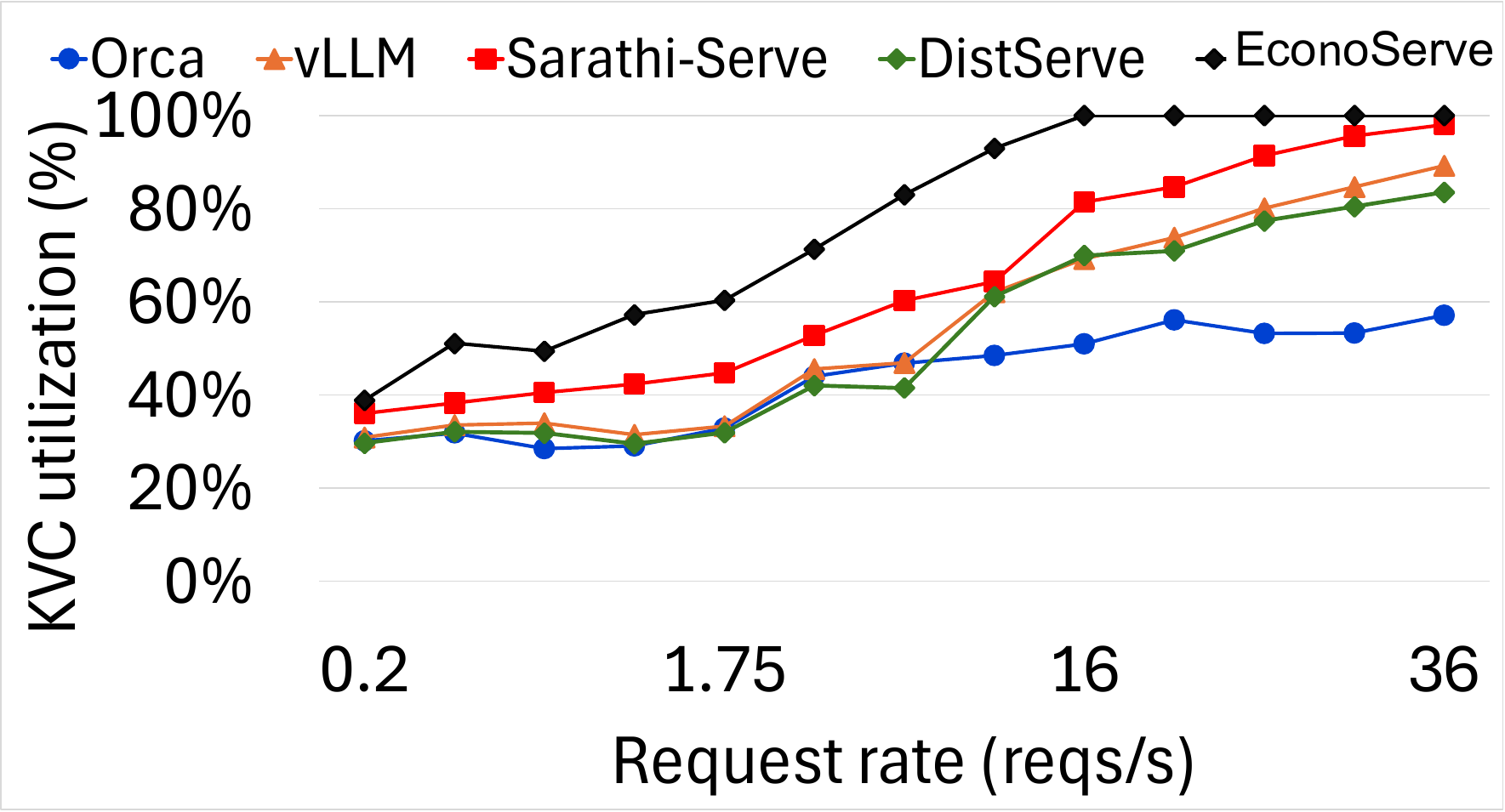} }}
    \hfill
    \subfloat[GPU utilization of OPT-13B.\vspace{-0.0in}\label{fig:gpu-13b}]{{\includegraphics[width=0.32\linewidth,height=0.112\textheight]{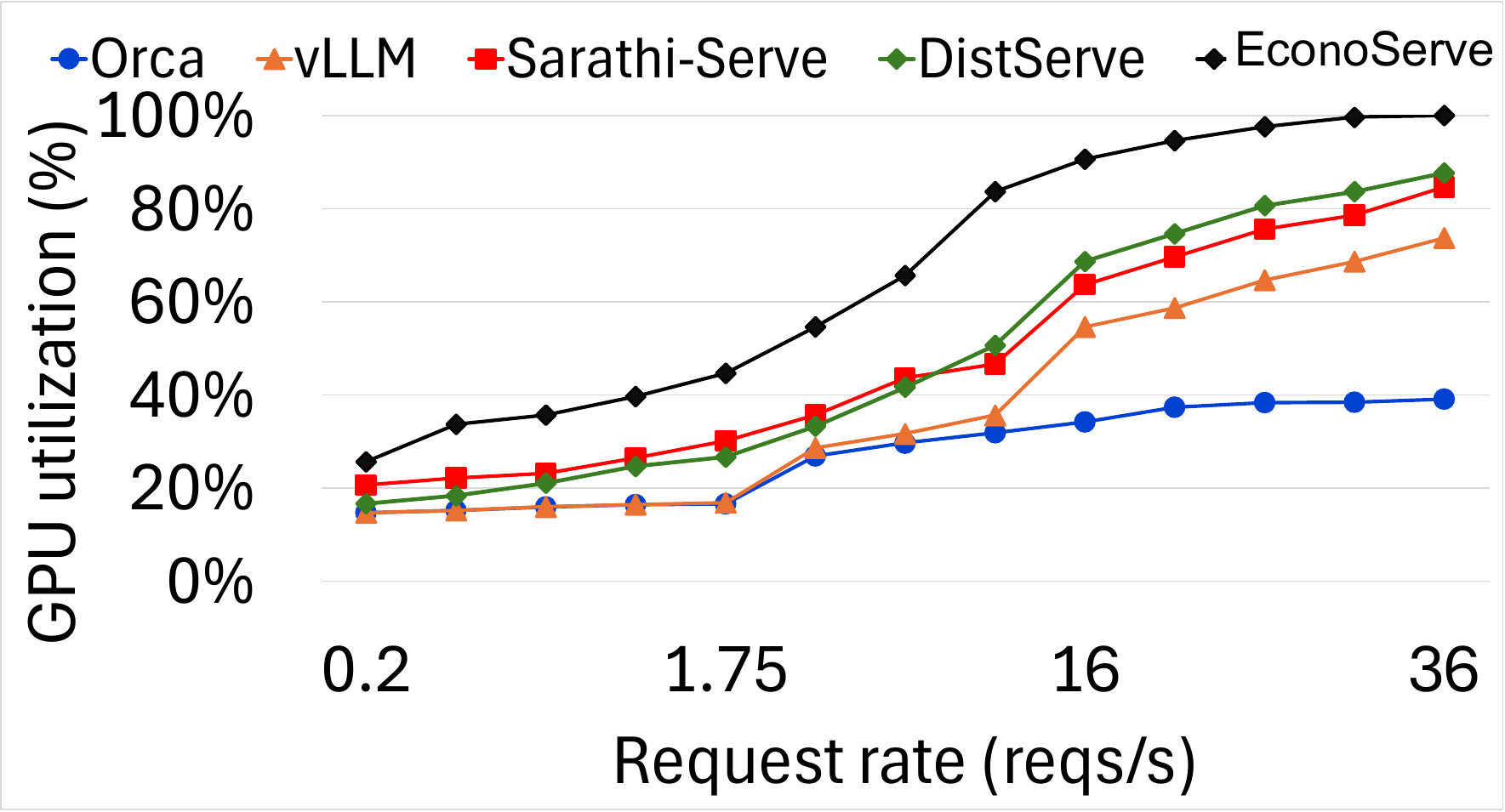} }}
    \hfill
    \subfloat[GPU utilization of OPT-175B.\vspace{-0.0in}\label{fig:gpu-175b}]{{\includegraphics[width=0.32\linewidth,height=0.112\textheight]{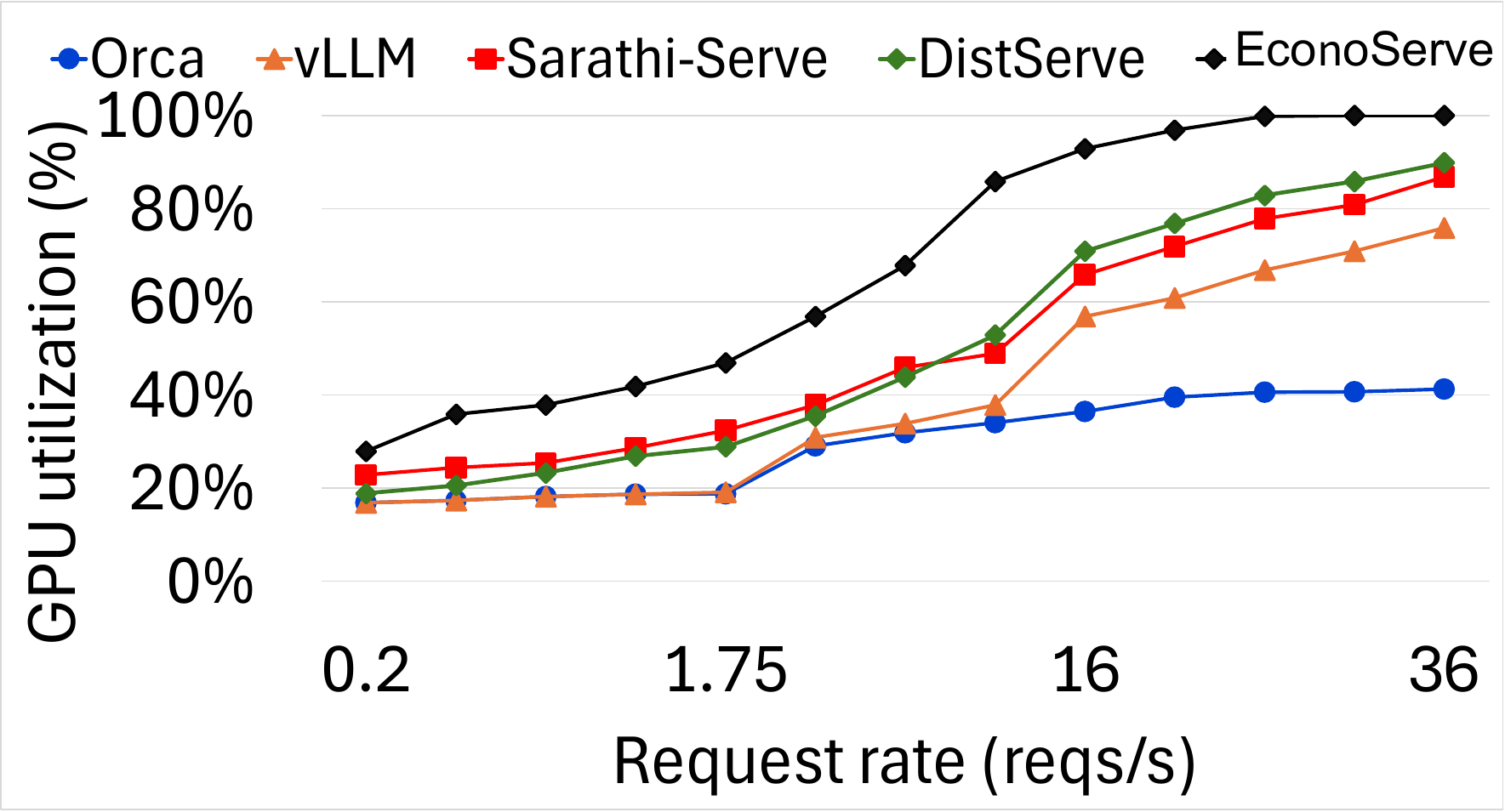} }}
    \hfill
    \subfloat[GPU utilization of Llama-33B.\vspace{-0.0in}\label{fig:gpu-llama}]{{\includegraphics[width=0.32\linewidth,height=0.11\textheight]{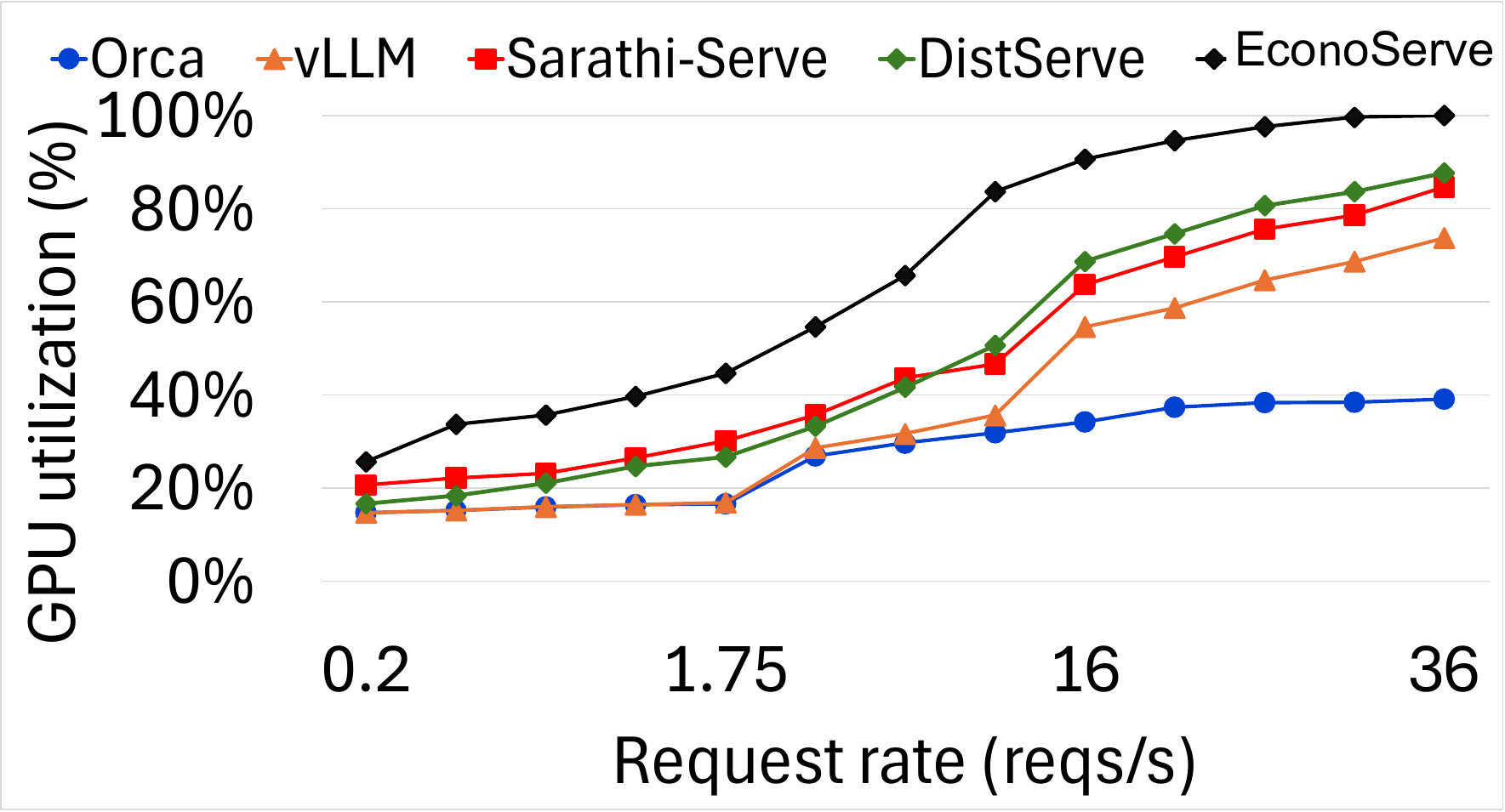} }}
    \hfill
    \vspace{-0.0in}
   \caption{\small{Overall performance comparison with varied request rates for all models on the three traces (DistServe utilizes twice as many GPUs compared to other methods).\vspace{-0.0in}}}%
    \label{fig:prompt-seq-vllm}
\end{figure*}}

\DEL{{\color{red}Figure~\ref{fig:prompt-methods} shows the performance of \sys for the OPT-13B model. First, Figures~\ref{fig:exp-1-a} to~\ref{fig:exp-1-b} show the JCT of the model for the three datasets, respectively. \sys has \%, \% , and \% lower JCT than the Sarathi-Serve, vLLM, and ORCA, respectively. \sys has \% and \% lower preemption time than the Sarathi-Serve and vLLM, respectively. \sys performs $\times$ better than the ORCA. \sys decouples prompts and GTs so that prompts can always be processed in time without the need to wait in the } 

\noindent{\textbf{Performance Comparison.}} Figure~\ref{fig:exp-1} illustrates the average JCT per request, and breaks down the JCT into prompt processing time, token generation time, and prompt waiting time. 
We see that \sys reduces the JCT of \Orca and {MultiRes} by 88\% and 1.32$\times$. Specifically, it reduces \Orca and {MultiRes}'s prompt processing time by 6\% and 5\%, respectively, and reduces their token generation time by 46\% and 1.7$\times$, respectively.  \sys reduces the waiting time of \Orca and {MultiRes} by 4.6$\times$ and 2.6$\times$, respectively.  

The lower JCT of MultiRes compared to \Orca indicates the effectiveness of selecting prompts to fully utilize both GPU and memory to increase throughput and reduce JCT. Based on our proposed MultiRes, \sys further reduces JCT. \sys decouples prompts and GTs so that prompts can always be processed in time without the need to wait in the queue for a long time. 
It combines the prompts and GTs to fully utilize the GPU and KVC, increasing the throughput and reducing waiting time and JCT. In contrast, \Orca simply uses FCFS and max-allocation, so it cannot achieve high throughput and low JCT. MultiRes tries to find a prompt to more fully utilize both GPU and the KVC. It generates high scheduling time and may not fully utilize the two resources. The effectiveness of the decoupling is shown by the result that \sys-D reduces the JCT of \Orca and {MultiRes} by 41\% and 72\%, respectively. 

By grouping the same-RL GT together for execution, \sys reduces the scheduling time, thus reducing the token generation time. Indeed, the combination of decoupling and GT grouping method (\sys-DG) reduces it \Orca and {MultiRes} by 69\% and 1.07$\times$, respectively. 

With the prompt and GT selection method, \sys-DGS reduces the JCT of \Orca and {MultiRes} by 83\% and 1.23$\times$ respectively. The selection method considers the occupied KVC, the SLO and the prompt length for prompts (or predicted response length for GTs), thus it quickly finds requests to fully utilize the GPU and KVC resources, reducing scheduling time and waiting time.  

Further, \sys has its novel KVC pipelining method to fully utilize idle allocated KVC to increase the throughput and reduce waiting time and JCT. This effect is shown by the result that \sys-DGS has a 5.3\% higher JCT than \sys. 

\sys-D reduces the JCT of \Orca by 5.4s, \sys-DG further reduces it by 2.21s, \sys-DGS further reduces it by 0.81s, and then \sys further reduces it by 0.26s. The results show that the decoupling method is the most effective one, followed by the selection method, and the same-RL GT grouping and the KVC pipeline have similar effects on reducing the JCT.

\DEL{\begin{figure*}[!t]
\centering
       \subfloat[Response latency (i.e., JCT).\vspace{-0.0in}\label{fig:exp-1-b}]{{\includegraphics[width=0.23\linewidth,height=0.112\textheight]{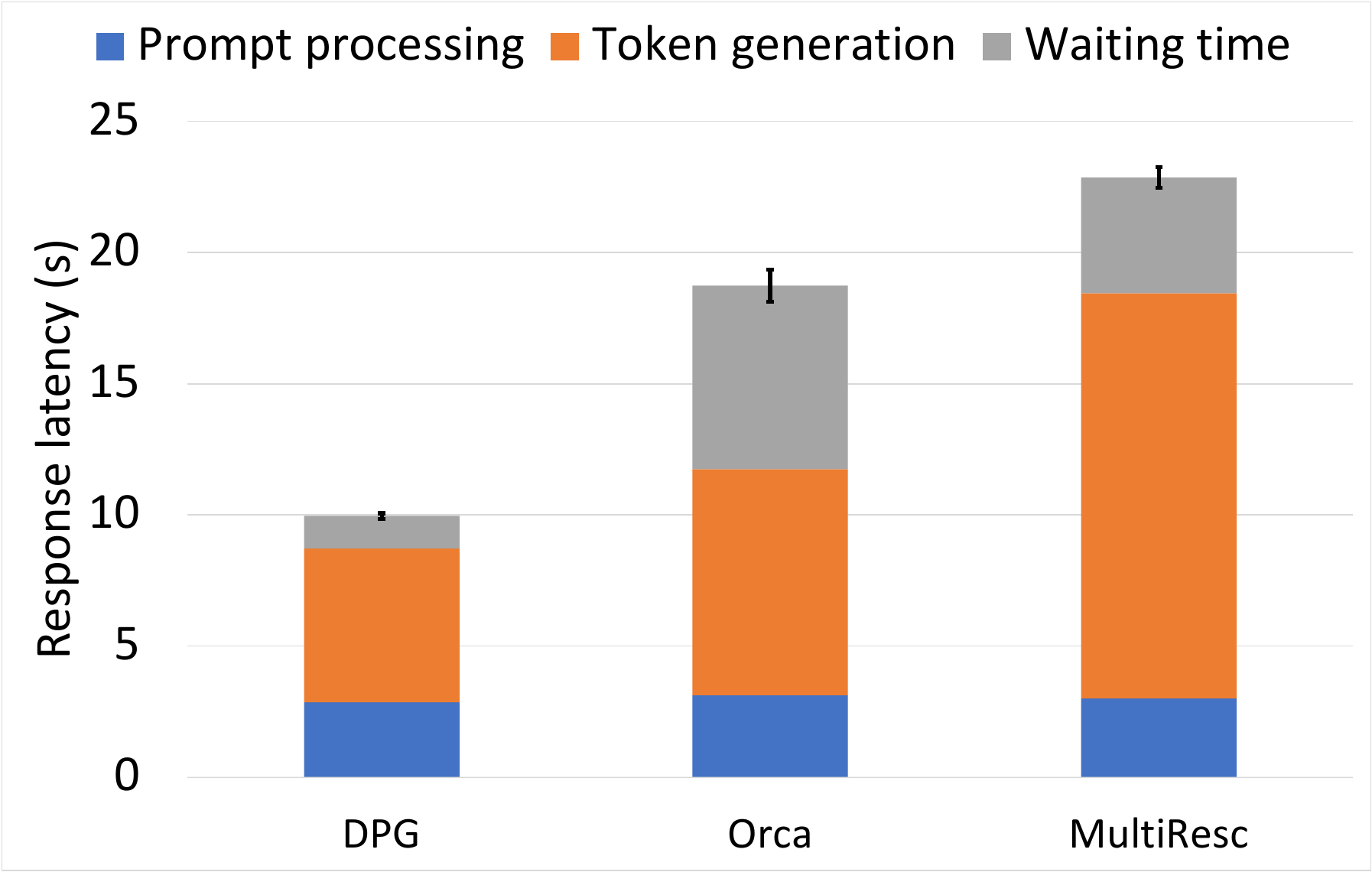} }}
    \hfill
    \subfloat[Iteration time.\vspace{-0.0in}\label{fig:exp-2-b}]{{\includegraphics[width=0.23\linewidth,height=0.112\textheight]{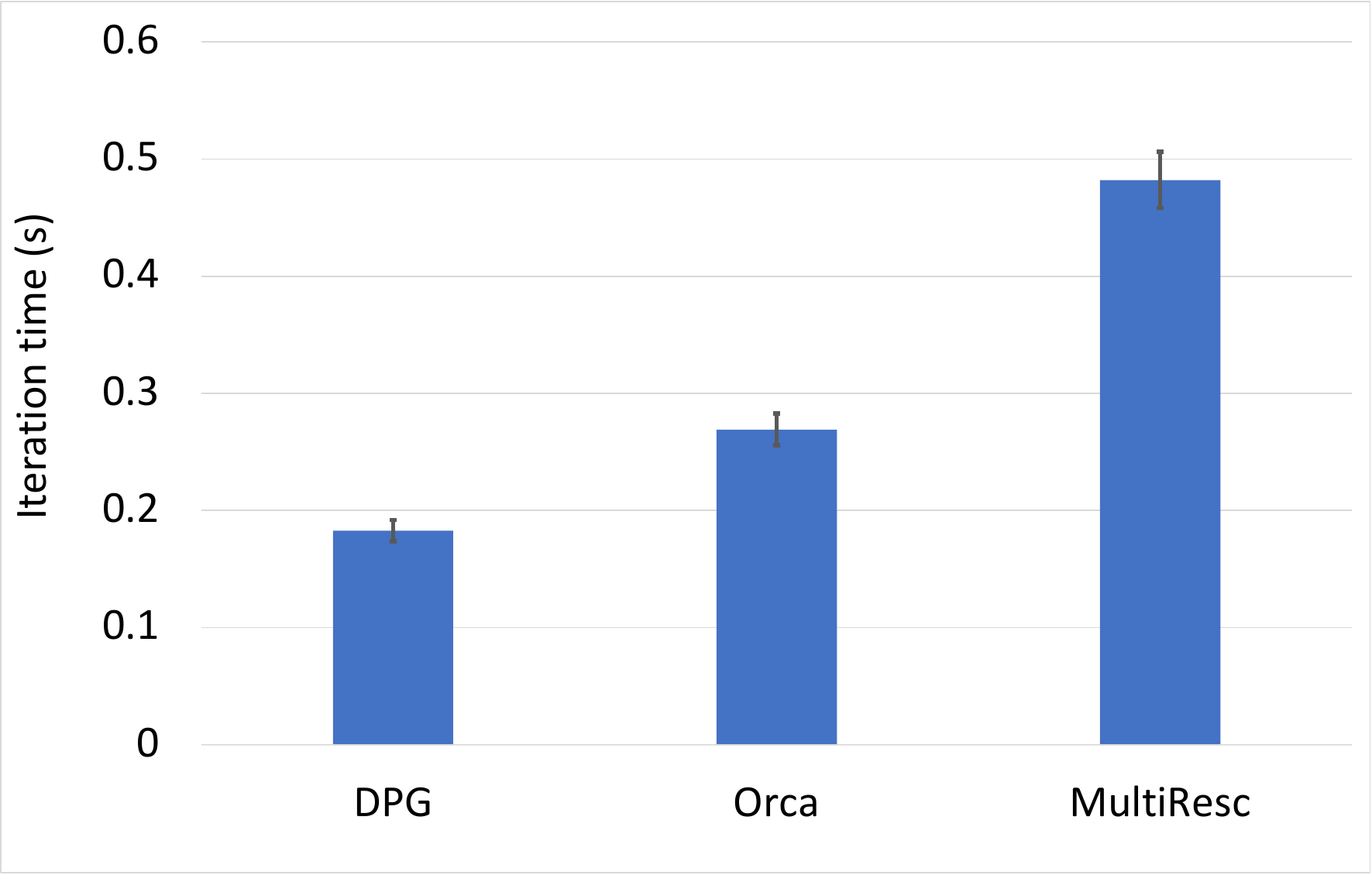} }}
    \hfill
    \subfloat[SLO.\vspace{-0.0in}\label{fig:exp-3-b}]{{\includegraphics[width=0.23\linewidth,height=0.112\textheight]{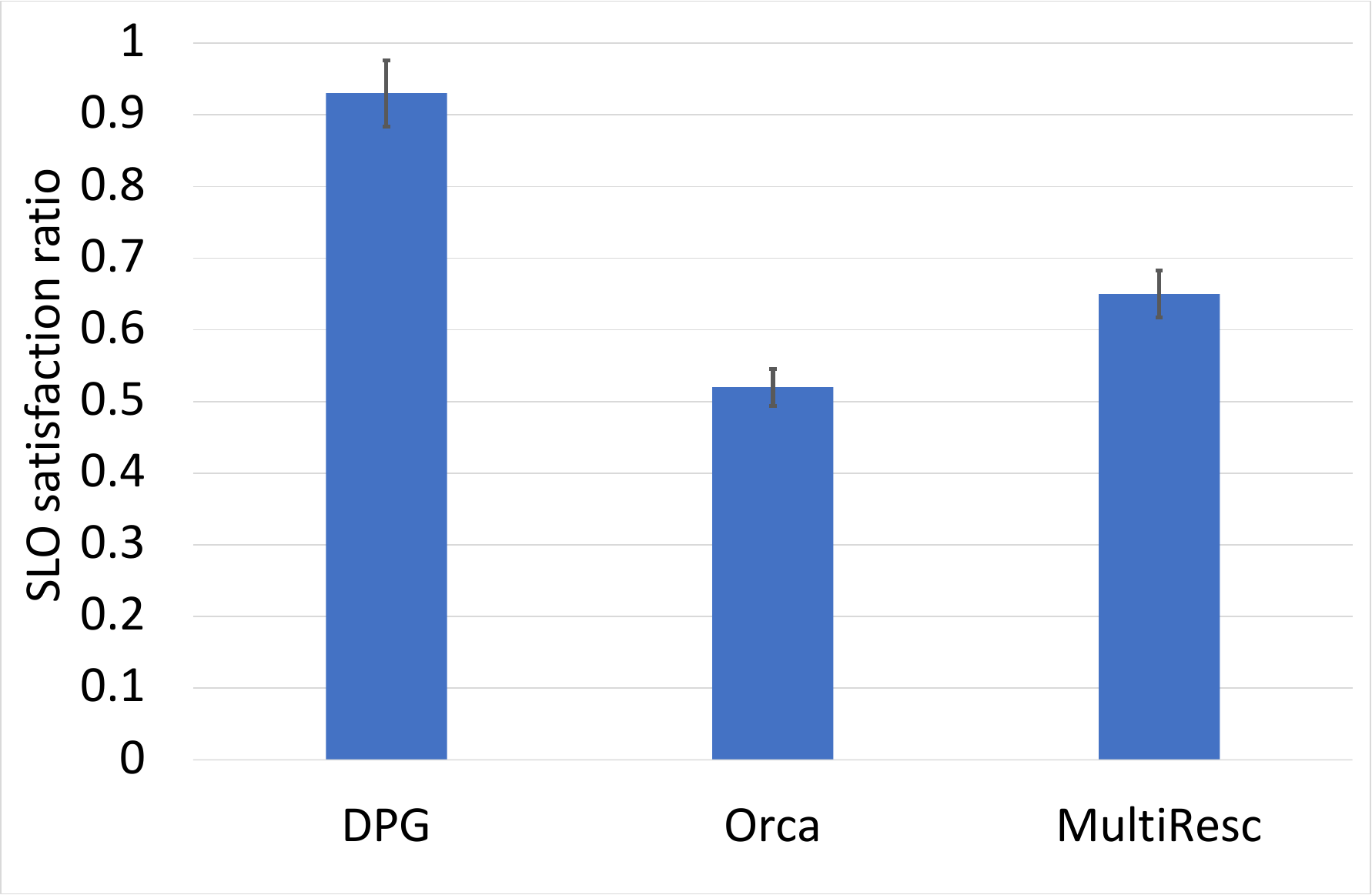} }}
    \hfill
\subfloat[Throughput.\vspace{-0.0in}\label{fig:exp-5-b}]{{\includegraphics[width=0.23\linewidth,height=0.112\textheight]{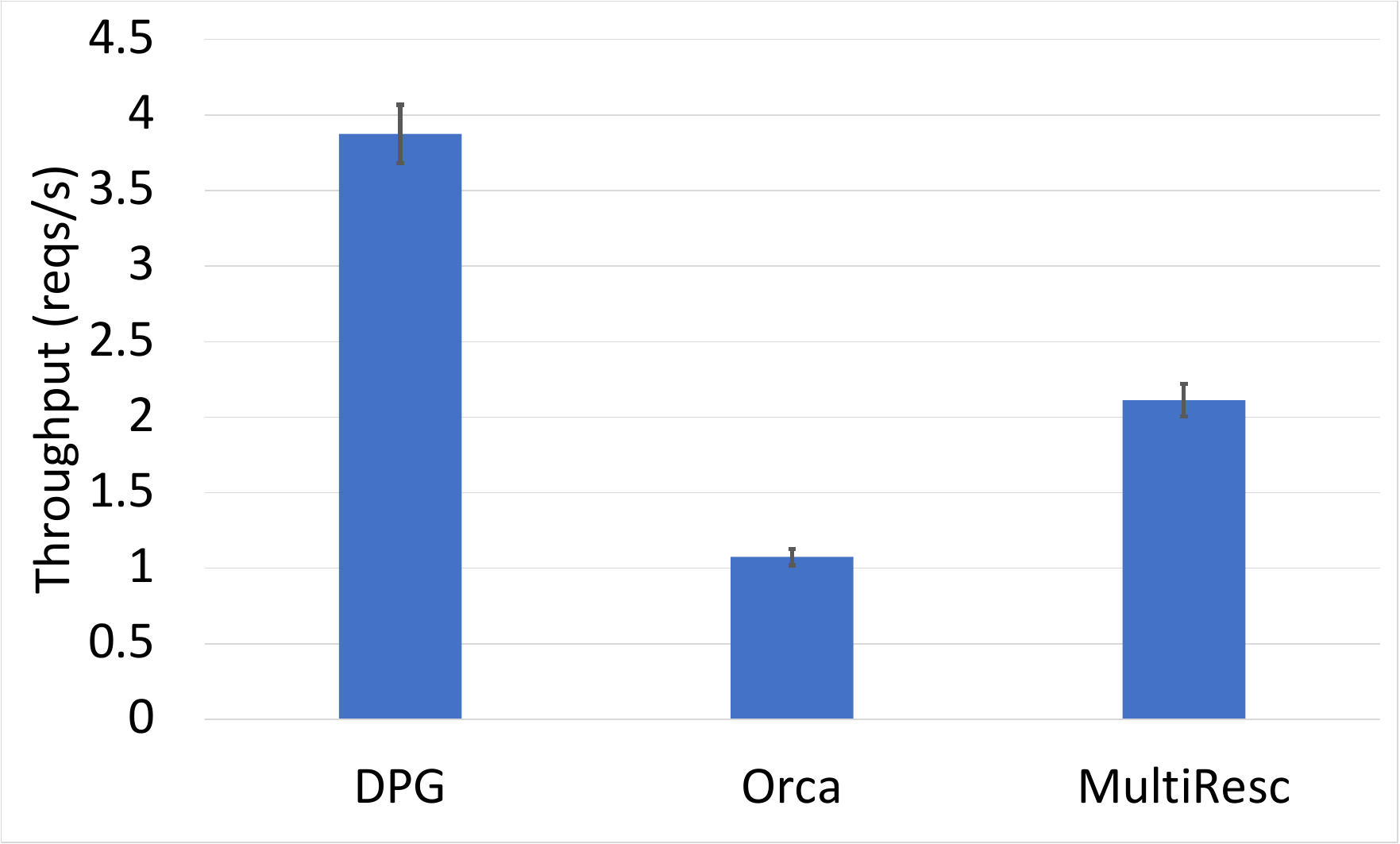} }}
    
    \hfill
    \subfloat[GPU utilization.\vspace{-0.0in}\label{fig:exp-4-b}]{{\includegraphics[width=0.23\linewidth,height=0.112\textheight]{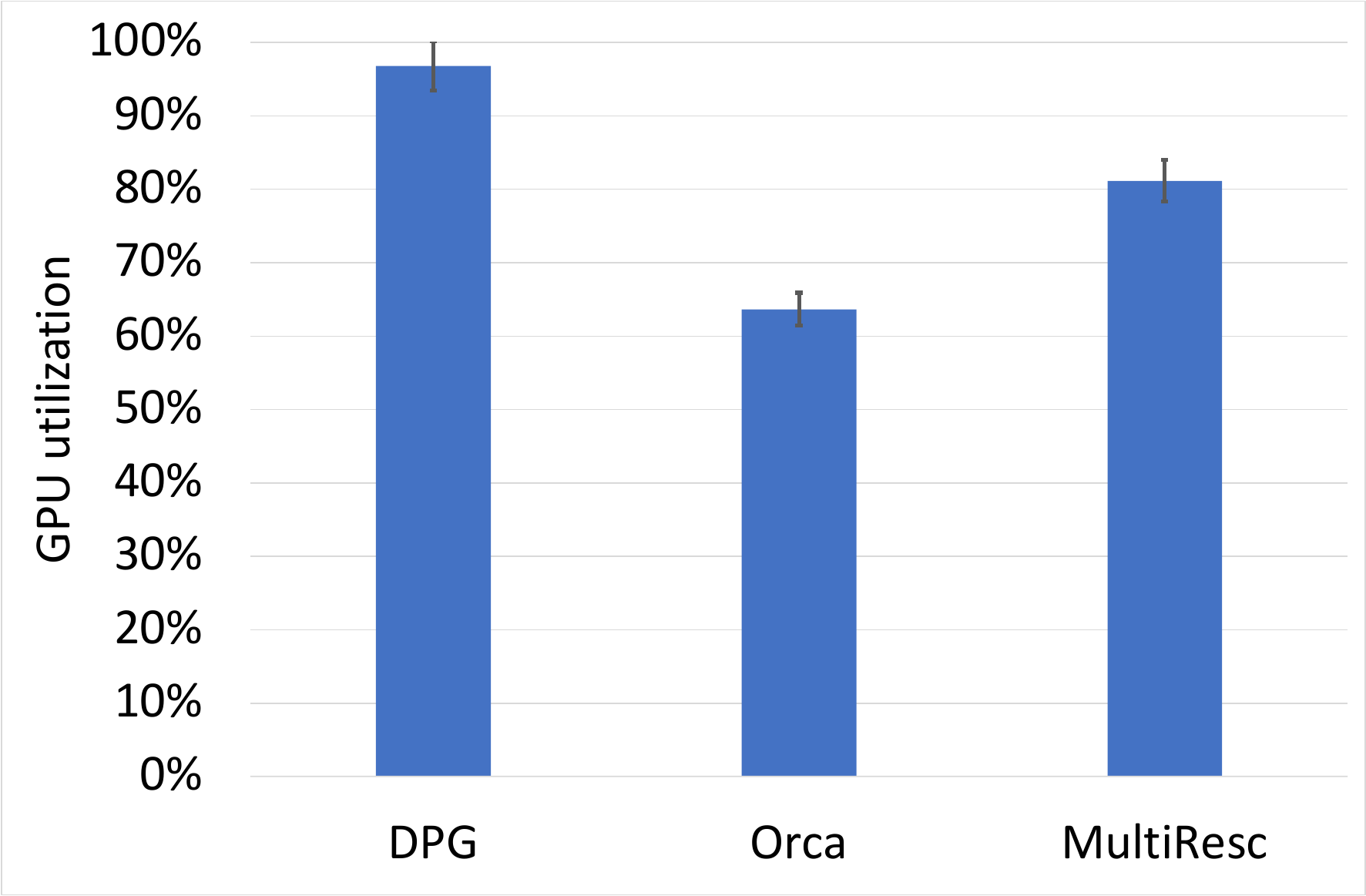} }}
    \hfill
        \subfloat[Memory utilization.\vspace{-0.0in}\label{fig:exp-6-b}]{{\includegraphics[width=0.23\linewidth,height=0.112\textheight]{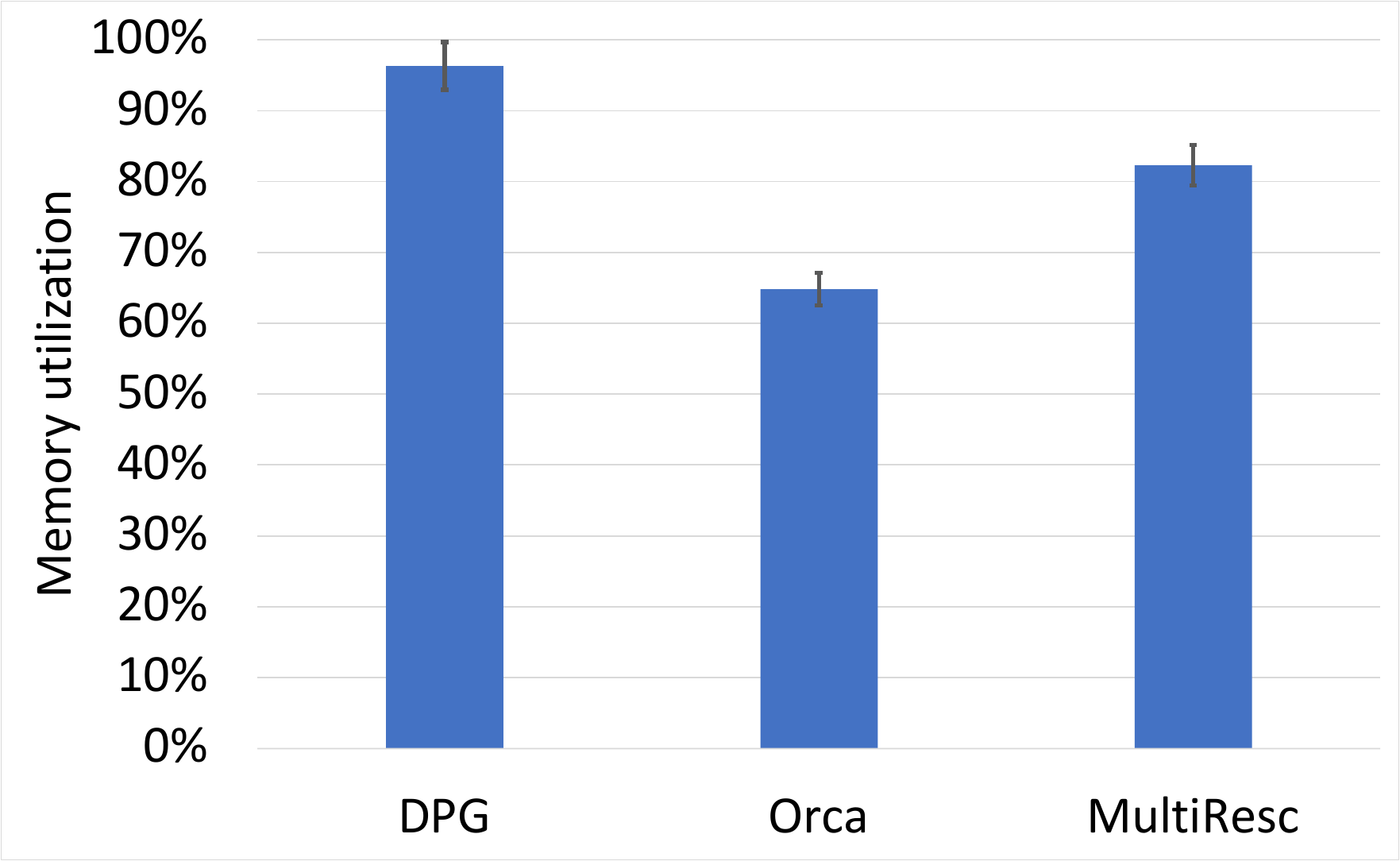} }}
    \hfill
    \subfloat[Time overhead.\vspace{-0.0in}\label{fig:exp-7-b}]{{\includegraphics[width=0.23\linewidth,height=0.112\textheight]{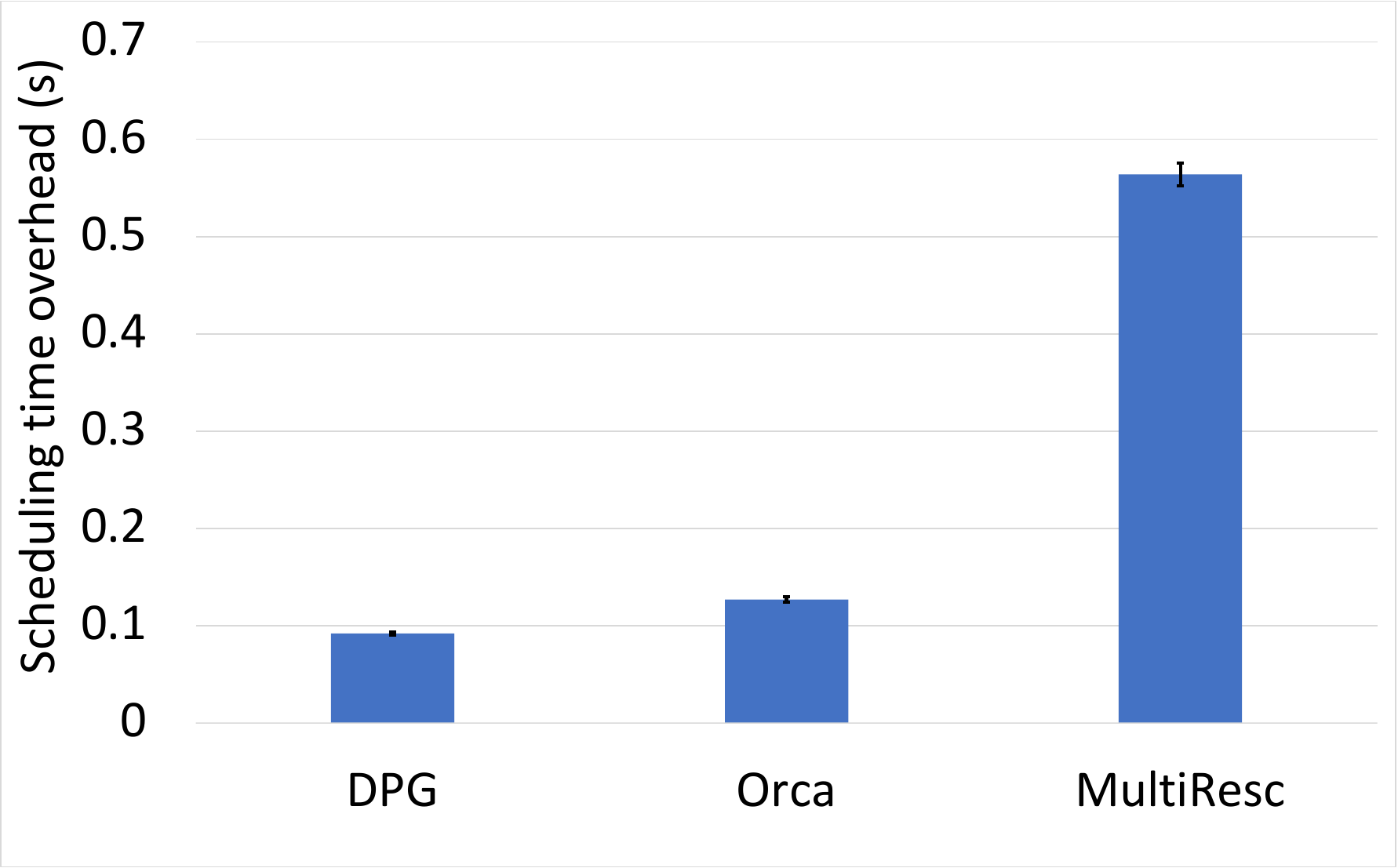} }}
    \hfill
    \subfloat[Workload processing time.\vspace{-0.0in}\label{fig:exp-8-b}]{{\includegraphics[width=0.23\linewidth,height=0.112\textheight]{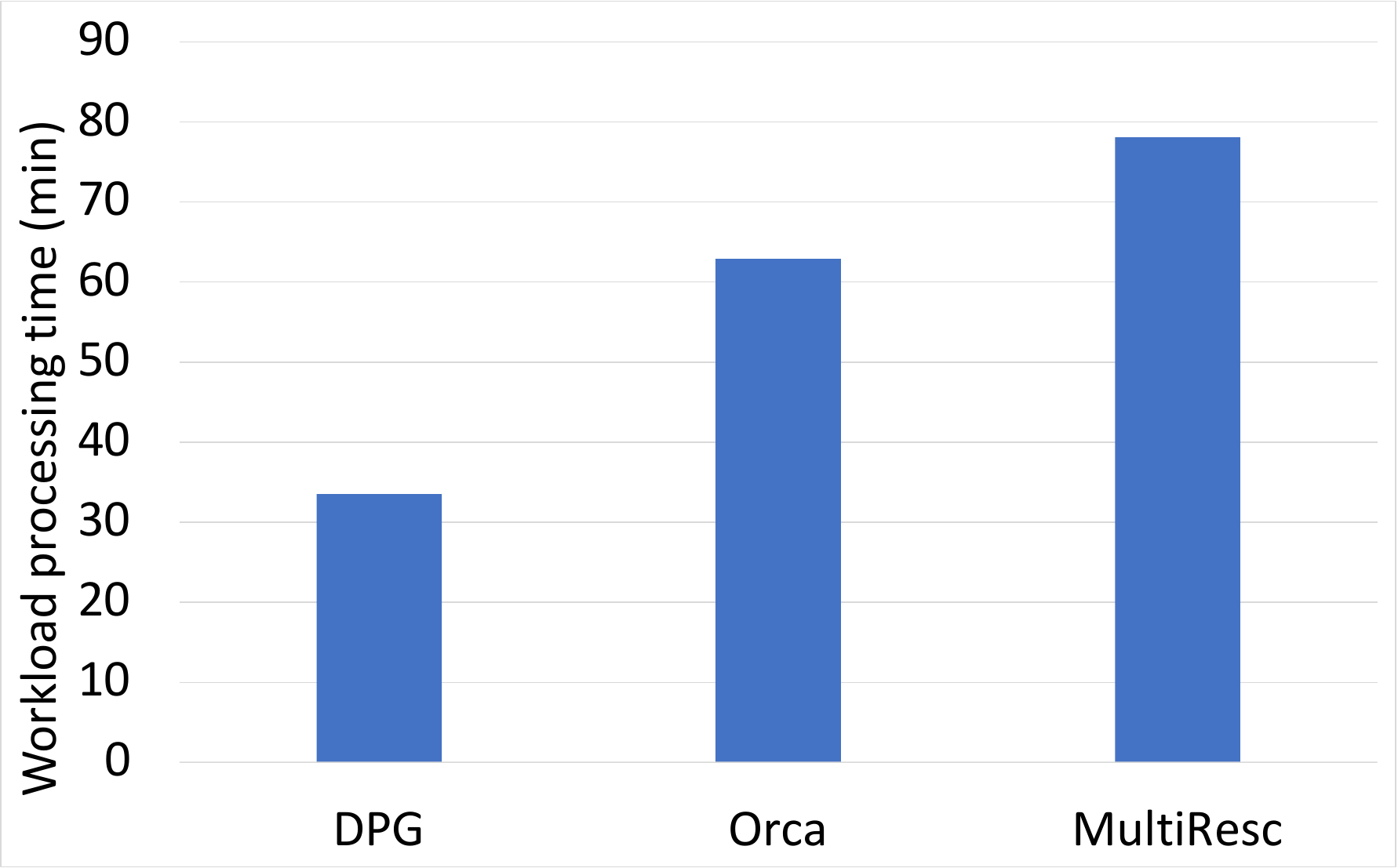} }}
    \hfill
    \vspace{-0.0in}
   \caption{\small{Performance of different methods under bursty request arrival rate.\vspace{-0.0in}}}%
    \label{fig:prompt-methods-burstiness}
\end{figure*}}

\DEL{The results show the effectiveness of each individual method in reducing the JCT. 
D: 1
G: 1.04 
S: 1.20
pipeline: 1.31
the above 4 values are comparable to show the effect of each method
5.39681	0.1268	0.446467	0.124363
Among the individual methods, ???? 
\sys-D compare to \Orca 1.08X, 1.08
\sys-DG compare to \Orca 1.13X, 0.05
 \sys-DGS compare to \Orca 1.35X, 0.22

{\color{red} \sys-DG further improves \sys-D by 4\%, \sys-DGS further improves \sys-DG by 16\%, and \sys improves \sys-DGS by 11\% for \Orca.}

As a result, \sys significantly reduces the JCT of \Orca and MultiRes.}

\DEL{No. you need to compare how much D improves Orca, then how much DG further improves, how much DGS further improves, and how much DPS further improves. Is it 1.06, 0.18, 0.18, 0.18?}

Figure~\ref{fig:exp-2} shows the average iteration time of different methods. 
\sys reduces the iteration time of \Orca and {MultiRes} by 46\% and 1.7$\times$, respectively. \sys-D has 15\% higher iteration time than \Orca, 
\sys-DG  and \sys-DGS reduce the average iteration time of \Orca by 15\%, and 28\%, respectively. \sys-D, \sys-DG, and \sys-DGS reduce that of {MultiRes} by 56\%, 1.11$\times$, and 1.28$\times$, respectively. 
\sys-D has a longer scheduling time than \Orca since it selects tasks sequentially from two queues to fully utilize GPU and KVC. Grouping same-RL GTs to run significantly reduces the scheduling time.  The reasons for other results are the same as explained above. 

Figure~\ref{fig:exp-3} shows the task SLO satisfaction ratio (SSR) of different methods. \sys improves the SSR of \Orca and {MultiRes} by 69\% and 35\%, respectively. \Orca does not consider SLO and just uses FCFS. {MultiRes} also does not consider SLO and just focuses on resource utilization. \sys tries to satisfy the SLOs of requests and prioritizes the requests with tighter SLOs in its scheduling and it greatly reduces the waiting time, thus achieving higher SSR. We also see that \sys-D, \sys-DG, \sys-DGS increase the SSR of \Orca by 40\%, 42\% and 59\%, respectively. They increase that of {MultiRes} by 19\%, 20\% and 35\%, respectively. As more individual methods are added, the SSR increases. 
This is because more fully utilizing the GPU and KVC from the decoupling, reducing scheduling time from the GT grouping, considering SLO and other factors in request selection from the queues, and pipelining the KVC use help reduce waiting time and satisfy SLOs.


Figure~\ref{fig:exp-5} shows the throughput of all the methods. \sys improves the throughput of \Orca and {MultiRes} by 1.9$\times$ and 48\%, respectively. We also see that \sys-D, \sys-DG, and \sys-DGS increase the throughput of \Orca by 1.42$\times$, 1.57$\times$, and 1.75$\times$, respectively. They increase those of MultiRes by 22\%, 30\%, and 39\%, respectively. As more individual methods are added, the throughput increases. Figure~\ref{fig:exp-4} shows the GPU utilization of all the methods. \sys improves the GPU utilization of \Orca and {MultiRes} by 53\% and 28\%. We also see that \sys-D, \sys-DG, \sys-DGS increase the GPU utilization of \Orca by 34\%, 37\% and 41\%, respectively. they increase the GPU utilization of {MultiRes} by 5\%, 7\%, and 8\%, respectively. The reasons of the results are the same as the above. \Orca's low throughput and GPU utilization is due to its inherent problem caused by max-allocation and hence limited batch size. {MultiRes} improves the throughput of \Orca at the cost of high scheduling time.

Memory utilization is the percentage of GPU memory used by a method as measured by GPUstat~\cite{gpustat}. 
Figure~\ref{fig:exp-6} shows the memory utilization of all the methods. \sys improves the memory utilization of \Orca and {MultiRes} by 48\% and 17\%, respectively. We also see that \sys-D, \sys-DG, \sys-DGS increase the memory utilization of \Orca by 29\%, 33\% and 34\%, respectively. They increase the memory utilization of {MultiRes} by 2.3\%, 4.9\% and 5.9\%, respectively. The reasons of the results are the same as the above.

\sys's online time overhead is for picking up requests from the two queues. Figure~\ref{fig:exp-7} shows the average online time overhead per request of all the methods. \sys, \Orca, {MultiRes} and \Orca produce 0.092s (1.23\% of JCT), 0.127s, and 0.564s online time overhead. \Orca simply uses FCFS at every iteration. Our proposed {MultiRes} finds a prompt among all prompts that can more fully utilize the available GPU and KVC resources. So it generates higher time overhead than \Orca, though it improves the throughput and response latency of \Orca. \sys-D, \sys-DG, and \sys-DGS cause 0.289s, 0.0898s, and 0.0905s time overhead, respectively. The GT grouping method reduces the scheduling time, thus reducing the online time overhead, and the KVC pipelining method causes a certain additional time overhead due to the selection of hosted tasks. 


\DEL{The additional offline time in \sys-DGS compared to DG is caused by the ordering the requests by SLO, prompt length and KV-cache.}

Table~\ref{tab:offline-time} shows the offline time of \sys and its component methods, averaged by the number of requests. The offline time overhead of \sys is for predicting response lengths in \sys-D, additionally grouping same-RL GTs in \sys-DG, and additionally ordering requests in two queues in \sys-DGS. Each method causes a certain offline time overhead.

Figure~\ref{fig:exp-8} shows the workload processing time for all the 200 requests. \sys reduces the workload processing time of \Orca and {MultiRes} by 58\% and 48\%, respectively. We also see that \sys-D, \sys-DG, \sys-DGS reduce that of \Orca by 51\%, 52\% and 57\%, respectively. They reduce the workload processing time of {MultiRes} by 40\%, 41\% and 46\%, respectively. The reasons of the results are the same as the above.
All the figures show that the results do not vary greatly. \looseness=-1

\noindent{\textbf{Resilience to High Arrival Rates.}}
To measure the resilience of \sys to high request arrival patterns, we changed the arrival rate of the requests to 8 requests per second, measured all the above metrics, and plotted them in Figure~\ref{fig:prompt-methods-burstiness}.  
Compared to Figure~\ref{fig:prompt-methods}, using (metric of high arrival rate - metric of normal arrival rate)/metric of normal arrival rate, \sys, for a higher request rate, increases JCT by 6.8\%, increases iteration time by 0.168\%, reduces SSR by 1.12\%, increases throughput by 13\%, increases GPU utilization by 3\%, increases memory utilization by 1.6\%, increases the workload processing time by 2.6\%. The online time overhead per request remains the same.

(SSR of high arrival rate - SSR of normal arrival rate)/SSR of normal arrival rate.

In this bursty case, compared to \Orca, \sys reduces JCT 82\%, reduces iteration time by 43\%, increases SSR by 
66\%, increases throughput by 1.91$\times$, increases GPU utilization by 58\%, increases memory utilization by 60\%, reduces workload processing time by 42\%, respectively. As a result, compared to the non-bursty case, \sys exhibits 16\%, 5\% and 2\% higher improvement in throughput, GPU utilization and memory utilization, respectively, but exhibits 6\%,3\%, 3\%, 6\% lower improvement in JCT, iteration time, SSR, and workload processing time, respectively. The bursty case gives an opportunity to further improve resource utilization and throughput but reduces the latency due to resource competition. The results show that \sys is resilient to the burstiness of the requests.}


\DEL{\begin{table}[]
\centering
\caption{Offline time for methods.}
\label{tab:offline-time}
\resizebox{\columnwidth}{!}{%
\begin{tabular}{|l|l|l|l|l|}
\hline
Method & \sys & \sys-D & \sys-DG & \sys-DGS \\ \hline
Offline time (s) & 0.478 & 0.244 & 0.301 & 0.344 \\ \hline
\end{tabular}%
}
\end{table}}


\noindent{\textbf{Overall Performance Comparison.}} 
Figures~\ref{fig:exp-13-s}-\ref{fig:exp-l3-al} show the normalized latency of the systems versus the request rate. Note that while DistServe utilizes twice as many GPUs as other methods, we aim to demonstrate that \sys can still outperform DistServe despite using significantly fewer GPU resources. A high-throughput serving system should retain low normalized latency against high request rates. We now compare the request rates that the methods can sustain while maintaining similar latencies. On ShareGPT, \sys can sustain 2.5-4$\times$ higher request rates than vLLM, 22-64$\times$ than \Orca, 1.25-2.25$\times$ than Sarathi-Serve, and 1.01-1.33$\times$ than DistServe. On BookCorpus, \sys achieves 2.5-2.8$\times$ higher rates than vLLM, 12.5-13$\times$ than \Orca, 1.88-2.33$\times$ than Sarathi-Serve, and 1.01-1.25$\times$ than DistServe. On Alpaca, \sys sustains 1.13-2.14$\times$ higher rates than vLLM, 5.6-9$\times$ than \Orca, 1.2-1.24$\times$ than Sarathi-Serve, and 1.1-1.13$\times$ than DistServe. \sys's advantage on Alpaca is less pronounced because it contains short sequences, allowing for more requests to be batched and processed quickly.

\DEL{On average, \sys shows the highest improvement for the ShareGPT dataset, followed by the BookCorpus, and the Alpaca, respectively. \sys can handle the ShareGPT the best, because it can efficiently manage the memory and batch more requests for ShareGPT's mixed sequence length. \sys has a relatively lower improvement for BookCorpus because of its high sequence length, not many requests can be batched together at the same time. Alpaca shows the least improvement because of its short sequence the requests can be processed quickly by all methods.}


\DEL{For OPT-13B, \sys can sustain 2.33$\times$, 2.25$\times$ and 1.2$\times$ higher request rates compared to Sarathi-Serve for ShareGPT, BookCorpus, and Alpaca, respectively, For OPT-175B, \sys can sustain 2.25$\times$, 1.88$\times$ and 1.11$\times$ higher request rates compared to Sarathi-Serve for ShareGPT, BookCorpus, and Alpaca, respectively. for Llama, \sys can sustain 2.18$\times$, 2$\times$ and 1.2$\times$ higher request rates than Sarathi-Serve for ShareGPT, BookCorpus, and Alpaca, }

\begin{figure*}[t]
\centering
    \subfloat[OPT-13B on ShareGPT.\vspace{-0.0in}\label{fig:exp-13-s-c}]{{\includegraphics[width=0.32\linewidth,height=0.11\textheight]{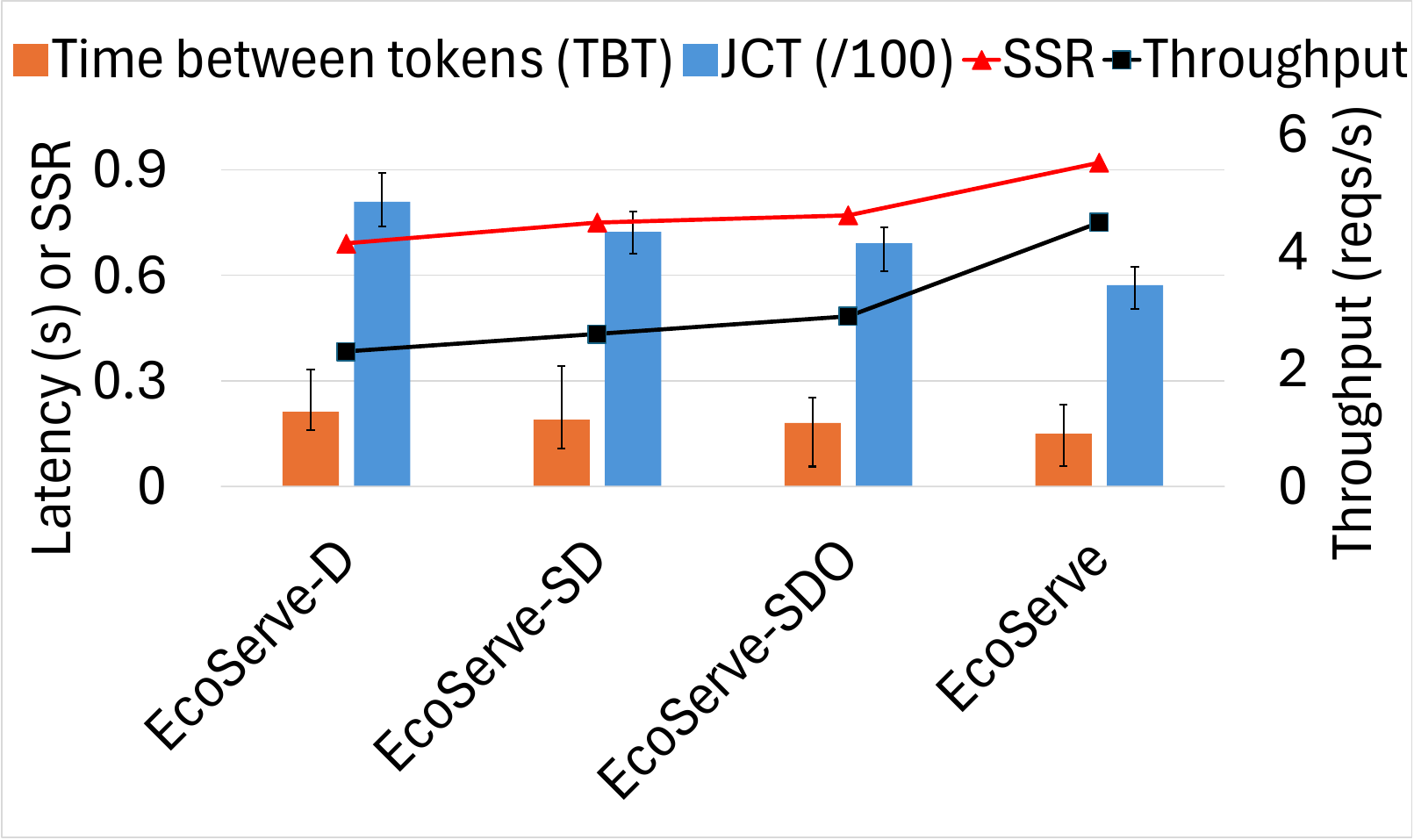} }}
    \hfill
    \subfloat[OPT-175B on ShareGPT.\vspace{-0.0in}\label{fig:exp-175-s-c}]{{\includegraphics[width=0.32\linewidth,height=0.11\textheight]{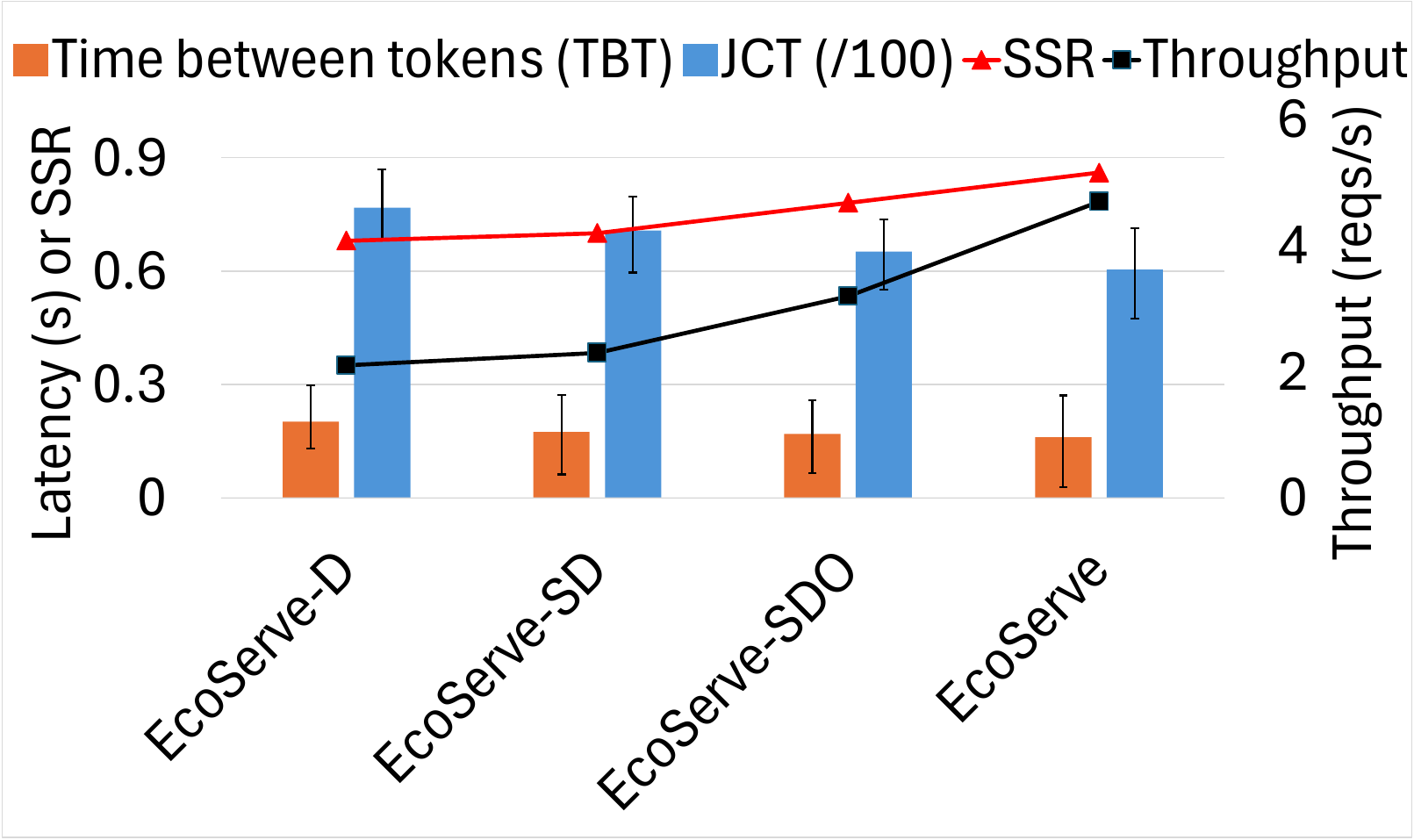} }}
    \hfill
    \subfloat[Llama-33B on ShareGPT.\vspace{-0.0in}\label{fig:exp-l3-s}]{{\includegraphics[width=0.32\linewidth,height=0.11\textheight]{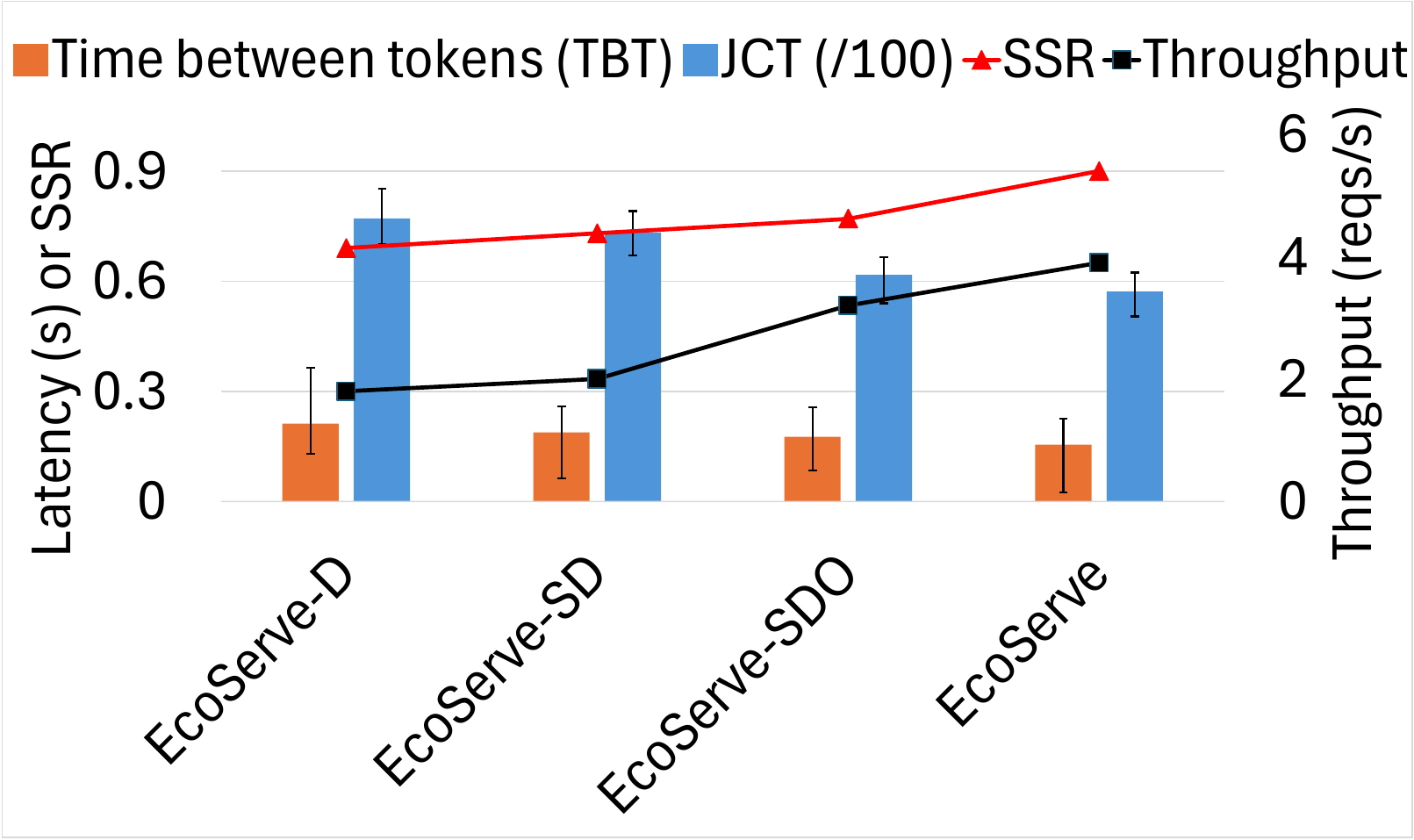} }}
    \hfill
     \subfloat[OPT-13B with BookCorpus.\vspace{-0.0in}\label{fig:exp-13-b}]{{\includegraphics[width=0.32\linewidth,height=0.11\textheight]{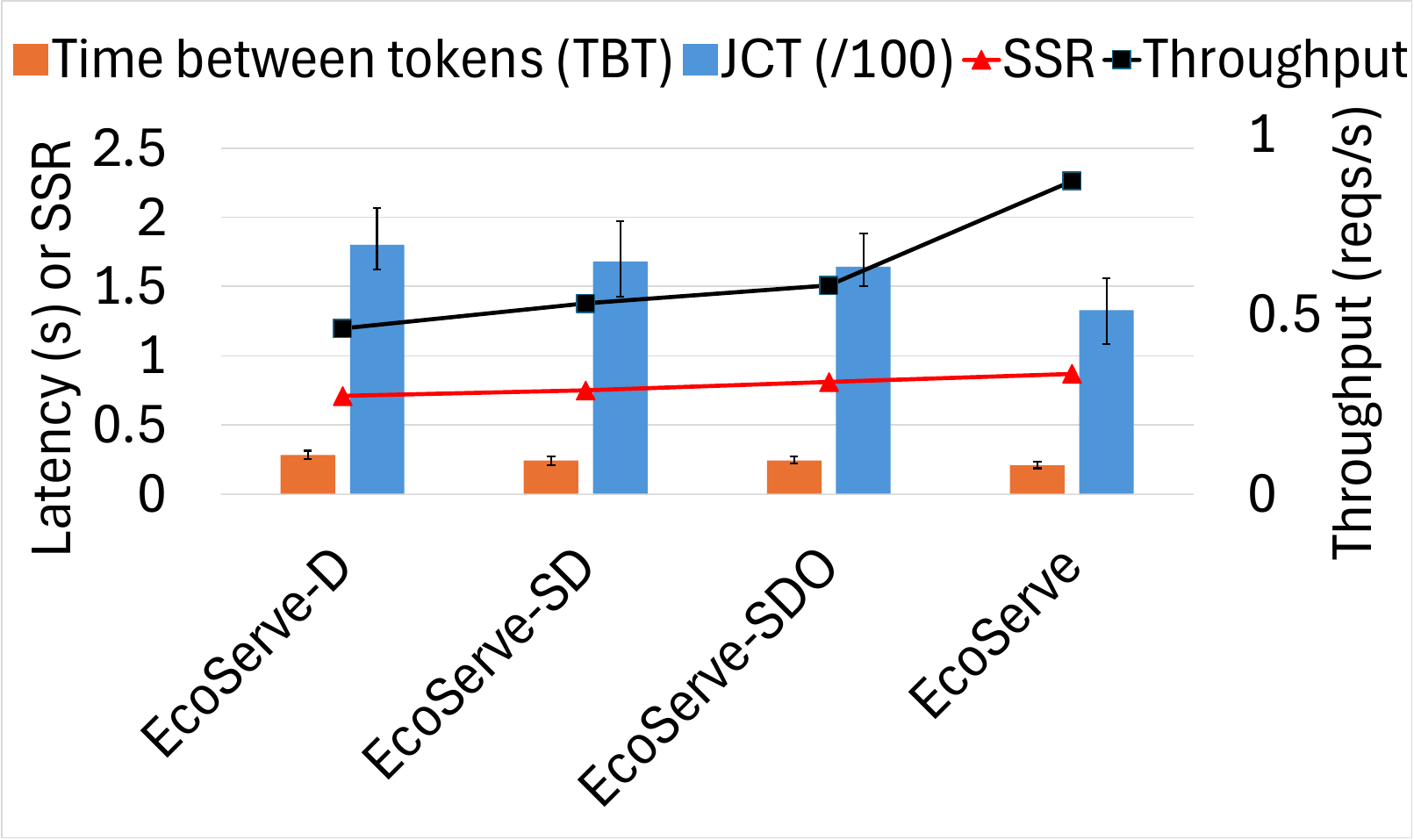} }}
    \hfill
    \subfloat[OPT-175B with BookCorpus.\vspace{-0.0in}\label{fig:exp-175-b-c}]{{\includegraphics[width=0.32\linewidth,height=0.11\textheight]{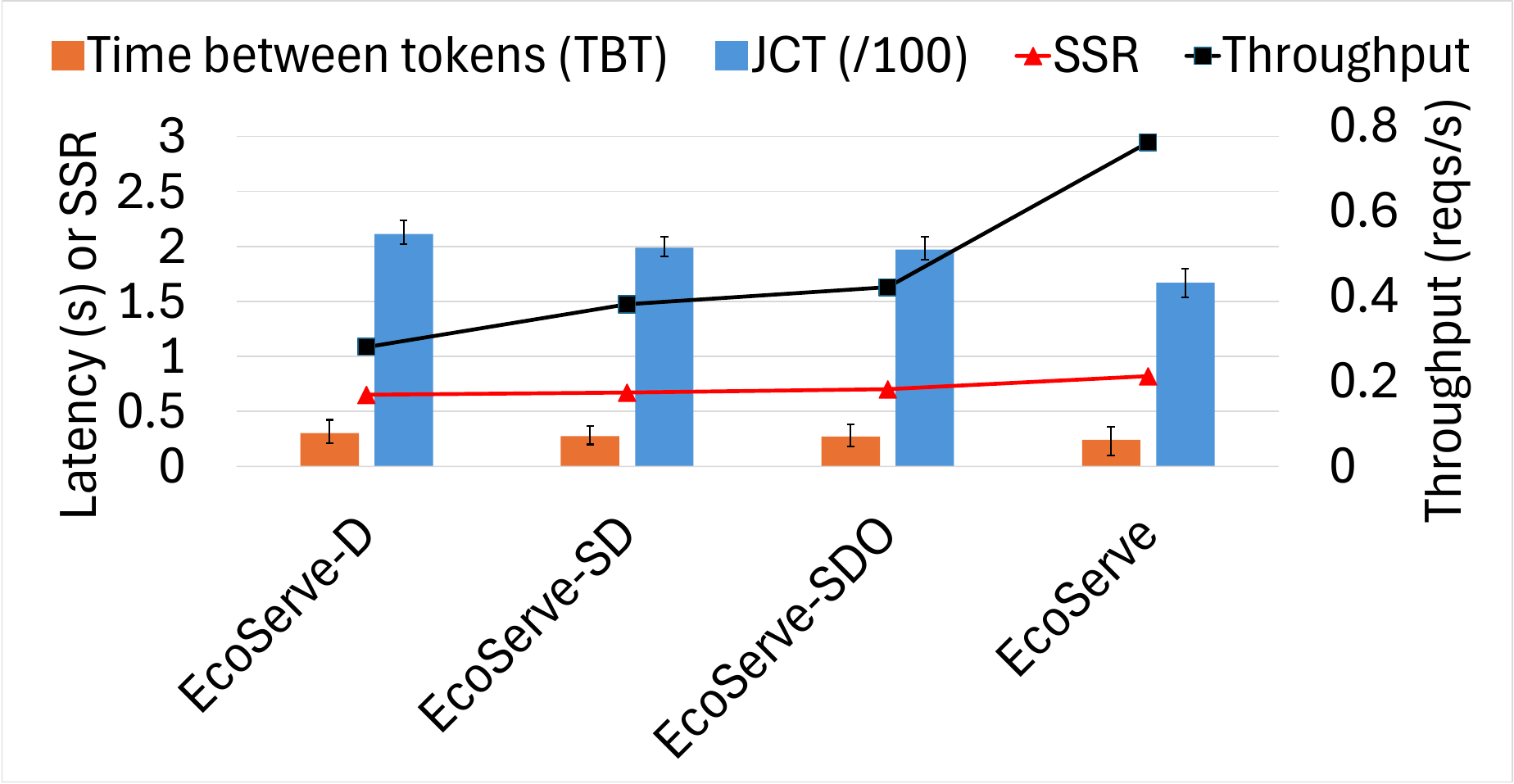} }}
    \hfill
    \subfloat[Llama-33B with BookCorpus.\vspace{-0.0in}\label{fig:exp-3}]{{\includegraphics[width=0.32\linewidth,height=0.11\textheight]{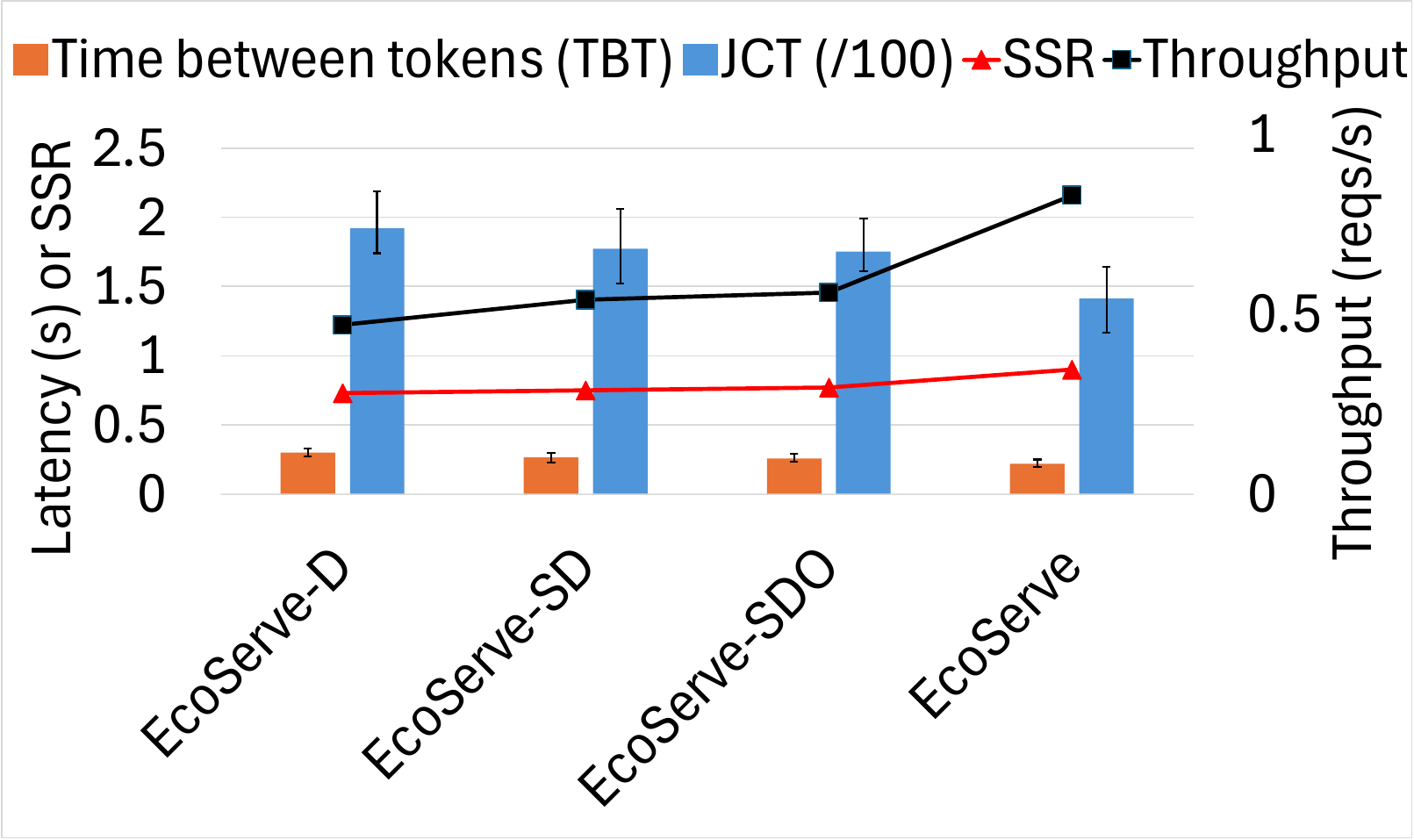} }}
    \hfill
    \subfloat[OPT-13B on Alpaca.\vspace{-0.0in}\label{fig:exp-13-a-c}]{{\includegraphics[width=0.32\linewidth,height=0.11\textheight]{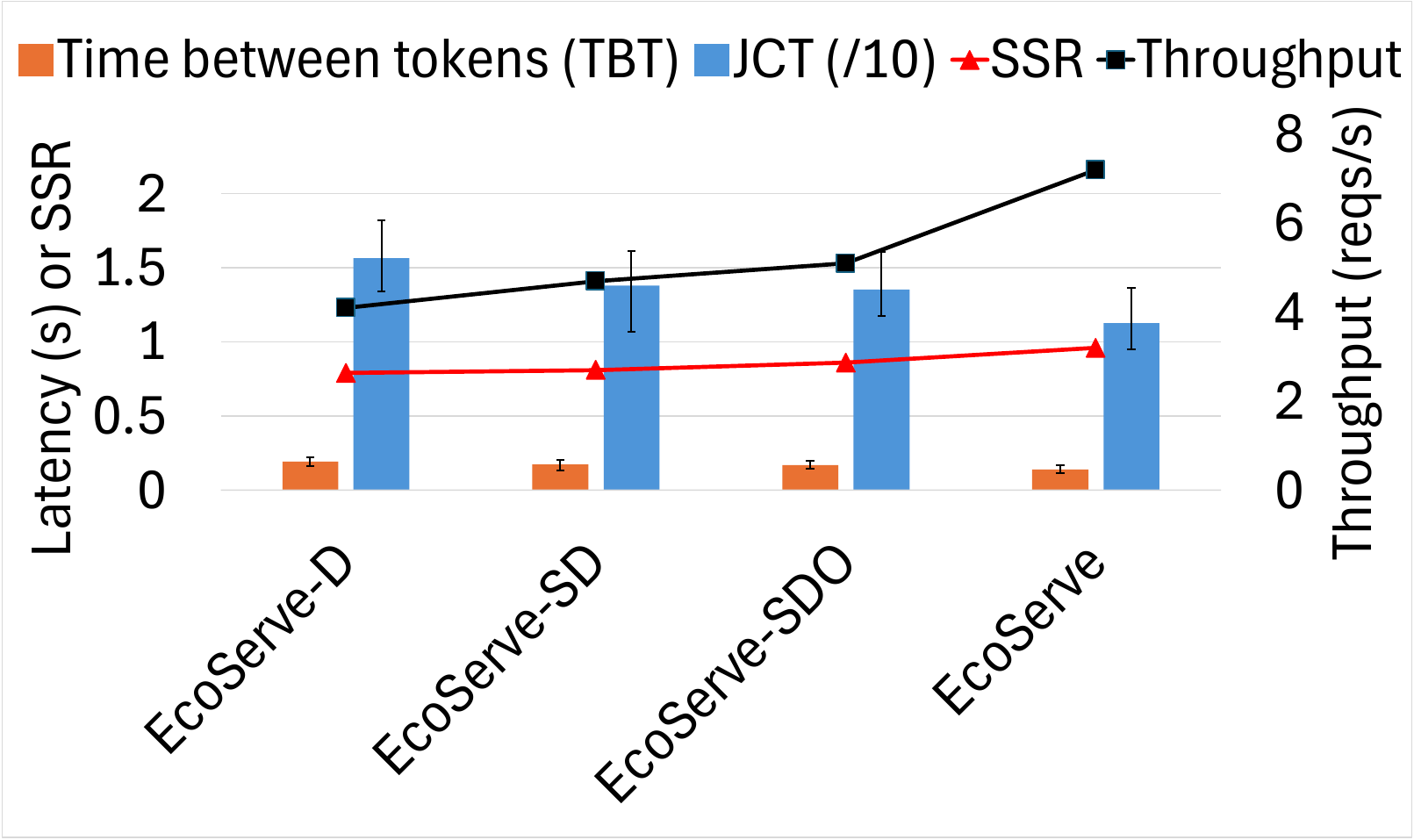} }}
    \hfill
    \subfloat[OPT-175B on Alpaca.\vspace{-0.0in}\label{fig:exp-175-a-c}]{{\includegraphics[width=0.32\linewidth,height=0.11\textheight]{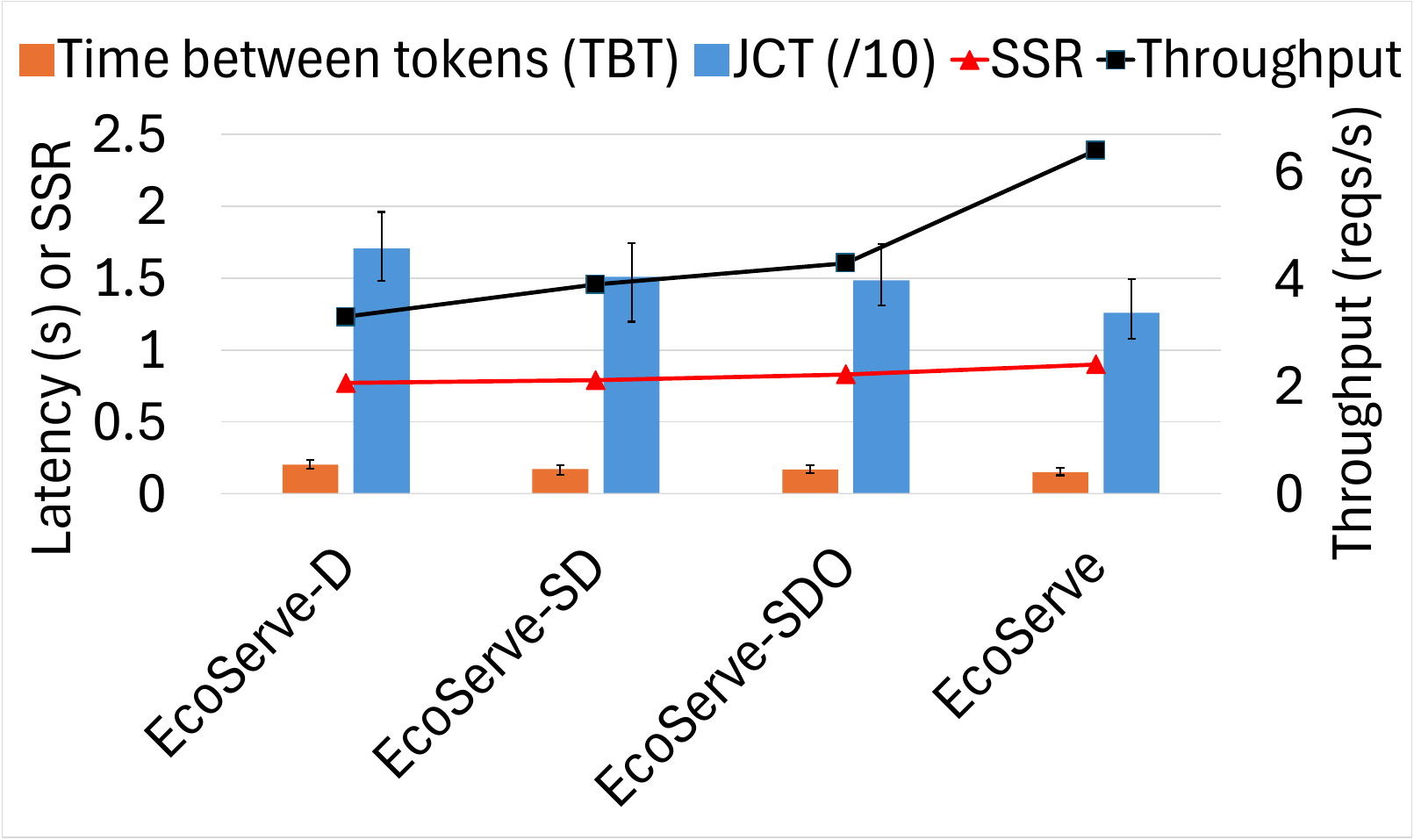} }}
    \hfill
    \subfloat[Llama-33B on Alpaca.\vspace{-0.0in}\label{fig:exp-l3-a-c}]{{\includegraphics[width=0.32\linewidth,height=0.11\textheight]{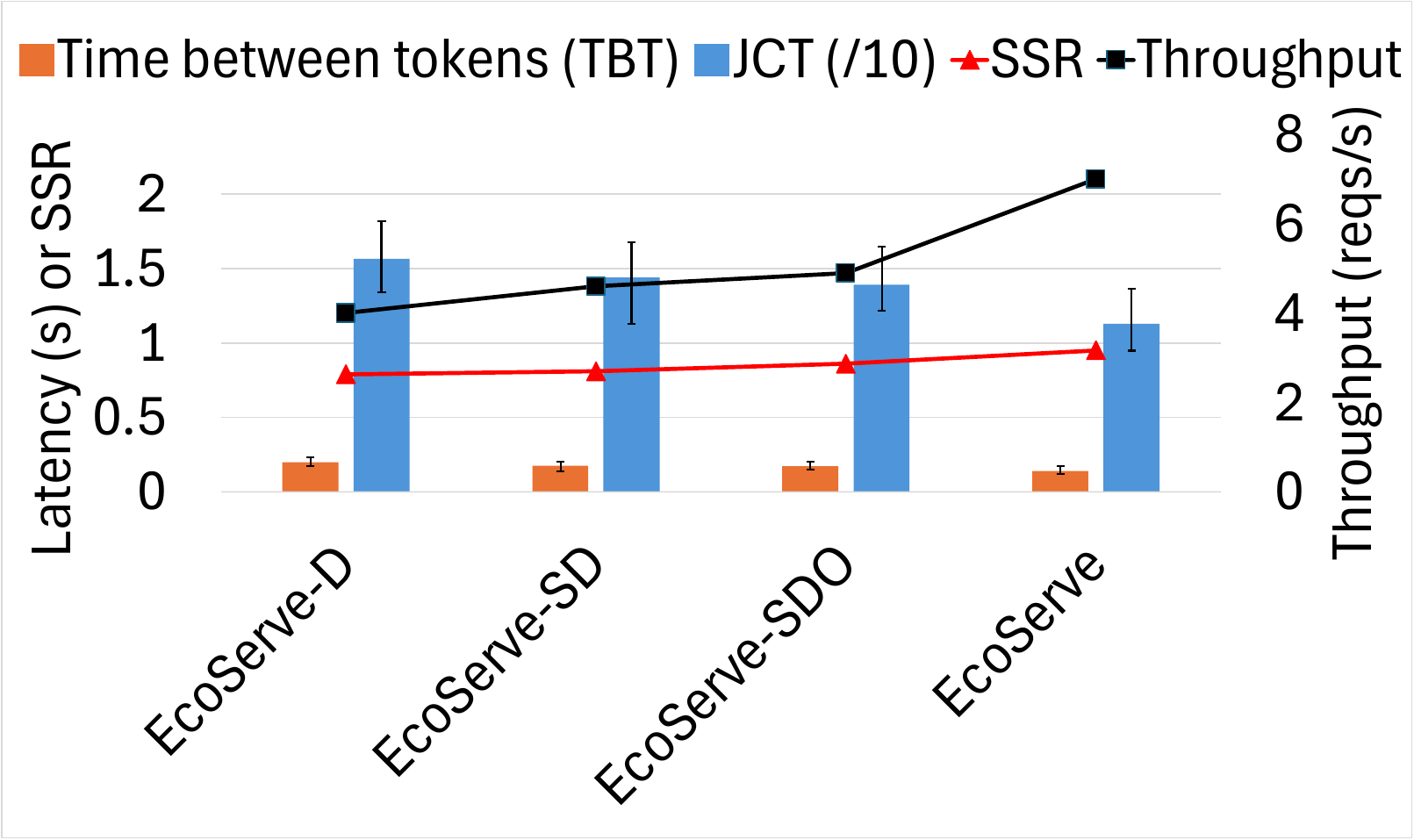} }}
    \hfill
    \vspace{-0.15in}
   \caption{\small{Ablation study of \sys for all models on the three traces.}\vspace{-0.2in}}%
    \label{fig:synd-comp-exp}
\end{figure*}

For OPT-13B, \sys can sustain 1.01-1.33$\times$ higher request rates than DistServe, 1.20-2.33$\times$ than Sarathi-Serve, 1.28-2.33$\times$ than vLLM and 3.5-25$\times$ than \Orca, respectively. 
For OPT-175B, \sys can sustain 1.20-1.50$\times$ higher request rates than DistServe, 1.20-2.25$\times$ than Sarathi-Serve, 1.28-2$\times$ than vLLM, and 2.8-22$\times$ than \Orca, respectively. Finally, for Llama, \sys can sustain 1.01-1.33$\times$ higher request rate than DistServe, 1.2-2.18$\times$ than Sarathi-Serve, 2.4-4$\times$ than vLLM, and 3-64$\times$ than \Orca, respectively. 
\sys outperforms other methods because it can more fully utilize KVC and GPU, avoid KVC allocation failures, and reduce scheduling time. \sys's time-synced batching helps reduce scheduling time and its decoupling method helps fully utilize the dual-resources. Further, its exact-allocation and padding strategy, KVC reservation and offload-free preemption help avoid KVC allocation failures or mitigate its adverse impact, while its KVC pipelining method helps improve KVC utilization. Moreover, its ordering method helps release occupied KVC earlier, expedite finding requests to fully utilize the dual-resources and satisfy SLO requirements. in contrast, vLLM does not aim to fully utilize GPU while Sarathi-Serve does not aim to fully utilize KVC, and both generate many KVC allocation failures due to block-allocation. \sys greatly improves \Orca, which uses max-allocation and a fixed small batch size.





\looseness=-1


Figures~\ref{fig:ssr-13b}-\ref{fig:ssr-llama} show the SLO satisfaction ratio (SSR) for each model for the three datasets, respectively. Compared to DistServe, Sarathi-Serve, vLLM, and Orca, \sys has 5\%, 42\%, 83\%, and 93\% higher SSR on average across the datasets in OPT-13B, 9\%, 47\%, 77\%, and 1.06$\times$   higher SSR in OPT-175B, and has 7\%, 41\%, 80\%, and 91\% higher SSR in Llama. \sys has only 2\%, 2.8\%, and 4\% lower SSR for the three models compared to Oracle. 

\begin{figure*}[]
\centering
    \subfloat[Scheduling overhead of OPT-13B.\vspace{-0.0in}\label{fig:sc0}]{{\includegraphics[width=0.32\linewidth,height=0.11\textheight]{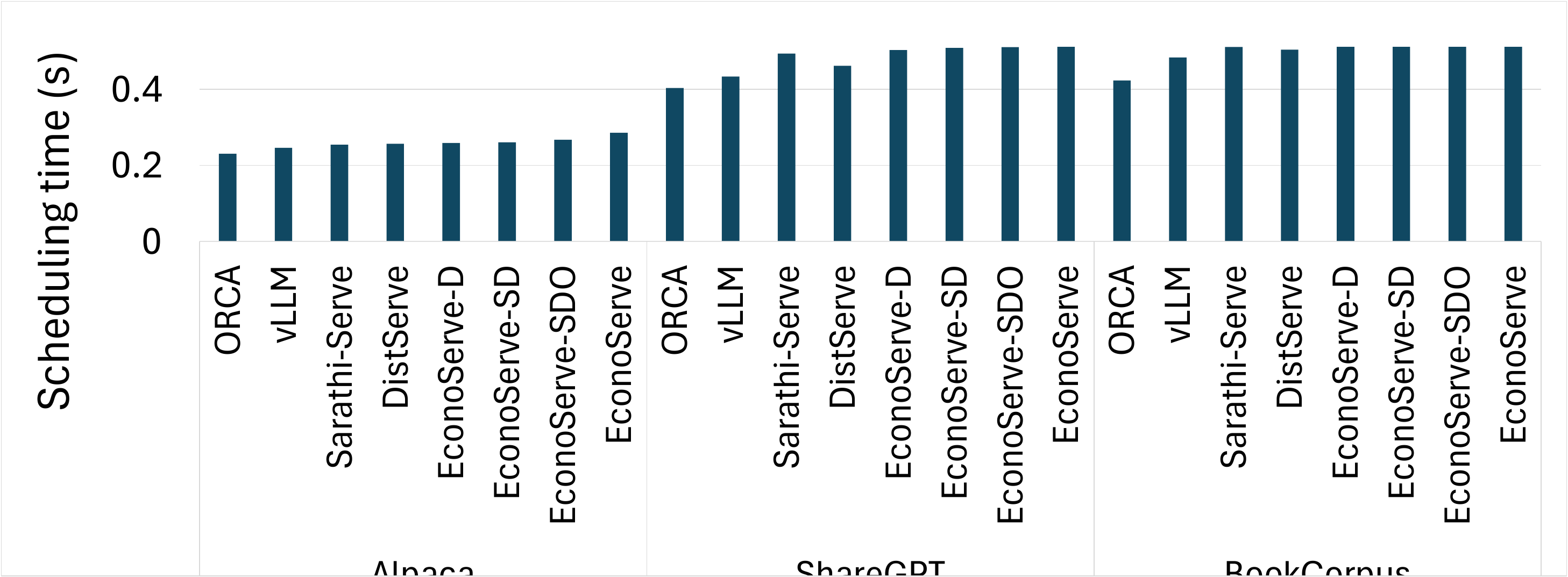} }}
    \hfill
    \subfloat[Scheduling overhead of OPT-175B.\vspace{-0.0in}\label{fig:sc1}]{{\includegraphics[width=0.32\linewidth,height=0.11\textheight]{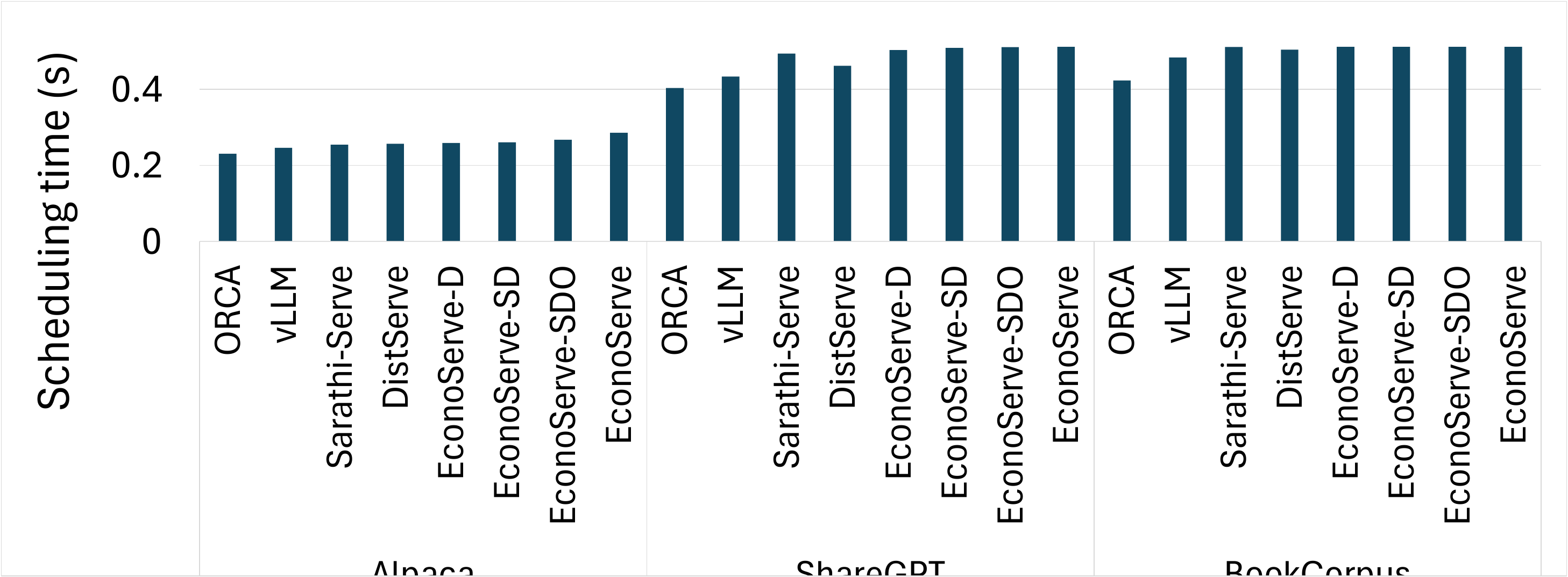} }}
    \hfill
    \subfloat[Scheduling overhead of Llama-33B.\vspace{-0.0in}\label{fig:sc2}]{{\includegraphics[width=0.32\linewidth,height=0.11\textheight]{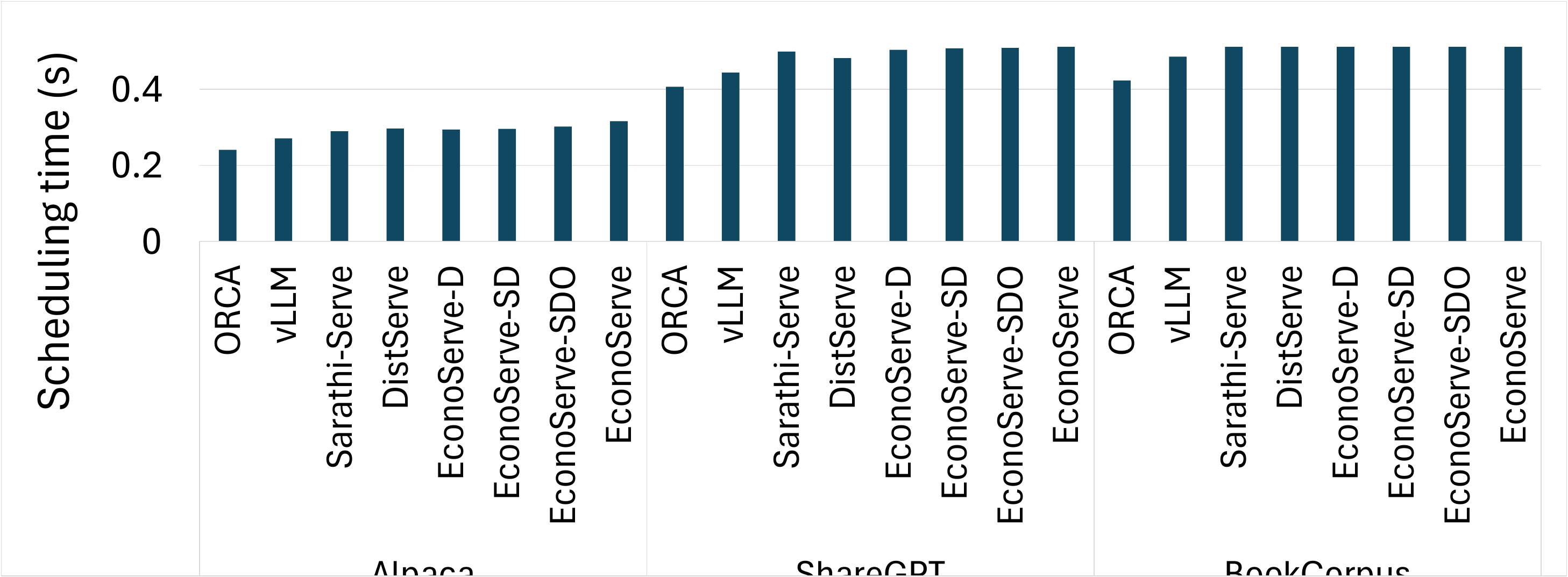} }}
    \hfill
    \vspace{-0.1in}
   \caption{\small{Scheduling overhead of \sys for all models on the three traces.}\vspace{-0.2in}}%
    \label{fig:synd-schedule}
\end{figure*}

In the following, we report the average GPU/KVC utilization of two GPUs for DistServe. We varied the request rate within the [0.2, 36] range with step size from the sequence \{0.3, 0.5, 0.5, 0.25, 0.25, 2, 4, 8, 8, 2, 2\}.  Figures~\ref{fig:kvc-13b}-\ref{fig:kvc-llama} show the average KVC utilization versus the request rate for ShareGPT, respectively. Compared to DistServe, Sarathi-Serve, vLLM, and Orca, \sys achieves on average 29\%, 18\%, 27\%, and 39\% higher KVC utilization for OPT-13B, 22\%, 12\%, 21\%, and 32\% for OPT-175B, and 25\%, 13\%, 22\%, and 34\% for Llama. 
Figures~\ref{fig:gpu-13b}-\ref{fig:gpu-llama} show the average GPU utilization for each model versus the request rate for the ShareGPT, respectively.  Compared to  DistServe, Sarathi-Serve, vLLM, and Orca, \sys has 16\%, 19\%, 28\%, and 39\% higher GPU utilization on average across the datasets in OPT-13B, 12\%, 17\%, 22\%, and 31\% for OPT-175B, and has 14\%, 17\%, 26\%, and 35\% for Llama. The reasons for \sys's higher resource utilization are explained earlier. 

\DEL{
\sys reduces the number of re-computations and swapping caused by the preemptions compared to these methods. Finally, }

\DEL{Figure~\ref{fig:prompt-seq-vllm-comp} shows the comparative performance of the individual components of \sys. Among the components, \sys-D can sustain the same request rate as of Sarathi-Serve for the OPT-13B model on the BookCorpus dataset for the same latency. For the Alpaca and the ShareGPT dataset,
\sys-D can sustain 1.07$\times$ and 1.5$\times$ request rates, respectively. The \sys-SD and \sys-SDO can handle  1.33$\times$-1.5$\times$ higher request rates than the Sarathi-Serve for the three datasets. \sys-P can handle 1.14$\times$-1.88$\times$ higher request rates than the Sarathi-Serve. For the OPT-175B model, \sys-D, \sys-SD, and \sys-SDO can sustain 1.07$\times$-1.5 $\times$ higher request rates than the Sarathi-Serve, while \sys-P can handle 1.14$\times$-1.88$\times$ higher request rates than the Sarathi-Serve. For the same attained request rate, \sys-P achieves 24\%-38\% lower latency than the other components, \sys-D, \sys-SD, \sys-SDO. The overall order of performance considering both the throughput and latency is as follows: \sys-P> \sys-SDO > \sys-SD >\sys-D.}



\noindent{\textbf{Resource-efficiency in Comparison to DistServe.}} We compared \sys with DistServe regarding the number of GPUs used at different request rates on the ShareGPT dataset across homogeneous, heterogeneous, and large-scale simulation settings, as shown in Figure~\ref{fig:num-of-gpus}. 

For the homogeneous settings, we used two A100 machines as described in~\ref{sec:analysis} and selected two proximity-close GPUs for the prefill and decode phases of each request.
When choosing a GPU on a machine, we prioritized the one with the shortest request queue. For each request rate, we determined the average goodput (requests meeting the SSR per second) for DistServe. We then iterated from 1 to the total number of GPUs to determine the minimum required by \sys to achieve the same goodput. Compared to DistServe, \sys uses 73\%, 58\%, and 67\% fewer GPUs for OPT-13B, OPT-175B, and Llama-33B, respectively. For the heterogeneous settings, we used one A100 with 8GPUs for decoding and one H100 with 4 GPUs (80 GB each) for prefilling.  Here, \sys uses 61\%, 51\%, and 57\% fewer GPUs for OPT-13B, OPT-175B, and Llama-33B, respectively. Finally, in a large-scale scenario using the Vidur Simulator~\cite{Agrawal2024VidurAL} with 4000 GPUs for the Meta-Llama-3-8B model (2000 GPUs for prefilling and 2000 for decoding), \sys uses 77\% fewer GPUs compared to DistServe.

\noindent{\textbf{Ablation Study.}} Figures~\ref{fig:exp-13-s-c}-\ref{fig:exp-l3-a-c} illustrate the performance of individual components of \sys in terms of average JCT, Time Between Tokens (TBT), SSR and throughput. In the figure, ``$/ 10$'' means that the figure divides the results of the method by 10 to make the figure visible. 



On average, across all datasets and all models, \sys-D, \sys-SD, \sys-SDO achieve 19\%, 7\%, 29\%, higher JCT than \sys, respectively.
This indicates that Decoupling, Synced batching, Ordering, and KVCPipe reduce JCT by 28\%, 19\%, 7\%, and 29\%, respectively. 
The individual methods reduce TBT by 17\%, 5\%, and 21\%, respectively. \sys has an average TBT of 0.141s with 5th and 95th percentile values of 0.129s and 0.221s. 
This indicates that GT queuing won't greatly increase their TBTs. 
Furthermore, \sys-D, \sys-SD, \sys-SDO has 16\%, 21\%, and 17\% lower SSR than \sys respectively. 
This implies that these individual methods have 36\%, 17\%, and 62\%, lower throughput than \sys, respectively. 
Ordering is less effective at improving JCT and TBT compared to other methods, but it improves SSR by considering SLO in queue ordering.\looseness=-1 

\DEL{??vLLM’s
advantage over Orca (Oracle) and Orca (Pow2) is less pronounced. This is because the model and server configuration
for OPT-175B (Table 1) allows for large GPU memory space
available to store KV cache, while the Alpaca dataset has
short sequences. In this setup, Orca (Oracle) and Orca (Pow2)
can also batch a large number of requests despite the inefficiencies in their memory management. As a result, the
performance of the systems becomes compute-bound rather
than memory-bound.}

\DEL{\noindent{\textbf{Ablation Study.}} Figures~\ref{fig:exp-13-s-c}-\ref{fig:exp-l3-a-c} illustrate the performance of individual components of \sys in terms of average JCT, Time Between Tokens (TBT), SSR and throughput. Across all three models and datasets, on average, \sys-D, \sys-SD, \sys-SDO, and \sys achieve 28\%, 47\%, 54\%, and 83\% lower JCT than vLLM, respectively. This indicates that Decoupling, Synced batching, Ordering, and KVCPipe reduce JCT by 28\%, 19\%, 7\%, and 29\%, respectively. Additionally, \sys-D, \sys-SD, \sys-SDO, and \sys achieve 27\%, 44\%, 49\%, and 70\% lower TBT than vLLM, respectively. This suggests that these methods reduce TBT by 27\%, 17\%, 5\%, and 21\%, respectively. \sys has an average TBT of 0.141s with 5th and 95th percentile values of 0.129s and 0.221s. 
This indicates that queued GTs won't greatly increase their TBTs. Furthermore, \sys-D, \sys-SD, \sys-SDO, and \sys achieve 26\%, 36\%, 57\%, and 78\% lower SSR than vLLM, respectively. This implies that these methods increase SSR by 26\%, 10\%\sh{this should be higher}, 21\%, and 21\%, respectively. \sys-D, \sys-SD, \sys-SDO, and \sys achieve 78\%, 1.04$\times$, 1.21$\times$, and 1.83$\times$ higher throughput than vLLM, respectively. This implies that these methods increase throughput by 78\%, 36\%, 17\%, and 62\%, respectively. 
Ordering is less effective at improving JCT and TBT compared to other methods, but it significantly improves SSR by considering SLO in queue ordering.
}


\DEL{Similarly, \sys-P, \sys-SDO, \sys-SD, \sys-D have  9\%,  24\%, 27\%, 35\% higher TBT than \sys, respectively. For the throughput, the order of the performance is as follows: \sys-D<\sys-SD<\sys-SDO<\sys-P<\sys. \sys-P, \sys-SDO, \sys-SD, and \sys-D, have 9\%, 78\%, 95\% and 2.15$\times$ lower throughput than \sys, respectively. \sys-P performs better than the others because the pipeline scheme reduces both the prompt and GT waiting time compared to other methods, reflected in the overall latency. \sys-SDO performs better than the \sys-SD in terms of latency although achieving the same throughput because the ordering method on top of the selection method considers the occupied KVC, the SLO, and the prompt length for prompts (or predicted response length for GTs). Thus, it quickly finds requests to fully utilize the GPU and KVC resources, reducing scheduling and waiting times. By grouping the requests with the same RL, \sys-SD reduces the overall token generation time, while \sys-D reduces the waiting time by decoupling the prompt and token generation.}

\DEL{{\color{blue} Across all models and datasets, \sys-P, \sys-SDO, \sys-SD, \sys-D have  11\%,  27\%, 29\%, 38\% higher JCT than \sys, respectively. Similarly, \sys-P, \sys-SDO, \sys-SD, \sys-D have  9\%,  24\%, 27\%, 35\% higher TBT than \sys, respectively. For the throughput, the order of the performance is as follows: \sys-D<\sys-SD<\sys-SDO<\sys-P<\sys. \sys-P, \sys-SDO, \sys-SD, and \sys-D, have 9\%, 78\%, 95\% and 2.15$\times$ lower throughput than \sys, respectively.} \sys-P performs better than the others because the pipeline scheme reduces both the prompt and GT waiting time compared to other methods, reflected in the overall latency. \sys-SDO performs better than the \sys-SD in terms of latency although achieving the same throughput because the ordering method on top of the selection method considers the occupied KVC, the SLO, and the prompt length for prompts (or predicted response length for GTs). Thus it quickly finds requests to fully utilize the GPU and KVC resources, reducing scheduling and waiting times. By grouping the requests with the same RL, \sys-SD reduces the overall token generation time, while \sys-D reduces the waiting time by decoupling the prompt and token generation.}

\noindent\textbf{Scheduling Time Overhead.} Figures~\ref{fig:sc0}-\ref{fig:sc2} show the total time overhead of different methods. \sys have 2.93\% higher time overhead than vLLM. However, \sys achieves much higher performance in other metrics than vLLM, as explained above. The time overhead of the components of \sys contributes to its overhead. \sys-D, \sys-SD, and \sys-SDO have 2.05\%, 2.16\%, and 2.36\% higher scheduling overhead than vLLM. \Orca has 2.15\% lower, DistServe 1.53\% higher, and Sarathi-Serve 1.84\% higher scheduling overhead than vLLM. DistServe's overhead stems from searching for parallelism strategies for prefill and decode instances, while Sarathi-Serve's arises from locating the necessary chunk for prefill. vLLM has slightly higher overhead than Orca due to block allocation. 



\begin{figure}[t]
\centering
    \subfloat[SLO scale.\vspace{-0.0in}\label{fig:expind-1-20-1}]{{\includegraphics[width=0.48\linewidth,height=0.11\textheight]{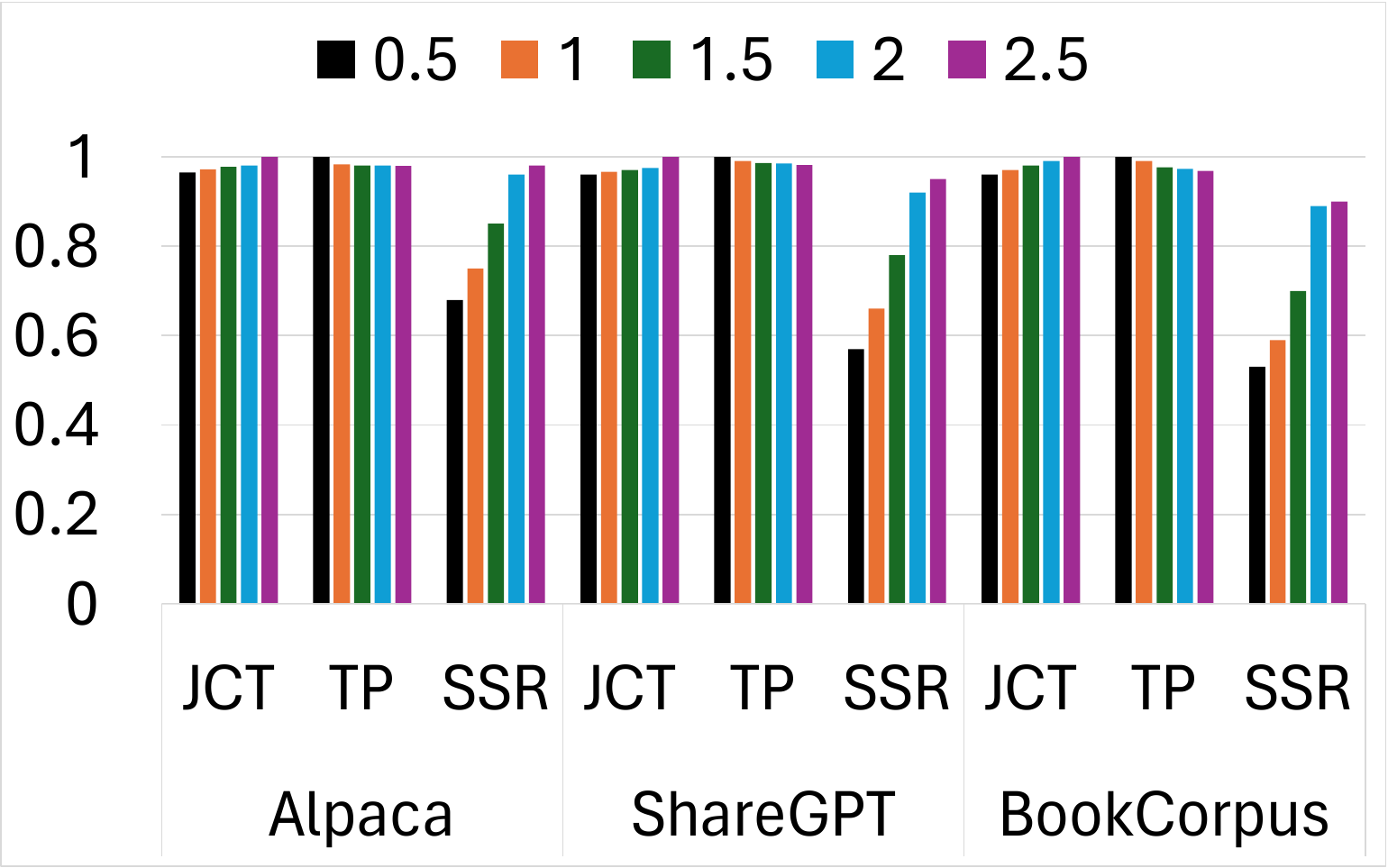} }}
    \hfill    \subfloat[Padding ratio.\vspace{-0.0in}\label{fig:expind-2-20-b}]{{\includegraphics[width=0.48\linewidth,height=0.11\textheight]{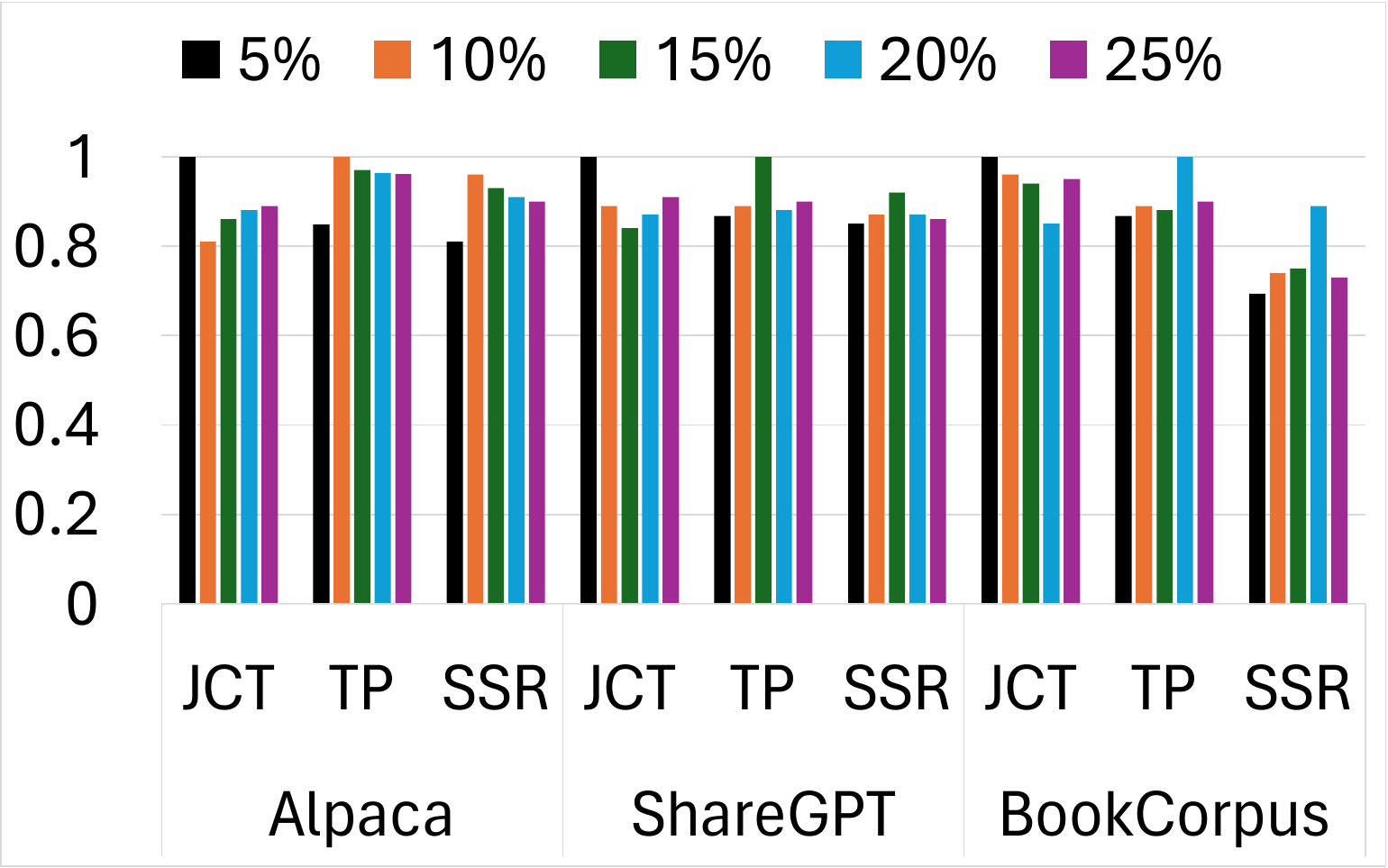} }}
    \hfill
    \subfloat[Reserved KVC.\vspace{-0.0in}\label{fig:expind-3-20-c}]{{\includegraphics[width=0.48\linewidth,height=0.11\textheight]{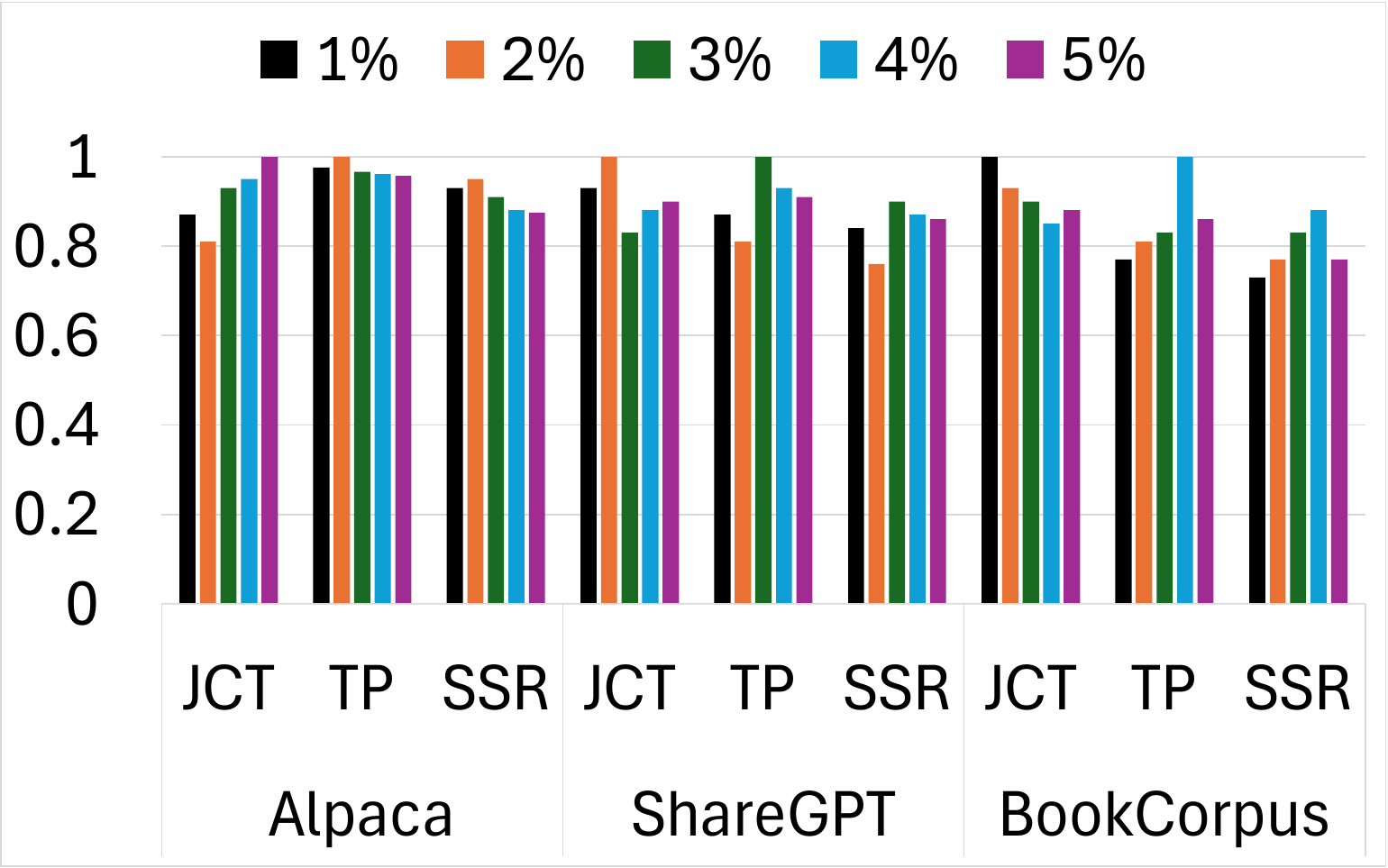} }}
    \hfill
    \subfloat[Buffer size/KVC size.\vspace{-0.0in}\label{fig:expind-3-20e}]{{\includegraphics[width=0.48\linewidth,height=0.11\textheight]{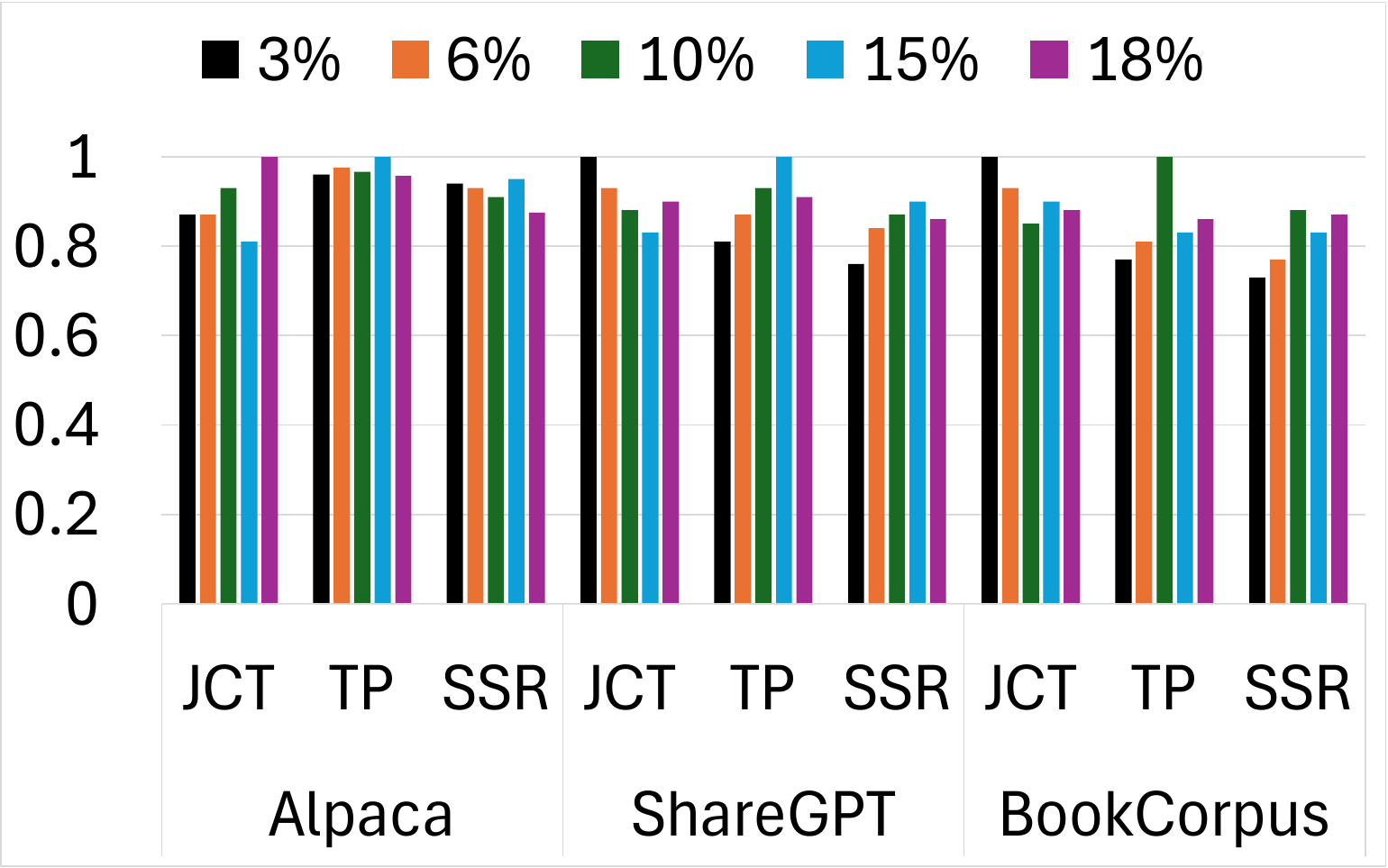} }}
    \hfill
 \DEL{   \subfloat[Buffer\sh{cannot infinitely decrease}.\vspace{-0.0in}\label{fig:expind-3-20e}]{{\includegraphics[width=0.325\linewidth,height=0.135\textheight]{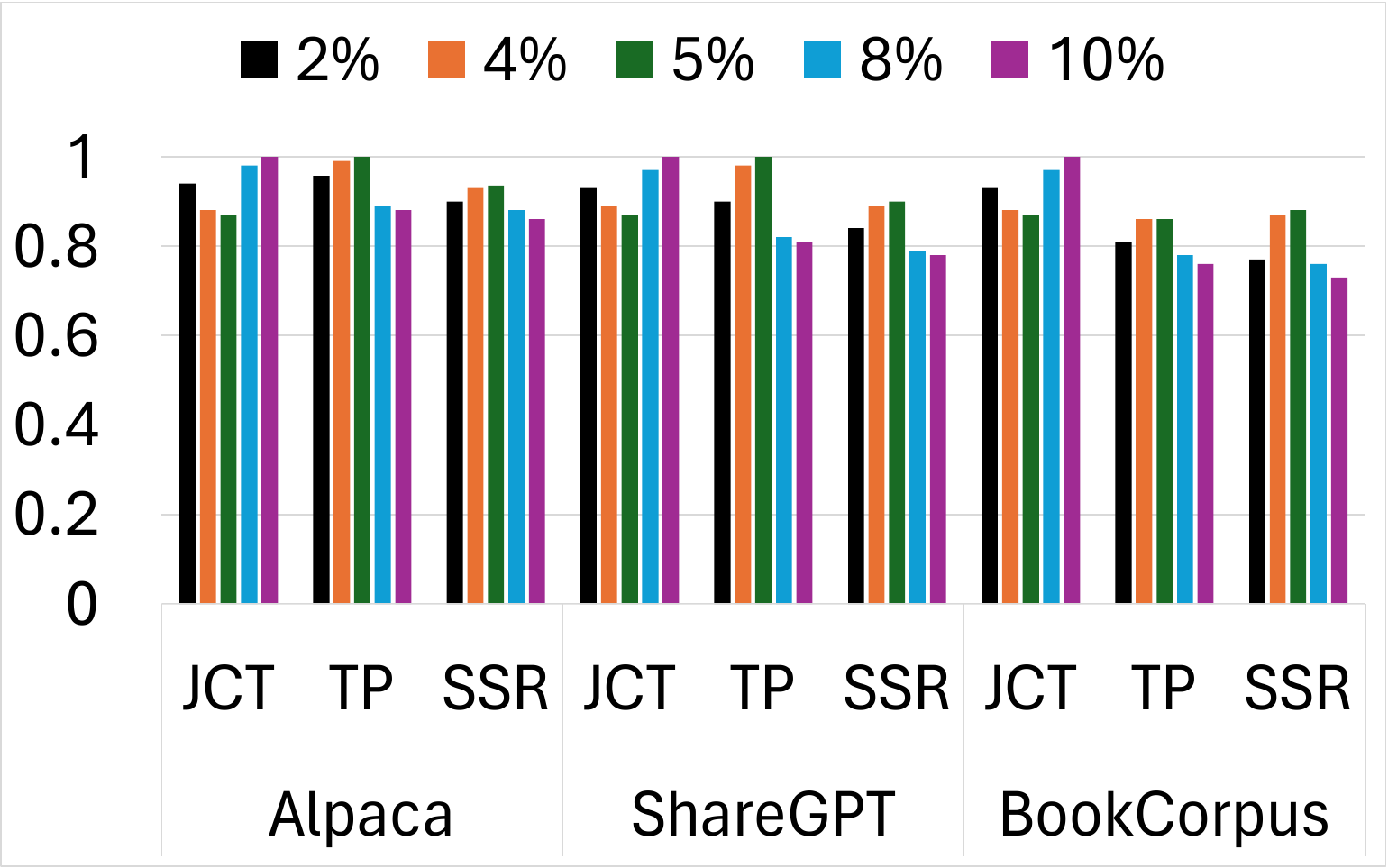} }}}
    \vspace{-0.25in}
   \caption{\small{Effect of factors for OPT-13B on the three traces. \vspace{-0.3in}}}%
    \label{fig:ind-factors-13b}
\end{figure}


\noindent{\textbf{Sensitivity Testing.}} 
Figure~\ref{fig:ind-factors-13b} shows the normalized performance of JCT, throughput (TP), and SSR by the maximum performance value with the different settings of each influencing factor for OPT-13B. 
Figure~\ref{fig:expind-1-20-1} shows that as the SLO scale increases from 0.5 to 2.5, the JCT and throughput are not greatly affected, but SSR increases by 23\% on average for the three datasets. This is because the SLO-scale only affects \sys's ordering method. 
BookCorpus has lower SSR than other datasets due to its longer sequence lengths and longer JCTs. Figure~\ref{fig:expind-2-20-b} shows that the three datasets have the best performance when the padding ratio equals to 10\%, 15\%, and 20\%, respectively. The JCT follows the same trend as in Figure~\ref{fig:response-latency-buffer} due to the same reasons explained. Accordingly, SSR and throughput first increase and then decrease.


\DEL{As \sys's queue ordering method considers requests' SLOs in order, and the priority relationship between requests will not be changed much, so the throughput and JCT are not greatly affected. We still observe slight increase in JCT and slight decrease in throughput. This is because looser SLO makes some requests wait longer in the queue. The }



\DEL{As the padding increases, JCT first decreases and the SSR and throughput increase, and then JCT increases, and the SSR and throughput decrease.\looseness=-1}  

Figure~\ref{fig:expind-3-20-c} shows that the best reserved KVC percentage for Alpaca, ShareGPT and BookCorpus is 2\%, 3\%, and 4\%, respectively. The general trend of the datasets on reserved KVC is similar to that of the padding ratio for the same reasons. Figure~\ref{fig:expind-3-20e}  demonstrates that as the buffer percentage increases, the performance improves initially but then decreases. The best buffer percentage for Alpaca, ShareGPT and BookCorpus is 15\%, 15\%, and 10\%, respectively.

\DEL{Figure~\ref{fig:expind-3-20e} illustrates the performance of \sys when varying the buffer in KVC pipelining from the set {3\%, 6\%, 10\%, 15\%, 18\%}.  }

\DEL{\begin{wrapfigure}{c}{4cm}\vspace{-0.0in}
    \centering
\includegraphics[width=0.459\columnwidth,height=0.135\textheight]{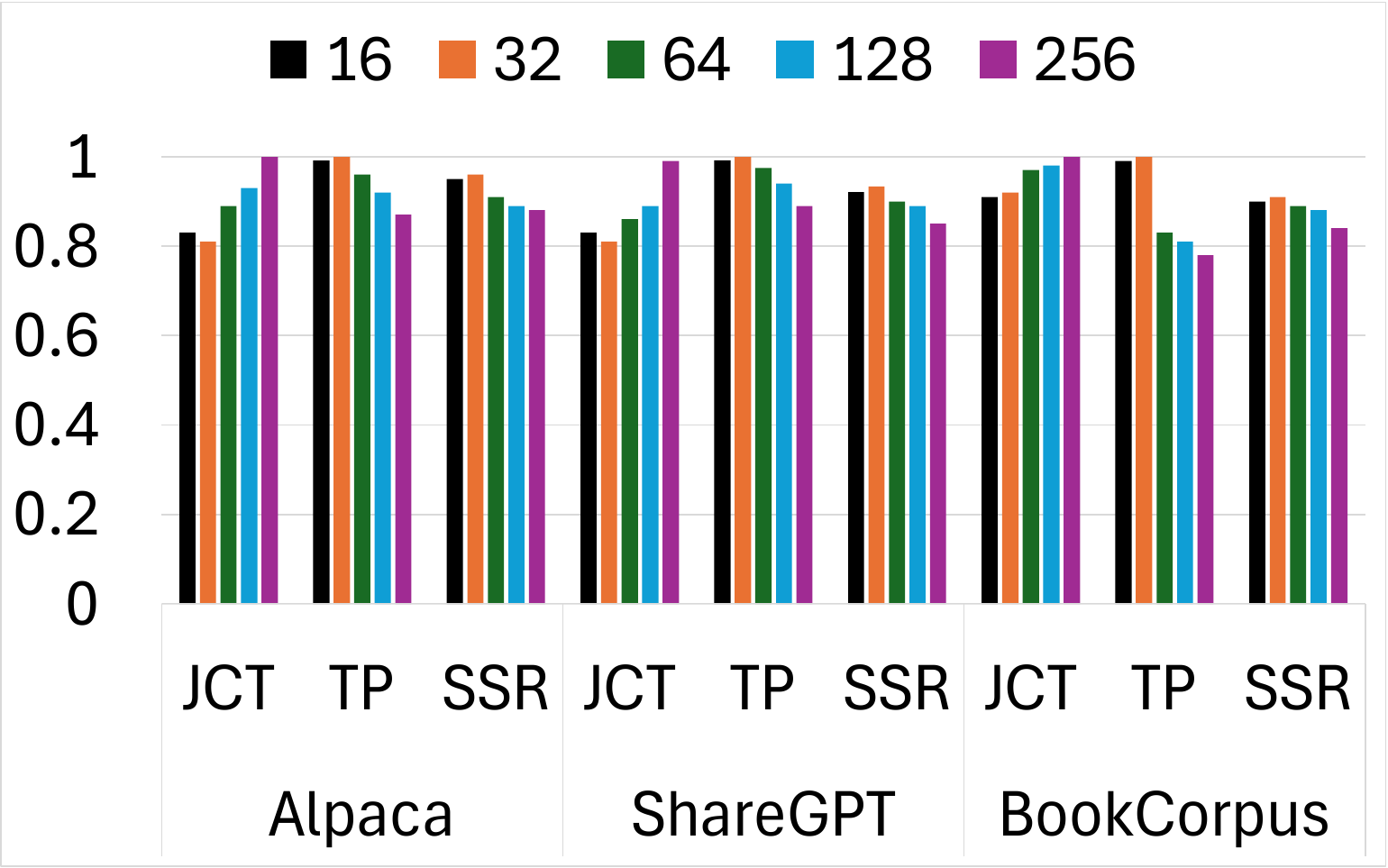}
    \caption{\small{Effect of block size for OPT-13B on the three traces.}}
    \label{fig:block-size}\vspace{-0.12in}
\end{wrapfigure}  
}


\DEL{Figure~\ref{fig:block-size} illustrates the normalized performance when varying the block size. For all three datasets, block sizes 16 and 32 perform well, with block size 32 showing approximately 1-3\% better performance across all metrics. However, performance deteriorates as the block size increases. The finding is consistent with that in \cite{vllm}.} 


\vspace{-0.1in}
\section{Related Work}
\vspace{-0.1in}
\label{sec:related-work}


\DEL{Initially, LLM inference employs request-level scheduling, in which all requests within a batch are processed collectively until they are completed~\cite{fang2021deployment,46801}. \Orca~\cite{280922} integrates the scheduler and the execution engine backend and overcomes the limitations of request-level scheduling with  two proposed techniques: iteration-level scheduling and selective batching. However, \Orca still lacks in terms of GPU under-utilization. 
Wu \emph{et al.}~\cite{Wu2023FastDI} proposed FastServe that uses preemptive scheduling to minimize JCT with a novel skip-join multi-level feedback queue scheduler. 
Sheng \emph{et al.}~\cite{sheng2023high} proposed FlexGen that solves a linear programming problem and searches for efficient patterns to store and access tensors across memory hierarchy. 
{\color{red} Zheng~\emph{et al.}~\cite{Radix-attention} proposed a technique for KVC reuse during runtime in the scenario of multiple LLM calls. Instead of discarding the KVC after finishing a generation request, the approach retains the KVC for both prompts and generation results in a radix tree for multiple requests. Liu~\emph{et al.}~\cite{liu2023scissorhands} proposed maintaining the memory usage of KVC under a fixed budget by only storing pivotal tokens based on the attention scores.} Sheng~\emph{et al.}~\cite{Fairness-OSDI-2024} proposed a scheduler that considers fairness between users. }

Initially, LLM inference utilizes request-level scheduling, where all requests within a batch are processed collectively until completion~\cite{fang2021deployment,46801}. 
To overcome its drawbacks of long response time, \Orca~\cite{280922} uses iteration-level scheduling and max-allocation but it generates GPU under-utilization. To address this, vLLM~\cite{vllm} uses a block-based approach. Jin \emph{et al.}~\cite{jin2023s}  proposed $S^3$, which predicts the output length and allocates the required memories to the request. Zheng \emph{et al.}~\cite{Zheng2023ResponseLP} proposed a system that predicts the output length of an input and then batches queries with similar response lengths. In contrast to \cite{Zheng2023ResponseLP}, which aims to prevent delaying faster requests in the request-level scheduling, we use a similar method for a different purpose -- reducing scheduling time at the iteration-level scheduling. Wu \emph{et al.}~\cite{Wu2023FastDI} proposed FastServe, employing preemptive scheduling to minimize JCT. 
Sheng\emph{et al.}~\cite{Fairness-OSDI-2024} considered achieving fairness in scheduling.  Sun \emph{et al.}~\cite{298685} proposed Llumnix, which reschedules requests across multiple model instances to improve tail latencies during inference. Some methods~\cite{patel2024splitwise,zhong2024distserve,hu2024inference} decouple the prefill and decode phases, running them on separate machines or GPUs. However, employing two model replicas across two GPUs for each inference job is resource-inefficient and adds latency due to KV value transfer. Compared to existing schedulers, \sys distinguishes itself by 
fully utilizing the dual resources at each GPU at each iteration while preventing KVC allocation failures. 

\DEL{to fully utilize both GPU and KVC resources. It also has KVC pipelining utilization to enhance KVC utilization and strategies such as padding on predicted RL, offload-free preemption, and using reserved KVC to mitigate adverse effects from KVC allocation failures.}



Several other approaches aim to enhance LLM performance. Sheng \emph{et al.}~\cite{flexgen} introduced FlexGen for memory hierarchy optimization. Oh \emph{et al.}~\cite{oh} proposed \emph{ExeGPT}, which finds the optimal execution configuration, including batch sizes and partial tensor parallelism, to maximize throughput while meeting latency constraints. 
Zheng~\emph{et al.}~\cite{zheng2024sglang} proposed storing KVC in a radix tree for multiple requests. Liu~\emph{et al.} proposed storing pivotal tokens in KVC based on attention scores~\cite{liu2023scissorhands} and reducing computational cost by processing only relevant parts of the input context~\cite{pmlr-v202-liu23am}. 
These methods can be employed in \sys to improve performance.



\DEL{a KV cache management framework tailored for long-text generation, which prefetches only the essential KV cache entries for computing the subsequent attention layer in the Transformer.}
\DEL{Compiler systems like~\cite{pit,280848} address dynamic sparsity in LLM models by constructing GPU-efficient tiles to increase GPU utilization and low memory wastage.}

\DEL{\noindent{\textbf{Scheduling of Transformer Models.}}
\Orca~\cite{280922} integrates the scheduler and the execution engine backend (system that executes the scheduled request) and overcomes the limitations of request-level scheduling with two proposed techniques: iteration-level scheduling and selective batching. However, \Orca still lacks in the terms of GPU under-utilization and meeting the latency SLOs. Gao \emph{et al.}~\cite{batchmaker} proposed \emph{Batchmaker} that divides the model graph for processing the request into RNN cells, and schedules execution at the level of cells. 
However, it does not handle the Transformer models. Romero \emph{et al.}~\cite{infaas} proposed INFaaS to automate the model selection, deployment, and serving process. 
Besides, there are systems, such as Clockwork~\cite{clockwork} and
Shepherd~\cite{shepherd}, which focuses on serving relatively small models like ResNet in a cluster and supports latency-aware resource allocation to maximize the overall throughput. Jin \emph{et al.}~\cite{jin2023s}  overcome the issue of GPU under-utilization by proposing $S^3$, which predicts the output sequence length. Then, it schedules generation queries based on the prediction by allocating the required memories to the request. Wu \emph{et al.}~\cite{Wu2023FastDI} proposed FastServe that uses preemptive scheduling to minimize JCT with a novel skip-join multi-level feedback queue scheduler. 
Zheng \emph{et al.}~\cite{Zheng2023ResponseLP} proposed a system that predicts the output sequence length of an input sequence and then leverages this information to gather queries with similar response lengths into micro-batches in order to reduce computational cost. Sheng \emph{et al.}~\cite{sheng2023high} proposed FlexGen that solves a linear programming problem and searches for efficient patterns to store and access tensors across memory hierarchy. Compared to existing work, \sys stands out with its innovative approach of decoupling prompt and GT processing to fully utilize both GPU and memory resources, reduce scheduling time, and optimize utilization of allocated but unused KVC. {\color{red} Zheng~\emph{et al.}~\cite{zheng2024sglang} proposed a technique for KVC reuse during runtime in the scenario of multiple LLM calls. Instead of discarding the KVC after finishing a generation request, the approach retains the KVC for both prompts and generation results in a radix tree for multiple requests. Liu~\emph{et al.}~\cite{liu2023scissorhands} proposed maintaining the memory usage of KVC under a fixed budget by only storing pivotal tokens based on the attention scores.} Sheng~\emph{et al.}~\cite{Fairness-OSDI-2024} proposed a scheduler that considers fairness between users. 



\noindent\textbf{Processors of Transformer Models.}
Several specialized inference systems have been developed targeting the transformer models. FasterTransformer~\cite{fastertransformer} and  TurboTransformers~\cite{fang-turbo} are such examples. These systems behave as a backend execution engine or model processor of existing scheduling systems such as Nvidia Triton Inference Server~\cite{fang2021deployment} and TensorFlow Serving~\cite{46801}. The systems use request-level scheduling as the scheduling mechanism.
While there are other systems like DeepSpeed-FastGen~\cite{deepspeed-fastgen}, DeepSpeed-Inference~\cite{deepspeed-inf}, HuggingFace-TGI~\cite{tgi}, Megatron-LM~\cite{megatron} for backend execution of inference requests. Deepspeed-MII~\cite{mii} has also emerged as a scheduling system with similar facilities as Triton.}


\DEL{
\section{Limitations and Future Work}
\squishlist


\item 
We will investigate enhancing the accuracy of RL prediction, reducing prediction overhead, and finding effective solutions for underestimation.

\item We will explore how to make use of the allocated KVC resulting from overestimation. 



\item We plan to investigate whether \sys would lead to imbalances in processing speeds of PTs and GTs, and how it affects the overall performance. 



\DEL{\item When a prompt task transitions to a GT, its prompt tokens occupy the KVC. Our research will investigate whether we can predict if a prompt's GT will be processed soon, and if not, we postpone the prompt's processing. This approach aims to reduce the KVC occupation of prompts' tokens.}

\item We will explore how to hide scheduling time by parallelizing it with batch execution. \DEL{It could be realized if the RL prediction is always accurate.} 

\DEL{\item In this paper, we focus on handling response length under-prediction. How to utilize the allocated KVC due to over-prediction remains as our future work. 

\item Though reserving a KVC pool can handle the lack of allocated KVC problem in the KVC pipelining method, it still wastes certain KVC space. We will study how to further improve the accuracy of response length prediction, and also explore more effective solutions for under-prediction.   

\item We will study how the different request arrival rates affect the performance of \sys. 

\item We will study whether selecting prompts to fully utilize GPU and selecting GTs to fully utilize KVC will lead to slow processing of one type of tasks or 
imbalanced waiting queue lengths between the two queues. How other factors such as the request arrival rate and batch size affect such imbalance. 

\item In the case of imbalanced waiting queues, we may not find enough prompts or GTs to fully utilize GPU or enough GTs to fully utilize the KVC sometimes. To address this problem, we will study whether and how we can predict the incoming requests to proactively make a schedule to avoid the imbalance. 

\item Once a prompt task is changed to GT, its prompt tokens occupy the KVC. We will study if and how we can predict whether a prompt's GT will be processed soon, and if not, we postpone processing the prompt. This way, we can reduce the cache occupation of prompts. Here, we will jointly consider the time-tolerance features of requests. 
}


\squishend
}

\vspace{-0.1in}
\section{Conclusion}
\vspace{-0.1in}
To improve throughput and SLO fulfillment performance of LLM inference, leveraging our findings from trace-based experimental analysis, we propose the \sys system, incorporating three key methods: 1) \emph{KVCPipe}, 2) \emph{SyncDecouple} and 3) \emph{Ordering}.
Our trace-based real experiments demonstrate that \sys exhibits superior performance in comparison with the state-of-the-art. 
We will investigate enhancing the accuracy of RL prediction and reducing prediction overhead. We also plan to investigate whether \sys would lead to imbalances in processing speeds of PTs and GTs, and how it affects the performance. 

\looseness=-1

\DEL{@@check arxiv to 
@@ 1, 2, 6 need to add proc. change journal to proc.
@@12 
@@14
@@17
@@21
@@22 proc
@@26
@@27 proc
@@29 double check -- done}

\DEL{improves the state-of-the-art method by up to 2.33$\times$ in throughput while maintaining similar latencies and improves the SLO satisfaction ratio by up to 78\%. Each method contributes effectively to achieving the goal. Our future work is outlined above.}
\DEL{In the context of LLM inference, where GPU utilization and KVC limitations pose challenges, maximizing GPU and KVC utilizations and preventing KVC failures while minimizing scheduling time are paramount for enhancing throughput and JCT. However, previous LLM job schedulers fail to tackle these challenges simultaneously. To address these challenges, leveraging our findings from trace-based experimental analysis, we propose the \sys system, employing three key methods: 1) Time-Synced Batching with Decoupled Prompt and
GT Processing (\emph{SyncDecouple}), 2) Prompt and GT Queue Ordering (\emph{Ordering}), 3) Pipelining KVC Utilization (\emph{KVCPipe}). 
Our experiments on a real testbed demonstrate that \sys improves the state-of-the-art method by up to 2.33$\times$ in throughput while maintaining similar latencies and improves the SLO satisfaction ratio by up to 78\%. Each method contributes effectively to achieving the goal. Our future work is outlined above.}
\bibliographystyle{unsrt}
\bibliography{nlp,myBib,myBib-2}

\DEL{{\color{gray}??KVC utilization of MultiRes should not be so high. In your old results, it is similar as \sys without KVC pipelining}

?? Why is \sys-DG KVC utilization higher than \sys-D? {\color{gray}WHy MultiRes has a higher utilization of \sys-D}

??Each component should have significant improvement over the other.  Pipleining should have 47\% improvement over the \sys-DGS

04/12/2024
{\color{gray}??Figure 1(a) Why SyncCoupled and others to the right have higher scheduling time? -fixed
?? Figure 1(b) SyncCoupled should be 0
?? Figure preempting time should be lower}
?? Figure 5(c), that's not right 
{\color{gray}?? Figure 6, change it to unused, put it with the CDF of GT}
??: for all the methods in fig 3, need to measure and draw
a fig, X=different percent values of requests in a batch com-
pleted after an iteration, Y=the percent of such iterations
over all iterations
{\color{gray}?? need to check the preemption time of the vllm - fixed, got higher
?? Why syncfastgen has higher unused kvcache?
?? why gpuutil. of sync fastgen is high?}

??check all recent conferences and SGLang paper to add new references

??vLLM also use a medium-size model (3 models together in exp). We will need to include the results of such a model.

??do not forget that "We will add results for Oracle that pre-know the response length."

??add experimental results for different SLO scales (maybe 1-2 fig)

??in exp. do not measure memory utilization, measure KVC utilization.

??Can you tell me what fig you will draw for the following comment: "Why is it reasonable to assume that there will be several other requests with the same request lengths present, given that a server can only be processing a handful requests at a time? Why doesn't skew for request lengths build up over time? Unless a large queue is built up to select from (which I suspect is the case, explained below) I would imagine that each request "group" would have only one or very few requests. It might be good to add more detailed evaluation results on this."
??In the analysis, X=\# of requests in one GT group in a batch, Y: CDF of \% of groups during whole exp., together with the equal RL fig.
\newpage

The average PT is  1.414, The ave iteration time or token generation time is  0.1443s

GT waiting time will be : 0.1124s

then around how many percent of 2nd tokens meet SLO: its output time<=request submission time+prompt SLO+generation SLo. or simply the gap between 1st token and 2nd token <=0.1875. NOw it is 0.1124+0.1443=0.25.

Tanmoy Sen 4:23 PM
For the first case, the SLO satisfaction percentage is : 96.3%
For the second case , the SLO satisfaction percentage is : 85.6%

we use the first case in the figs. use one fig to show the SLO satisfication ratio for the 2nd token--add this ?? comment to overleaf. }


\DEL{\appendix

\begin{appendices}
\section{More Experimental Analysis Results.}\label{sec:app}
\begin{figure*}[h]
\centering
  \DEL{  \subfloat[GPU utilization. 
    \vspace{-0.0in}\label{fig:gpu-schedulers1}]{{\includegraphics[width=0.32\linewidth,height=0.112\textheight]{NewFigs/gpu-util-analysis-opt-13b.pdf} }}
    \hfill}
    \DEL{\subfloat[Throughput.\vspace{-0.0in}\label{fig:throughput-schedulers1}]{{\includegraphics[width=0.32\linewidth,height=0.112\textheight]{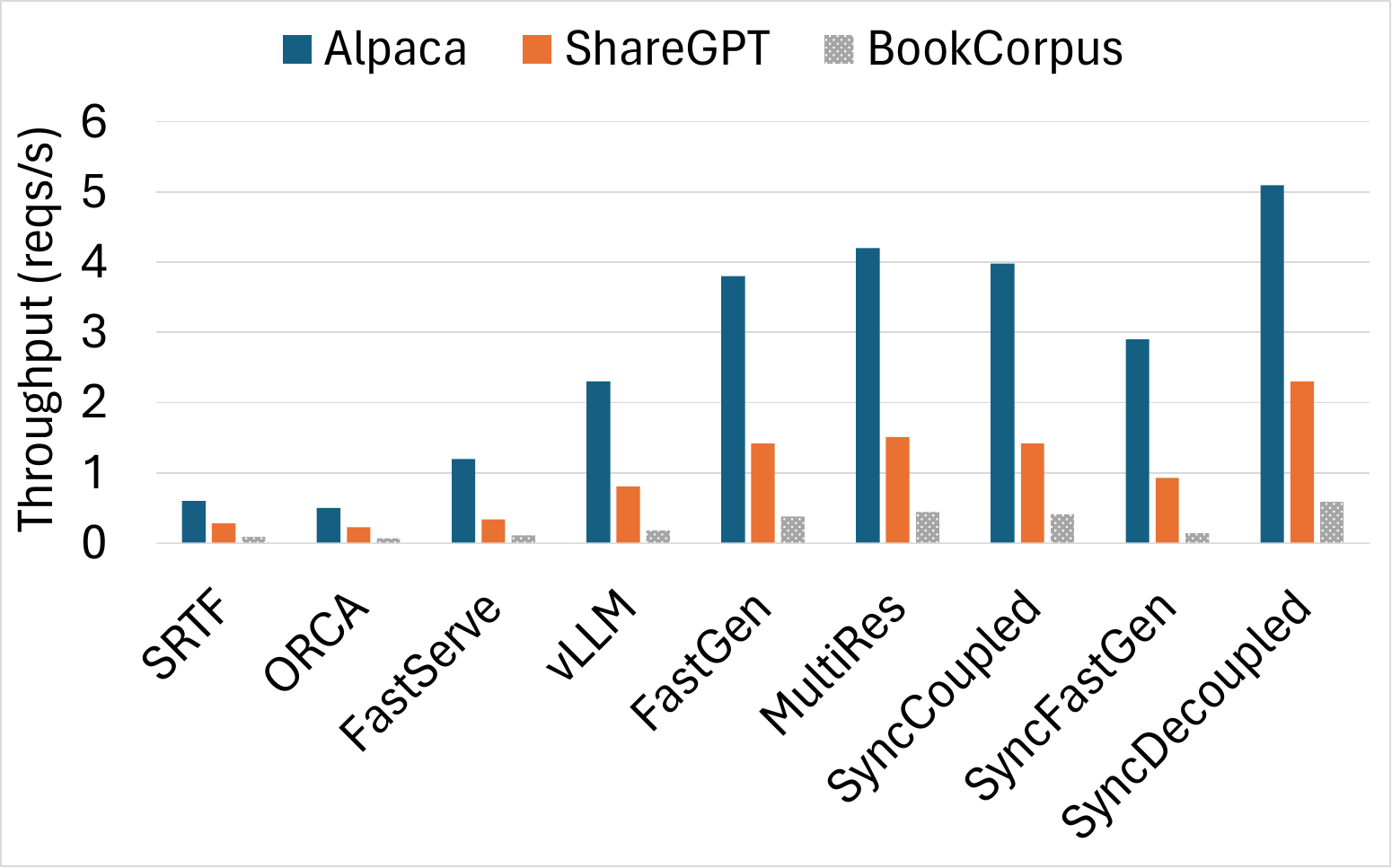} }}
    \hfill
    \DEL{\subfloat[TG latency.\vspace{-0.0in}\label{fig:tgl-schedulers1}]{{\includegraphics[width=0.32\linewidth,height=0.112\textheight]{NewFigs/tgl.pdf} }}
    \hfill}
    \subfloat[KVC utilization.\vspace{-0.0in}\label{fig:KVC-schdulers1}]{{\includegraphics[width=0.32\linewidth,height=0.112\textheight]{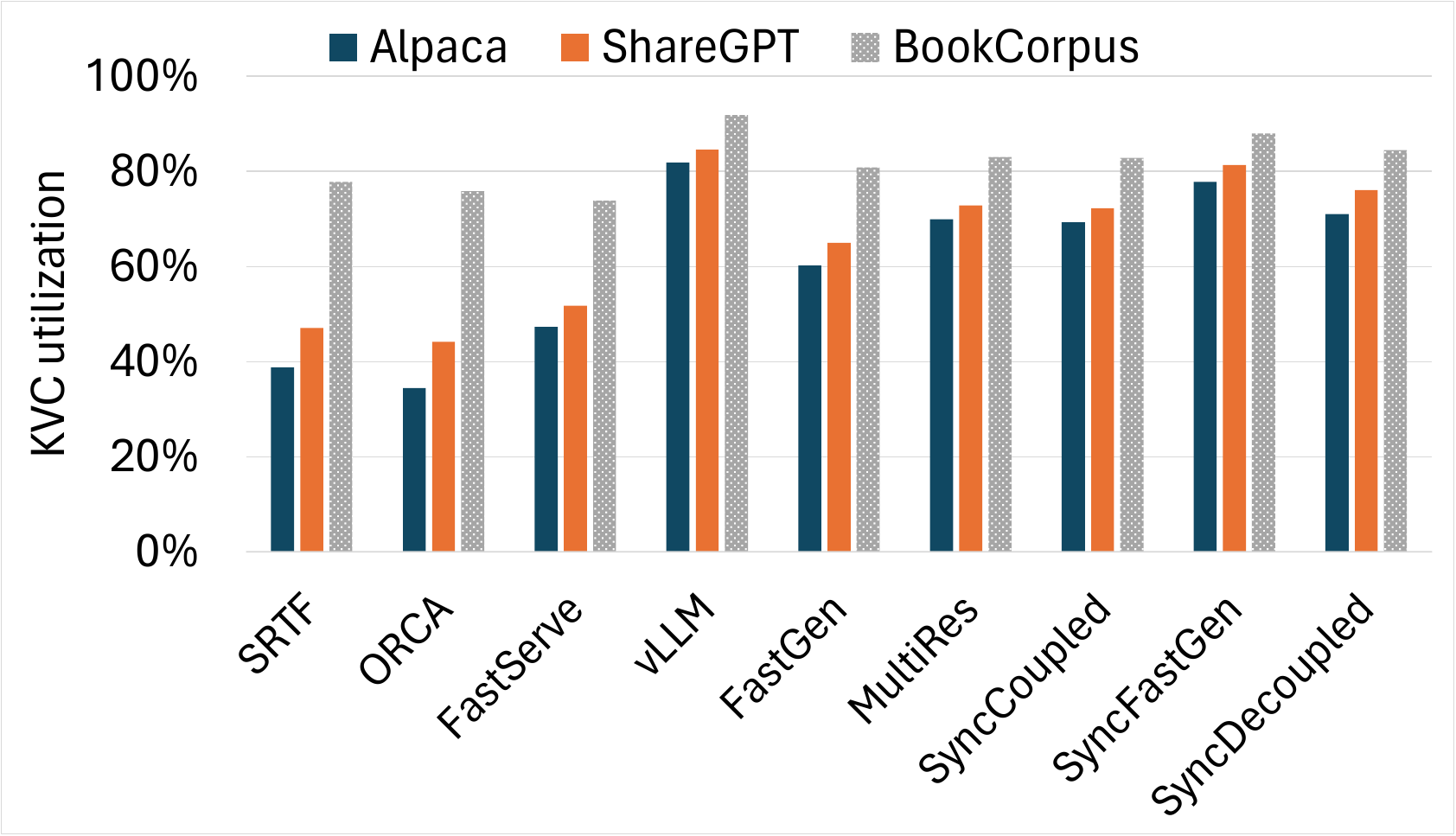} }}
    \hfill
    \subfloat[Forward size.\vspace{-0.0in}\label{fig:forwardsize-schedulers1}]{{\includegraphics[width=0.32\linewidth,height=0.112\textheight]{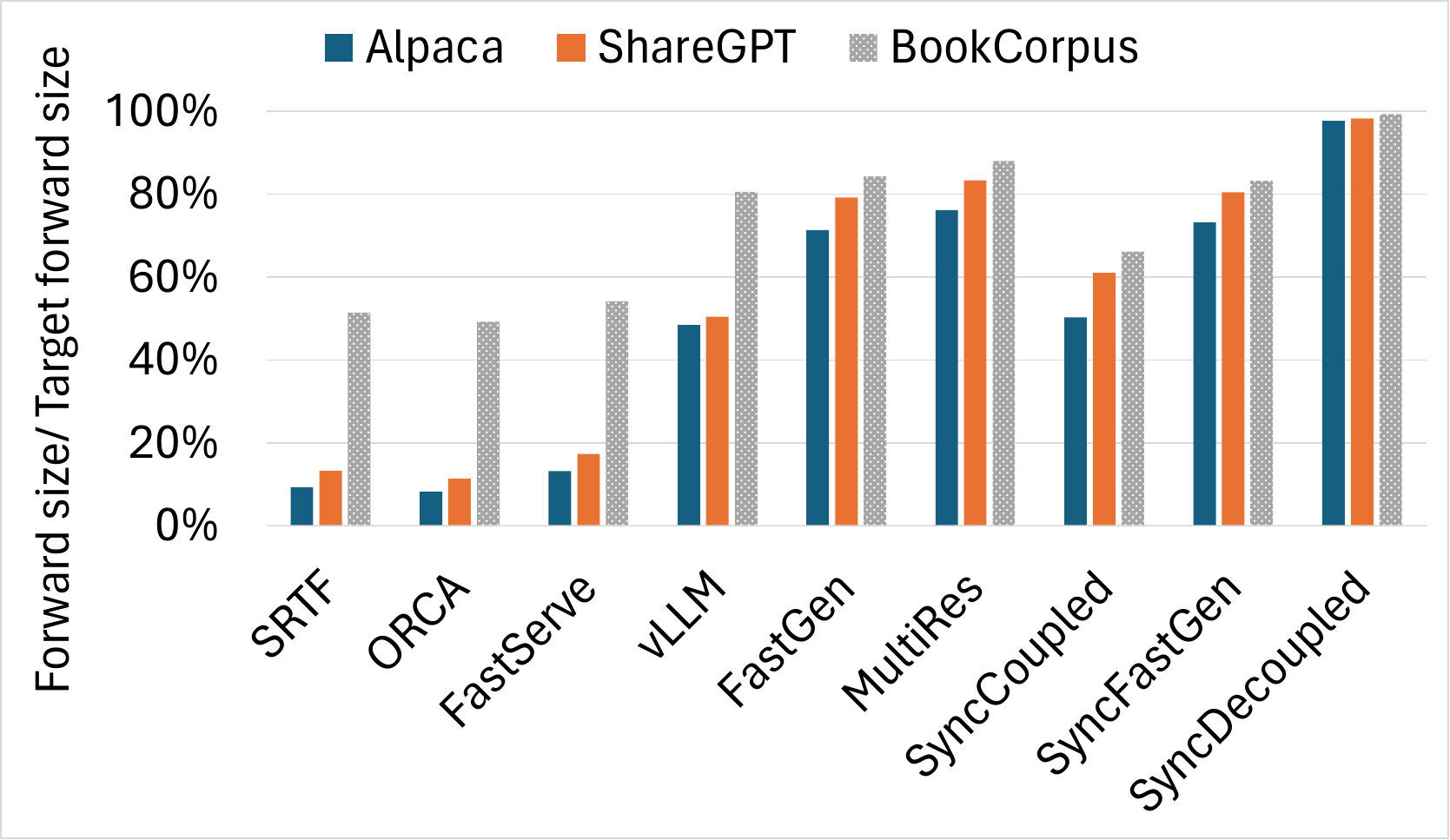} }}
    \hfill
    \subfloat[KVC allocation failure.\vspace{-0.0in}\label{fig:alloc-schedulers1}]{{\includegraphics[width=0.32\linewidth,height=0.112\textheight]{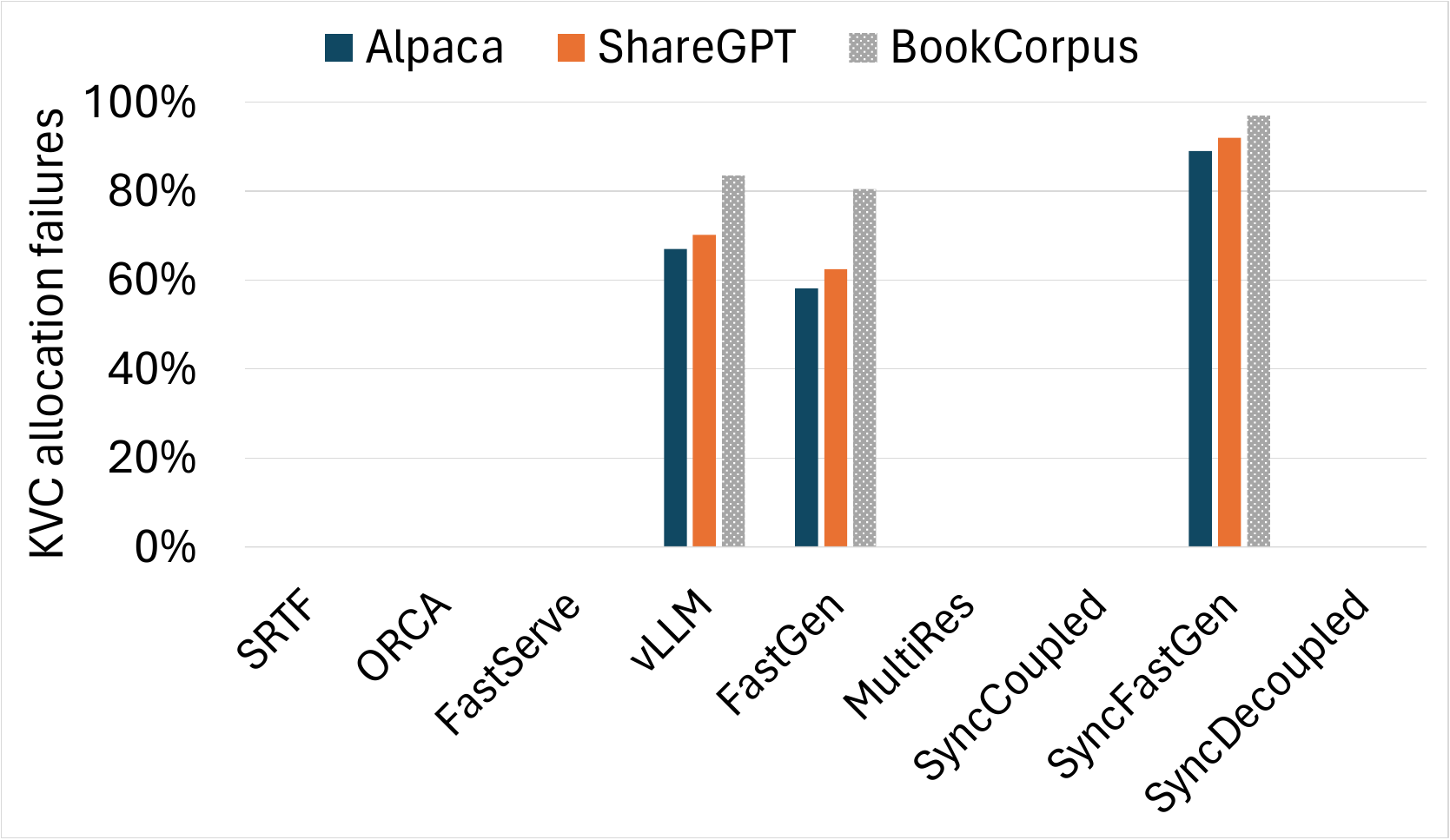}  }}
    \hfill}
    \subfloat[JCT for Alpaca.\vspace{-0.0in}\label{fig:overhead-schedulers1}]{{\includegraphics[width=0.32\linewidth,height=0.112\textheight]{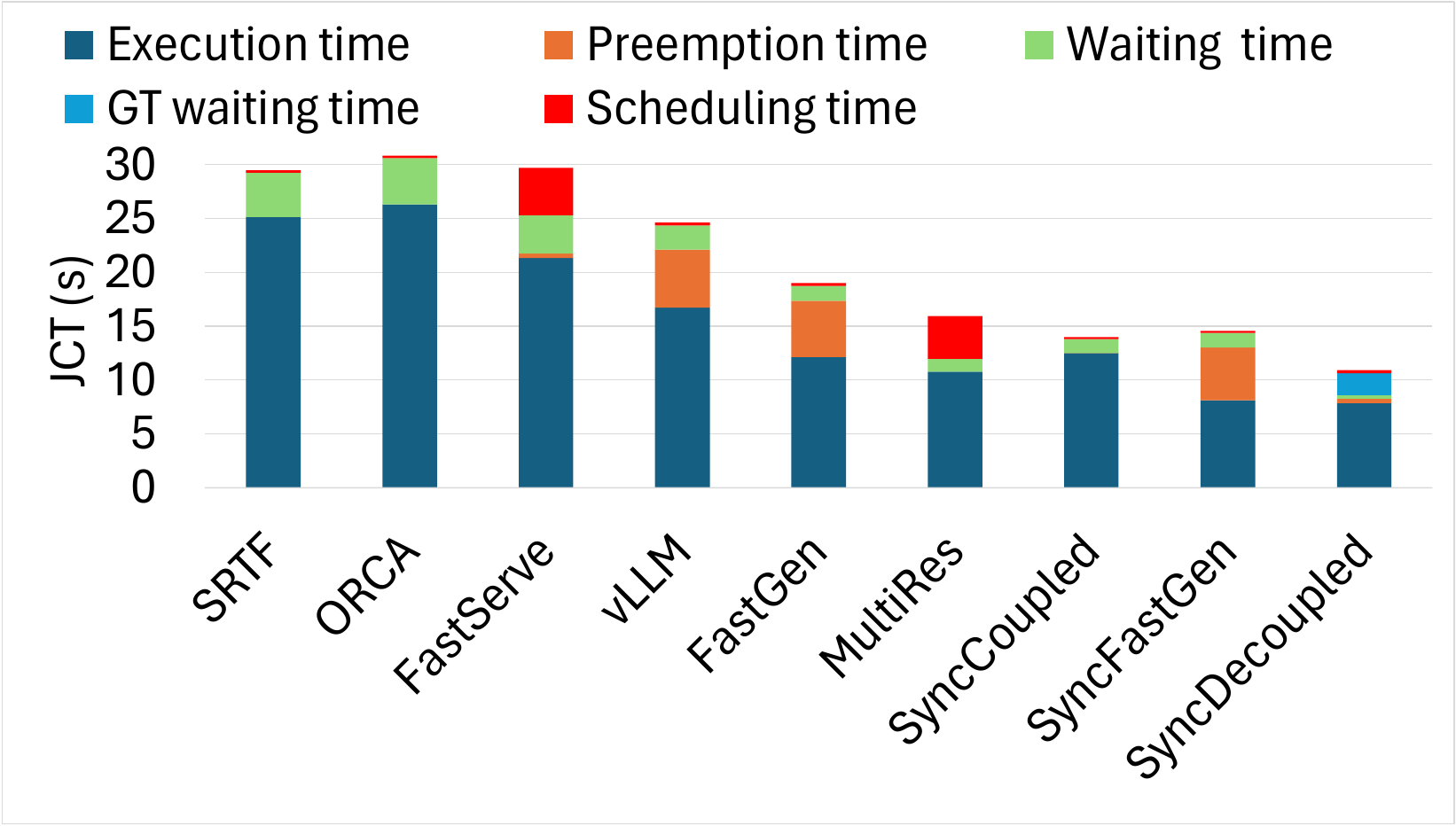}  }}
    \hfill
  { \subfloat[JCT for ShareGPT.\vspace{-0.0in}\label{fig:overhead-schedulers1}]{{\includegraphics[width=0.32\linewidth,height=0.112\textheight]{NewFigs/jct-sharegpt-13b-analysis-fin.pdf}  }}
    \hfill
    \subfloat[JCT for BookCorpus.\vspace{-0.0in}\label{fig:overhead-schedulers-b1}]{{\includegraphics[width=0.32\linewidth,height=0.112\textheight]{NewFigs/jct-bookcorpus-13b-analysis-fin}}}
    \hfill}
    \DEL{\subfloat[Scheduling time.\vspace{-0.0in}\label{fig:overhead-schedulers1}]{{\includegraphics[width=0.32\linewidth,height=0.112\textheight]{NewFigs/sched-all.pdf}}}
    \hfill}
    \subfloat[Distribution of iterations with a certain num. of requests completed after an iteration for Alpaca.\vspace{-0.0in}\label{fig:req-schedulers1}]{{\includegraphics[width=0.32\linewidth,height=0.112\textheight]{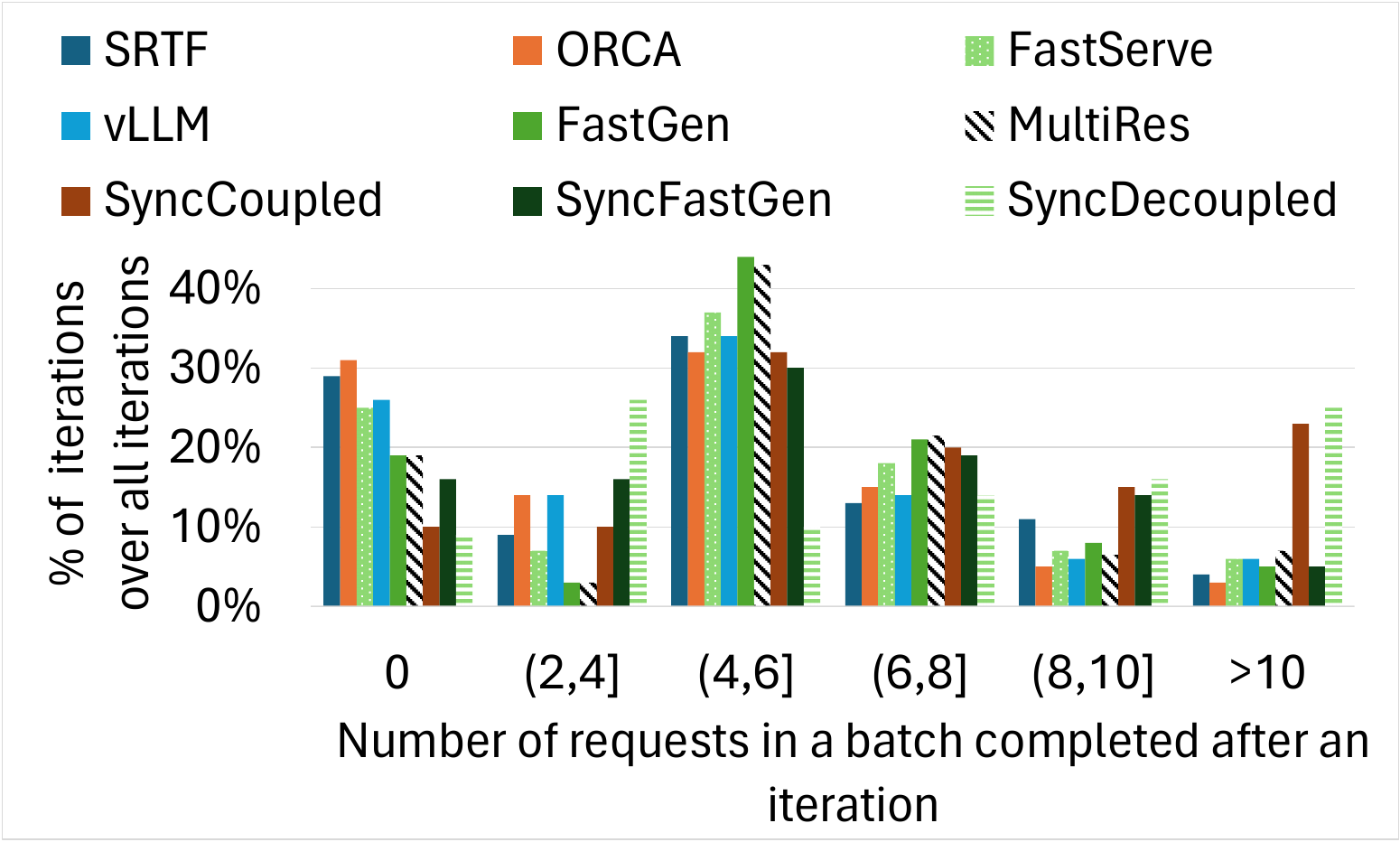}}}
    \hfill
   {\subfloat[Distribution of iterations with a certain num. of requests completed after an iteration for   ShareGPT.\vspace{-0.0in}\label{fig:req-schedulers-sharegpt1}]{{\includegraphics[width=0.32\linewidth,height=0.112\textheight]{NewFigs/iterations-vs-requests-sharegpt-up.pdf}}}
    \hfill
     \subfloat[Distribution of iterations with a certain num. of requests completed after an iteration for BookCorpus.\vspace{-0.0in}\label{fig:req-schedulers-bookcprpus1}]{{\includegraphics[width=0.32\linewidth,height=0.112\textheight]{NewFigs/iterations-vs-requests-bookcorpus-up.pdf}}}
    \hfill}
    \vspace{-0.2in}
   \caption{{Comparison of different schedulers.
\vspace{-0.1in}}}%
    \label{fig:schedulers-measurement1}
\end{figure*}


\begin{figure*}[h]
\centering
    \subfloat[SLO scale.\vspace{-0.0in}\label{fig:expind-1-21-a}]{{\includegraphics[width=0.33\linewidth,height=0.112\textheight]{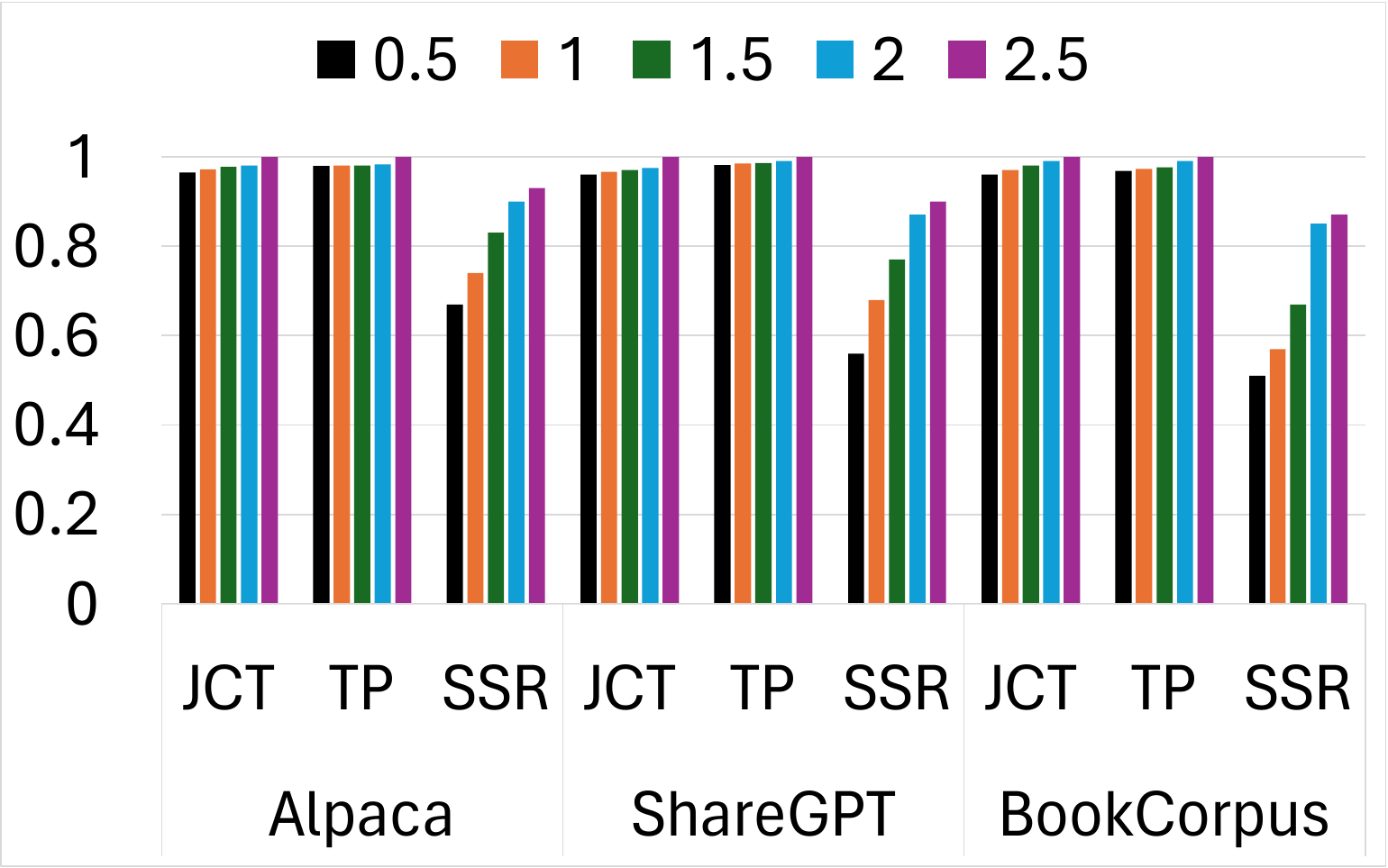} }}
    \hfill
    \subfloat[Padding.\vspace{-0.0in}\label{fig:expind-2-21-b}]{{\includegraphics[width=0.32\linewidth,height=0.112\textheight]{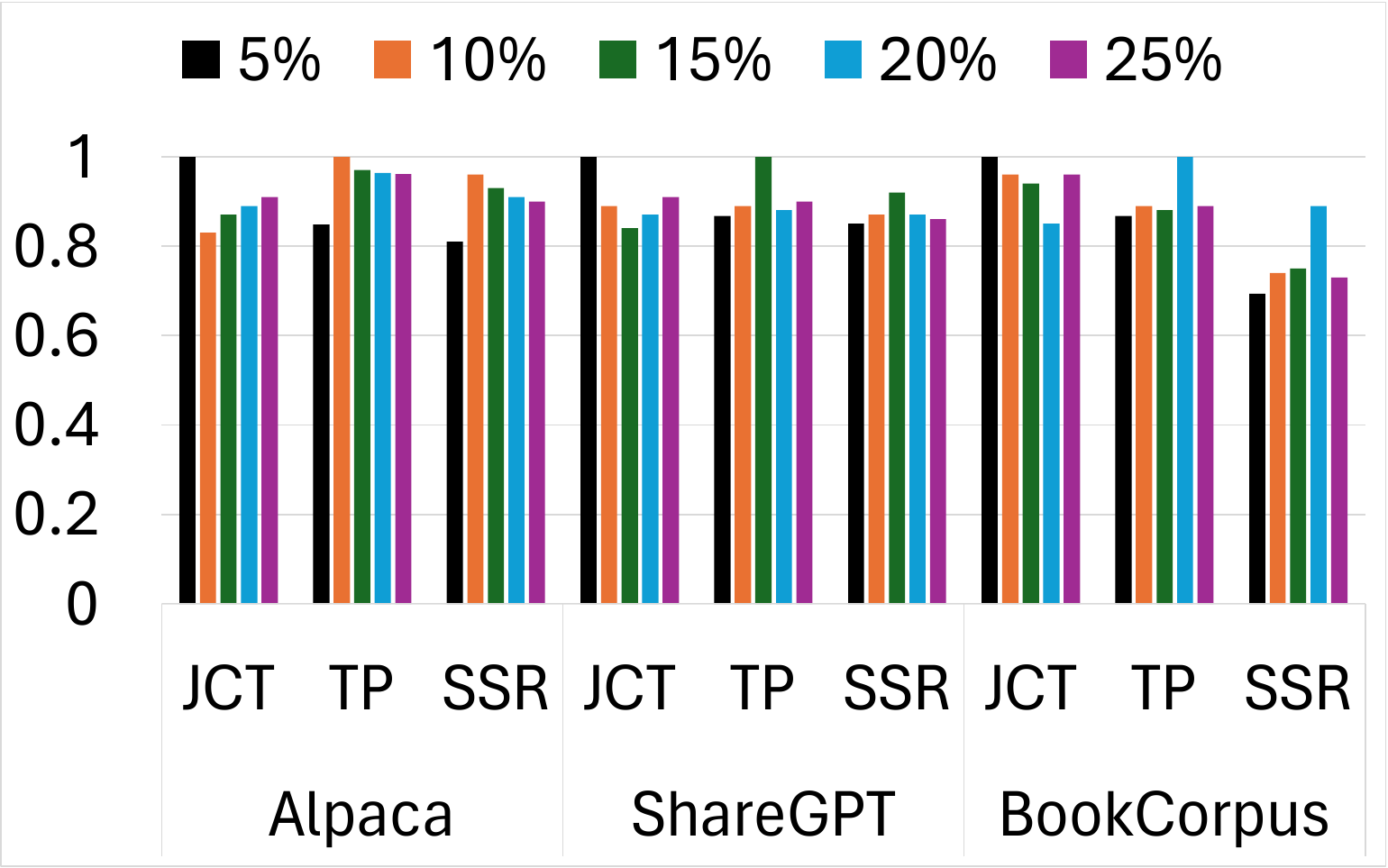} }}
    \hfill
    \subfloat[Reserved KVC.\vspace{-0.0in}\label{fig:expind-3-221-c}]{{\includegraphics[width=0.32\linewidth,height=0.112\textheight]{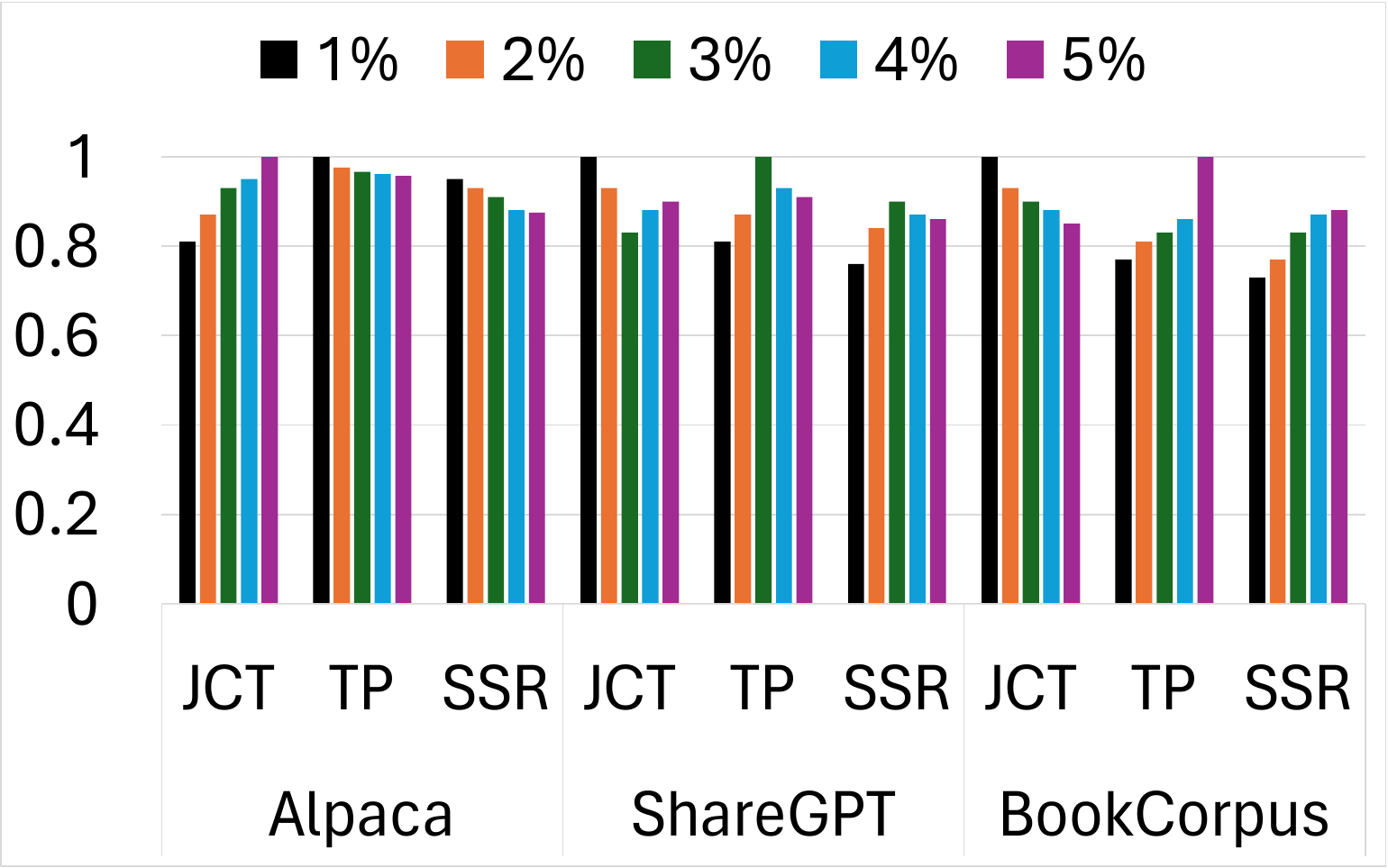} }}
    \hfill
    \subfloat[Block size.\vspace{-0.0in}\label{fig:expind-3-21d}]{{\includegraphics[width=0.32\linewidth,height=0.112\textheight]{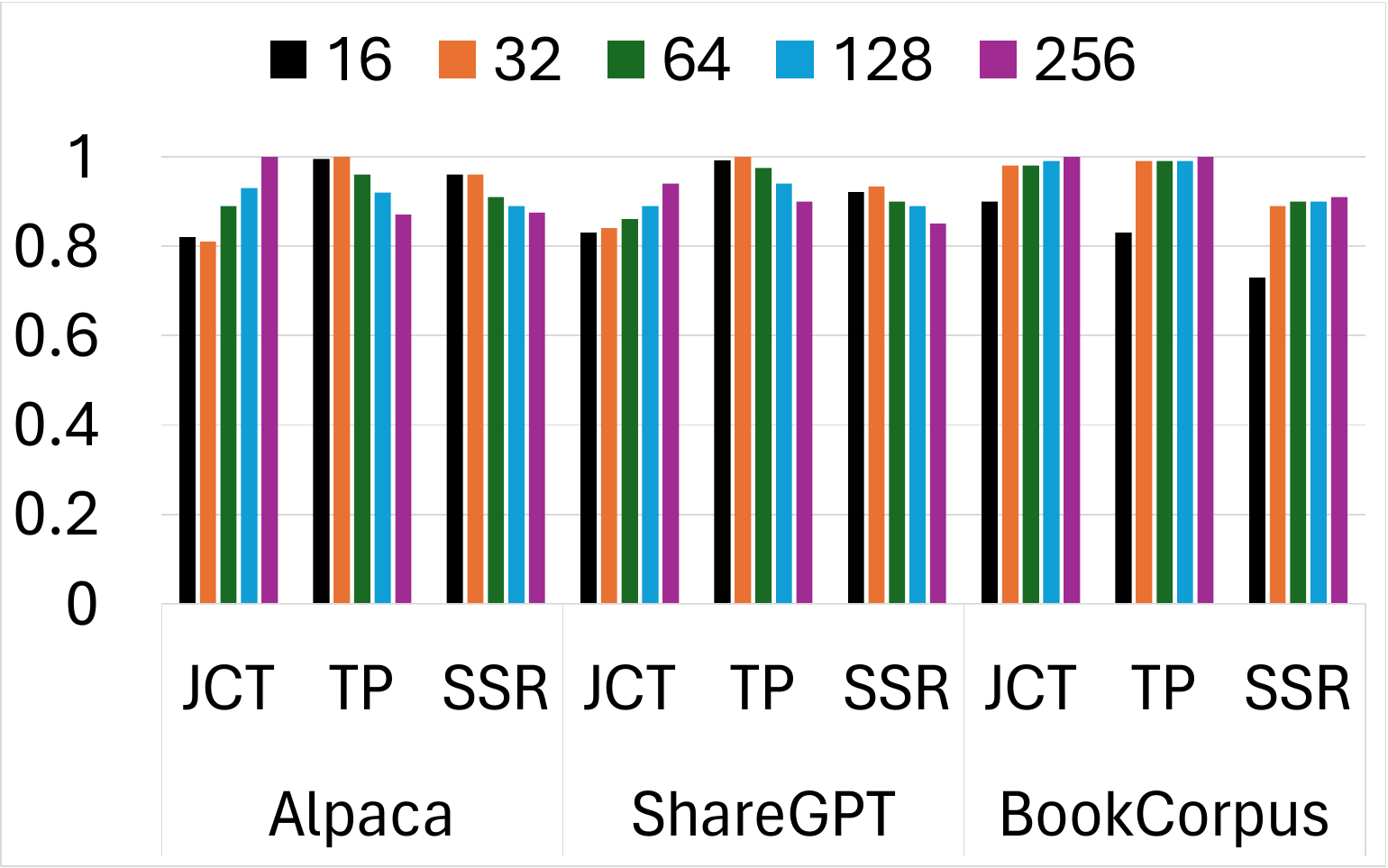} }}
    \hfill
    \subfloat[Buffer size/KVC size.\vspace{-0.0in}\label{fig:expind-3-21-e}]{{\includegraphics[width=0.32\linewidth,height=0.112\textheight]{newaspfig/21-c-up copy.pdf} }}
    \hfill
    \vspace{-0.0in}
   \caption{\small{Effect of individual factors for the OPT-175B on the three traces. \vspace{-0.0in}}}%
    \label{fig:ind-factors-175b}
\end{figure*}

\begin{figure*}[h]
\centering
    \subfloat[SLO scale.\vspace{-0.0in}\label{fig:expind-1-22a}]{{\includegraphics[width=0.32\linewidth,height=0.112\textheight]{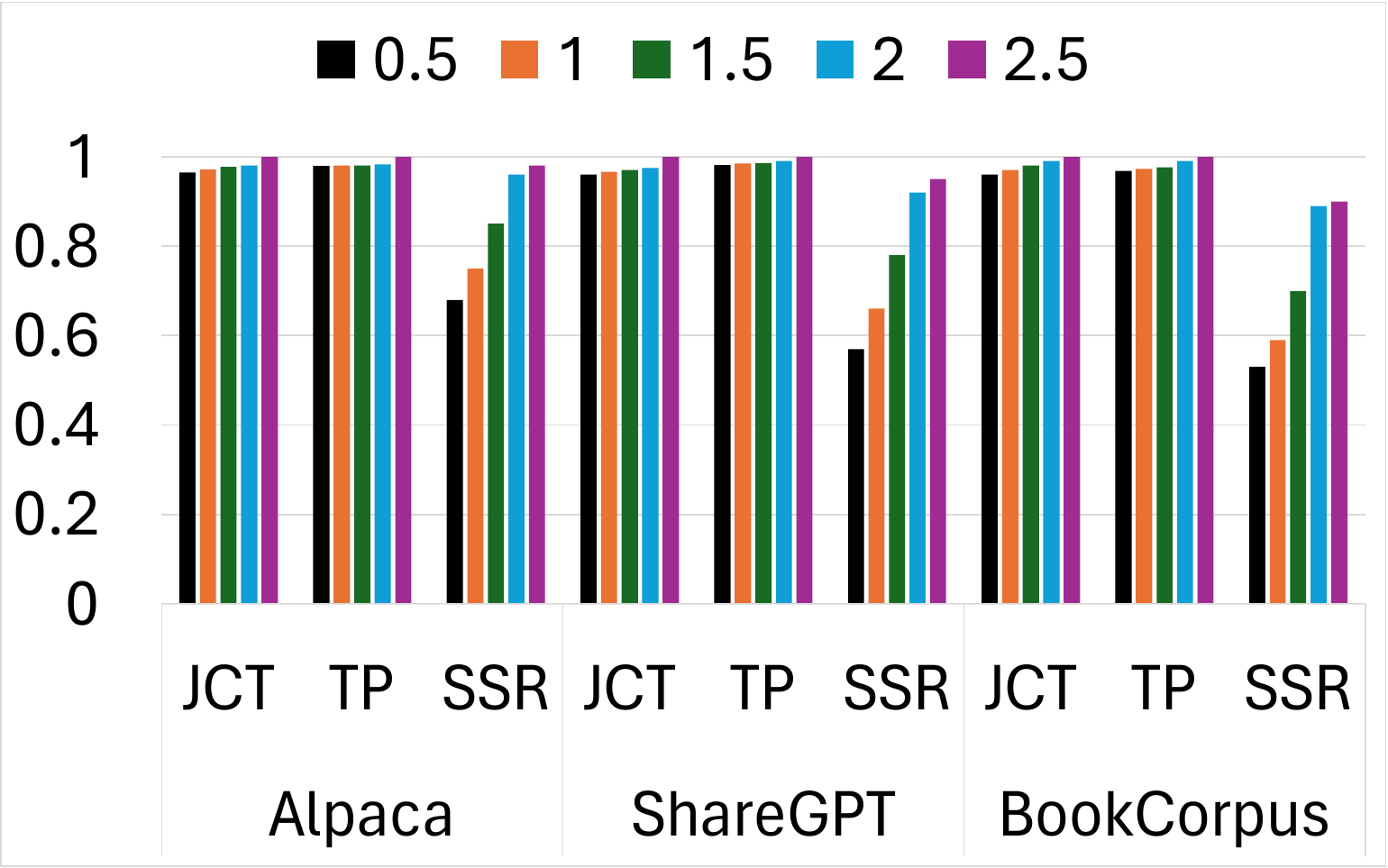} }}
    \hfill
\subfloat[Padding.\vspace{-0.0in}\label{fig:expind-2-22b}]{{\includegraphics[width=0.32\linewidth,height=0.112\textheight]{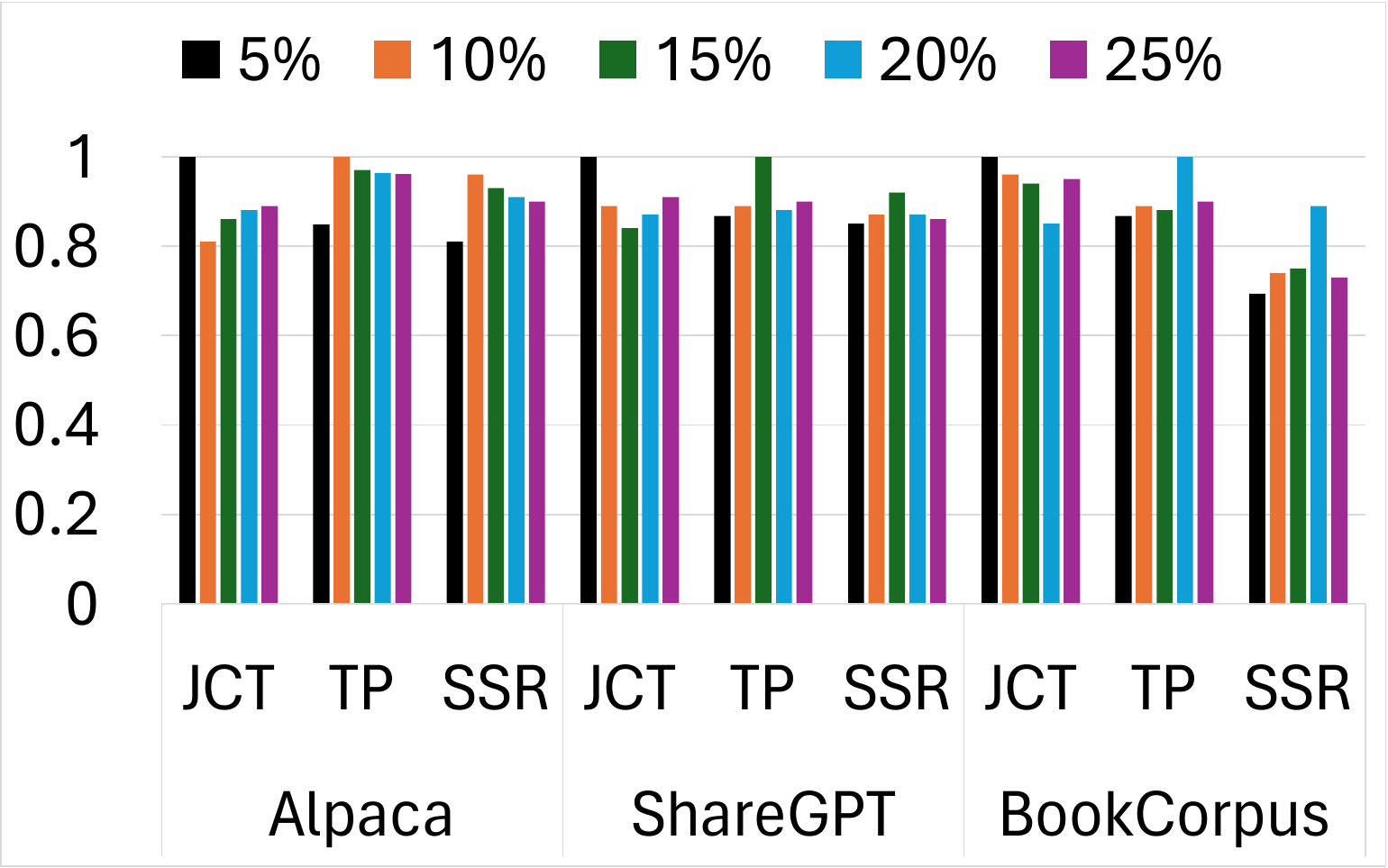} }}
\hfill
\subfloat[Reserved KVC.\vspace{-0.0in}\label{fig:expind-3-22c}]{{\includegraphics[width=0.32\linewidth,height=0.112\textheight]{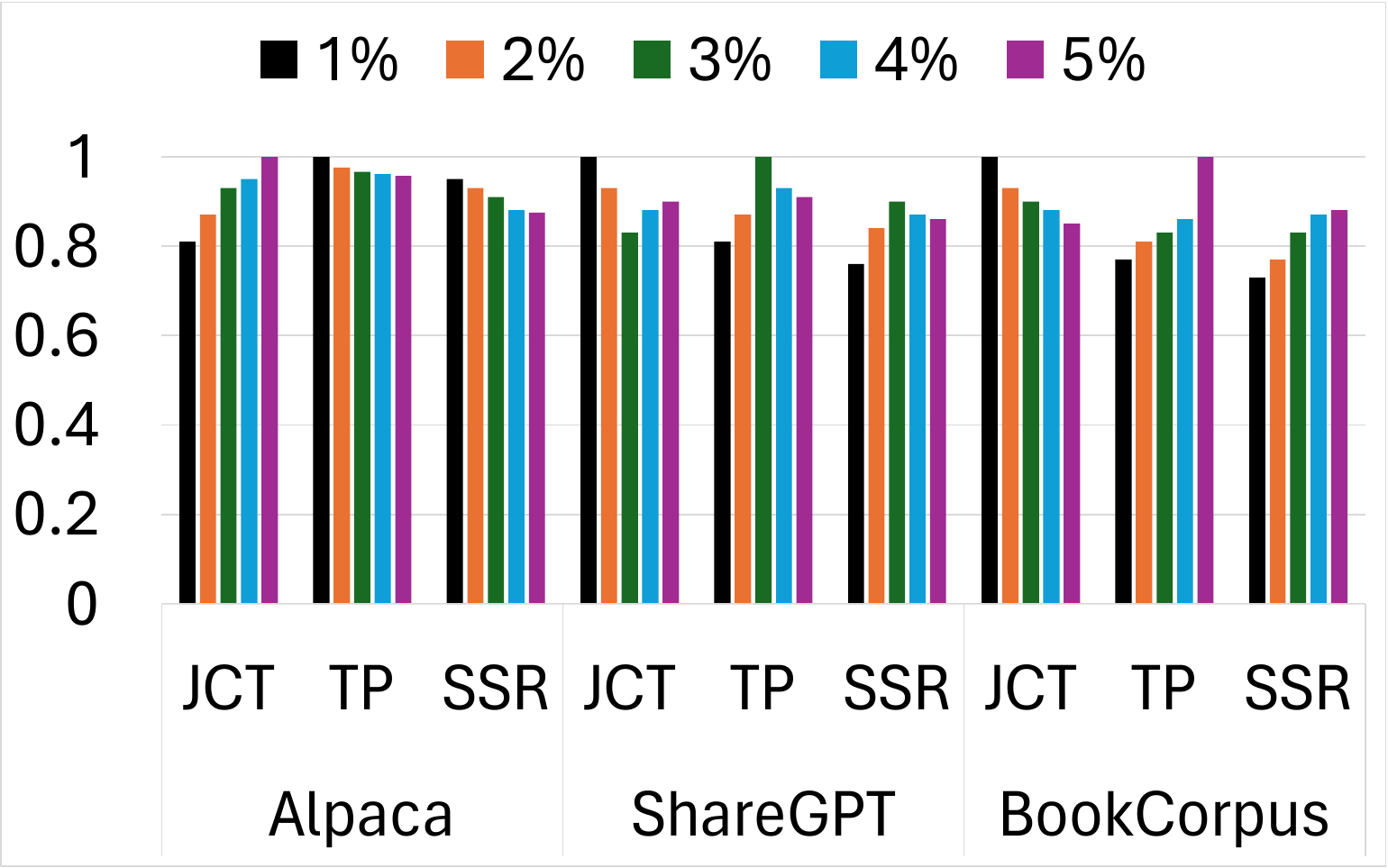} }}
 \hfill
\subfloat[Block size.\vspace{-0.0in}\label{fig:expind-3-22d}]{{\includegraphics[width=0.32\linewidth,height=0.112\textheight]{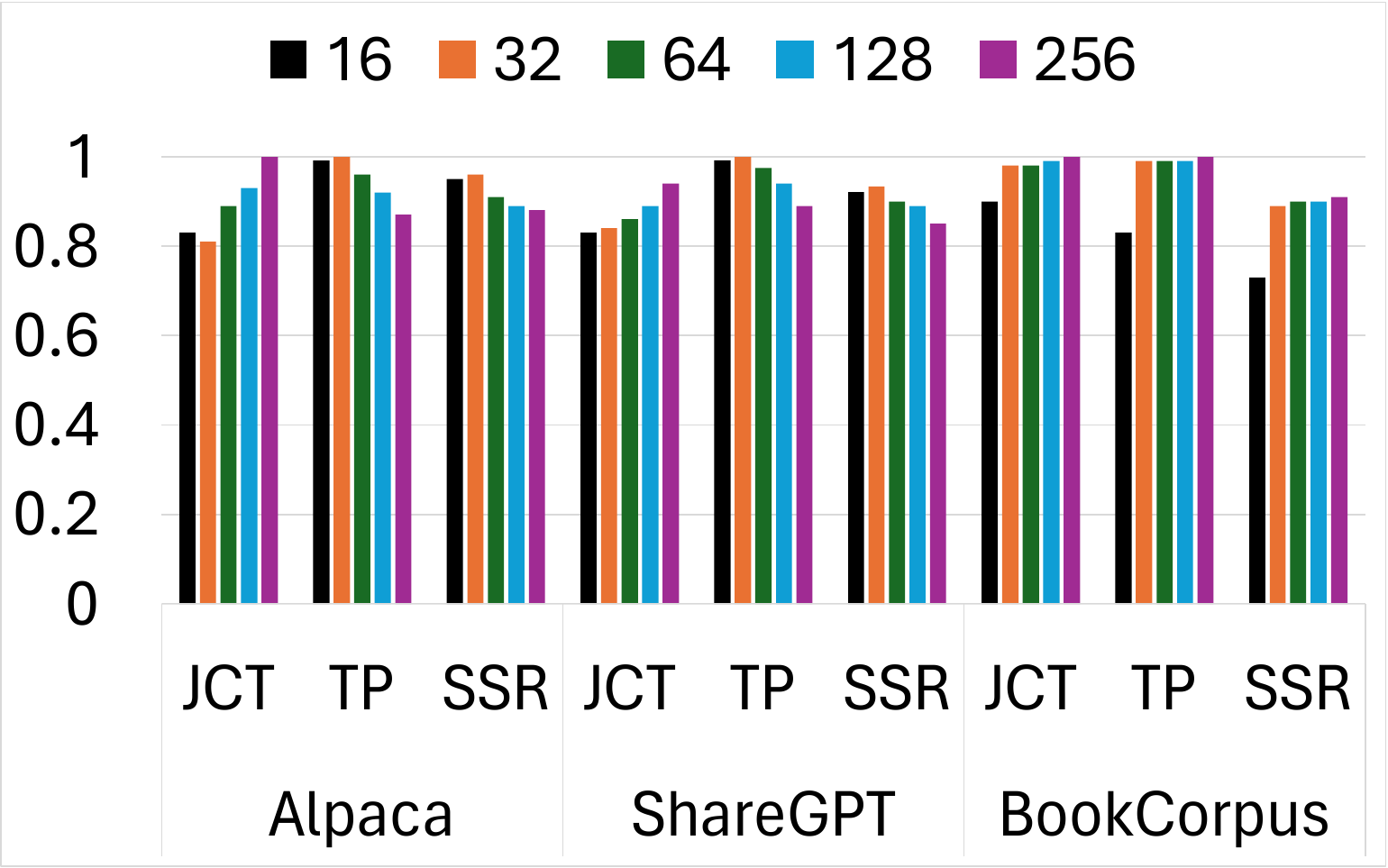} }}
    \hfill \subfloat[Buffer size/KVC size.\vspace{-0.0in}\label{fig:expind-3--22e}]{{\includegraphics[width=0.32\linewidth,height=0.112\textheight]{newaspfig/22-c-up copy.pdf} }}
    \hfill
    \vspace{-0.0in}
   \caption{\small{Effect of individual factors for the Llama on the three traces }. \vspace{-0.0in}}%
    \label{fig:ind-factors-llama}
\end{figure*}

\DEL{\begin{figure*}[t]
\centering
    \subfloat[SLO scale, vary from (0.5 to 2.5 with 0.5 increase.\vspace{-0.0in}\label{fig:expind-1}]{{\includegraphics[width=0.23\linewidth,height=0.112\textheight]{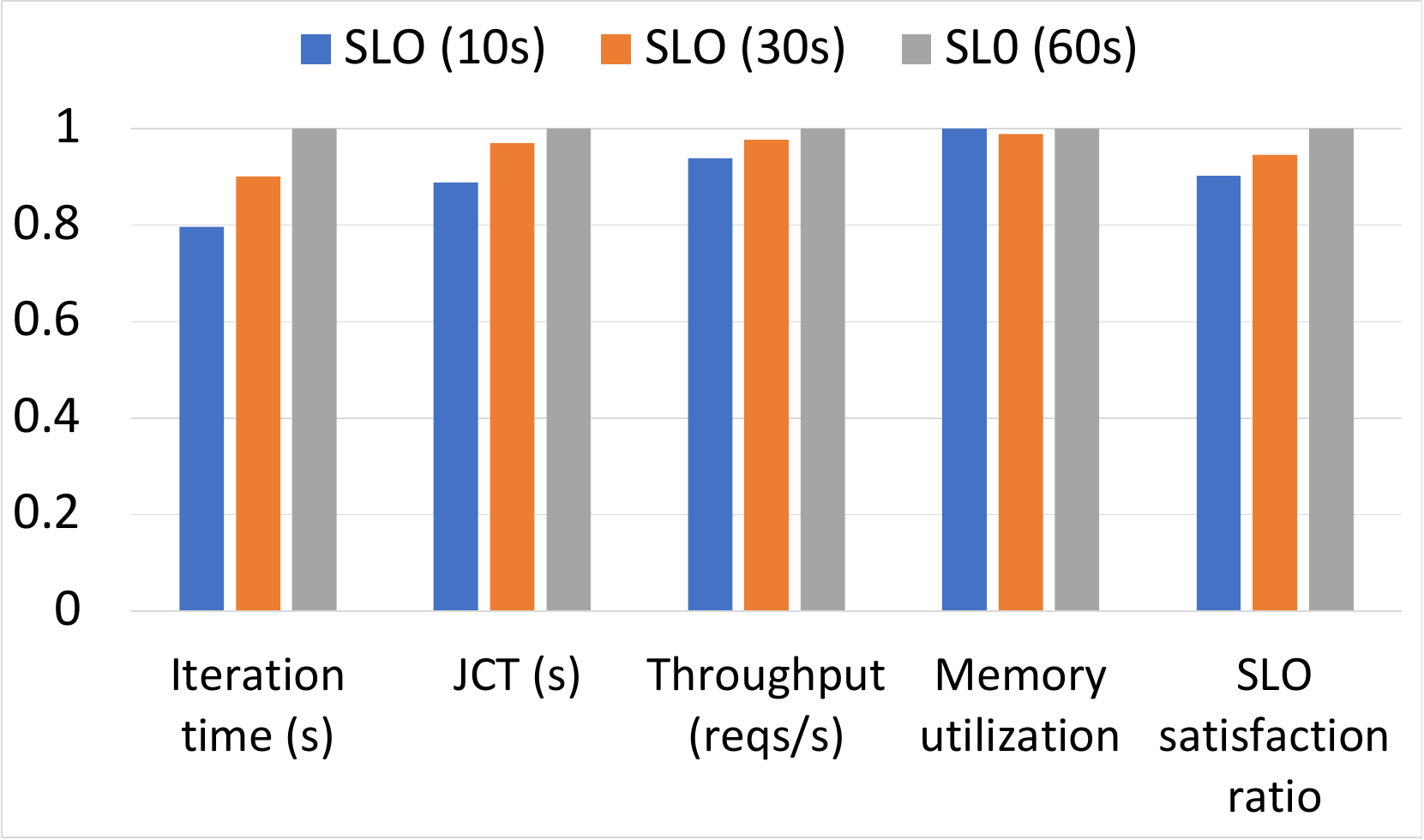} }}
    \subfloat[Padding (vary from 5\% to 25\%, plot normalized latency).\vspace{-0.0in}\label{fig:expind-2}]{{\includegraphics[width=0.23\linewidth,height=0.112\textheight]{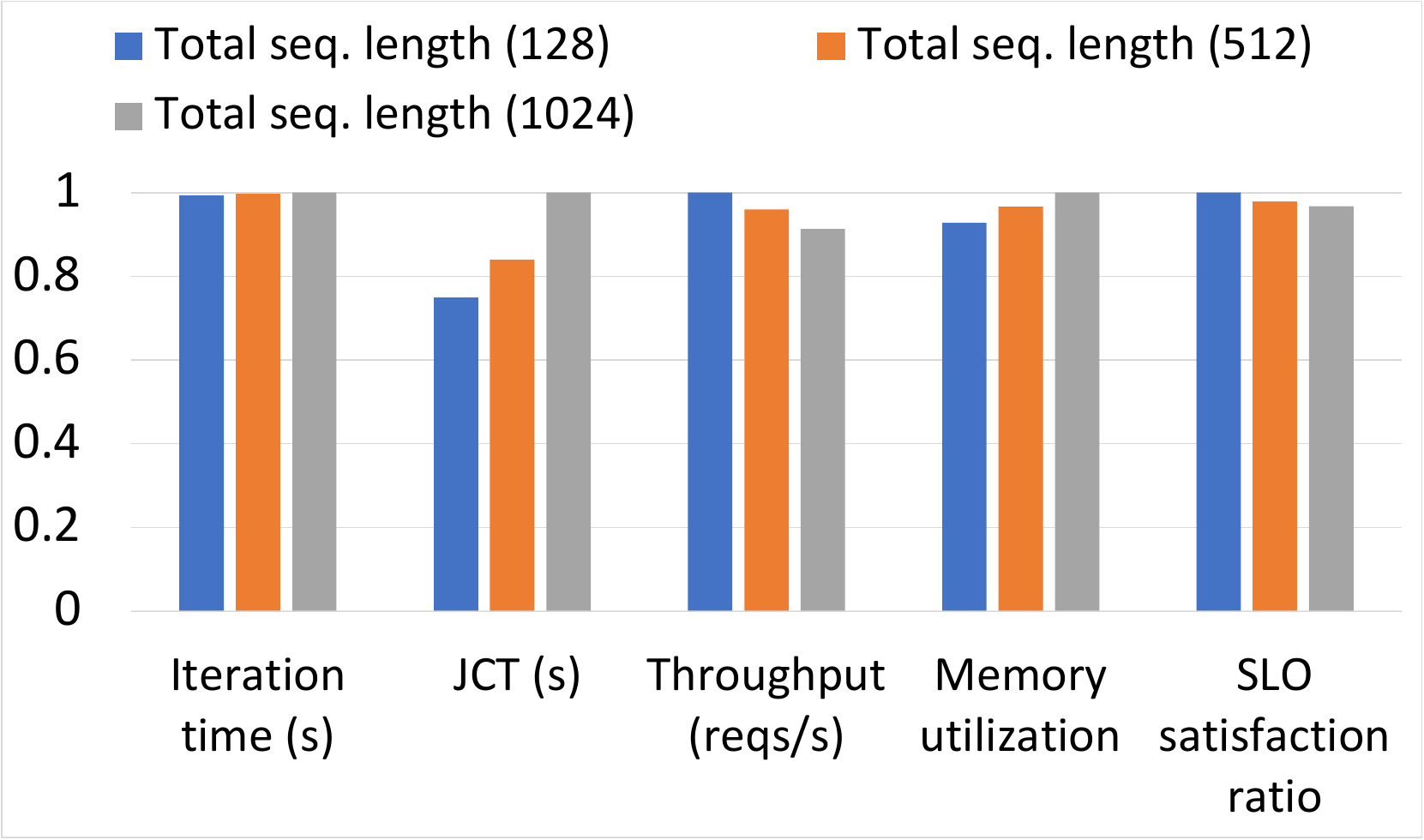} }}
    \subfloat[Reserved KVC (vary from 64 to 512 with 2X increase, plot normalzied latency).\vspace{-0.0in}\label{fig:expind-3}]{{\includegraphics[width=0.23\linewidth,height=0.112\textheight]{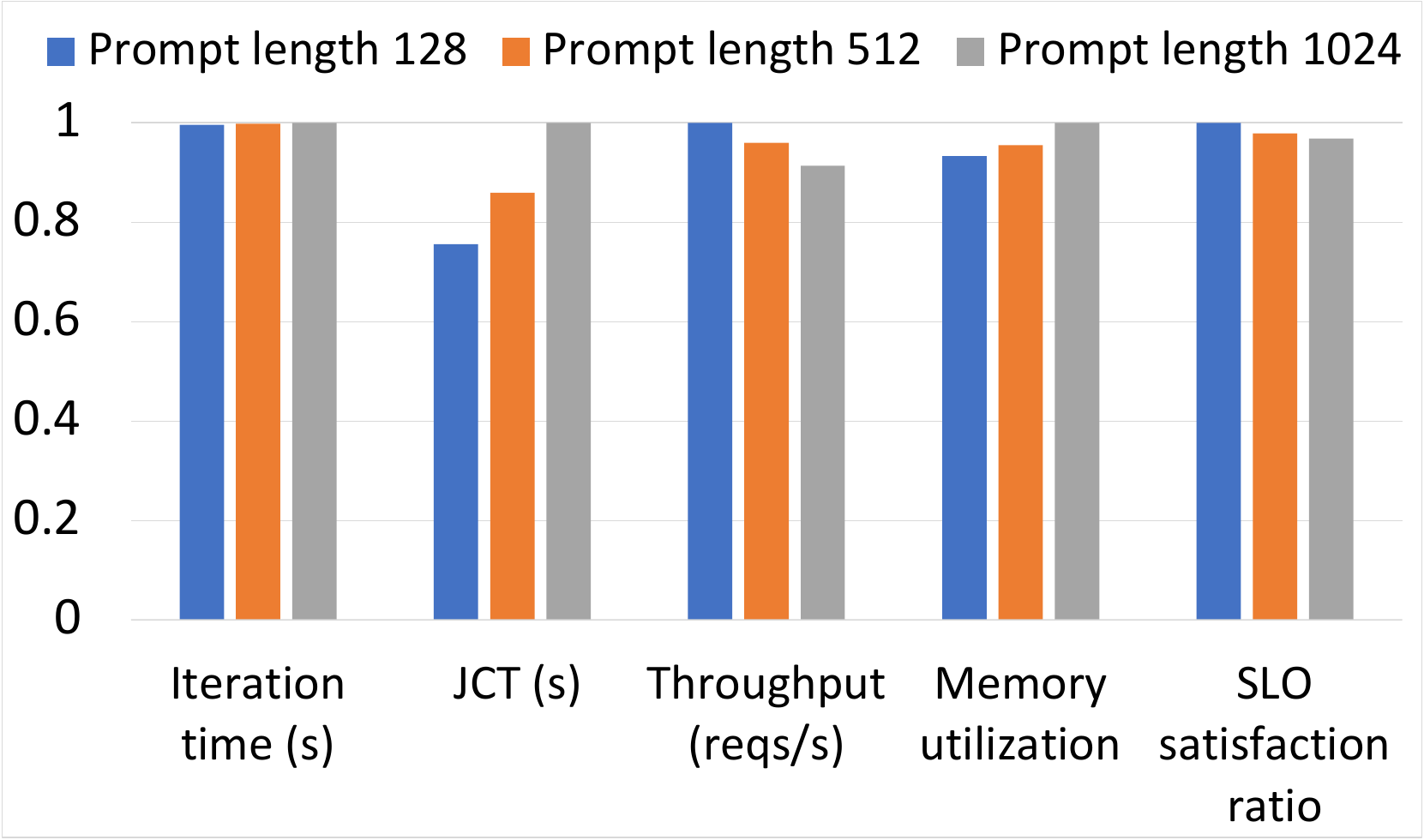} }}
    \subfloat[Block size (vary from 16 to 256 with 2X increase, plot normalized latency).\vspace{-0.0in}\label{fig:expind-3}]{{\includegraphics[width=0.23\linewidth,height=0.112\textheight]{Fig/prompt-ind-norm.pdf} }}
    \subfloat[Buffer (vary from 16 to 256 with 2X increase, plot normalized latency).\vspace{-0.0in}\label{fig:expind-3}]{{\includegraphics[width=0.23\linewidth,height=0.112\textheight]{Fig/prompt-ind-norm.pdf} }}
    \vspace{-0.0in}
   \caption{\small{Effect of individual factors for the OPT-175B on the three traces. \vspace{-0.0in}}}%
    \label{fig:ind-factors}
\end{figure*}

\begin{figure*}[t]
\centering
    \subfloat[SLO scale, vary from (0.5 to 2.5 with 0.5 increase.\vspace{-0.0in}\label{fig:expind-1}]{{\includegraphics[width=0.23\linewidth,height=0.112\textheight]{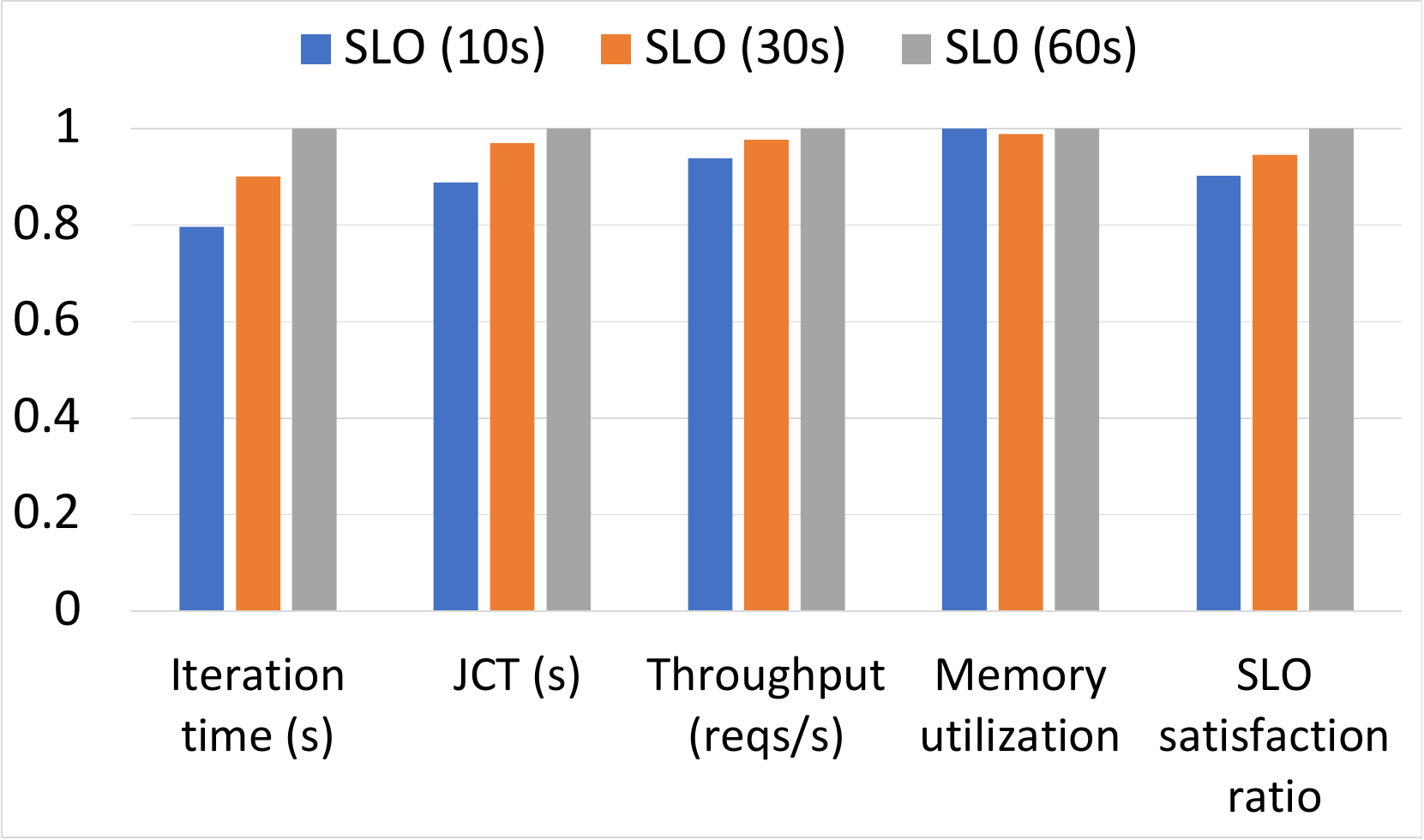} }}
    \subfloat[Padding (vary from 5\% to 25\%, plot normalized latency).\vspace{-0.0in}\label{fig:expind-2}]{{\includegraphics[width=0.23\linewidth,height=0.112\textheight]{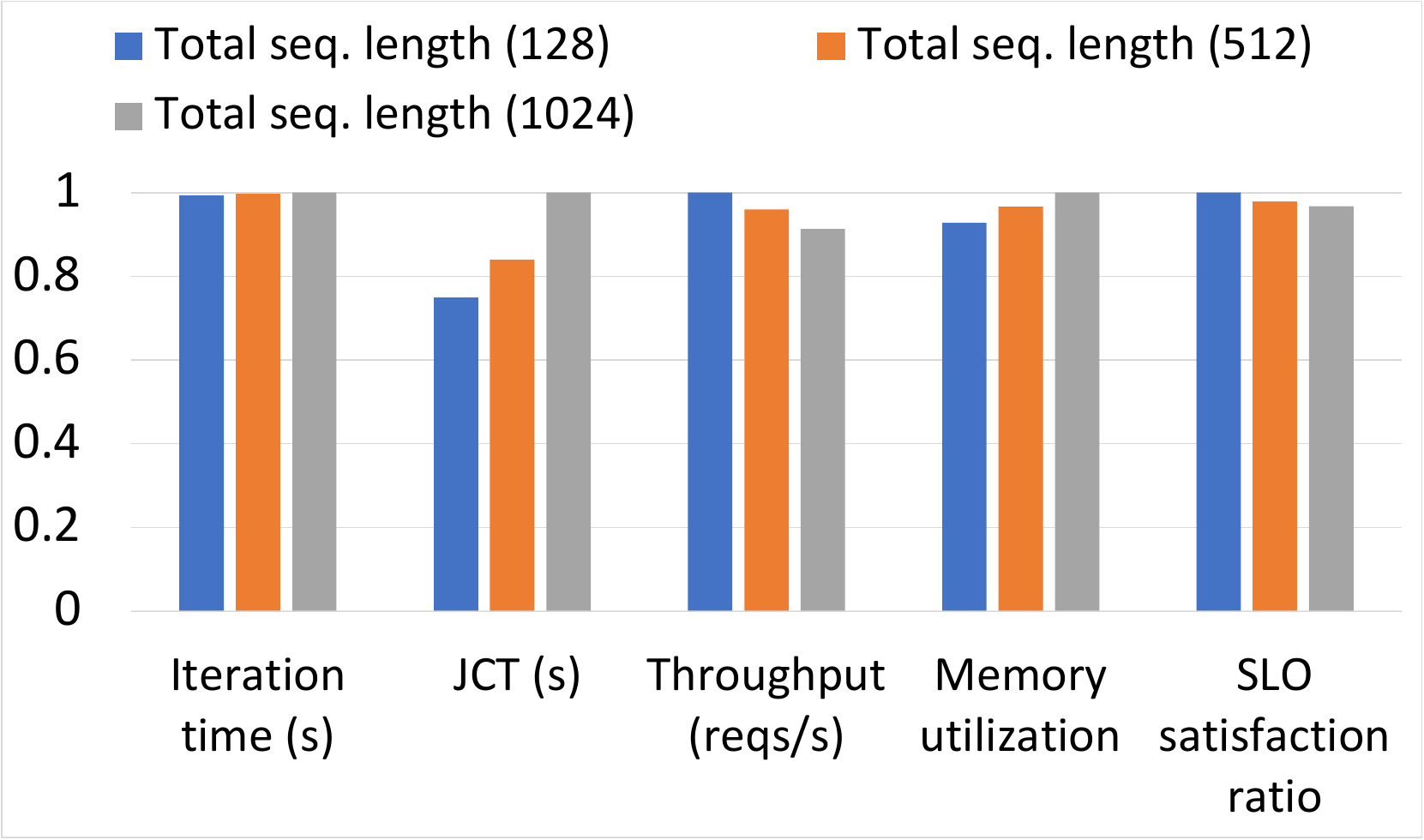} }}
    \subfloat[Reserved KVC (vary from 64 to 512 with 2X increase, plot normalzied latency).\vspace{-0.0in}\label{fig:expind-3}]{{\includegraphics[width=0.23\linewidth,height=0.112\textheight]{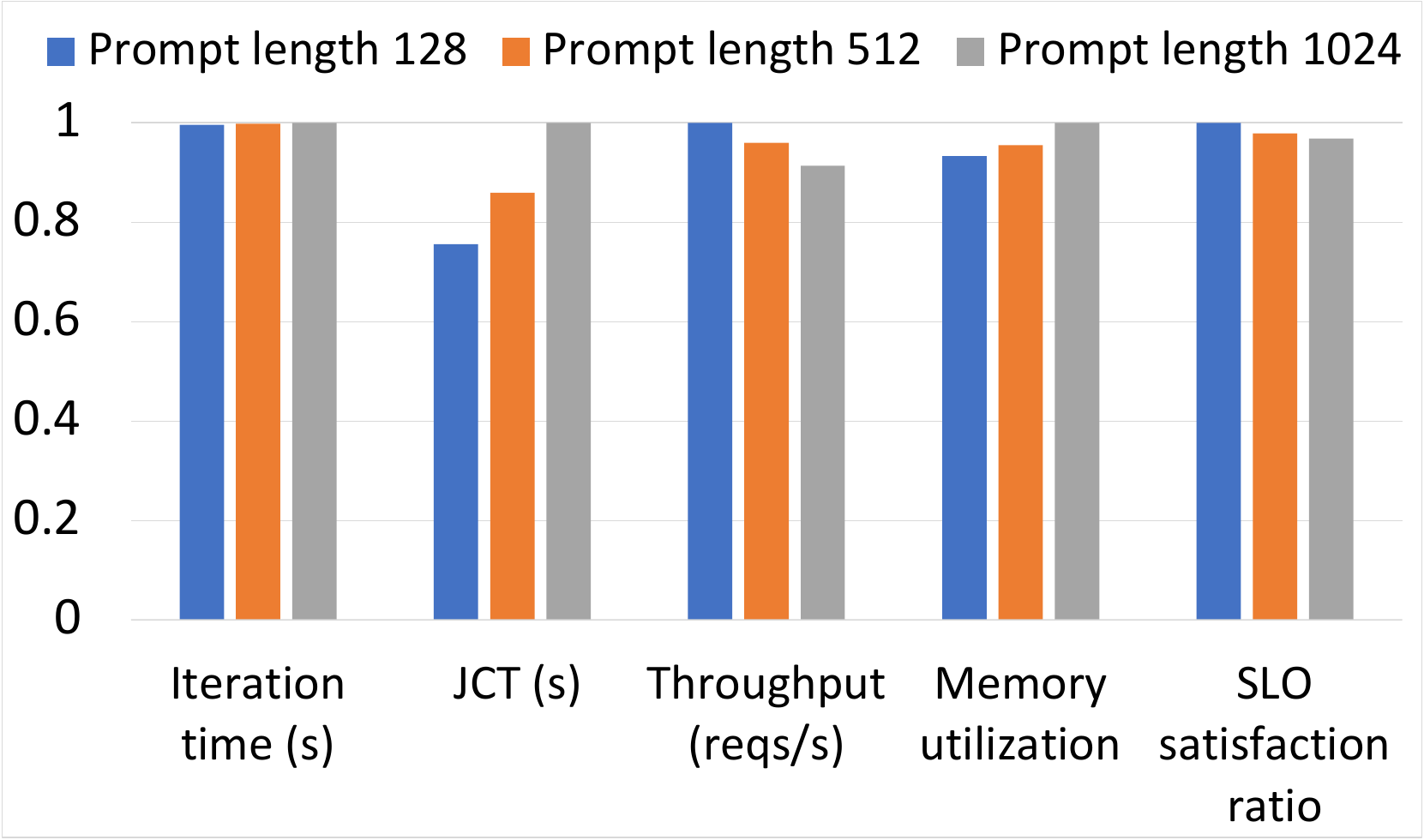} }}
    \subfloat[Block size (vary from 16 to 256 with 2X increase, plot normalized latency).\vspace{-0.0in}\label{fig:expind-3}]{{\includegraphics[width=0.23\linewidth,height=0.112\textheight]{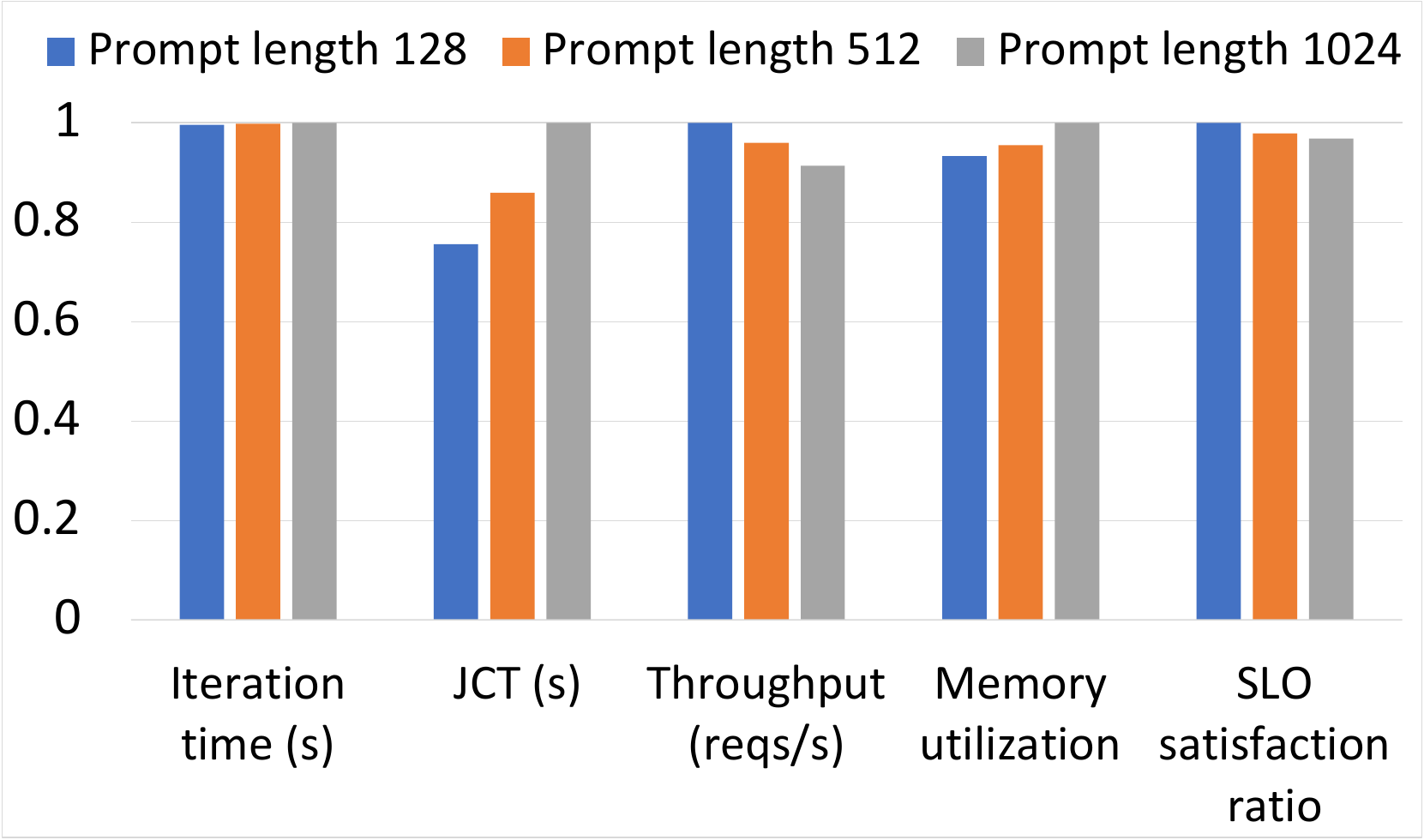} }}
    \subfloat[Buffer (vary from 16 to 256 with 2X increase, plot normalized latency).\vspace{-0.0in}\label{fig:expind-3}]{{\includegraphics[width=0.23\linewidth,height=0.112\textheight]{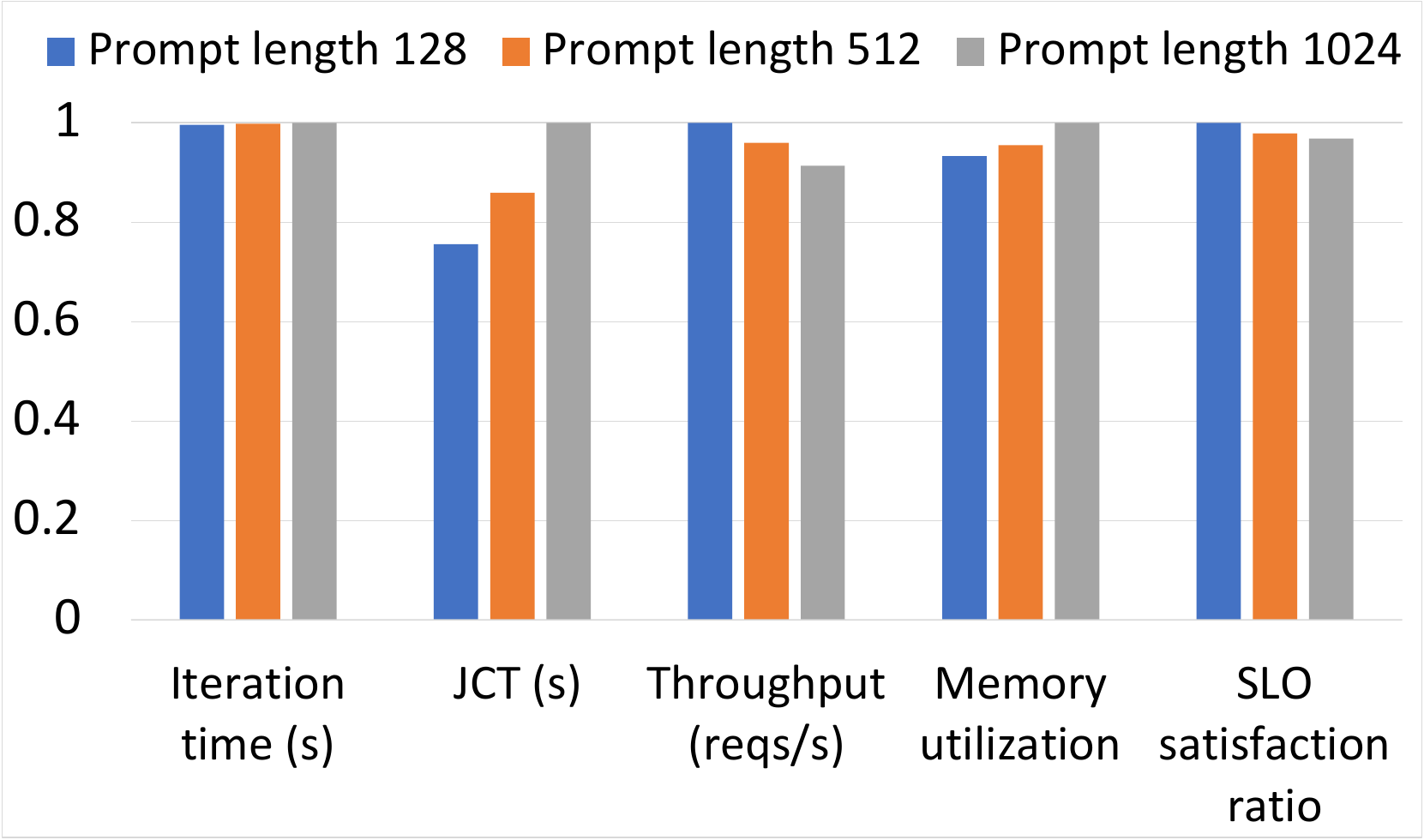} }}
    \vspace{-0.0in}
   \caption{\small{Effect of individual factors for the Llama on the three traces}. \vspace{-0.0in}}}%
    \label{fig:ind-factors}
\end{figure*}
\end{appendices}
}



\end{document}